\documentclass[sigconf, nonacm]{acmart}

\AtBeginDocument{%
	}

\setcopyright{acmcopyright}
\copyrightyear{2023}
\acmYear{2023}
\acmDOI{XXXXXXX.XXXXXXX}

\acmConference[CCS '23]{}{November 26--30,
	2023}{Copenhagen, Denmark}
\acmPrice{11.00}
\acmISBN{978-1-4503-XXXX-X/18/06}

\usepackage{stfloats}

\usepackage{amsmath, bm, amsfonts, amsthm}
\usepackage{algorithmic}
\usepackage{graphicx}
\usepackage{subcaption}
\usepackage{textcomp}
\usepackage{xcolor}
\usepackage[normalem]{ulem}
\usepackage{enumitem}

\usepackage{msc}
\usepackage{spi}
\usepackage{mathtools}

\usepackage{dashbox}

\usepackage{pdflscape}

\usepackage{tikz}

\newtheorem{definition}{Definition}
\newtheorem{scheme}{Scheme}
\newtheorem{corollary}{Corollary}

\newtheorem{innercustomthm}{Theorem}
\newenvironment{customthm}[1]
{\renewcommand\theinnercustomthm{#1}\innercustomthm}
{\endinnercustomthm}

\newtheorem{hypothesis}{Hypothesis}

\newcommand{\SY}[1]{{\color{purple} SY: #1}}

\newcommand{\SM}[1]{{\color{cyan} SM: #1}}

\newcommand{\change}[1]{#1}  %Uncomment to make changes invisible
\newcommand{\delete}[1]{}  %Uncomment to make deletions invisible
\newcommand{\after}[1]{#1}

\newcommand*\squeezespaces[1]{% %% <- #1 is a number between 0 and 1
	\thickmuskip=\scalemuskip{\thickmuskip}{#1}%
	\medmuskip=\scalemuskip{\medmuskip}{#1}%
	\thinmuskip=\scalemuskip{\thinmuskip}{#1}%
	\nulldelimiterspace=#1\nulldelimiterspace
	\scriptspace=#1\scriptspace
}
\newcommand*\scalemuskip[2]{%
	\muexpr #1*\numexpr\dimexpr#2pt\relax\relax/65536\relax
}

\newcommand{\card}{\textit{C}}
\newcommand{\terminal}{\textit{T}}
\newcommand{\cloud}{\textit{B}}
\newcommand{\paysys}{\textit{PaySys}}
\newcommand{\bankc}{\textit{B\card}}
\newcommand{\bankt}{\textit{B\terminal}}
\newcommand{\psu}{\textsf{Unlinkable}}
\newcommand{\hi}{\texttt{hi}}
\newcommand{\lo}{\texttt{lo}}
\newcommand{\on}{\texttt{on}}
\newcommand{\of}{\texttt{of}}
\newcommand{\gen}{\mathfrak{g}}
\newcommand{\pkg}[1]{\smult{#1}{\gen}}

\newcommand{\opin}{\ensuremath{\textit{opin}}}
\newcommand{\out}{\ensuremath{\textit{out}}}

\usepackage{pifont}
\newcommand{\cmark}{\ding{51}}%
\newcommand{\xmark}{\ding{55}}%

\newcommand{\protocol}{\text{U}\text{TX}}
\newcommand{\protocolbf}{\text{U}\text{TX}}

\newcommand{\mult}[2]{\ensuremath{#1 \cdot #2}}
\newcommand{\smult}[2]{\mathopen\phi\left(#1, #2\right)}
\newcommand{\checksig}[2]{\mathopen\texttt{check}\left(#2, #1\right)}
\newcommand{\checksigv}[2]{\mathopen\texttt{vcheck}\left(#2, #1\right)}
\newcommand{\sig}[2]{\mathopen\texttt{sig}\left(#2, #1\right)}
\newcommand{\sigv}[2]{\mathopen\texttt{vsig}\left(#2, #1\right)}

\newcommand{\h}[1]{\texttt{h}(#1)}

\newcommand{\proj}[2]{\mathopen\texttt{p}_\texttt{#1}\left(#2\right)}
\newcommand{\nlist}[1]{\langle#1\rangle}
\newcommand{\parenth}[1]{\left({#1}\right)}

\newcommand\frsh{\#}
\newcommand\fresh{\frsh{}}

\newcommand{\nw}{\mathopen\nu}
\newcommand{\lett}{\texttt{let}}
\newcommand{\inn}{\texttt{in}}
\newcommand{\ifff}{\texttt{if}}

\newcommand{\thenn}{\texttt{then}}
\newcommand{\elsee}{\texttt{else}}

\newcommand{\bef}[1]{\emph{\textbf{#1}}}

\newcommand{\mk}{\ensuremath{mk}}
\newcommand{\bt}{\ensuremath{b_t}}
\newcommand{\fail}{\ensuremath{\perp}}

\newcommand{\crec}{\ensuremath{\textit{R}}}

\newcommand{\ctaili}{\ensuremath{\mathcal{C}^\textit{Impl}}}
\newcommand{\ctails}{\ensuremath{\mathcal{C}^\textit{Spec}}}

\newcommand{\ctail}{\ensuremath{\mathcal{C}}}
\newcommand{\ttonhi}{\ensuremath{\mathcal{TONH}}}
\newcommand{\ttofhi}{\ensuremath{\mathcal{TOFH}}}
\newcommand{\ttlo}{\ensuremath{\mathcal{TLO}}}
\newcommand{\btail}{\ensuremath{\mathcal{B}}}

\newcommand{\tparamsmm}{\ensuremath{\vec{\psi}^\texttt{MM}}}
\newcommand{\cparams}{\ensuremath{\vec{\chi}}}
\newcommand{\tparams}{\ensuremath{\vec{\psi}}}
\newcommand{\bparams}{\ensuremath{\vec{\omega}}}
\newcommand{\user}{\ensuremath{\textit{user}}}
\newcommand{\ch}{\ensuremath{\textit{ch}}}
\newcommand{\kbt}{\ensuremath{\textit{kbt}}}

\newcommand{\specc}[1]{#1_{\text{spec}}}
\newcommand{\spec}[1]{\textit{#1}_{\text{spec}}}
\newcommand{\impl}[1]{\textit{#1}_{\text{impl}}}
\newcommand{\impll}[1]{#1_{\text{impl}}}

\newcommand{\ind}{\ensuremath{\textit{ind}}}

\newcommand{\key}[1]{\mathopen\textit{k}_{#1}}
\newcommand{\pks}[1]{\mathopen\texttt{pk}\left(#1\right)}
\newcommand{\pkv}[1]{\mathopen\texttt{vpk}\left(#1\right)}

\newcommand{\event}[2]{\mathopen\texttt{ev:}\texttt{#1}\left(#2\right)}

\newcommand{\shrug}[1][]{%
	\begin{tikzpicture}[baseline,x=0.8\ht\strutbox,y=0.8\ht\strutbox,line width=0.125ex,#1]
		\def\arm{(-2.5,0.95) to (-2,0.95) (-1.9,1) to (-1.5,0) (-1.35,0) to (-0.8,0)};
		\draw \arm;
		\draw[xscale=-1] \arm;
		\def\headpart{(0.6,0) arc[start angle=-40, end angle=40,x radius=0.6,y radius=0.8]};
		\draw \headpart;
		\draw[xscale=-1] \headpart;
		\def\eye{(-0.075,0.15) .. controls (0.02,0) .. (0.075,-0.15)};
		\draw[shift={(-0.3,0.8)}] \eye;
		\draw[shift={(0,0.85)}] \eye;
		\draw (-0.1,0.2) to [out=15,in=-100] (0.4,0.95); 
\end{tikzpicture}}

\begin{document}

	\title{Provably Unlinkable Smart Card-based Payments}
	
		\author{Sergiu Bursuc, Ross Horne$^\dagger$, Sjouke Mauw, Semen Yurkov$^\ddagger$}
	\orcid{0000-0002-0409-5735}
	\affiliation{%
		\institution{University of Luxembourg, Esch-sur-Alzette, Luxembourg}
		\country{}
		\city{}
		%		\department{Department of Computer Science}
		\thanks{$^{\ddagger}$Semen Yurkov is supported by the Luxembourg National Research Fund through grant PRIDE15/10621687/SPsquared.}
	}
	\affiliation{%
			\institution{$^\dagger$University of Strathclyde, Glasgow, UK}
			%	\department{Computer \& Information Sciences}
			\city{}
			\country{}
		}
	\email{sergiu.bursuc, sjouke.mauw, semen.yurkov@uni.lu}
	\email{ross.horne@strath.ac.uk}

 \renewcommand{\shortauthors}{Sergiu Bursuc, Ross Horne, Sjouke Mauw, and Semen Yurkov}
	
	\begin{abstract} 
	The most prevalent smart card-based payment method, EMV, currently
	offers no privacy to its users. Transaction details and the card
	number are sent in cleartext, enabling the profiling and tracking
	of cardholders. Since public awareness of privacy issues is growing and legislation, such as GDPR, 
	%		aiming to regulate unwanted data collection, 
	is emerging, we believe it is necessary to investigate the possibility of making payments anonymous and unlinkable 
	%		towards eavesdroppers and terminals 
	without compromising essential security guarantees and functional
	properties of EMV. This paper draws attention to trade-offs
	between functional and privacy requirements in the design of such
	a protocol. We present the $\protocolbf$ protocol -- an enhanced
	payment protocol satisfying such requirements, and we formally certify key security and privacy properties using techniques based on the 
	applied $\pi$-calculus. 
	%		This paper explains the design of the $\protocolbf$ protocol, an enhanced payment protocol satisfying such requirements, and formally certifies key security and privacy properties using techniques based on the applied pi-calculus. 
	%    \SM{Better to avoid special characters in an abstract, because it will often be used as a separate text, so let's call it `pi-calculus''.}
	%RH: Lets keep $\pi$ for the $\pi$-calculus in the abstract for consistency.		
\end{abstract}
	

%\ccsdesc[500]{Security and privacy~Formal security models}

%\keywords{security analysis, formal methods, payment protocols}
	
	\maketitle
	
	%\SM{I like the very compact and still expressive name UPay. Still a
%short search indicates that this name has already been taken for
%various banking related applications or protocols. Is this a risk? I
%know that every short combination of letters already has many
%meanings, so there is probably no real reason to abandon the name UPay.}
\section{Introduction}

%\SY{Copyright and ACM reference format footnotes are suppressed.}

%\SY{track: formal methods: no auth -- just text (remove 2nd column), passive/active privacy}

%\SY{in the introduction explain what offline transaction is after fig 1}

%\SY{rerun the code without the terminal's check}

%\SY{software update}

%\SY{asynchronious update, unlinkable payments could coexist with normal card} downgrade attacks!

%\SY{Small comments that are not addressed: Reviewer2: p1 public transport(s), p3, coarse identities, p5, more info on verheul signarures, p5 ``is publicly known for the past few months''}
%
%\SY{say in intro about the attacker model and the scope of the formal analysis}
%
%\SY{and about the tools?}

As a payment method, EMV came into place in the mid-1990s to replace magstripe cards as they are incapable of computation and easy to clone. The EMV standard~\cite{emv} is a series of documents that specify how exactly payments should be done with the main focus on card-terminal communication. This specification is quite flexible -- only minimal requirements must be respected, so it is up to the payment system that implements EMV which additional options to include. Hence, the standard describes not a single protocol, but a whole variety of configurations. It was shown several times that some configurations are not secure~\cite{murdoch2010drimer, jorge2020emv, drimer2007keep, radu2022practical}, thus to achieve the primary goal of EMV, the safety of money, one should carefully select a secure configuration.
%one should still be careful.
%\SM{``still be careful'' is a bit unclear. Why still? Careful about
%what?}

On the other hand, currently, the privacy of payments is not an explicit requirement of EMV. To this day the communication between the card and the terminal is not encrypted. Valuable sensitive data such as the card number PAN (Primary Account Number), the amount, the country code, and the time, are exposed and an attacker eavesdropping on wireless communications can profile cardholders engaged in transactions.
%\before{In addition, an active attacker can 
%	power up a contactless card without a cardholder being aware using an antenna~\cite{habraken2015dolron}, e.g., at a doorway or by a seat on public transport. A powered-up card is ready to start an EMV session and to present its PAN -- a strong form of identity. This enables an attacker to silently track the movements of anyone who holds a payment card, and even without a genuine EMV transaction involved. Hence, active attackers pretending to be honest terminals and capable of starting communicating with the card should be part of the threat model when privacy is among the concerns.}
\after{In addition, the card presents its PAN -- a strong form of identity\delete{,} \change{--} to \emph{any} device that asks. Nearby smartphones supporting NFC and antennas~\cite{habraken2015dolron}, installed, e.g., at a doorway or by a seat on public transport, are examples of active attackers that can power up cards without cardholders being aware. After being powered-up, a card engages in what it thinks is a legitimate EMV session during which the PAN is transmitted. This enables an attacker to track the movements of anyone who holds a payment card by forcing the card to run a session and obtaining the card's permanent identity, even \emph{without a genuine EMV transaction involved}. Hence, active attackers capable of initiating communication with cards using an unauthorised device should be part of the threat model when privacy is among our concerns.}

Our position is that unwanted data collection should be mitigated at the protocol level since legal sanctions are not enough to ensure privacy -- we have examples of their violation~\cite{germanpolice, rsf}. EMVCo, a consortium of payment processing companies that develops the EMV standard, \delete{are} \change{is} aware that privacy issues are present in EMV and have proposed in the next generation of EMV \after{(EMV 2nd Gen)} to encrypt communications 
%\before{by employing the BDH protocol, an authenticated key establishment protocol~\cite{rfc} that protects from eavesdropping} 
\after{between the card and the terminal~\cite{overview} by running an authenticated key establishment before exchanging sensitive data. Obviously, a na{\i}ve solution to employ the standard Diffie-Hellman (DH) would not solve the tracking problem described above since the card's permanent identity, its public key, involved in the handshake allows both eavesdroppers and active attackers to identify the same card through different sessions. To mitigate that, EMVCo developed a blinded version of the DH protocol, BDH~\cite{rfc}, where in each session the card's public key is blinded with a fresh scalar, making eavesdroppers locked out from subsequent communication.}
However, even in the presence of encryption, the problem of active attackers persists: the card still sends its \after{unblinded} signed public key, a strong form of identity, to the terminal allowing the attacker to trace the card.

%\SM{...allowing the attacker to trace the card. or: ...breaking unlinkability.}

%\SY{Scheme explaining key establishment followed by data exchange OR table. Naive solution BDH plus EMV. DH plus EMV, BDH plus EMV etc}

%\SY{title provably unlinkable smar tcard-based payments}

The recently proposed UBDH~\cite{horne2022csf} protocol, an \emph{unlinkable} version of the BDH protocol, where an attacker cannot link key establishment sessions with the same card,
%\SM{Explain already here that U stands for unlinkable, so it's a variation of the BDH protocol proposed by EMV that achieves unlinkability (for the key establishment phase).}
is an example of an authenticated key establishment protocol that satisfies both the initial EMV privacy goals~\cite{rfc}, and rules out active attackers. The essence of UBDH is that the public key of the card appears to be fresh in each session, yet a terminal can still authenticate that the card was issued by a recognised payment system. \after{The following table summarises the privacy level each key establishment mechanism achieves. 
	
\noindent
\begin{tabular}{ |p{0.28\columnwidth}||p{0.29\columnwidth}|p{0.29\columnwidth}|  }
	%	\hline
	%	\multicolumn{4}{|c|}{Operations} \\
	\hline
	&passive privacy&active privacy\\
	\hline
	DH &\multicolumn{1}{|c|}{\xmark}& \multicolumn{1}{|c|}{\xmark}\\
	\hline
	BDH &\multicolumn{1}{|c|}{\cmark}& \multicolumn{1}{|c|}{\xmark}\\
	\hline
	UBDH&\multicolumn{1}{|c|}{\cmark}& \multicolumn{1}{|c|}{\cmark}\\
	\hline
\end{tabular}
}

%\SM{Maybe we can mention here already that this is implemented using ``Verheul
%signatures''. That will already prep the mind of the reader when
%introducing them in Section 4.1.}
%\SY{We decided this is a distraction and we point to a paper where Verheul is a crucial. We think For this paper Verheul is not specific}
According to the proposal of EMVCo~\cite{rfc, overview}, an EMV 2nd Gen transaction would consist of a key establishment phase followed by a data exchange.
%\SM{Here we need to be as clear as possible, and start with repeating
%that BDH and UBDH only concern the Key Establishment phase. So
%preserving privacy in this phase may be guaranteed, but it also needs
%to be guaranteed in the data exchange phase (and in their
%composition). This is the main concern addressed in this paper. If I
%remember correctly, not fully understanding this was one of the
%sources of concern for the reviewers.
%Would it be possible to express this a bit sharper until the end of
%this paragraph?}
Hence we need to consider unlinkability of the full protocol, as an active attacker in the second phase could gather
the information allowing to link payment sessions even if the first
phase, key agreement, is unlinkable. If we simply follow what EMV
offers now, this information includes the card's explicit identity PAN
that the payment system uses to route payments through the network.
\after{The table below presents the degrees of privacy obtained by combining a key establishment with the default EMV data exchange.

\noindent
\begin{tabular}{ |p{0.28\columnwidth}||p{0.29\columnwidth}|p{0.29\columnwidth}|  }
	%	\hline
	%	\multicolumn{4}{|c|}{Operations} \\
	\hline
	&passive privacy&active privacy\\
	\hline
	EMV &\multicolumn{1}{|c|}{\xmark}& \multicolumn{1}{|c|}{\xmark}\\
	\hline
	BDH + EMV &\multicolumn{1}{|c|}{\cmark}& \multicolumn{1}{|c|}{\xmark}\\
	\hline
	UBDH + EMV &\multicolumn{1}{|c|}{\cmark}& \multicolumn{1}{|c|}{\xmark}\\
	\hline
	UBDH + ? = \protocol &\multicolumn{1}{|c|}{\cmark}& \multicolumn{1}{|c|}{\cmark}\\
	%	\hline
	%	Our objective, \protocol &\multicolumn{1}{|c|}{\cmark}& \multicolumn{1}{|c|}{\cmark}\\
	\hline
\end{tabular}
}

%\SB{Note: the overall protocol is not a simple sequential composition of UBDH + UTX', since some other protocol elements (e.g. the blinding scalar) from UBDH are reused in addition to the established key for UTX'. I think the last line in the table could be "Our objective" and in text we could give some intuition on why we need to carefully design the interaction between the key establishment phase relying on UBDH and the transaction confirmation phase. }

\after{
While there is no privacy in cleartext EMV, encrypting EMV by running BDH or UBDH as the first step does not help achieve an unlinkable protocol where an active attacker cannot link payment sessions, thereby tracing the cardholder. The fact that EMVCo officially abandoned efforts on EMV 2nd Gen to enhance privacy in 2019~\cite{emv2019statement} also emphasises the need for a newly designed protocol (called UTX in the table) to meet future privacy demands. 

%The last line reflects the fact that an incremental advance over neither UBDH nor EMV would lead to an unlinkable protocol where an active attacker cannot link sessions and, thereby, trace the cardholder. The fact that EMVCo officially abandoned efforts on EMV 2nd Gen to enhance privacy in 2019~\cite{emv2019statement} also emphasises the need for a newly designed protocol (called UTX in the table) to meet future privacy demands. 

To the best of our knowledge, no existing solutions satisfy the basic functional and security requirements of EMV while relying exclusively on the computational resources of a smart card and being unlinkable at the same time. Mobile wallet apps like Apple Pay~\cite{applepay} protect the card number from being revealed by replacing it with a permanent Device Account Number (DAN) stored in the device (e.g. the smartphone). The DAN is exposed to an active attacker in the same way the PAN is exposed in a traditional EMV transaction\footnote{\footnotesize However, an additional layer of security is provided in this case since the device should be ready for communication, e.g. unblocked with the proper app running, etc.}. At the same time, anonymous credential (AC) schemes~\cite{idemix1, verheul2001self} are a popular way for establishing unlinkability, e.g. in the context of anonymous access to online services. Some AC schemes have been effectively implemented on smart cards~\cite{vullers2013aplar, batina2010developing}. In principle, an AC scheme could be employed to prove the legitimacy of the card to the terminal without revealing any identifying information. However, the full functionality of an EMV-like transaction requires a much richer functionality. For example, the parties need to agree on the parameters of the transaction, the terminal may need to verify the user PIN, and the bank needs to check that the payment request comes from a valid interaction with the corresponding card. AC schemes can be augmented with attributes that can be used to encode a richer functionality (e.g. attesting that the card is still valid at a certain date). However, such extensions typically rely on zero-knowledge proofs, that we aim to avoid since they would introduce too much overhead for a payment smart card. Furthermore, the design question remains, i.e. how to adapt an AC scheme for use in a larger payment system. In this paper, we demonstrate that a protocol with the desired functional, security and privacy requirements can be designed based on a particular and simple instance of anonymous credentials, namely self-blindable certificates~\cite{verheul2001self}. We discuss some deployment questions at the end of the paper and argue that our protocol could be implemented with minimal overhead on current smart cards.}

%\SM{The last two sentences are long/complicated and maybe ungrammatical.}
%}

%\SY{AC to smartcards} 

The main contributions of the paper are as follows. 
\change{
	\begin{itemize}
		\item \textit{A non-trivial threat model.} We build on recent work~\cite{horne2022csf} that explains why active attackers pose a real threat for contactless payments and how an appropriate Dolev-Yao model~\cite{dolev1983onthesec} fully accounts for them. A key novelty of our model is that we account for both honest and dishonest terminals, but in very different ways. Attackers impersonating terminals not requiring the PIN are implicitly accounted for in the Dolev-Yao model. In contrast, honest terminals requiring the PIN are explicitly represented as processes.
				
		\item \textit{Requirements for privacy-preserving card payments.} From EMV we extract functional and security requirements. For privacy requirements we extract from the EMV 2nd Gen draft~\cite{emv2019statement} an unlinkability requirement and clarify it with respect to our threat model. 
		
		\item \textit{A new payment protocol.} We design a non-trivial protocol that we argue is feasible to implement since it uses standard components that respect limited computational resources of the card. The assemblage, however, is unique. We also explain that new demands imposed by the protocol on infrastructure may be handled by sofware updates for the existing EMV infrastructure.
		
		\item \textit{A proof that the protocol satisfies our requirements.} Notably, unlinkability is proven directly using state-of-the-art bisimulation techniques and does not make use of tools. Our experiments show that our particular combination of protocol and threat model is not yet in scope of current tools.
		
	\end{itemize}
	
%	Firstly, we consider a peculiar attacker that the card cannot distinguish from honest terminals. Since the card is by nature passive, waiting to be activated, any device can engage a card in a contactless session. However, when the PIN is requested the cardholder is aware of the reason it is requested and enters their PIN only into terminals at the legitimate points of sale, thence we assume that an attacker cannot silently obtain the PIN. This attacker is fully covered by the Dolvev-Yao model~\cite{dolev1983onthesec}.

}

\delete{We identify the requirements for an unlinkable payment protocol, discuss how exactly the information that enables linking payment sessions with the same card can be hidden from the terminal, and present a protocol that we prove to satisfy these privacy requirements while preserving key requirements of EMV. }
%while providing stronger security and privacy guarantees. 
%\RH{Is this necessary? you have offline+PIN} Unless specified otherwise, we consider contactless online transactions, as nowadays they prevail~\cite{forbes}. 
%\before{The protocol we introduce respects the current flow of EMV and its
%infrastructure, supports contactless online transactions, as nowadays
%they prevail~\cite{forbes}, as well as offline transactions requiring a PIN.}
\delete{The protocol we introduce respects the current flow of EMV and its infrastructure. It supports cardholder verification via PIN, contactless online transactions, as nowadays they prevail~\cite{forbes}, as well as offline transactions when the terminal cannot immediately connect to its acquiring bank. }
%\change{Third, we conduct a small case study of the capabilities of state-of-the-art automatic protocol verifiers capable of proving unlinkability.}

%We explain the current EMV infrastructure and privacy issues in Section~\ref{sec:background}. 
%However, we stress that in this work we build a new protocol from scratch. 
We begin by presenting a design space where 
we determine the requirements of an unlinkable payment protocol in Section~\ref{sec:building},
and draw attention to trade-offs between functional and privacy requirements.
We then present an unlinkable payment protocol $\protocol$ in Section~\ref{sec:protocoltop}, and provide formal analysis in Section~\ref{sec:analysistop}.

	\section{Design space for unlinkable transactions} \label{sec:building}
%\section{Towards unlinkable smartcard-based payments}

In this section, we explore the design space for a privacy-preserving payment protocol.
This top-level design space is narrowed down in later sections to guide the design of our proposed protocol.
We explain the architecture of a payment system that should be respected,
and emphasise the functional, security and privacy requirements. 
%We conclude the section by discussing 
%explain techniques and measures that support these requirements, and 
%the appropriate attacker and trust models.

%In this section, we describe several domains to cover when designing a privacy-preserving payment protocol. We start by describing the agents involved in the protocol and briefly introduce the flow of a typical transaction. Then we discuss the desired requirements, list the techniques that support them, and look at different trust and attacker models.

%\SM{The reader may expect a linear story on what is needed. However
%what we offer is an exploration of the options that we see when designing the
%protocol. We must stress this here, because the reader will be lost if
%he starts reading this section with the wrong mindset.}

%The proposed protocol in Section~\ref{sec:protocol} satisfies the majority of the requirements and whenever we intentionally skip a requirement, we indicate that.  

\subsection{EMV infrastructure}

\begin{figure}[h]
	\resizebox{.65\linewidth}{!}{\includegraphics[width=\linewidth]{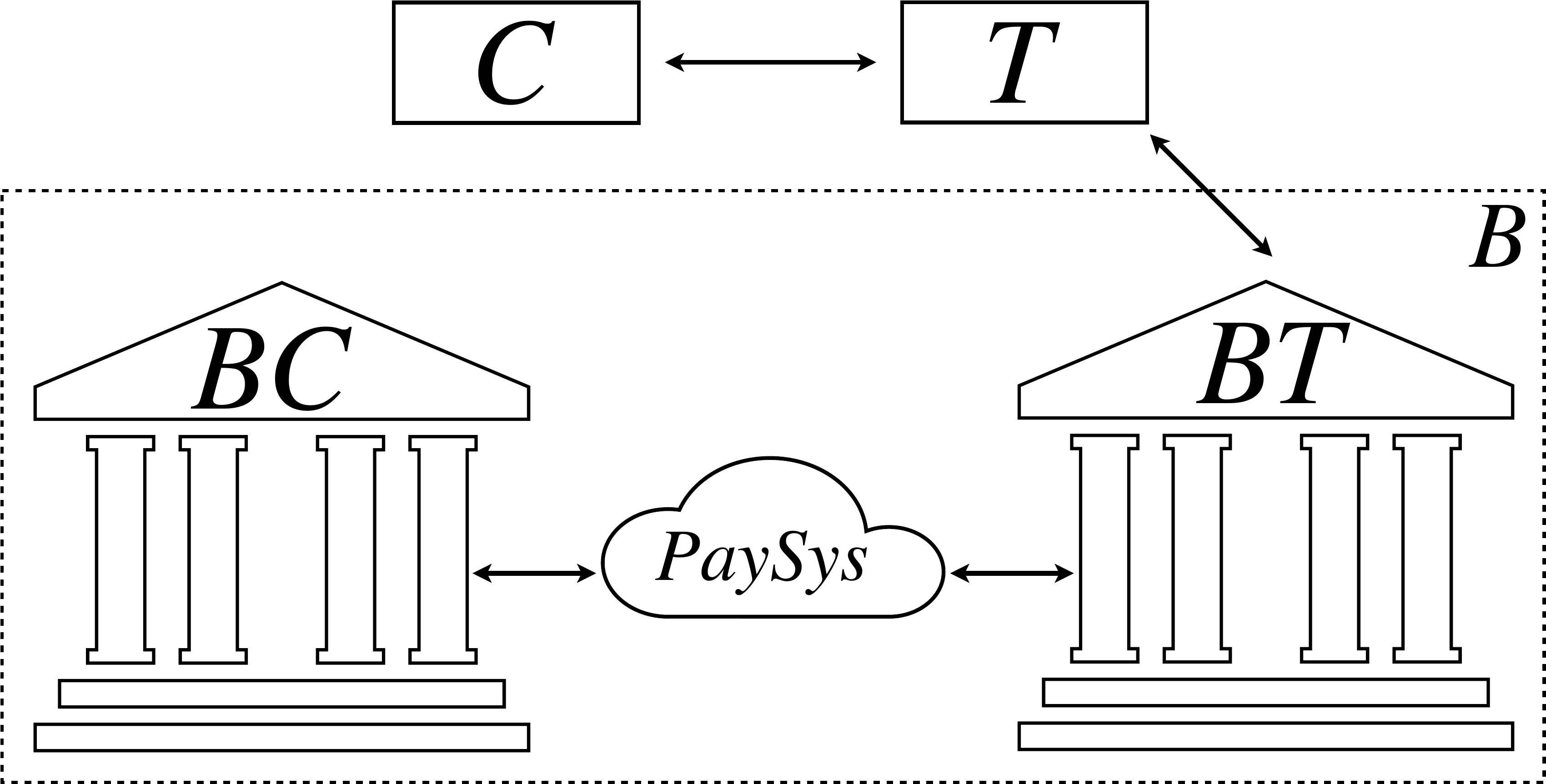}}
	\centering
	\caption{Payment architecture.} 
	\label{fig:network}
\end{figure}

We present an overview of the payment infrastructure, \change{assumed by the current EMV standard,} in Figure~\ref{fig:network}. The card $\card$ is manufactured by the \textit{issuing bank} $\bankc$ in collaboration with the payment system $\paysys$ (e.g. Visa or Amex).
%\SM{We haven't introduced PaySys yet.}
The terminal $\terminal$ is connected to an \textit{acquiring bank} $\bankt$ supporting $\paysys$ that processes payments on behalf of the terminal. 
The acquiring bank processes payments by connecting to the $\paysys$ network that exchanges messages between banks.

A successful run of the protocol results in the generation of an \emph{Application Cryptogram} $\textsf{AC}$ by $\card$. $\textsf{AC}$ is eventually sent by $\terminal$ to $\bankt$, either before of after the payment is approved by the terminal, depending on whether the payment is online or offline, respectively. The issuing bank $\bankc$ receiving $\textsf{AC}$, decides to decline or accept the transaction, and replies with the appropriate message. 
\delete{This infrastructure respects the current EMV standard and the exact processing method on the bank's side is not mandatory and ``issuers may decide to adopt other methods''\mbox{~\cite[Book~2,~Section~8]{emv}.}}

In this paper, we are concerned about hiding the information about the card from the terminal. Thus, when modelling the system, 
we merge $\bankt$, $\paysys$, and $\bankc$ into a single agent $\cloud$, modelling their common interface with the terminal when processing payments, as indicated in Figure~\ref{fig:network}. \change{This is consistent with the fact, that EMV currently does not enforce any exact processing method on the bank's side, i.e. the standard contains an example while ``issuers may decide to adopt other methods''~\cite[Book~2,~Section~8]{emv}.}

\subsection{Requirements for unlinkable payments}\label{sec:requirements}
An unlinkable protocol should satisfy three types of requirements: functional, security and privacy requirements. \delete{Functional and security requirements are extracted from the current EMV specification. Some security requirements are strengthened, and privacy requirements that were not previously present in EMV are introduced.} \change{We extract functional and security requirements from the current EMV specification, strengthen some security requirements, and introduce privacy requirements not previously present in EMV.}

\subsubsection{Functional requirements}\label{sec:functionalrequirements}
%Recall that the agents involved in the transaction are the card $\card$, the terminal $\terminal$, and the bank $\cloud$, the endpoint that the terminal is connected to. 

We consider smart card-based payments, hence we rely only on the
computational resources of the smart card and the terminal.
Devices like smartphones that can establish direct
communication between the card and the bank are excluded from the discussion in this paper. We also prohibit indirect card-bank communication by means of, e.g.\ synchronised clocks since the card has no long-term power source.

%and can be powered up only during the payment protocol session.

%Third-party devices like smartphones are excluded from the consideration. Such devices are superior to the cards in terms of computational power, are connected to the internet, and hence can communicate with the bank directly. In that case, privacy-preserving protocols, such as Revolut and Apple Pay are already available. We also prohibit indirect card-bank communication by the means of, e.g. synchronised clocks since the card can be powered up only during the payment protocol session.  

The card should use \delete{ECC-based cryptography} \change{Elliptic-Curve Cryptography (ECC)}, as already required for the new iteration of the EMV standard~\cite{rfc}. Since, currently, the card must be present within the reader's field for at most 500ms~\cite{emvcontactless}, computationally-heavy  general-purpose zero-knowledge proofs 
%or anonymous credentials 
are out of scope. 

%A study of the users' behaviour is necessary to reconsider this time limit. 
%\SM{What is the purpose of this opinion on the time limit? Will it
%come back later in the paper?}

The protocol should support contact and contactless transactions. For the purpose of this analysis we consider the PIN as the only  cardholder verification method and the PIN is always required for high-value transactions. Hardware solutions that might help to replace the PIN
%e.g.\ cards with built-in fingerprint
%sensors~\cite{idex} 
are beyond the scope of this work.
%\SM{What is the reason for the footnote? Is that information relevant
%for the reader?} 

Cards can optionally support offline transactions which carry two risks resulting in the terminal not being paid (when $\textsf{AC}$ is finally processed by the bank): either there is not enough money in the cardholder's account, or the card is blocked, e.g. reported as stolen. If offline transactions are supported, the insurance policy must cover these risks.

% see  Section~\ref{sec:policies} for details. 

%Moreover, we require that the card can only be blocked on the backend when the issuing bank includes it in the list of suspended cards.

%Being tied to card payments, there are natural restrictions on the payment protocol, i.e. there are no synchronised clocks between $\card$ and $\cloud$.

%present the description of our design space. We list here the main requirements for our payment protocol, present different attacker and trust models, and discuss major attacks and countermeasures. The same information is presented in the form of the list of nodes is presented in Appendix~\ref{a:list}.

%Delegation: the card is authenticated by the terminal, and the validity date is checked by the terminal; Does the BT checks anything?; the BC simply verifies the cryptogram. 

\subsubsection{Security requirements} \label{sec:securityrequirements}

Recall that some configurations of EMV have been shown to be insecure. The primary security goals we extract from good configurations of EMV are the following authentication and secrecy properties.

\begin{itemize}
	\item $\terminal$ must be sure that the presented card is a legitimate card that was issued by the $\paysys$ that $\terminal$ supports and that $\card$ is not expired. 
	\item If \delete{$\cloud$} \change{the bank} accepts the transaction, then $\terminal$, $\card$, and \delete{$\cloud$} \change{the bank} must agree on the transaction.
	\item Keys for message authentication and PIN are secret.
\end{itemize}

Notice that the card does not authenticate the terminal. The reason
is, in the philosophy of the EMV standard, that the payment system
allows anyone to manufacture terminals. We strengthen these requirements by assuring the card that if the cryptogram is processed, then it is processed by a legitimate bank.

%to check that the terminal is connected to a legitimate bank.

In addition to the requirements extracted from EMV above we introduce
the additional requirement that the application cryptogram $\textsf{AC}$ must be secret. This is in line with the proposal of secret channel establishment~\cite{rfc}, where a session-specific secret channel was introduced to protect all messages between the card and the terminal from eavesdroppers.
Currently, the communication between the card and the terminal is in cleartext, and the $\textsf{AC}$, that contains transaction details, is always exposed.
%Optionally, relay-resistance methods are put in place for contactless transactions.
Formal security definitions reflecting these requirements are introduced in Section~\ref{sec:securityanalysis} where we present the analysis of our proposal for a protocol.

%: the terminal authenticates the card, it checks that the card is valid at the moment of the transaction, all three parties agree on transaction details, sensitive data is secret, and, for contactless transactions, relay-resistance methods are put in place.

%\begin{itemize}
%\item The terminal authenticates the card.
%\item The terminal checks that the card is valid at the moment of the transaction. 
%\item All three parties agree on transaction details.
%\item Critical data, such as keys, PIN, and $\textsf{AC}$ is secret.
%\item Relay-resistance methods are put in place for contactless transactions.
%\end{itemize}

%The third requires further explanation. If the terminal accepts an offline transaction and issues a receipt, the bank (later) should always accept the application cryptogram for this transaction, hence \emph{later} checks must always go through for offline transactions. For an online transaction whenever the terminal believes that either the card or the bank was running the protocol processing the transaction, the card or the bank was indeed running the protocol processing this transaction. The same should hold for the bank's belief about the card and the terminal. We can not ensure this property considering the card's belief, since the default scenario in this paper is contactless payments. The authorisation response time limit currently is 5 seconds for Europe~\cite{find_the_source}, and it is unrealistic to expect the card's presence within the terminal's vicinity for such a long period: without the round trip from the card to the bank, the property can not hold. 

\subsubsection{Privacy requirements}\label{sec:privacyreq}

As mentioned in the introduction and expanded upon next, currently no
privacy properties are preserved by EMV. 
%A payment protocol that is designed to satisfy privacy must nonetheless also preserve the initial functional and security goals.
The privacy property we aim for in this paper is \emph{unlinkability}.
Unlinkability is standardised in the Common Criteria for Information Technology Security Evaluation ISO 15408~\cite{cc},
as ensuring that two uses of a card cannot be linked.
ISO standard 15408 also covers anonymity.
Unlinkability is stronger in the sense that, if two sessions are not anonymous then they can be linked, but the converse does not hold.
This explains why unlinkability is a suitable benchmark for privacy.

For this initial discussion, we give an intuitive scheme for defining
unlinkability. A formal definition is presented in
Section~\ref{sec:unlinkanalysis}, where we prove that the protocol we introduce in later sections satisfies unlinkability.
\begin{scheme}\emph{(unlinkability)}\label{def:unlink}
	Transactions are unlinkable if an attacker cannot distinguish between a system where a card can participate in multiple transactions and another system where a card can participate in at most one transaction.
\end{scheme}

\hyphenation{life-span}
Let us reflect on the above scheme. The former system represents a real-world scenario where the card is issued and within its lifespan can participate in several protocol sessions. The latter system is an idealised situation, where cards are disposed of after each transaction and can participate in one payment session at most, hence sessions are trivially unlinkable. Whenever, with respect to all attack strategies, there is no distinction between the two scenarios for a given payment protocol, \delete{unlinkability must hold also for the real-world scenario.} \change{such a protocol is unlinkable.} Guaranteeing this property without compromising the aforementioned security and privacy requirements is our primary challenge.

We explain that unlinkability cannot hold in all contexts, if we aim to fulfil also our functional and security requirements.
As mentioned above, two sessions that are not anonymous can be linked. Therefore, to achieve unlinkability, certainly any identity unique either to the card or the cardholder must never be revealed to an attacker. We call such identities \emph{strong} and they include the cardholder's name, the PAN, the card's public key, and any signature on the data specific to the card.

On the other hand, even if strong identities were protected,
\emph{coarse} identities, that are common to a group of cards, may enable tracking of groups of cardholders.
%while any strong identity is still hidden.
Coarse identities include the payment system, the validity date, the format of transaction data, and other implementation-specific features. Some coarse identities are inevitably exposed as a consequence of the requirements in Sections~\ref{sec:functionalrequirements},~\ref{sec:securityrequirements}. For instance, the terminal needs to know which payment system the card uses to authenticate the card, and needs to
be able to distinguish between valid and expired cards.
Other coarse identities include the network traffic response times,
% , since the response time from the bank, 
which may reveal information about whether the card belongs to a local or foreign bank.

Coarse identities can be combined to \textit{fingerprint} a card.
Thus we are obliged to accept that unlinkability can only be achieved up to their fingerprint,
that is, we can link two sessions with the same fingerprint only.
However, we require that this fingerprint is minimised,
thereby limiting the capability of an attacker to perform unauthorised profiling of cardholders and their behaviours.

	\section{The $\protocolbf$ protocol} \label{sec:protocoltop}

%\SM{I like the acronym UTX and I didn't find any collision with existing
%systems. Let's tell the reader what it means (Unlinkable Transaction Exchange,
%or something similar?)}
In this section, we introduce the $\protocol$ (Unlinkable Transactions) protocol that satisfies the security and privacy requirements introduced in Section~\ref{sec:requirements}. We pay particular attention to minimising the fingerprint given by the coarse identities thereby maximising unlinkability. 
%Only the fact that the card supports unlinkable payments, and, possibly, the PIN is revealed to honest terminals in the $\protocol$ protocol. 
We start by discussing the initialisation phase, then we introduce the message theory representing cryptographic primitives employed in the protocol. We then explain the key distribution between the participants of the protocol. 
%how cards are issued and clarify how the payment system connects an acquiring bank to the network, and how the acquiring bank connects the terminal. 
Finally, we thoroughly explain transactions that can either be offline, online, high, or low-value. 

%For low-value transactions We \SY{require} the terminal to send the PIN to the card or to the bank for high-value offline or online transactions respectively. 
%i.e. how cards are issued and what information participants of the protocol store.

\subsection{Application selection} \label{sec:init}

\begin{figure}[h]
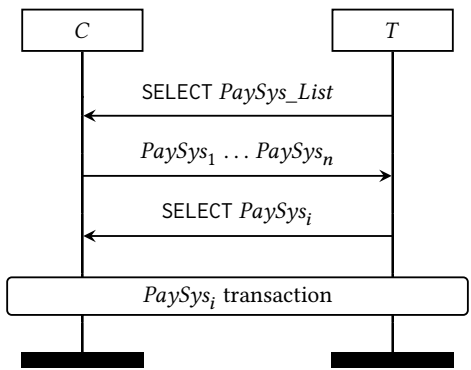

	\[
	\setmsckeyword{}
	\drawframe{no}
	\setlength{\topheaddist}{1pt}
	\setlength{\bottomfootdist}{1pt}
	\begin{msc}[instance distance=2.5cm]{}
		\declinst{c}{\parbox[c]{3.9cm}}{$C$}
		\declinst{t}{}{$T$}
		\nextlevel[0.5]
		\mess{\texttt{SELECT} \paysys\textunderscore \textit{List}}{t}{c}
		\nextlevel[1.6]
		\mess{$\text{\paysys}_1$ $\hdots$ $\text{\paysys}_n$}{c}{t}
		\nextlevel[1.6]
		\mess{\texttt{SELECT} $\paysys_i$}{t}{c}
		\nextlevel[1.1]
		\referencestart{r}{\\[-4pt] $\paysys_i$ transaction}{c}{t}
		\nextlevel[1.0]
		\referenceend{r}
		\nextlevel[0.2]
	\end{msc}
	\]
	\caption{Payment System Selection.} \label{fig:selectpaysys}
\end{figure}

The card can generally support several payment methods, or, in EMV lingo,
\emph{applications}. In Fig.~\ref{fig:selectpaysys} we schematically show how
the terminal currently selects the application. First, the terminal asks the
card to send the list of supported applications, then the card provides the
list, and the terminal selects one (possibly with the help of the cardholder).
Knowing the payment system, the terminal can select the appropriate public key
to authenticate the data on the card. Notice that the list of payment
applications is a coarse identity of the card even if this list consists of a
single application, since it can still be distinguished from other cards.

In order to avoid a coarse identity being exposed at this point,
 we design the protocol such that the card presents a list comprising a single element, $\psu$.
This means that a group of payment systems agree to provide privacy-preserving payments using the name $\psu$ for the respective application. 
Terminals, thus, should also be upgraded to support $\psu$ in order to accept unlinkable payments, before such cards are rolled out.
An alternative is to allow each payment system to provide their own unlinkable application, and to tolerate that the payment system becomes part of the coarse identity of the card. Our analysis covers both choices.

\subsection{Keys required to set up $\psu$}\label{sec:keys}
Here we explain who generates and holds keys and signatures involved in the $\protocol$ protocol. An authority, who is either a payment system or a delegate acting on behalf of a group of payment systems, produces signatures involved in the protocol using 
two types of signing keys. Firstly, a secret key $s$, is used to produce \emph{certificates for banks}, which are kept by the terminal and used by the card to check that the terminal is connected to a legitimate bank. Secondly, a list of secret keys $\chi_{\texttt{MM}}$ is maintained for each new calendar month.
%that are used to sign all cards implementing $\psu$.
%\SM{What does ``common'' mean? It may lead the reader to think that
%every card has all these secret keys.}
They are used by the authority upon request from the payment system to generate \emph{month certificates} unique to each card supporting $\psu$ for every month the card is valid.
%for each new calendar month.
%\SM{Are these lists unique for each card? Or generic?}
A card valid for five years would store $61$ such month certificates,
% where $0 \leq \texttt{MM} \leq 60$,
that the terminal checks to be sure that the card is valid at the month of a particular purchase. 
The public key for checking month certificates is broadcast to terminals from the first of every month.

%Whether the card is currently valid is a coarse identity of the card that the terminal requires.
We take care to prohibit an attacker from learning the expiry or the issuing month, which would allow many cards to be distinguished.
To do so, we introduce the following pointer mechanism. 
The card maintains a pointer to the most recent month certificate that
has been used in response to a legitimate request by the terminal.
%\SM{By whom?}
When the terminal asks the card to show the certificate for the month, the card compares the pointer with the received month. If the received month is greater than what the pointer references, the card advances the pointer to this month
%\SM{Why ``next'' and not the received month? The received month might
%not be the next month, but a later month. (Because the card hasn't
%been used for a few months.)}
and shows the respective certificate. If either the received
month coincides with or is one month behind the pointer, the card
simply shows the certificate for this month and the pointer remains untouched.
Otherwise, if the month requested is older than two months the card
terminates the session. A terminal cannot request a month in the
future, assuming that the public keys for verification are carefully managed, such that they are never released in advance.

%\SM{Wouldn't this lead to a DOS/desynchronisation attack if an adversarial
%terminal inquires about a month in the far future? In that case, the card updates
%to that month and will not be able to make any purchases until that month.}
%\SM{When reading the protocol below a bit closer, I think this problem
%has been anticipated, but please confirm that there is no problem.}

We allow a window of two months, to allow time for offline terminals to eventually receive the most recent public key for the month.
For this reason a new card valid for 60 months is loaded with 61 month certificates with a pointer referencing the issuing month. That way a newly issued card cannot be distinguished from cards already in circulation as it is ready to present the certificate for the month prior to the month in which it was issued.
Thus, the only coarse identities revealed are whether the card is outdated or has not been used since the beginning of the month.
%That way a new card can present an outdated certificate, while the terminal cannot ask for an older one. 

\subsection{Message theory}\label{sec:messagetheory}

We now introduce cryptographic primitives employed by the $\protocol$ protocol. Since later in Section~\ref{sec:analysistop} we reason about $\protocol$ symbolically and assume perfect cryptography, low-level details such as ECC domain parameters are out of scope. In particular, \delete{we assume the use of authenticated encryption, such that the message integrity is guaranteed.} \change{we assume the use of an encryption scheme that guarantees message integrity.} Fig.~\ref{fig:syntax} presents the message theory 
%for the $\protocol$ protocol 
that consists of the syntax, that defines messages agents can form, and the equational theory $E$, that axiomatises cryptographic operations.

\begin{figure}[h]
	\[\arraycolsep=0.5pt
	\begin{gathered}
		\begin{array}{c}
			\begin{array}{rlr}
				M, N \Coloneqq
				& \gen & \mbox{DH group generator (constant)} \\
				\small|& x & \mbox{variable} \\
				\small|& \mult{M}{N} & \mbox{multiplication} \\
				\small|& \smult{M}{N} & \mbox{scalar multiplication} \\
				\small|& \hash{M} & \mbox{hash} \\
				\small|& \enc{M}{N} & \mbox{symmetric encryption} \\
				\small|& \nlist{M_1,\hdots,M_k} & \mbox{$n$-tuple} \\
				\small|& \pks{M} & \mbox{public key} \\
				\small|& \sig{N}{M} & \mbox{signature} \\
				\small|& \pkv{M} & \mbox{Verheul public key} \\
				\small|& \sigv{N}{M} & \mbox{Verheul signature} \\
%			\small|& \aenc{M}{N} & \mbox{asymmetric encryption} \\
				\small|& \checksig{N}{M} & \mbox{check signature} \\
				\small|& \checksigv{N}{M} & \mbox{check Verheul signature} \\
				\small|& \proj{i}{N} & \mbox{$i$th projection} \\
				\small|& \dec{N}{M} & \mbox{symmetric decryption} \\
%			\small|& \adec{N}{M} & \mbox{asymmetric decryption} \\
%			\small|& \texttt{on} & \mbox{online transaction tag} \\
				\small|& \fail, \texttt{ok}, \texttt{accept}, \texttt{auth}, \texttt{lo},~\texttt{hi} & \mbox{constants} 
%				\\
%				\small|& \texttt{ok} & \mbox{PIN success reply} \\
%				\small|& \texttt{accept} & \mbox{bank success reply} \\
%				\small|& \texttt{auth} & \mbox{transaction authorisation} \\
%				\small|& \texttt{lo},~\texttt{hi} & \mbox{low and high amount} \\
%				\small|& out, term, bank, card & \mbox{chanels}
			\end{array} 
		\end{array}
	\end{gathered}
	\]
		\[
	\begin{gathered}
		\begin{array}{c}
			\begin{array}{ll}
				\mult{M}{N} =_{E} \mult{N}{M} \\[1pt]
				\mult{\parenth{\mult{M}{N}}}{K} =_{E} \mult{M}{\parenth{\mult{N}{K}}} \\[1pt]
				\smult{\mult{M}{N}}{K} =_{E} \smult{M}{\smult{N}{K}} \\[1pt]
				\proj{i}{\nlist{M_1,\hdots,M_k}} =_{E} M_i \\[1pt]
				\dec{K}{\enc{M}{K}} =_{E} M \\[1pt]
%			\adec{K}{\aenc{M}{K}} =_{E} M \\[1pt]
				\checksig{\pks{K}}{\sig{K}{M}} =_{E} M \\[1pt]
				\checksigv{\pkv{K}}{\sigv{K}{M}} =_{E} M \\[1pt]
				\smult{M}{\sigv{K}{N}} =_E \sigv{K}{\smult{M}{N}}
			\end{array}
		\end{array}
	\end{gathered}
	\]
	\caption{$\protocol$ message theory.} 
	\label{fig:syntax}
\end{figure}

%\SM{Isn't it more common to use $\pi_i$ for projection, instead of $p_i$?}
%\SY{I don't know how to adress this :) When I replace it with $pi$ I don't like how formulas become + there is a confusion with $pi$-calculust. For now I leave it as is}

The message theory admits operations for ECC\delete{-based cryptography}, i.e.
multiplication between two field elements (scalars), and multiplication
between a scalar and an element of the DH group. Whenever we say that ``a message is blinded with a scalar'', we mean multiplication by that scalar. Next, we include a standard set of cryptographic operations such as hashing, symmetric key cryptography, $n$-tuples, and generic digital signatures. Finally, we introduce the Verheul signature scheme~\cite{verheul2001self}, which is invariant under blinding of the message-signature pair 
%with the same scalar 
(hence can appear as ``new'' in each session). This scheme \delete{uses} \change{supports} ECC\delete{-based cryptography} and has been demonstrated to work sufficiently fast on smart cards~\cite{batina2010developing}. We also define several constants employed in $\protocol$.

The equational theory $E$ captures the two types of multiplication and contains conventional destructor functions: decryption, projection, and two versions of signature verification. A digital signature is successfully verified whenever the message corresponds to the message extracted from the signature by applying the appropriate check function. Notice that the last equation ensures that if the function $\checksigv{\cdot}{\cdot}$ is applied to the signature, blinded with some scalar and the matching Verheul public key, it returns the message, blinded with the same scalar.

%In the description of the protocol in Section~\ref{sec:protocol} we employ a hash-based message authentication code (HMAC) mechanism to ensure the authenticity and integrity of a cryptogram. We abbreviate $\hmac{N}{M} \coloneqq \nlist{M, \hash{\nlist{M, N}}}$, hence do not include the $\hmac{\_}{\_}$ operation in the message theory. The verification requires the message $N$: one computes the hash of the first element of $\hmac{N}{M}$ together with $N$ and compares it with the second element.

\subsection{Before running the protocol: the setup}\label{sec:setup}

Before describing the protocol we explain how the payment system issues a card in collaboration with the issuing bank, how the acquiring bank joins the payment system, and how the terminal connects to the acquiring bank. In the next section, where we describe the transaction, we collapse the payment system, the issuing bank, and the acquiring bank into a single agent.

% Though we abstract away the issuing bank, the payment system, and
%the acquiring bank as the bank agent \cloud{} to present and later model our protocol, however, we consider them separately to describe thoroughly this setup phase.
%\SM{I don't understand the previous sentence. Maybe we could split it in two
%separate sentences.}

%We have abstracted away the issuing bank, the payment system, and the acquiring bank as the bank agent \cloud to model our protocol, however, we consider them separately to describe thoroughly the setup phase. In what follows we assume a \emph{trusted} delegate that can produce signatures on behalf of the payment systems providing unlinkable payments. The delegate uses two types of signing keys: $s$, to produce the \emph{bank's certificate} kept by the terminal that the card checks to be sure that the terminal is connected to the legitimate bank; and keys $\chi_{\texttt{MM}}$, where $0 \leq \texttt{MM} \leq 60$ to produce the \emph{month certificates} kept by the card that the terminal checks to be sure that the card is valid at the month of the purchase. 

\subsubsection{Issuing a card} \label{sec:issuing}
Here we outline how a card could be manufactured involving \change{a} signing authority that payment systems could share as explained in Section~\ref{sec:keys}. 

To issue a card, the payment system generates a new card's private key $c$, computes the card'\delete{a}\change{s} public key $\pkg{c}$, and 
%\SM{Maybe we could briefly explain the purpose of the blinding scalar.
%Who is not supposed to see what?}
asks the signing authority to generate the following list of month Verheul signatures $\{\nlist{\texttt{MM}, \sigv{\chi_\texttt{MM}}{\pkg{c}}}\}_{\texttt{MM}=0}^{60}$ which it loads to the card together with $\pks{s}$, $c$, $\pkg{c}$, PAN, and PIN.
%It also sets the counter ATC to 0. 
Then the card is sent to the issuing bank together with $\pkg{c}$, PAN, and PIN; the bank generates and loads to the card a new master key $\mk$, and finally sends the card to the user. Since no one should ever have access to $c$ except the card, we assume the \delete{issuing bank} \change{payment system} never shares or stores $c$. 
%Notice that this scheme prevents the delegate to issue cards alone, without the payment system.

\subsubsection{The keys used by the terminal to connect to the payment system} \label{sec:connectingterminal}

To allow an acquiring bank supporting the payment system to process payments in the month $\texttt{MM}$, 
the authority knowing $s$ issues a certificate of the form $\nlist{\nlist{\texttt{MM}, \pkg{\bt}}, \sig{s}{\nlist{\texttt{MM}, \pkg{\bt}}}}$ to each acquiring bank, where $b_t$ is the private key of the bank. In turn, the acquiring bank loads the terminal with both this data and a symmetric key $\kbt$ used for secure communication between the terminal and the bank. The terminal presents the bank's certificate at each run of the protocol. As explained in Section~\ref{sec:keys}, the terminal and the bank must update the month key certificate and the month validation key regularly without being offline for more than two months.

First\delete{ly}\change{,} we explain why the month $\texttt{MM}$ is signed. Recall the card points to the most recent month it has seen. Hence, if this month requested by the terminal is the month pointed to by the card, or the month before, it is safe to reveal that it is valid for either of these two months. The signature $\sig{s}{\nlist{\texttt{MM}, \pkg{\bt}}}$ containing the month $\texttt{MM}$ is required in the situation where the next month is requested, in which case this signature serves as proof to the card that the next month has arrived. This prevents attackers learning whether the card is valid next month, and also avoids the pointer being advanced too quickly thereby invalidating the card in the current month.  
%Firstly, this certificate enables the card to verify that $\texttt{MM}$ is authentic, hence to invoke the appropriate month signature $\sigv{\chi_\texttt{MM}}{\pkg{c}}$ when proving that it is valid in the month $\texttt{MM}$. 
Notice\delete{,} that $\pkv{\chi_{\texttt{MM}}}$ is publicly known for the past few months and could have been transferred by the terminal to the card and used by the card to check whether a request for the next month is valid. However, since checking Verheul signatures is too expensive for the card, we avoid using keys $\pkv{\chi_{\texttt{MM}}}$, and instead only check the certificate $\nlist{\nlist{\texttt{MM}, \pkg{\bt}}, \sig{s}{\nlist{\texttt{MM}, \pkg{\bt}}}}$ against the generic $\pks{s}$ already present in the card which can employ a more efficient signature scheme since it does not need to support blinding. 

Second\delete{ly}, the bank's certificate enables the card to verify that $\pkg{\bt}$ is a public key for a legitimate bank connected to the payment system providing $\psu$, hence it can safely use $\pkg{\bt}$ to encrypt the application cryptogram at the end of the transaction. This signature helps to avoid the situation when an attacker introduces their own public key and thereby can look inside the cryptogram to gather sensitive information including the PAN.

%We describe this in detail in the next section.

\delete{In principle, the signature on the month and the signature on the bank's public key could be separate, yet it is efficient to transmit them in a single message. Moreover, the bank's public key introduces certain padding to small and publically known constants $\texttt{MM}$ representing months. If a bank requires multiple keys, multiple certificates could be produced.} \change{It is efficient to transmit the month and the bank's public key in a single message, however, in principle, the signatures on each could be separate. In this case, to prevent offline guessing attacks, the payment system should introduce certain padding to small and publicly known constants $\texttt{MM}$ representing months. If a bank requires multiple keys, the payment system could produce multiple certificates.}  

%as verifying that this is the correct Verheul public key for the month $\texttt{MM}$ would require expensive on-card computation. Instead the card verifies the signature using one generic signature $\pks{s}$. 

%$\texttt{MM}$ and $\pkg{\bt}$ are authentic using $\pks{s}$. The card uses hence to invoke the appropriate month signature $\sigv{\chi_\texttt{MM}}{\pkg{c}}$ when proving that it is valid in the month $\texttt{MM}$, and to safely use $\pkg{\bt}$ to encrypt the application cryptogram at the end of the transaction. We describe this in detail in the next section. 
%The purpose of associating the month and the bank's public key is to assure the card that the terminal has been connected to a legitimate bank in the last two months, and that the card can use this public key to encrypt the application cryptogram at the end of the transaction.
%This bank certificate should not be confused with the list of month signatures on the card described above. 
%Finally $\nlist{\nlist{\texttt{MM}, \pks{\bt}}, \sig{s}{\nlist{\texttt{MM}, \pks{\bt}}}}$ is transferred to the bank together with the current month validation key $\pkv{\chi_\texttt{MM}}$. In turn, the acquiring bank loads the terminal with this data, and a symmetric key $\kbt$ used for secure communication between them. 

The secure channel between the bank and the terminal modelled here as a symmetric key $\kbt$ could be established by other means, which is consistent with EMV as it is not specified.

\subsection{The $\protocol$ transaction} \label{sec:protocol}
\begin{figure*}
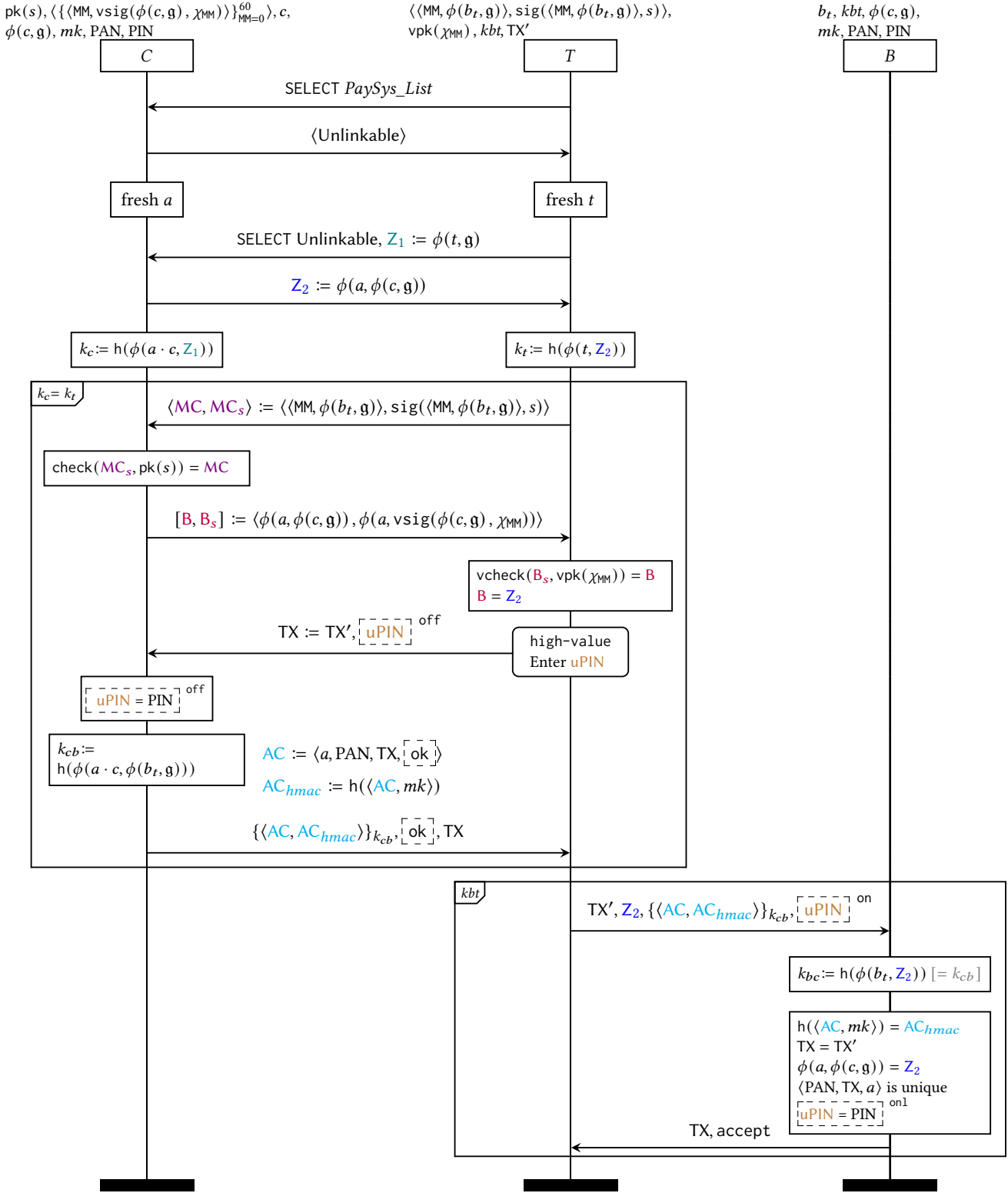

	\[
	\setmsckeyword{}
	\drawframe{no}
	\setlength{\topheaddist}{1pt}
	\setlength{\bottomfootdist}{1pt}
	\begin{msc}{}
		\mscset{instance distance=5.7cm}
		\declinst{c}{\raisebox{0.12cm}{\parbox[c]{4.9cm}{\small $\pks{s}$, $\nlist{\{\nlist{\texttt{MM}, \sigv{\chi_\texttt{MM}}{\pkg{c}}}\}_{\texttt{MM}=0}^{60}}$, $c$, $\pkg{c}$, $\mk$, PAN, PIN}}}{$C$}
		\declinst{t}{\raisebox{0.12cm}{\parbox[c]{5.6cm}{\small $\nlist{\nlist{\texttt{MM}, \pkg{\bt}}, \sig{s}{\nlist{\texttt{MM}, \pkg{\bt}}}}, \\ \pkv{\chi_{\texttt{MM}}}, \kbt, \textsf{TX}'$}}}{$T$}
		\mscset{instance distance=3.9cm}
		\declinst{cloud}{\raisebox{0.12cm}{\parbox[c]{2.5cm}{\small $\bt$, $\kbt$, $\pkg{c}$, \\ $\mk$, PAN, PIN}}}{$B$}
		\mess{\texttt{SELECT} \paysys\textunderscore \textit{List}}{t}{c}
		\nextlevel[1.6]
		\mess{$\nlist{\psu}$}{c}{t}
		\nextlevel[-2.5]
		\action*{fresh $a$}{c}
		\action*{fresh $t$}{t}
		\nextlevel[2.6]
		\mess{\texttt{SELECT} $\psu$, $\textcolor{teal}{\textsf{Z}_1} \coloneqq \smult{t}{\gen}$}{t}{c}
		\nextlevel[1.6]
		\mess{$\textcolor{blue}{\textsf{Z}_2} \coloneqq \smult{a}{\pkg{c}}$}{c}{t}
		\nextlevel[1]
		\action*{\small $\key{c} \coloneqq \hash{\smult{\mult{a}{c}}{\textcolor{teal}{\textsf{Z}_1}}}$}{c}
		\action*{\small $\key{t} \coloneqq \hash{\smult{t}{\textcolor{blue}{\textsf{Z}_2}}}$}{t}	
		\nextlevel[1.7]
		\inlinestart[2][2]{exp1}{\footnotesize $\key{c} = \key{t}$}{c}{t}
		\nextlevel[1.5]
		\mess{$\nlist{\textcolor{violet}{\textsf{MC}}, \textcolor{violet}{\textsf{MC}_s}} \coloneqq \nlist{\nlist{\texttt{MM}, \pkg{\bt}}, \sig{s}{\nlist{\texttt{MM}, \pkg{\bt}}}}$}{t}{c}
%		\mess{$\nlist{\textcolor{violet}{\textsf{MC}}, \textcolor{violet}{\textsf{MC}_s}} \coloneqq \nlist{\nlist{\texttt{MM}, \pkg{\bt}}, \sig{s}{\textcolor{violet}{\textsf{MC}}}}$}{t}{c}
%			\mess{\begin{minipage}[t]{0.48\textwidth} \raisebox{0.3cm}{ \hspace*{2.45cm}{ $\textcolor{violet}{\textsf{MC}} \coloneqq \nlist{\texttt{MM}, \pkg{\bt}}$}}  \\ \raisebox{0.3cm}{ \hspace*{2.45cm}{ $\textcolor{violet}{\textsf{MC}_{s}} \coloneqq \sig{s}{\textcolor{violet}{\textsf{MC}}}$}} \\[3pt] \hspace*{2.45cm} { $\nlist{\textcolor{violet}{\textsf{MC}}, \textcolor{violet}{\textsf{MC}_s}}$} \end{minipage}}{t}{c}	
		\nextlevel[0.9]
		\action*{\small \parbox[c]{3.28cm}{$\checksig{\pks{s}}{\textcolor{violet}{\textsf{MC}_s}} = \textcolor{violet}{\textsf{MC}}$}}{c}
		\nextlevel[3]	
		\mess{$\left[\textcolor{purple}{\textsf{B}}, \textcolor{purple}{\textsf{B}_s}\right] \coloneqq \nlist{\smult{a}{\pkg{c}}, \smult{a}{\sigv{\chi_{\texttt{MM}}}{\pkg{c}}}}$}{c}{t}
		\nextlevel[0.8]
		\action*{\small \parbox[c]{3.24cm}{$\checksigv{\pkv{\chi_{\texttt{MM}}}}{\textcolor{purple}{\textsf{B}_s}} = \textcolor{purple}{\textsf{B}}$ \\ $\textcolor{purple}{\textsf{B}} = \textcolor{blue}{\textsf{Z}_2}$}}{t}
		\nextlevel[2.3]
		\referencestart{ref1}{\raisebox{-0.33cm}{\hspace*{0.3cm}\parbox[c]{1.9cm}{\small $\texttt{high-value}$ \\ \ Enter \textcolor{brown}{\textsf{uPIN}}}}}{t}{t}
		\nextlevel[0.9]
%		\mess{\begin{minipage}[t]{0.48\textwidth} \raisebox{0.3cm}{\hspace*{1.1cm}{\scriptsize $\textsf{TX} \coloneqq \nlist{\text{CCY}, \text{AMT}, \text{DATE}} \eqqcolon \textsf{TX}'$}}  \\ \hspace*{3.6cm} {\small $\textsf{TX}, \dbox{\textcolor{brown}{\textsf{uPIN}}}$} \end{minipage}}{ref1left}{c}
			\mess{\begin{minipage}[t]{0.48\textwidth} \hspace*{3.3cm} {$\textsf{TX} \coloneqq \textsf{TX}', \dbox{\textcolor{brown}{\textsf{uPIN}}}^\texttt{ off}$} \end{minipage}}{ref1left}{c}
		\nextlevel[0.8]
		\referenceend{ref1}
		\action*{\small \parbox[c]{2.0cm}{$\hspace*{-0.06cm}\dbox{\textcolor{brown}{\ \textsf{uPIN}} = \text{PIN}}^\texttt{ off}$}}{c}
		\nextlevel[2.0]
		\action*{\small \parbox[c]{3.1cm}{$\key{cb} \coloneqq \hash{\smult{\mult{a}{c}}{\pkg{b_t}}}$}}{c}
		\nextlevel[4.1]
		\mess{\begin{minipage}[t]{0.48\textwidth} \raisebox{0.3cm}{ \hspace*{2.45cm}{ $\textcolor{cyan}{\textsf{AC}} \coloneqq \nlist{a, \text{PAN}, \textsf{TX}, \dbox{\texttt{ok}}}$}}  \\ \raisebox{0.3cm}{ \hspace*{2.45cm}{ $\textcolor{cyan}{\textsf{AC}_{hmac}} \coloneqq \hash{\nlist{\textcolor{cyan}{\textsf{AC}}, mk}}$}} \\[3pt] \hspace*{2.26cm} { $\enc{\nlist{\textcolor{cyan}{\textsf{AC}}, \textcolor{cyan}{\textsf{AC}_{hmac}}}}{\key{cb}}, \dbox{\texttt{ok}}\, , \textsf{TX}$} \end{minipage}}{c}{t}	
%		\mess{\begin{minipage}[t]{0.48\textwidth} \raisebox{0.3cm}{ \hspace*{0.57cm}{\small $\nlist{\textcolor{cyan}{\textsf{AC}}, \textcolor{cyan}{\textsf{AC}_{hmac}}} \coloneqq \nlist{\nlist{a, \text{PAN}, \textsf{TX}, \dbox{\texttt{ok}}}, \hash{\nlist{\textcolor{cyan}{\textsf{AC}}, mk}}}$}}  \\ \hspace*{2.45cm} { $\enc{\nlist{\textcolor{cyan}{\textsf{AC}}, \textcolor{cyan}{\textsf{AC}_{hmac}}}}{\key{cb}}, \dbox{\texttt{ok}}$} \end{minipage}}{c}{t}
		\nextlevel[0.5]
		\inlineend{exp1}
		\nextlevel[0.5]
		\inlinestart[2][2]{exp2}{\footnotesize $\kbt$}{t}{cloud}
		\nextlevel[1.7]
		\mess{$\textsf{TX}', \textcolor{blue}{\textsf{Z}_2}, \enc{\nlist{\textcolor{cyan}{\textsf{AC}}, \textcolor{cyan}{\textsf{AC}_{hmac}}}}{\key{cb}}, \dbox{\textcolor{brown}{\textsf{uPIN}}}^\texttt{ on}$}{t}{cloud}
		\nextlevel[0.9]
		\action*{\small \parbox[c]{3.2cm}{$\key{bc} \coloneqq \hash{\smult{\bt}{\textcolor{blue}{\textsf{Z}_2}}} \textcolor{gray}{[= \key{cb}]}
		$}}{cloud}
		\nextlevel[1.9]
		\action*{\small \parbox[c]{3.2cm}{$\hash{\nlist{\textcolor{cyan}{\textsf{AC}}, \mk}} = \textcolor{cyan}{\textsf{AC}_{hmac}} \\ \textsf{TX} = \textsf{TX}' \\ \smult{a}{\pkg{c}} = \textcolor{blue}{\textsf{Z}_2} \\ \nlist{\text{PAN}, \textsf{TX}, a} \text{ is unique} \\ \dbox{\hspace*{-0.15cm}\textcolor{brown}{\ \textsf{uPIN}} = \text{PIN}}^\texttt{ onl}$}}{cloud}
		\nextlevel[4.77]
		\mess{$\textsf{TX}, \texttt{accept}$}{cloud}{t}
		\nextlevel[0.37]
		\inlineend{exp2}
	\end{msc}
	\]
	\caption{The $\protocol$ protocol. Offline and online high-value modes are annotated as $\texttt{off}$ and $\texttt{on}$ respectively.}
	\label{fig:protocol}
\end{figure*}

We introduce online and offline modes of the $\protocol$ protocol in Fig.~\ref{fig:protocol}. The PIN is asked for in high-value purchases. 
%i.e. whenever the transaction amount AMT exceeds the threshold value \textsl{LIM}. 
In the offline mode, the PIN is sent to the card. As the PIN must be transferred to the card, and the card cannot leave the session until the PIN is entered, high-value offline transactions are always performed as a contact payment. In online mode, the PIN is not sent to the card, instead it is sent to the bank together with the application cryptogram. Parts of the protocol involving the PIN check are indicated by dashed lines and annotated as $\texttt{off}$ and  $\texttt{on}$ indicating these two modes of operation. In Fig.~\ref{fig:protocol} the two messages exchanged between the terminal and the bank are either executed during the transaction (online mode) or postponed to the moment when the terminal goes online to upload collected cryptograms and, optionally, to update its bank's certificate (offline mode). 

\subsubsection{Initialisation} When the card is close enough to the terminal, it is powered up, and the terminal asks which payment methods the card supports by issuing the $\texttt{SELECT}$ command. The card supporting unlinkable payments, replies with a singleton list containing only $\psu$, as explained in Section~\ref{sec:init} The terminal then selects this payment method and sends to the card the ephemeral public key $\pkg{t}$. The card in response sends to the terminal $\smult{a}{\pkg{c}}$, which is its public key, blinded with a fresh scalar $a$. After that the card and the terminal establish the symmetric session key $\key{c} \coloneqq \hash{\smult{\mult{a}{c}}{\pkg{t}}} =_E \hash{\smult{t}{\smult{a}{\pkg{c}}}} \eqqcolon \key{t}$ which they use \textit{to encrypt all further communications}. \delete{We represent that all communications between the card and the terminal are encrypted by putting the box around the messages they exchange.} \change{In Fig.~\ref{fig:protocol}, phases of the protocol that are encrypted are represented by a box with a label in the top-left corner indicating the encryption key.}

A \emph{passive eavesdropper} \delete{who only listens and intercepts messages} \change{who only observes messages} is now locked out from the session since it has no access to the derived key. 
\delete{On the other hand}\change{However}, an \emph{active attacker}
can choose their own public key and engage in the handshake.
\delete{so we must}\change{We will} explain below how \delete{such an active attacker is} \change{active attacks are} mitigated. The only information about the card exposed at this point is the fact that the card supports the application $\psu$. 

%which can interact with the card and use its own public key to and acquire the card's public key link transactions can not do so since the public key of the card in each session is freshly blinded, hence is distinct and cannot be used to link transactions. 

\subsubsection{Validity check}\label{sec:mutualauth} 
 After the secret key is established, the card presents evidence that it is valid. To do so, firstly, the terminal sends to the card $\nlist{\nlist{\texttt{MM}, \pkg{\bt}}, \sig{s}{\nlist{\texttt{MM}, \pkg{\bt}}}}$, the current bank's certificate.
%that contains the current month $\texttt{MM}$, the threshold limit for PIN-less low-value transactions $\textsl{LIM}$, and the acquiring bank's public key $\pks{\bt}$. 
The card verifies this certificate against the public key $\pks{s}$, hence believes that this terminal is connected to a legitimate acquiring bank, and that $\texttt{MM}$, and $\pkg{\bt}$ are authentic.

Having received this legitimate request to show the month certificate corresponding to $\texttt{MM}$, the card updates its pointer, leaves it untouched or, aborts the transaction as described in Section~\ref{sec:keys}.
%compares its pointer to the
%last shown month Verheul signature with the newly received month. If
%the received month is greater than what the pointer references, the card
%advances the pointer to the signature of month $\texttt{MM}$.
%If the received month coincides with the card's local view on $\texttt{MM}$, the pointer remains untouched. If the received $\texttt{MM}$ is one month behind the card's, the card does not update the pointer and decides that this transaction is offline. As mentioned in Section~\ref{sec:connectingterminal} in this work we tolerate terminals that haven't been online for one month at most. If the certificate is too old, the card terminates the session. 
After the decision about the pointer has been made, the card 
%increments the transaction counter ATC, 
blinds the appropriate month Verheul signature $\sigv{\chi_{\texttt{MM}}}{\pkg{c}}$ 
%(it is always what the received $\texttt{MM}$ points at) 
with a the scalar $a$, and sends to the terminal the following blinded pair $\nlist{\smult{a}{\pkg{c}}, \smult{a}{\sigv{\chi_{\texttt{MM}}}{\pkg{c}}}}$. 

The terminal verifies this blinded message-signature pair against the
current month Verheul public key $\pkv{\chi_{\texttt{MM}}}$ and
additionally checks that the first element of the received pair
coincides with the card's blinded public key used to establish a
session key. This check ensures that the terminal is still
communicating with the same card \change{and prevents the construction of fake cards loaded with previously exposed blinded message-signature pairs}.

Since both elements of the message coming from the card at this stage are freshly blinded, as for the session key, they are distinct in each session, hence the terminal cannot use it to reidentify the card in future sessions by simply requesting the same month. 
At this point in the protocol the card exposes that it is valid at the month $\texttt{MM}$ (since the key $\pkv{\chi_{\texttt{MM}}}$ fits) which is not a coarse card's identity, as all other cards that have not yet expired and support unlinkable payments, expose the same information. 

%To prevent the terminal from learning the issuing month of the card we restrict offline
%transactions to one month behind at most, and introduce a pointer mechanism that forbids the terminal to feed too old certificates to the card. Finally, the terminal cannot even learn whether the card is \emph{new} since at the month of issue the card carries the signature valid for the previous month (hence the total amount is sixty-one) and is ready to complete offline transactions with outdated terminals.

\subsubsection{Cardholder verification (high-value)}
In case of a high-value \emph{offline} transaction, the terminal asks the cardholder to enter the PIN and sends the entered number $\textsf{uPIN}$ to the card together with the transaction details. If this input matches the actual card's PIN, the card includes the $\texttt{ok}$ message both in the reply to the terminal and in the cryptogram to indicate to the issuing bank that the PIN has been successfully verified on the card's side. Otherwise, the card includes the $\fail$ message in the reply and in the cryptogram, which the terminal has to send to the bank anyway to log failed attempts to enter the PIN for auditing purposes. In case of a high-value \emph{online} transaction, the terminal also asks the cardholder to enter the PIN but instead keeps it and sends it to the acquiring bank together with the cryptogram. 

\subsubsection{Cryptogram generation}\label{sec:acgen}
The terminal sends to the card the transaction details 
%$\squeezespaces{1}\textsf{TX} \coloneqq \nlist{\text{CCY}, \text{AMT}, \text{DATE}}$ 
$\squeezespaces{1}\textsf{TX}'$ 
comprising the currency, the amount, and the date; and either $\fail$ (when the transaction is low-value), or, the entered $\textsf{uPIN}$ (when the transaction is high-value offline). The card computes $\squeezespaces{1}\key{cb} \coloneqq \hash{\smult{\mult{a}{c}}{\pkg{b_t}}}$, which serves as a symmetric session key between the card and the acquiring bank for this transaction only. Then the card generates one of the cryptograms. 
\begin{itemize}
	\item $\squeezespaces{1}\textsf{AC}^\texttt{\fail} \coloneqq \nlist{a, \text{PAN}, \textsf{TX}}$ if no \textsf{uPIN} has been received.
	\item $\squeezespaces{1}\textsf{AC}^{\texttt{ok}} \coloneqq \nlist{a, \text{PAN}, \textsf{TX}, \texttt{ok}}$ if the received \textsf{uPIN} is correct.
	\item $\squeezespaces{1}\textsf{AC}^{\texttt{no}} \coloneqq \nlist{a, \text{PAN}, \textsf{TX},\texttt{no}}$ otherwise.
\end{itemize}

Finally, the card uses the master key $\mk$ that has already been shared between the card and the issuing bank to compute hash-based message authentication code of the form $\hash{\nlist{\textsf{AC}, \mk}}$ and replies respectively with one the following messages to the terminal.

\begin{itemize}
	\item $\squeezespaces{1}\nlist{\enc{\nlist{\textsf{AC}^{\fail}, \hash{\nlist{\textsf{AC}^\fail, mk}}}}{\key{cb}}, \fail, \textsf{TX}}$ \vspace{2pt}
	\item $\squeezespaces{1}\nlist{\enc{\nlist{\textsf{AC}^{\texttt{ok}}, \hash{\nlist{\textsf{AC}^{\texttt{ok}}, mk}}}}{\key{cb}}, \texttt{ok}, \textsf{TX}}$ \vspace{2pt}
	\item $\squeezespaces{1}\nlist{\enc{\nlist{\textsf{AC}^{\texttt{no}}, \hash{\nlist{\textsf{AC}^{\texttt{no}}, mk}}}}{\key{cb}}, \texttt{no}, \textsf{TX}}$
\end{itemize}

Each of these messages correspond\change{s} to the cryptograms described and contain\change{s} additional information \change{on} whether the PIN was successfully verified by the card ($\texttt{ok}$ entry), or the PIN verification has failed ($\texttt{no}$ entry) because the terminal cannot open the cryptogram encrypted for the acquiring bank.

Notice that the card includes \delete{a} \change{the} nonce $a$ in each of the cryptograms to make it unique per session. The fact that the same $a$ is used for blinding the card's public key at the initialisation step allows the bank to strongly connect the cryptogram to the current session, thereby avoiding the cryptogram being replayed in other sessions.
Although a trusted terminal is already assured that a valid card generated the cryptogram in the current session, it is beneficial for the bank to also check this. This is because the bank may not fully trust \change{the} terminal to be implemented correctly in which case, if the terminal fails to authenticate the card properly as described in Section~\ref{sec:mutualauth}, the terminal cannot be reimbursed for the cryptogram generated by an honest card in another session and replayed in a session with an unauthenticated device posing as a card. Therefore $\protocol$ ensures recent aliveness of the card from the perspective of the bank even in the presence of compromised terminals.

\subsubsection{Transaction authorisation}
In the final stage of the protocol the terminal asks the bank to authorise the payment. The terminal uses the pre-established secret key $\kbt$ that is shared with the acquiring bank to send the following.
\begin{itemize}
	\item The transaction details $\textsf{TX}'$.
	\item The blinded card's public key $\textsf{Z}_2 \coloneqq \smult{a}{\pkg{c}}$.
	\item The encrypted cryptogram of one of the three types described above that it has received from the card.
	\item The user-entered PIN $\textsf{uPIN}$ in case the transaction is high-value online, or the message $\fail$ otherwise.
\end{itemize}

%the transaction details $\textsf{TX}'$, the blinded card's public key $\textsf{Z}_2 \coloneqq \smult{a}{\pkg{c}}$, the encrypted cryptogram received from the card and of one of the three types described above,
%and the user-entered PIN $\textsf{uPIN}$ in
%case the transaction is high-value online (or the message $\fail$ otherwise). 

Recall that $B$ in Fig.~\ref{fig:protocol} represents both the acquiring and the issuing banks. 
The acquiring 
bank uses its private key $b_t$ and the received card's blinded public key $\textsf{Z}_2$ to compute the symmetric key with the card $\key{bc} \coloneqq \hash{\smult{\bt}{\textsf{Z}_2}} = \hash{\smult{\bt}{\smult{a}{\pkg{c}}}}$ and to decrypt the cryptogram. 
%The payment network that connects acquiring and issuing banks can now use the $\text{PAN}$ from the cryptogram to route the cryptogram (unmodified, together with the information from the terminal listed above) to the correct issuing bank.
Internally to $B$, the acquiring bank uses the PAN from the decrypted cryptogram and forwards all the information received from terminal to the issuing bank. In turn, the issuing bank 
%After receiving the data from the terminal and decrypting the cryptogram, that includes the PAN, the 
%issuing 
determines $\mk$, $\pkg{c}$, and the $\text{PIN}$ corresponding to the PAN received and performs the following.

\begin{itemize}
	\item It checks that the first element of the cryptogram hashed with $mk$ equals the second element, making sure the cryptogram is authentic.
%	using the master key $\mk$: $\hash{\nlist{\textsf{AC}, \mk}} = \textsf{AC}_{hmac}$.
	\item It checks that the transaction details $\textsf{TX}'$ received from the terminal match the transaction details from the cryptogram: $\textsf{TX}'=\textsf{TX}$
	\item 
%	Receiving the PAN as part of the cryptogram, it takes the corresponding card's public key $\pkg{c}$ and 
 It checks that the blinding factor $a$ from the cryptogram multiplied by the card's public key $\pkg{c}$ matches the blinded public key $\textsf{Z}_2$ received from the terminal: $\squeezespaces{0.3}\smult{a}{\pkg{c}} = \textsf{Z}_2$.
	\item It checks the transaction history of the card and ensures that
  the received $a$ has not been used for an identical transaction,
  hence preventing a replay of the cryptogram. This replaces the transaction counter ATC from the EMV standard.
%  \footnote{\scriptsize In the EMV standard currently, this check is optimised by using the transaction counter $\text{ATC}$ on the card which increases at the beginning of each session.}. 
	\item If the transaction value is high, the bank checks if the $\texttt{ok}$ tag is present in the cryptogram and proceeds with the reply, otherwise, if the $\texttt{ok}$ tag is not present, the bank checks if the received \textsf{uPIN} matches the card's PIN: $\textsf{uPIN} = \text{PIN}$ and proceeds with the reply.
\end{itemize}

If the above is successful, the terminal receives the reply message $\nlist{\text{TX}, \texttt{accept}}$ encrypted with $\kbt$.

%Notice that since the acquiring bank obtains the $\text{PAN}$ from the cryptogram, the payment system that connects it to the issuing bank can use the $\text{PAN}$ to route the cryptogram (unmodified, together with the information from the terminal listed above). 

Notice that in $\protocol$ the payment system still uses the PAN to route payments between acquiring and issuing banks, however, it is now hidden from the terminal in contrast to the current EMV standard, where it is exposed. The main changes to the infrastructure to roll out $\protocol$ are as follows. The acquiring bank requires a key for decrypting the cryptogram. The issuing bank \delete{requires} \change{is required} to ensure itself that the nonce from the cryptogram is tied to the legitimate card-terminal session. In addition a substantial update is needed for public key infrastructure explained in Sections~\ref{sec:keys},~\ref{sec:issuing}, and~\ref{sec:connectingterminal}.

%our $\protocol$ protocol it is now hidden from the terminal in contrast to the current EMV. 

	\section{Unlinkability and security analysis} \label{sec:analysistop}
\def\specTonhiSimpleEv{
	\begingroup
	\addtolength{\jot}{1.3pt}
	\begin{equation}
		\squeezespaces{0.5}
		\notag
		\begin{aligned}
			& \nw \, \textsf{TXdata}. \\ 
			& \lett \, \textsf{TX} \coloneqq \nlist{\textsf{TXdata}, \hi} \; \inn \\
			& \nw t. \lett z_1 \coloneqq \pkg{t} \inn \; \\ & \cout{\ch}{z_1}. \\
			& \cin{\ch}{z_2}. \\
			& \lett \, \key{t} \coloneqq \hash{\smult{t}{z_2}} \inn \\
			& \cout{\ch}{\enc{\textsf{crt}}{\key{t}}}. \\
			& \cin{\ch}{n}.\\
			& \lett \, \nlist{\textsf{B}, \textsf{B}_s} \coloneqq \dec{n}{\key{t}} \; \inn \\
			& \ifff \ \checksigv{pk_\texttt{MM}}{\textsf{B}_s} = \textsf{B} \ \thenn \\
			& \ifff \ \textsf{B} = z_2 \ \thenn \\
			& \cin{user}{\textsf{uPIN}}.\\
			& \lett \, \textit{etx} = \enc{\nlist{\textsf{TX}, \fail}}{\key{t}} \; \inn \\
			& \cout{\ch}{\textit{etx}}. \\
			& \cin{\ch}{y}. \\
			& \lett \, \nlist{\textsf{EHAC}, \textsf{pinV}, \textsf{tx}} \coloneqq \dec{y}{\key{t}} \; \inn \\
			& \ifff \ \textsf{tx} = \textsf{TX} \ \thenn \\
			& \event{\textcolor{blue}{TComC}}{z_1, z_2, \enc{\textsf{crt}}{\key{t}}, n, \textit{etx}, y} \\
			& \lett \, \textit{req} = \enc{\nlist{\textsf{TX}, z_2, \textsf{EHAC}, \textsf{uPIN}}}{\kbt} \; \inn \\
			& \event{\textcolor{violet}{TRunBC}}{\textit{req}, z_1, z_2, \enc{\textsf{crt}}{\key{t}}, n, \textit{etx}, y}\\
			& \cout{\ch}{\textit{req}}. \\
			& \cin{\ch}{r}. \\
			& \ifff \; \dec{r}{\kbt} = \nlist{\textsf{TX},\texttt{rtype}} \ \thenn \\
			& \quad \event{\textcolor{red}{TComBC}}{\textit{req}, r, z_1, z_2, \enc{\textsf{crt}}{\key{t}}, n, \textit{etx}, y} \\ 
			&  \quad \ifff \; \texttt{rtype} = \texttt{accept} \;\thenn\\ 
			& 	\quad \quad \event{\textcolor{brown}{TAccept}}{\kbt,\textsf{TX}} \\
			&   \quad \quad \cout{\ch}{\texttt{auth}} \\
		\end{aligned}
	\end{equation}
	\endgroup
}

\def\specCsimpleEv{
	\begingroup
	\addtolength{\jot}{1.3pt}
	\begin{equation}
		\squeezespaces{0.5}
		\notag
		\begin{aligned}
			& \cin{\ch}{z_1}. \\
			& \nw a.\; \lett z_2 \coloneqq \smult{a}{\pkg{c}} \; \inn \\ 
			&\cout{\ch}{z_2}.\\
			& \lett \, \key{c} \coloneqq \hash{\smult{\mult{a}{c}}{z_1}} \inn \\
			& \cin{\ch}{m}. \ \ \textcolor{purple}{*}\\ 
			& \lett \, \nlist{\nlist{\textsf{MM},y_B}, \textsf{MC}_s} \coloneqq \dec{m}{\key{c}}  \inn \\
			&  \ifff \ \checksig{pk_s}{\textsf{MC}_s} = \nlist{\textsf{MM},y_B} \ \thenn \\
			& \lett \, \textit{emc} \coloneqq \enc{\nlist{\smult{a}{\pkg{c}}, \smult{a}{\texttt{vsig}_\texttt{MM}}}}{\key{c}} \; \inn \\
			& \cout{\ch}{\textit{emc}}.\\
			& \cin{\ch}{x}.\\
			% low-value online, low-value offline, high-value online
			& \lett \, \nlist{\textsf{TX},\textsf{uPin}} \coloneqq \dec{x}{\key{c}} \; \inn\\
			& \lett \, \textsf{AC}^\fail \coloneqq \nlist{a, \text{PAN}, \textsf{TX}} \; \inn\\
			& \lett \, \textsf{AC}^{\texttt{ok}} \coloneqq \nlist{a, \text{PAN}, \textsf{TX}, \texttt{ok}} \; \inn\\
			& \lett \, \textsf{AC}^{\texttt{no}} \coloneqq \nlist{a, \text{PAN}, \textsf{TX}, \texttt{no}} \; \inn\\
			& \lett \, \key{cb} \coloneqq \hash{\smult{\mult{a}{c}}{y_B}} \inn \\
			& \ifff \ \textsf{uPin} = \fail \ \thenn \\ 
			%		& \quad \event{CAccept}{\sf{PAN,TX}}.\\
			& \quad \lett \, \textsf{HAC} \coloneqq \nlist{\textsf{AC}^{\fail},\hash{\nlist{\textsf{AC}^{\fail}, \mk}}} \; \inn \\
			& \quad \lett \, \textit{eac} \coloneqq \enc{\nlist{\enc{\textsf{HAC}}{\key{cb}}, \fail, \textsf{TX}}}{\key{c}} \; \inn \\
			& \quad \event{\textcolor{teal}{CRunB}}{\textit{eac}} \\
			& \quad \event{\textcolor{blue}{CR}\textcolor{red}{u}\textcolor{violet}{n}}{z_1, z_2, m, \textit{emc}, x, \textit{eac}} \\
			& \quad \cout{\ch}{\textit{eac}} \\
			& \elsee \, \ifff \ \textsf{uPin} =\text{PIN} \ \thenn \\	
			& \quad \lett \, \textsf{HAC} \coloneqq  \nlist{\textsf{AC}^{\texttt{ok}},\h{\nlist{\textsf{AC}^{\texttt{ok}}, \mk}}} \; \inn \\
			& \quad \lett \, \textit{eac} \coloneqq \enc{\nlist{\enc{\textsf{HAC}}{\key{cb}}, \texttt{ok}, \textsf{TX}}}{\key{c}} \; \inn \\
			& \quad \event{\textcolor{teal}{CRunB}}{\enc{\textsf{HAC}}{\key{cb}}} \\
			& \quad \event{\textcolor{blue}{CR}\textcolor{red}{u}\textcolor{violet}{n}}{z_1, z_2, m, \textit{emc}, x, \textit{eac}} \\
			%		& \quad \event{CAccept}{\sf{PAN,TX}}. \\
			& \quad \cout{\ch}{\textit{eac}} \\
			& \elsee \, \\
			& \quad \lett \, \textsf{HAC} \coloneqq \nlist{\textsf{AC}^{\texttt{no}}, \h{\nlist{\textsf{AC}^{\texttt{no}}, \mk}}} \; \inn \\
			& \quad \lett \, \textit{eac} \coloneqq \enc{\nlist{\enc{\textsf{HAC}}{\key{cb}}, \texttt{no}, \textsf{TX}}}{\key{c}} \; \inn \\
			& \quad \event{\textcolor{teal}{CRunB}}{\enc{\textsf{HAC}}{\key{cb}}} \\
			& \quad \event{\textcolor{blue}{CR}\textcolor{red}{u}\textcolor{violet}{n}}{z_1, z_2, m,\textit{emc}, x, \textit{eac}} \\
			&\quad \cout{\ch}{\textit{eac}}
		\end{aligned}
	\end{equation}
	\endgroup
}

\def\specBsimpleEv{
		\begingroup
	\addtolength{\jot}{1.3pt}
	\begin{equation}
		\squeezespaces{0.5}
		\notag
		\begin{aligned}
			& \cin{\ch}{x}.\\
			& \lett \, \mbox{\squeezespaces{1}$\nlist{\textsf{TX}', z_2, \textsf{EAC}, \textsf{uPIN}} \coloneqq \dec{x}{\kbt}$} \\
			%		& \event{BRunning}{\kbt,\textsf{TX'}}. \\
			& \lett \, \key{bc} \coloneqq \hash{\smult{b_t}{z_2}} \inn \\
			& \lett \, \mbox{\squeezespaces{0.5}$\nlist{\textsf{AC}, \textsf{AC}_{hmac}} \coloneqq \dec{\textsf{EAC}}{\key{bc}}$} \inn \\
			& \lett \, \nlist{x_a, \text{PAN}, \textsf{TX}, \textsf{pinV}} = \textsf{AC} \; \inn \\
			%			& \cin{si}{=\text{PAN},\text{PIN}, mk, pk_{c}}. \\
			&\cin{\nlist{si, \text{PAN}}}{\text{PIN}, mk, pk_{c}}. \\
			& \ifff \ \hash{\nlist{\textsf{AC}, mk}} = \textsf{AC}_{hmac} \ \thenn \\
			& \ifff \ \textsf{TX} = \textsf{TX'} \ \thenn \\
			& \ifff \ \smult{x_a}{pk_{c}}= z_2 \\
			%	& \lett \, r \coloneqq \nlist{\textsf{TX'}, \texttt{accept}} \inn \\
			% low-value offline, low-value online
			& \lett \ \nlist{\textsf{TXdata,TXtype}}  \coloneqq \textsf{TX'} \ \inn \\
			& \ifff \; \textsf{TXtype}=\lo \; \thenn \\
			& \quad \event{\textcolor{teal}{BComC}}{\textsf{EAC}} \\
			& \quad \event{\textcolor{red}{BRunT}}{x, \enc{\nlist{\textsf{TX'}, \texttt{accept}}}{\kbt}} \\
			%		& \quad\event{BAccept}{\textsf{PAN,TX}} \\
			& \quad \event{\textcolor{violet}{BComTC}}{x} \\
			& \quad\cout{\ch}{\enc{\nlist{\textsf{TX'}, \texttt{accept}}}{\kbt}} \\
			& \elsee \; \ifff \; \textsf{TXtype} = \hi \ \thenn \\
			% high-value offline		
			& \quad \mbox{\squeezespaces{1}$\ifff \ (\textsf{pinV} = \texttt{ok}) \vee (\textsf{uPIN} = \text{PIN}) \ \thenn$} \\ 
			%		& \quad \quad \event{BAccept}{\text{PAN},\textsf{TX}}. \\
			& \quad \quad \event{\textcolor{teal}{BComC}}{\textsf{EAC}} \\
			& \quad \quad \event{\textcolor{red}{BRunT}}{x, \enc{\nlist{\textsf{TX}', \texttt{accept}}}{\kbt}} \\
			& \quad \quad \event{\textcolor{violet}{BComTC}}{x} \\
			& \quad \quad \cout{\ch}{\enc{\nlist{\textsf{TX}', \texttt{accept}}}{\kbt}} \\
			% high-value online	
			%& \quad \elsee \ \ifff \ \textsf{uPIN} = \text{PIN} \ \thenn \\
			%& \quad \quad \event{BAccept}{\textsf{PAN,TX}} \\
			%& \quad \quad \cout{\ch}{\enc{\textsf{TX}', \texttt{accept}}{\kbt}} \\
			& \quad \elsee \\
			& \quad \quad \event{\textcolor{brown}{BReject}}{\kbt, \textsf{TX'}} \\
			& \quad \quad \event{\textcolor{teal}{BComC}}{\textsf{EAC}} \\
			& \quad \quad \event{\textcolor{red}{BRunT}}{x, \enc{\nlist{\textsf{TX'}, \texttt{reject}}}{\kbt}} \\
			& \quad \quad \event{\textcolor{violet}{BComTC}}{x} \\
			& \quad \quad  \cout{\ch}{\enc{\nlist{\textsf{TX'}, \texttt{reject}}}{\kbt}}. \\ 
			%		&  \quad \quad \event{BReject}{\textsf{PAN,TX}}. \\
		\end{aligned}
	\end{equation}
	\endgroup
}

\def\specTonhiSimple{
\begin{equation}
	\squeezespaces{1}
		\notag
		\begin{aligned}
%	T_{\texttt{onhi}}(&\user, \ch, pk_\texttt{MM}, \textsf{crt}, \kbt)  \triangleq \\
& \nw \, \textsf{TXdata}. \\ 
& \lett \, \textsf{TX} \coloneqq \nlist{\textsf{TXdata}, \hi} \; \inn \\
& \nw t. \lett \, z_1 \coloneqq \pkg{t} \inn \; \\ & \cout{\ch}{z_1}. \\
& \cin{\ch}{z_2}. \\
& \lett \, \key{t} \coloneqq \hash{\smult{t}{z_2}} \inn \\
%		& \event{TRunning}{\kbt,\key{t},z_1,z_2, \textsf{TX}}. \\
& \cout{\ch}{\enc{\textsf{crt}}{\key{t}}}. \\
& \cin{\ch}{n}.\\
& \lett \, \nlist{\textsf{B}, \textsf{B}_s} \coloneqq \dec{n}{\key{t}} \; \inn \\
& \ifff \ \checksigv{pk_\texttt{MM}}{\textsf{B}_s} = \textsf{B} \ \thenn \\
& \ifff \ \textsf{B} = z_2 \ \thenn \\
%& \event{TCommit}{\key{t},z_1,z_2, \textsf{TX}}. \\
& \cin{user}{\textsf{uPIN}}.\\
& \cout{\ch}{\enc{\nlist{\textsf{TX}, \fail}}{\key{t}}}. \\
& \cin{\ch}{y}. \\
%& \event{\textcolor{blue}{TCommitWithC}}{z_1, z_2, \enc{\textsf{crt}}{\key{t}}, n, \enc{\nlist{\textsf{TX}, \fail}}{\key{t}}, y} \\
%& \lett \, \textit{req} = \enc{\nlist{\textsf{TX}, z_2, \dec{y}{\key{t}}, \textsf{uPIN}}}{\kbt} \; \inn \\
%& \event{\textcolor{violet}{TRunningWithBC}}{\textit{req}, z_1, z_2, \enc{\textsf{crt}}{\key{t}}, n, \enc{\nlist{\textsf{TX}, \fail}}{\key{t}}, y}\\
& \cout{\ch}{\enc{\nlist{\textsf{TX}, z_2, \dec{y}{\key{t}}, \textsf{uPIN}}}{\kbt}}. \\
& \cin{\ch}{r}. \\
& \ifff \; \dec{r}{\kbt} = \nlist{\textsf{TX},\texttt{rtype}} \ \thenn \\
%& \quad \event{\textcolor{red}{TCommitWithBC}}{\textit{req}, r, z_1, z_2, \enc{\textsf{crt}}{\key{t}}, n, \enc{\nlist{\textsf{TX}, \fail}}{\key{t}}, y} \\ 
%		& \quad \event{TCommit}{\kbt,\key{t},z_1,z_2, \textsf{TX}}. \\
&  \quad \ifff \; \texttt{rtype} = \texttt{accept} \;\thenn\\ 
%		& \quad \quad  \; \event{TAccept}{\kbt,\TX}. \\
%& 	\quad \quad \event{\textcolor{brown}{TAccept}}{\kbt,\textsf{TX}} \\
&   \quad \quad \cout{\ch}{\texttt{auth}} \\
		\end{aligned}
	\end{equation}
}

\def\specCsimple{
	\begin{equation}
		\squeezespaces{1}
		\notag
		\begin{aligned}
%			C(& \ch, c, pk_s, \texttt{vsig}_\texttt{MM}, \text{PAN}, \mk, \text{PIN}) \triangleq \\ 
			& \cin{\ch}{z_1}. \\
			& \nw a.\lett \, z_2 \coloneqq \smult{a}{\pkg{c}} \; \inn \\ 
			&\cout{\ch}{z_2}.\\
			& \lett \, \key{c} \coloneqq \hash{\smult{\mult{a}{c}}{z_1}} \inn \\
			%		& \event{CRunning}{\text{PAN},\key{c},z_1,z_2}. \\
			& \cin{\ch}{m}.\\ 
			& \lett \, \nlist{\nlist{\textsf{MM},y_B}, \textsf{MC}_s} \coloneqq \dec{m}{\key{c}}  \inn \\
			&  \ifff \ \checksig{pk_s}{\textsf{MC}_s} = \nlist{\textsf{MM},y_B} \ \thenn \\
			%	& \ifff \ \proj{1}{\textsf{MC}} =\texttt{MM} \ \thenn \\
%			& \lett \, \textit{emc} = \enc{\nlist{\smult{a}{\pkg{c}}, \smult{a}{\texttt{vsig}_\texttt{MM}}}}{\key{c}}\; \inn \\
			& \cout{\ch}{\enc{\nlist{\smult{a}{\pkg{c}}, \smult{a}{\texttt{vsig}_\texttt{MM}}}}{\key{c}}}.\\
			& \cin{\ch}{x}.\\
			% low-value online, low-value offline, high-value online
			& \lett \, \nlist{\textsf{TX},\textsf{uPin}} \coloneqq \dec{x}{\key{c}} \; \inn\\
			& \lett \, \textsf{AC} \coloneqq \nlist{a, \text{PAN}, \textsf{TX}} \; \inn\\
			& \lett \, \textsf{AC}^{\texttt{ok}} \coloneqq \nlist{a, \text{PAN}, \textsf{TX}, \texttt{ok}} \; \inn\\
			& \lett \, \textsf{AC}^{\fail} \coloneqq \nlist{a, \text{PAN}, \textsf{TX}, \fail} \; \inn\\
			& \lett \, \key{cb} \coloneqq \hash{\smult{\mult{a}{c}}{y_B}} \inn \\
			& \ifff \ \textsf{uPin} = \fail \ \thenn \\ 
			%		& \quad \event{CAccept}{\sf{PAN,TX}}.\\
%			& \quad \lett \, \textit{eac} \coloneqq \enc{\enc{\nlist{\textsf{AC},\hash{\nlist{\textsf{AC}, \mk}}}}{\key{cb}}}{\key{c}} \; \inn \\
%			& \quad \event{\textcolor{teal}{CRunningWithB}}{\textit{eac}} \\
%			& \quad \event{\textcolor{blue}{CR}\textcolor{red}{unn}\textcolor{violet}{ing}}{z_1, z_2, m, \textit{emcert}, x, \textit{eac}} \\
			& \quad \cout{\ch}{\enc{\enc{\nlist{\textsf{AC},\hash{\nlist{\textsf{AC}, \mk}}}}{\key{cb}}}{\key{c}}} \\
			& \elsee \, \ifff \ \textsf{uPin} =\text{PIN} \ \thenn \\
%			& \quad \lett \, \textit{eac} \coloneqq \enc{\nlist{\enc{\nlist{\textsf{AC}^{\texttt{ok}},\h{\nlist{\textsf{AC}^{\texttt{ok}}, \mk}}}}{\key{cb}}, \texttt{ok}}}{\key{c}} \; \inn \\
%			& \quad \event{\textcolor{teal}{CRunningWithB}}{\enc{\nlist{\textsf{AC}^{\texttt{ok}},\h{\nlist{\textsf{AC}^{\texttt{ok}}, \mk}}}}{\key{cb}}} \\
%			& \quad \event{\textcolor{blue}{CR}\textcolor{red}{unn}\textcolor{violet}{ing}}{z_1, z_2, m, \textit{emcert}, x, \textit{eac}} \\
			%		& \quad \event{CAccept}{\sf{PAN,TX}}. \\
			& \quad \cout{\ch}{\enc{\nlist{\enc{\nlist{\textsf{AC}^{\texttt{ok}},\h{\nlist{\textsf{AC}^{\texttt{ok}}, \mk}}}}{\key{cb}}, \texttt{ok}}}{\key{c}}} \\
			& \elsee \, \\
%			& \quad \lett \, \textit{eac} \coloneqq \enc{\nlist{\enc{\nlist{\textsf{AC}^{\fail}, \h{\nlist{\textsf{AC}^{\fail}, \mk}}}}{\key{cb}}, \fail}}{\key{c}} \; \inn \\
%			& \quad \event{\textcolor{teal}{CRunningWithB}}{\enc{\nlist{\textsf{AC}^{\fail},\h{\nlist{\textsf{AC}^{\fail}, \mk}}}}{\key{cb}}} \\
%			& \quad \event{\textcolor{blue}{CR}\textcolor{red}{unn}\textcolor{violet}{ing}}{z_1, z_2, m,\textit{emcert}, x, \textit{eac}} \\
			&\quad \cout{\ch}{ \enc{\nlist{\enc{\nlist{\textsf{AC}^{\fail}, \h{\nlist{\textsf{AC}^{\fail}, \mk}}}}{\key{cb}}, \fail}}{\key{c}}}
		\end{aligned}
	\end{equation}
}

\def\specBsimple{
	\begin{equation}
		\squeezespaces{1}
		\notag
		\begin{aligned}
			%		B(&\ch, si, \kbt, b_t) \triangleq \\
			& \cin{\ch}{x}.\\
			& \lett \, \mbox{\squeezespaces{1}$\nlist{\textsf{TX}', z_2, \textsf{EAC}, \textsf{uPIN}} \coloneqq \dec{x}{\kbt}$} \\
			%		& \event{BRunning}{\kbt,\textsf{TX'}}. \\
			& \lett \, \key{bc} \coloneqq \hash{\smult{b_t}{z_2}} \inn \\
			& \lett \, \mbox{\squeezespaces{0.5}$\nlist{\textsf{AC}, \textsf{AC}_{hmac}} = \dec{\textsf{EAC}}{\key{bc}}$} \inn \\
			& \lett \, \nlist{x_a,\textsf{PAN,TX,pinV}} = \textsf{AC} \; \inn \\
			%			& \cin{si}{=\text{PAN},\text{PIN}, mk, pk_{c}}. \\
			&\cin{\nlist{si, \text{PAN}}}{\text{PIN}, mk, pk_{c}}. \\
			& \ifff \ \hash{\nlist{\textsf{AC}, mk}} = \textsf{AC}_{hmac} \ \thenn \\
			& \ifff \ \textsf{TX} = \textsf{TX'} \ \thenn \\
			%		& \event{BCommit}{\kbt,\textsf{TX}, \text{PAN}}. \\
			& \ifff \ \smult{x_a}{pk_{c}}= z_2 \\
			%	& \lett \, r \coloneqq \nlist{\textsf{TX'}, \texttt{accept}} \inn \\
			% low-value offline, low-value online
			& \lett \, \nlist{\textsf{TXdata, TXtype}}  \coloneqq \textsf{TX'} \ \inn \\
			& \ifff \; \textsf{TXtype}=\lo \; \thenn \\
			%	& \quad \event{\textcolor{teal}{BCommitWithC}}{\textsf{EAC}} \\
			%	& \quad \event{\textcolor{red}{BRunningWithT}}{x, \enc{\nlist{\textsf{TX'}, \texttt{accept}}}{\kbt}} \\
			%		& \quad\event{BAccept}{\textsf{PAN,TX}} \\
			%	& \quad \event{\textcolor{violet}{BCommitWithTC}}{x} \\
			& \quad\cout{\ch}{\enc{\nlist{\textsf{TX'}, \texttt{accept}}}{\kbt}} \\
			& \elsee \; \ifff \; \textsf{TXtype} = \hi \ \thenn \\
			% high-value offline		
			& \quad \mbox{\squeezespaces{1}$\ifff \ (\textsf{pinV} = \texttt{ok}) \vee (\textsf{uPIN} = \text{PIN}) \ \thenn$} \\ 
			%		& \quad \quad \event{BAccept}{\text{PAN},\textsf{TX}}. \\
			%	& \quad \quad \event{\textcolor{teal}{BCommitWithC}}{\textsf{EAC}} \\
			%	& \quad \quad \event{\textcolor{red}{BRunningWithT}}{x, \enc{\nlist{\textsf{TX}', \texttt{accept}}}{\kbt}} \\
			%	& \quad \quad \event{\textcolor{violet}{BCommitWithTC}}{x} \\
			& \quad \quad \cout{\ch}{\enc{\nlist{\textsf{TX}', \texttt{accept}}}{\kbt}}\\
			% high-value online	
			%& \quad \elsee \ \ifff \ \textsf{uPIN} = \text{PIN} \ \thenn \\
			%& \quad \quad \event{BAccept}{\textsf{PAN,TX}} \\
			%& \quad \quad \cout{\ch}{\enc{\textsf{TX}', \texttt{accept}}{\kbt}} \\
			& \quad \elsee \\
			%	& \quad \quad \event{\textcolor{brown}{BReject}}{\kbt, \textsf{TX'}} \\
			%	& \quad \quad \event{\textcolor{teal}{BCommitWithC}}{\textsf{EAC}} \\
			%	& \quad \quad \event{\textcolor{red}{BRunningWithT}}{x, \enc{\nlist{\textsf{TX'}, \texttt{reject}}}{\kbt}} \\
			%	& \quad \quad \event{\textcolor{violet}{BCommitWithTC}}{x} \\
			& \quad \quad  \cout{\ch}{\enc{\nlist{\textsf{TX'}, \texttt{reject}}}{\kbt}} \\ 
			%		&  \quad \quad \event{BReject}{\textsf{PAN,TX}}. \\
		\end{aligned}
	\end{equation}
}

\def\specC{
	\begin{equation}
		\squeezespaces{1}
		\notag
		\begin{aligned}
			C(&\ch, c, pk_s, \texttt{vsig}_\texttt{MM}, \text{PAN}, \mk, \text{PIN}) \triangleq \\
			& \cin{\ch}{z_1}.\\
			& \nw a.\cout{\ch}{\smult{a}{\pkg{c}}}.\\
			& \lett \, \key{c} \coloneqq \hash{\smult{\mult{a}{c}}{z_1}} \inn \\
			& \cin{\ch}{m}.\\ 
			& \lett \, \nlist{\textsf{MC}, \textsf{MC}_s} \coloneqq \dec{m}{\key{c}} \inn \\
			& \ifff \ \checksig{pk_s}{\textsf{MC}_s} = \textsf{MC} \ \thenn \\
			& \ifff \ \proj{1}{\textsf{MC}} =\texttt{MM} \ \thenn \\
			& \cout{\ch}{\enc{\nlist{\smult{a}{\pkg{c}}, \smult{a}{\texttt{vsig}_\texttt{MM}}}}{\key{c}}}.\\
			& \cin{\ch}{x}.\\
			% low-value online, low-value offline, high-value online
			& \lett \, \nlist{\textsf{TX},\textsf{uPin}} \coloneqq \dec{x}{\key{c}} \inn\\
			& \lett \, \textsf{AC} \coloneqq \nlist{a, \text{PAN}, \textsf{TX}} \inn\\
			& \lett \, \textsf{AC}^{\texttt{ok}} \coloneqq \nlist{a, \text{PAN}, \textsf{TX}, \texttt{ok}} \inn\\
			& \lett \, \textsf{AC}^{\fail} \coloneqq \nlist{a, \text{PAN}, \textsf{TX}, \fail} \inn\\
			& \lett \, \key{cb} \coloneqq \hash{\smult{\mult{a}{c}}{\proj{3}{\proj{1}{\dec{m}{\key{c}}}}}} \inn \\
			& \ifff \ \textsf{uPin} = \fail \ \thenn \\
%			& \event{accept}{ offlo, onlo, onhi} \\
			& \cout{\ch}{\enc{\enc{\nlist{\textsf{AC},\hash{\nlist{\textsf{AC}, \mk}}}}{\key{cb}}}{\key{c}}} \\
			& \ifff \ \textsf{uPin} =\text{PIN} \ \thenn \\
%			& \event{accept}{offhi} \\
			& \cout{\ch}{\enc{\nlist{\enc{\nlist{\textsf{AC}^{\texttt{ok}},\hash{\nlist{\textsf{AC}^{\texttt{ok}}, \mk}}}}{\key{cb}}, \texttt{ok}}}{\key{c}}} \\
			& \elsee \ \cout{\ch}{\enc{\nlist{\enc{\nlist{\textsf{AC}^{\fail},\hash{\nlist{\textsf{AC}^{\fail}, \mk}}}}{\key{cb}}, \fail}}{\key{c}}}
		\end{aligned}
	\end{equation}
}

\def\specTonhi{
	\begin{equation}
		\notag
		\begin{aligned}
			T_{\texttt{onhi}}(&\user, \ch, pk_\texttt{MM}, \textsf{crt}, \kbt) \triangleq \\
			& \nw \, \text{DATE}. \lett \, \textsf{TX} \coloneqq \nlist{\text{DATE}, \hi} \inn \\
			& \nw t.\cout{\ch}{\pkg{t}}. \\
			& \cin{\ch}{z_2}. \\
			& \lett \, \key{t} \coloneqq \hash{\smult{t}{z_2}} \inn \\
			& \cout{\ch}{\enc{\textsf{crt}}{\key{t}}}. \\
			& \cin{\ch}{n}.\\
			& \lett \, \nlist{\textsf{B}, \textsf{B}_s} \coloneqq \dec{n}{\key{t}} \inn \\
			& \ifff \ \checksigv{pk_\texttt{MM}}{\textsf{B}_s} = \textsf{B} \ \thenn \\
			& \ifff \ \textsf{B} = z_2 \ \thenn \\
			& \cin{user}{\textsf{uPIN}}.\\
			& \cout{\ch}{\enc{\nlist{\textsf{TX}, \fail}}{\key{t}}}. \\
			& \cin{\ch}{y}.\\
			& \cout{\ch}{\enc{\nlist{\textsf{TX}, z_2, \dec{y}{\key{t}}, \textsf{uPIN}}}{\kbt}}. \\
			& \cin{\ch}{r}. \\
			& \ifff \ \proj{1}{\dec{r}{\kbt}} = \textsf{TX} \ \thenn \\
			& \ifff \ \proj{2}{\dec{r}{\kbt}} = \texttt{accept} \ \thenn \\
			%			& \event{TAccept}{\kbt,\TX}; \\
			&  \cout{\ch}{\texttt{auth}} \\
		\end{aligned}
	\end{equation}
}

\def\specB{
	\begin{equation}
		\notag
		\begin{aligned}
		%	B(&\ch, si, \kbt, b_t) \triangleq \\
			& \cin{\ch}{x}.\\
			& \lett \, \nlist{\textsf{TX}', z_2, \textsf{EAC}, \textsf{uPIN}} \coloneqq \dec{x}{\kbt} \\
			& \lett \, \key{bc} \coloneqq \hash{\smult{b_t}{z_2}} \inn \\
			& \lett \, \nlist{\textsf{AC}, \textsf{AC}_{hmac}} = \dec{\textsf{EAC}}{\key{bc}} \inn \\
			& \lett \, \text{PAN} \coloneqq \proj{2}{\textsf{AC}} \\
			& \cin{\nlist{si, \text{PAN}}}{y}. \\
			& \lett \, \nlist{\text{PIN}, mk, pk_{c}} \coloneqq y \\
			& \ifff \ \hash{\nlist{\textsf{AC}, mk}} = \textsf{AC}_{hmac} \ \thenn \\
			& \ifff \ \proj{3}{\textsf{AC}} = \textsf{TX}' \ \thenn \\
			& \ifff \ \smult{\proj{1}{\textsf{AC}}}{pk_{c}}= z_2 \\
			& \lett \, r \coloneqq \nlist{\textsf{TX}', \texttt{accept}} \inn \\
			% low-value offline, low-value online
			& \ifff \ \proj{2}{TX'} = \lo \ \thenn \\
%			& \event{accept}{offlo, onlo} \\
			& \cout{\ch}{\enc{r}{\kbt}} \\
			& \ifff \ \proj{2}{\textsf{TX}'} = \hi \ \thenn \\
			% high-value offline		
			& \ifff \ \proj{4}{\textsf{AC}} = \texttt{ok} \ \thenn \\ 
%			& \event{accept}{offhi} \\
			& \cout{\ch}{\enc{r}{\kbt}} \\
			% high-value online	
			& \elsee \ \ifff \ \textsf{uPIN} = \text{PIN} \ \thenn \\
%			& \event{accept}{onhi} \\
			& \cout{\ch}{\enc{r}{\kbt}} \\
		\end{aligned}
	\end{equation}
}

\def\specRealIdeal{
\begin{figure*}
\begin{subfigure}[t]{0.45\textwidth}
\caption{The real protocol specification $\impl{UTX}$}
\specProtocolReal 
\end{subfigure} 
\begin{subfigure}[t]{0.45\textwidth}
\caption{The ideal unlinkable protocol specification $\spec{UTX}$}
\specProtocolIdeal 
\end{subfigure} 
\caption{Specifications for the real $\protocol$ protocol and its ideal unlinkable version}\label{fig:system}
\end{figure*}
}

\def\specProtocolReal{
	\begin{equation}
	\notag
	\begin{aligned}
		 &\nu \, \user, s, si, \chi_{\texttt{MM}}. \cout{out}{\pks{s}}.\cout{out}{\pkv{ \chi_{\texttt{MM}}}}.\Bigl( \\ 
		&
		\;\; 
		\begin{array}{l}
			\mathbf{\bang} \nu \text{PIN}, mk, c, \text{PAN}. \bigl(\\
			\quad \lett \, \textsf{crtC} \coloneqq  \sigv{\chi_{\texttt{MM}}}{\pkg{c}} \inn \\
			\quad 
			\begin{array}[t]{l}
			\;\; \textcolor{red}{\bang}\nw \ch.\cout{card}{\ch}.C(\ch, c, \pk{s}, \textsf{crtC}, \text{PAN}, mk, \text{PIN}) \\
			\cpar \; !\cout{\user}{\text{PIN}} \cpar \; ! \cout{\nlist{si, \text{PAN}}}{\nlist{\text{PIN}, mk, \pkg{c}}} \;\bigl) \; \cpar \; \\

				%					\textit{DB}(si, \text{PAN}, mk, \text{PIN}, \pkg{c})~\textcolor{green}{\Bigl)} \cpar
			\end{array}
			\\
			\nu b_t.\bang \nu \kbt. \textcolor{cyan}{\bigl(}
			\\
			\;\; 
			\begin{array}[t]{l}
				\nu \ch.\cout{bank}{\ch}.B(\ch, si, \kbt, b_t) \cpar \\
				\lett \, \textsf{crt} \coloneqq \nlist{\nlist{\texttt{MM}, \pkg{b_t}}, \sig{s}{\nlist{\texttt{MM}, \pkg{b_t}}}} \inn \\
				\nu \ch.\cout{term}{\ch}.T(user, \ch, \pkv{\chi_{\texttt{MM}}}, \textsf{crt}, \kbt)
				\textcolor{cyan}{\bigl)} ~\Bigl)
			\end{array}
		\end{array}
	\end{aligned}
\end{equation}
}

\def\specProtocolIdeal{
	\begin{equation}
	\notag
	\begin{aligned}
		 &\nu \, \user, s, si, \chi_{\texttt{MM}}. \cout{out}{\pks{s}}.\cout{out}{\pkv{ \chi_{\texttt{MM}}}}.\Bigl( \\ 
		&
		\;\; 
		\begin{array}{l}
			\mathbf{\bang} \nu \text{PIN}, mk, c, \text{PAN}. \bigl(\\
			\quad \lett \, \textsf{crtC} \coloneqq  \sigv{\chi_{\texttt{MM}}}{\pkg{c}} \inn \\
			\quad
			\begin{array}[t]{l}
			\;\; \nw \ch.\cout{card}{\ch}.C(\ch, c, \pk{s}, \textsf{crtC}, \text{PAN}, mk, \text{PIN})  \\
			\cpar \; !\cout{\user}{\text{PIN}} \cpar \; ! \cout{\nlist{si, \text{PAN}}}{\nlist{\text{PIN}, mk, \pkg{c}}} \;\bigl) \; \cpar \; \\

				%					\textit{DB}(si, \text{PAN}, mk, \text{PIN}, \pkg{c})~\textcolor{green}{\Bigl)} \cpar
			\end{array}
			\\
			\nu b_t.\bang \nu \kbt. \textcolor{cyan}{\bigl(}
			\\
			\;\; 
			\begin{array}[t]{l}
				\nu \ch.\cout{bank}{\ch}.B(\ch, si, \kbt, b_t) \cpar \\
				\lett \, \textsf{crt} \coloneqq \nlist{\nlist{\texttt{MM}, \pkg{b_t}}, \sig{s}{\nlist{\texttt{MM}, \pkg{b_t}}}} \inn \\
				\nu \ch.\cout{term}{\ch}.T(user, \ch, \pkv{\chi_{\texttt{MM}}}, \textsf{crt}, \kbt)
				\textcolor{cyan}{\bigl)} ~\Bigl)
			\end{array}
		\end{array}
	\end{aligned}
\end{equation}
}

We specify and verify our proposed protocol in a variant of the applied $\pi$-calculus~\cite{applied-pi}.
%\change{We specify and prove unlinkability of our proposed protocol in a variant of the applied $\pi$-calculus~\cite{applied-pi}. Security analysis is performed for a simplified version of $\protocol$ -- specifically, we restrict the security analysis to the case where all cards are synchronised to execute within the same month $\texttt{MM}$. The security analysis of the model admitting cards valid in different months as well as advancing pointers from one month to the next is left as future work.}
 In the formulation of the property of transaction unlinkability, we employ \emph{quasi-open bisimilarity}~\cite{HORNE2023113842} -- 
% \before{a notion of equivalence preserved in all contexts} 
an equivalence notion that is preserved in all contexts and captures an attacker capable of making dynamic decisions -- and its corresponding labelled transition system. For the properties that constitute payment security, we rely on the ProVerif tool and its notion of \emph{correspondence assertions} \delete{introduced in} \cite{blanchet2001efficient, blanchet-corresp}. We focus the analysis on the core component of our protocol, modelling its key agreement and transaction authorisation steps. We omit the application selection step as it involves only constant messages that are the same for all sessions. 
 
%While it would be desirable to also have a tool-based proof for unlinkability, state of the art tools like ProVerif and Tamarin cannot yet handle equivalence queries for message theories as complex as ours. %For correspondence assertions we still have to perform certain abstractions to the message theory in order to achieve termination in ProVerif, as we explain below. 

\subsection{Attacker model}\label{sec:attackermodel}
The attacker model we use for verification of the $\protocol$ protocol
is a Dolev-Yao attacker~\cite{dolev1983onthesec} who controls the
communications between the card, the terminal, and the bank. Such
attackers can intercept, block, modify, and inject
messages.
In the presence of contactless payments,
%\SM{I can't remember having read that we have a ``default'' scenario
%and that it's contactless.}
%it is natural to consider different attackers depending on how close they are to the card.
the Dolev-Yao attacker is particularly relevant since, within a range of 100cm, an attacker can power up the card and interact with it~\cite{habraken2015dolron}, explaining why we insist on this attacker model when verifying our protocol.
%since it covers both scenarios. 
The connection between the terminal and the bank is not necessarily secure and an attacker could manipulate this connection, e.g.\ cutting it and forcing the terminal to go offline. 
%We leave modeling such attackers' capabilities for future work.

%Furthermore, in addition to the network, we assume the attacker may also corrupt a subset of the payment terminals.
%\BEGIN{PREVIOUS VERSION}
%\before{We assume that cardholders only enter their PIN into honest terminals.
%In other words, the cardholder uses terminals at reputable points of sale that 
%observe adequate security measures, such as security cameras.}
%\after{We assume that active attackers, acting as terminals without the cardholder's awareness, cannot
%%\SM{will? can?}
%obtain the PIN. 
%%\SM{directly from the user?}
%%\SM{Should we make an explicit statement how realistic this assumption is?}
%In other words, we assume that cardholders only enter their PIN into terminals during the actual purchase at reputable points of sale that observe adequate security measures, such as security cameras. Since we aim at preserving the requirement that anyone can manufacture terminals, we admit that there is a risk of a malicious terminal being present to the user even in this situation.}
%\END{PREVIOUS VERSION}

%\SY{explain better honest terminals with PIN in the model. honest terminals do not require any special information to run the protocol (at least to communicate with the card) -- dishonest terminals can run in parallel to the specification, thanks to compositionality}

We assume that cardholders only enter their PIN into honest terminals. In other words, the cardholder uses terminals at reputable points of sale \delete{that observe adequate security measures, such as security cameras.} \change{in the process of a conscious purchase and never enters their PIN into random terminals that pop up on the street.} The properties of unlinkability and PIN secrecy are immediately compromised if the PIN is entered into a malicious terminal which reveals the PIN to attackers. If an attacker possesses a PIN, clearly the card can be stolen and then used for high-value purchases for which the PIN is required. While theft may be mitigated by cancelling cards, an attacker knowing the PIN may authorise high-value purchases \delete{via a relay attack} \change{by relaying the messages between an honest terminal and an honest card~\cite{relayorig}}, making it difficult for the cardholder to dispute the transaction, as legally a cardholder is always held liable for transactions authorised by a PIN; and hence the primary goal of the security of money in the account would be compromised.
Supposing that relay attacks were mitigated, an attacker knowing the PIN may still attack unlinkability as follows. For high-value transactions, it becomes possible for a terminal that remembers the PIN to track cards by the fact that the same PIN is used. 
%curious terminals
Moreover, even in a low-value contact scenario not requiring the PIN, the PIN can nonetheless be used to track specific individuals, since such terminals remembering PINs can run a fake session with a high-value amount requiring the PIN to be sent from the terminal to the card in order to check if it has already seen this card before processing the legitimate low-value transaction. \after{In contrast to the above, if an attacker is physically unable to perform contact transactions, low-value contactless payments are unlinkable even if the PIN is compromised. We analyse this case separately in Appendix~\ref{apx:lowval}.}
%by the compositionality we could strip away low-value contactless terminals
%Note that however, since high-value offline transactions require the card to be inserted in the terminal. The PIN cannot be used for tracking cards when a malicious terminal secretly attempts to run the session for a fake high-value offline transaction and to check if the leaked PIN is valid for a card. 
 
%\SM{A reviewer may be triggered by this assumption, and the previous sentence
%may be too implicit to take away their concern. We must make it explicit that
%you're definitely lost if you type your PIN in a malicious terminal, so the
%unlinkability problem is one of you minor concerns.}
%\SY{rewritten, to revise}

There are other attacker models.
%requires the protocol to be side-channel resistant.
%that is cryptographic operations and the transaction authorisation response time from the bank should reveal no information about the card. 
We could have verified with respect to a weaker distant attacker that operates
within a distance of 100cm to 20m from the card and can only eavesdrop on communications~\cite{novotny2008guerrieri, engelhardt2013pfeiffer}. This attacker would have been sufficient to establish privacy for the proposal already considered by EMVCo establishing a channel to encrypt regular EMV transactions~\cite{rfc}. 
Other attackers may attempt side-channel attacks by measuring execution time of cryptographic operations, or the response time from the bank, which is out of scope of our analysis.
       
\subsection{Formal specification of the protocol} 

%\begin{figure*} 
%\begin{subfigure}[t]{0.33\textwidth}
%\caption{\fbox{\squeezespaces{1.0}Card $C(\ch, c, pk_s, \texttt{vsig}_\texttt{MM}, \text{PAN}, \mk, \text{PIN})$}}
%\label{fig:card}
%\specCsimple
%\end{subfigure}
%\begin{subfigure}[t]{0.33\textwidth}
%\caption{\fbox{\squeezespaces{1.0}Terminal $T_{\texttt{onhi}}(\user, \ch, pk_\texttt{MM}, \textsf{crt}, \kbt)$}}
%\label{fig:tonhi}
%\specTonhiSimple
%\end{subfigure}
%\begin{subfigure}[t]{0.33\textwidth}
%\captionsetup{singlelinecheck = false, justification=justified}\caption{\fbox{\squeezespaces{1.0}Bank $B(\ch, si, \kbt, b_t)$}}
%\label{fig:b}
%\specBsimple
%\end{subfigure}
%\vspace{0.001cm}

\begin{figure*} 
	\begin{subfigure}[t]{0.33\textwidth}
		\caption{\fbox{\squeezespaces{1.0}Card $C(\ch, c, pk_s, \texttt{vsig}_\texttt{MM}, \text{PAN}, \mk, \text{PIN})$}}
		\label{fig:card}
		\specCsimpleEv
	\end{subfigure}
	\begin{subfigure}[t]{0.33\textwidth}
		\caption{\fbox{\squeezespaces{1.0}Terminal $T_{\texttt{onhi}}(\user, \ch, pk_\texttt{MM}, \textsf{crt}, \kbt)$}}
		\label{fig:tonhi}
		\specTonhiSimpleEv
	\end{subfigure}
	\begin{subfigure}[t]{0.33\textwidth}
		\captionsetup{singlelinecheck = false, justification=justified}\caption{\fbox{\squeezespaces{1.0}Bank $B(\ch, si, \kbt, b_t)$}}
		\label{fig:b}
		\specBsimpleEv
	\end{subfigure}
	\vspace{0.001cm}

In addition, there are processes $T_{\texttt{offhi}}$ and $T_{\texttt{lo}}$ defining the behavior for offline high-value and low-value transactions, respectively (presented in Appendix~\ref{sec:appx}). Moreover there is $T$ defined as $T_{\texttt{onhi}} + T_{\texttt{offhi}} + T_{\texttt{lo}}$. \change{The star} $\textcolor{purple}{*}$ \change{indicates at which point the card selects the appropriate month certificate to present (see Appendix~\ref{appx:months} for a larger specification making this choice explicit).}
\caption{Specifications for the three roles in the $\protocol$ protocol.}\label{fig:specification}
\end{figure*}

We use the applied $\pi$-calculus language~\cite{applied-pi} to
specify the formal model of the $\protocol$ protocol \change{where all
cards are synchronised to execute within the same month
$\texttt{MM}$.} In the essence of this formalism, we have processes
that can communicate by sending and receiving messages using channels.
We write $\cout{ch}{M}$ and $\cin{ch}{x}$ for sending the message $M$
or receiving the input $x$ on the channel $ch$, respectively. A
process can also generate private values (used e.g. for fresh secret
keys and nonces), written as $\nw a$, be replicated using the $\bang$
operator (allowing an unbounded number of its instances to execute),
and run in parallel with other processes using the $\cpar$ operator.
\delete{We} \change{In Fig.~\ref{fig:specification} we} have three
processes that model the execution of a session of our protocol by the
three roles in the UTX protocol: the terminal, the card, and the bank.
\change{Events, marked with \texttt{ev:}, will be used in the security
analysis and can be ignored until
Section~\ref{sec:securityanalysis}.}\delete{A}
\change{Fig.~\ref{fig:system} specifies the} top-level process
\change{that} expresses how these processes are assembled and
instantiated across multiple payment sessions in a full execution of
the protocol.

%\bef{The card process $C$} represents the execution of a payment session by a card: 
\subsubsection{The card process} $C$, described in Fig.~\ref{fig:card}, represents the execution of a payment session by a card.
$$\nw \ch.\cout{card}{\ch}.C(\ch, c, pk_s, \texttt{vsig}_\texttt{MM}, \text{PAN}, mk, \text{PIN})$$ 
It is parameterised by the session channel $ch$, the card's secret key $c$, the system-wide public key $pk_{s}$ used to check the bank's certificate $\textsf{crt}$ received from the terminal, the signature $\texttt{vsig}_\texttt{MM}$ on the card's public key for the current month (considering the currently valid month only simplifies the initial analysis), the card number $\text{PAN}$, and the $\text{PIN}$. First, the card establishes a key with the terminal, then checks the certificate of the terminal and sends back its own month certificate (comprising its public key and the corresponding Verheul signature) blinded with the scalar $a$ used in the shared key establishment. Using the data provided in the terminal's certificate, the card also generates $k_{cb}$, which is a fresh symmetric key to be used by the card to communicate securely with the bank (the terminal cannot obtain this key). Upon receiving the transaction details, the card decides as follows: if no $\text{PIN}$ has been provided or the corresponding PIN matches its own PIN, the card accepts the transaction and replies with the corresponding cryptogram. Otherwise, the rejection cryptogram $\textsf{AC}^{\texttt{no}}$ is generated and sent as reply to the terminal. 

%\bef{The terminal process $T$} is
\subsubsection{The terminal process} 
The modes in which a terminal can operate  are   
combined in a role $T$ defined as follows.
 $$\nu \ch.\cout{term}{\ch}.T(user, \ch, pk_\texttt{MM}, \textsf{crt}, \kbt)$$
$T$ is parametrised by the secret channel $user$ used to enter the $\text{PIN}$, the session channel $\ch$, the public key used for verifying the card certificate for the given month $pk_\texttt{MM}$, and the shared secret key between the terminal and the bank $\kbt$. 
%where $\ch$ identifies the communication channel used in that session, $user$ is a secret channel name used to enter the $\text{PIN}$, $pub_\MM$ is the public key used for verifying the card certificate for the given month, and $\kbt$ is a shared secret key between the terminal and the bank. 
To incorporate various operation modes for the terminal, we have three types of processes from which the terminal process $T$ is made of: the process for online high-value transactions $T_{\texttt{onhi}}$, for offline high-value transactions $T_{\texttt{offhi}}$, and for low-value transactions $T_{\texttt{lo}}$. 

Initially, each terminal proceeds with the key establishment phase with the card, sends its certificate, and checks the received month certificate. High-value terminals rely on the $\text{PIN}$ \delete{received from the $\user$ channel} \change{entered by the cardholder} to perform transaction authorisation. To represent the different types of transactions that can occur, we have constants $\lo$ and $\hi$ for low-value and high-value transactions respectively. 

The online high-value terminal process $T_{\texttt{onhi}}$ is given in Fig.~\ref{fig:tonhi}. Since the transaction is high-value, the $\text{PIN}$ is required and after the initialisation, the user enters the $\text{PIN}$ using the \change{private} channel $\user$, \change{which models that the PIN can only be entered into honest terminals}. Then the terminal sends the transaction details to the card, receives the application cryptogram in the response, and sends it to the bank together with the entered $\text{PIN}$. Since we are in the online mode, the terminal authorises the transaction only after receiving confirmation from the bank. In contrast, offline terminals authorise transactions right after receiving the reply from the card.

The offline high-value and low-value modes are similar, and their specifications appear in Appendix~\ref{sec:appx}. The offline high-value mode requires the terminal to send the entered PIN to the card since only the card can verify the PIN if the terminal is offline. Terminals operating in this mode accept transactions only if the $\texttt{ok}$ reply has been received from the card, however, regardless of the outcome, the cryptogram is always sent to the bank eventually.
%\delete{In the} \change{In} offline high-value \change{mode}, the user similarly enters the $\text{PIN}$ after the initialisation \change{into the terminal}, and the terminal sends this $\text{PIN}$ for verification to the card, since the \delete{transaction is} \change{only the card can verify PINs} offline. The terminal accepts the transaction only if it receives the $\texttt{ok}$ reply from the card. Regardless of \change{the} outcome, the terminal also sends the cryptogram to the bank. 
%The low-value terminal \change{behaviour} does not require the $\text{PIN}$\delete{.}\change{, hence the corresponding role specification $T_{\texttt{lo}}$ does not require that online and offline modes are distinguished.} \change{For High-Value terminal behaviours,} \delete{D}\change{d}epending on \change{whether} the mode\delete{,} \change{is} online or offline, the terminal authorises a transaction, respectively, after receiving the confirmation from the bank, or after receiving the application cryptogram from the card.
\change{Low-value transactions are PINless, hence the corresponding role specification $T_{\texttt{lo}}$ does not require that online and offline modes are distinguished.}

%\bef{The bank process $B$} connected to a terminal session identified by the shared key $\kbt$ is represented by $$\nu \ch.\cout{bank}{\ch}.B(\ch, si, \kbt, b_t)$$
\subsubsection{The bank process} $B$, specified in Fig.~\ref{fig:b}, that connects to a terminal session identified by the shared key $\kbt$ is represented as follows.
$$\nu \ch.\cout{bank}{\ch}.B(\ch, si, \kbt, b_t)$$
In addition to $\kbt$, its parameters are the session channel $\ch$, the system-wide channel $si$ that is used by the payment system to access the card database, and the bank's secret key $b_t$. We model each entry inserted into the card database using the instruction 
%$\bang \cout{si}{\text{PAN}, \text{PIN}, mk, \pkg{c}}$, 
$\bang \cout{\nlist{si, \text{PAN}}}{\nlist{\text{PIN}, mk, \pkg{c}}}$,
and the corresponding entry can be read by receiving a message on the channel consisting of the pair $\nlist{si, \text{PAN}}$ 
where the first component of the channel keeps the database private to the bank and the second component indicates the entry to look up.
After receiving a transaction request from a terminal, the bank derives the symmetric key with the card $k_{bc}$, obtains the $\text{PAN}$ from the 
%application 
cryptogram, and obtains the card's $\text{PIN}$, its master key $mk$, and the public key $\pkg{c}$ from the database channel $si$. The integrity of the cryptogram is then checked against the corresponding information from the database, taking into account the verification of the PIN if the transaction is high value. If all the checks are ok, the transaction is accepted, otherwise not; and in all cases, a confirmation message is sent in reply to the terminal. 

%\bef{The full protocol.}  
\subsubsection{The full protocol}
\specRealIdeal
To complete the specification, in
Fig.~\ref{fig:system} we present the full system, which operates as follows. At the start, the
system-wide parameters are generated and public data that includes the
system public key $\pks{s}$ and the month public key
$\pkv{\chi_{\texttt{MM}}}$ is announced on the public channel $out$. A
new card is issued by the generation of the card-specific parameters
$\text{PIN}$, $mk$, $c$, and $\text{PAN}$, and can participate in many
sessions, hence the red replication operator
``$\textcolor{red}{\bang}$''. Notice that together with the card the
system has a $\cout{\user}{\text{PIN}}$ process that models the user entering $\text{PIN}$  into a terminal on the channel $\user$ known only to the terminals; and the process $\cout{\nlist{si, \text{PAN}}}{\nlist{\text{PIN}, mk, \pkg{c}}}$ that models the entry into the card database that the bank can access to get the card's data. The bottom part of the figure specifies the back end of the system, i.e. the banks and the terminals. There is a system-wide secret key of the bank $b_t$ and a session-wise (hence the replication) symmetric key between the bank and the terminal $kbt$. Notice also that we are using public session channels $\ch$ to give an attacker the power to observe which agents are communicating.

\subsubsection{The Dolev-Yao model accounts for malicious terminals} \label{sec:dy}
\change{Terminals operated by attackers should be accounted for in our threat model, since, consistent with EMV, terminals are not authenticated by the card and hence can be implemented and operated by anyone. In our model, indeed, an attacker can impersonate a terminal, either up until the point when the PIN is requested, or, in modes where the PIN is never requested, proceed to obtain the encrypted application cryptogram produced by the card. To operate as a terminal, an attacker only needs the bank's certificate $\nlist{\nlist{\texttt{MM}, \pkg{\bt}}, \sig{s}{\nlist{\texttt{MM}, \pkg{\bt}}}}$ which is straightforward to obtain since an honest terminal gives away this certificate to anyone it communicates with. Indeed, a fake card can be used to obtain new monthly certificates even if authorities only distribute them to honest terminals.
Such a fake card would first engage in a Diffie-Hellman handshake with an honest terminal, which establishes a channel on which an attacker can receive the certificate currently loaded into the terminal. No knowledge of any private key is required to implement such fake cards.
This viable threat is accounted for in the proofs of unlinkability theorems in the next section.
}

\subsection{Unlinkability definition and analysis}\label{sec:unlinkanalysis}

%\SY{this discussion about the tools should be moved after the formal definition}

%\noindent
%\begin{tabular}{ |p{1.64cm}||p{1.25cm}|p{1.26cm}|p{1.25cm}|p{1.25cm}|}
%	%	\hline
%	%	\multicolumn{4}{|c|}{Operations} \\
%	\hline
%	Protocol&PV 2.4 diff&T1 diff &T2 diff (to remove)&T2 obs\\
%	\hline
%	BDH (no T) &\shrug&\textcolor{red}{TRUE} \xmark&\shrug&\shrug\\
%	\hline
%	BDH (full) &freezes&\textcolor{red}{TRUE} \xmark&freezes& \shrug (23h) \\
%	\hline
%	UBDH (no T) &\textcolor{blue}{TRUE} \cmark&\textcolor{blue}{TRUE} ?&\textcolor{blue}{TRUE} \cmark&\textcolor{blue}{TRUE} (time?) \cmark\\
%	\hline
%	UBDH (full) &freezes&\textcolor{blue}{TRUE} ? &freezes (compositionality makes a difference) &\textcolor{blue}{TRUE} \cmark (7h)\\
%	\hline
%	UTX &freezes&fatal error&freezes&freezes\\
%	\hline
%\end{tabular}

In this section, we clarify the informal definition of unlinkability given by Scheme~\ref{def:unlink} presented in Section~\ref{sec:privacyreq} and formally prove that $\protocol$ is unlinkable.
\change{We also present some variations on the unlinkability problem that show that 
unlinkability still holds even if certain marginal coarse identities are tolerated.}
%\change{We also provide evidence that $\protocol$ is beyond the capabilities of state-of-the-art tools that can prove privacy properties of security protocols.}

\subsubsection{The formal definition of unlinkability}\label{sec:resultunlink}
Recall that the core of the unlinkability scheme is the equivalence between the idealised and the real-world system. We define both in Fig.~\ref{fig:system}. Notice that in the system $\impl{UTX}$ defining the real-world scenario the card with the private key $c$ can participate in any number of sessions, while in the system $\spec{UTX}$ defining the idealised situation, the card can only participate in one session at most. The possibility 
%card defined by $\text{PIN}$, $mk$, $c$, and $\text{PAN}$ can only start one session, yet the possibility 
of entering the $\text{PIN}$ arbitrarily many times is given by the process $\bang \cout{\user}{\text{PIN}}$, and accessing the database 
in arbitrarily many bank-terminal sessions given by the process 
$\bang \cout{\nlist{si, \text{PAN}}}{\nlist{\text{PIN}, mk, \pkg{c}}}$, remains the same for both real and idealised worlds.

We are ready now to give the unlinkability definition. 

\begin{definition}\emph{(unlinkability)}\label{def:unlinkformal}
	We say that the payments are unlinkable if $\impl{UTX} \sim \spec{UTX}$, where $\sim$ is quasi-open bisimilarity.
\end{definition}

There is a difference with the definition of unlinkability for
key establishment considered in~\cite{horne2022csf} \after{(provided in Appendix~\ref{app:unlinkkey})}, 
%\after{There is a
%difference between the definition of unlinkability for key establishment considered in~\cite{horne2022csf} and the one given below}, 
where the terminal and the bank are deliberately omitted. The reason is that the key establishment in isolation\after{, i.e. the $\protocol$ protocol up to the \emph{Cardholder verification} phase,} requires no shared secret between the parties, yet to execute, for instance, a full high-value transaction, at least the $\text{PIN}$ is required to be shared between all three parties involved in the protocol.
In addition, to validate a transaction there is a secret $mk$ shared between the bank and the card,
meaning that, even if only transactions without the PIN are modelled, the bank and card must be explicitly modelled in a transaction.

Finally we are ready to formulate our first result.

%\begin{customthm}{1}The \begin{upshape}$\protocol$\end{upshape} protocol is unlinkable. \label{thm:link}
%\end{customthm}

\begin{customthm}{1}$\impl{UTX} \sim \spec{UTX}$. \label{thm:link}
\end{customthm}

The detailed proof of Theorem~\ref{thm:link} is given in Appendix~\ref{sec:appx},
%~\ref{app:proofunlink}, 
however, we give a proof sketch here. The key is to give a relation $\mathfrak{R}$ between processes representing states of the two worlds demonstrating that an attacker has no strategy allowing to distinguish between these two worlds. We form such a relation by pairing the appropriate states and checking that it satisfies the conditions for \change{a} quasi-open bisimulation. We pair the states based on the number of sessions \emph{started} with terminals, cards, and banks  and the respective stages of each session; and we ignore the number of exhausted processes that model entering the $\text{PIN}$ and accessing the database for card's details. Then we check that each possible transition that either world can make can be matched by the opposing world; that the resulting states are related by $\mathfrak{R}$, that any two related states are statically equivalent, i.e. indistinguishable by an attacker who can only observe which messages are on the network in this state; and finally, that $\mathfrak{R}$ is \emph{open}, i.e. there is no way for an attacker to distinguish between two worlds by manipulating free variables.

%\SM{We need to refer to \cite{horne2021quasi} again, otherwise thereader will be unhappy that we don't explain these notions.}

%\todo{Definition and proof sketch. Maybe point out the relation/difference with the previous unlinkability definition from CSF paper.}

\subsubsection{Unlinkability in the face of coarse identities}\label{sec:corollary}
\change{Below we justify the observation made in Section~\ref{sec:privacyreq} where we pointed out that unlinkability can only be achieved up to the fingerprint comprising the coarse identities of the card being revealed. We explain below how such coarse identities of the card can exist in the system without compromising unlinkability.}
	
\change{\emph{Signing authority.} We demonstrate that $\protocol$ is unlinkable even if an attacker can distinguish two cards that use different signing authorities. To do so, we exploit the fact that quasi-open bisimilarity}
\delete{The choice of quasi-open bisimilarity as the equivalence relation suitable for modelling privacy properties is justified in~\cite{HORNE2023113842}. Firstly, it captures a powerful realistic attacker and hence certainly takes into account all attacks ranged over by coarser equivalences. Secondly, it}
is a congruence~\cite{HORNE2023113842}, i.e. when a \emph{smaller system} satisfies unlinkability, then a \emph{larger system} containing the smaller one as a subsystem also satisfies unlinkability, i.e. process equivalence is preserved in any context. A context is a process ``with a hole'' such as $\mathcal{O}(\cdot) \coloneqq \bang(\cdot)$. \delete{Either the $\impl{UTX}$ or $\spec{UTX}$ being put in $\mathcal{O}$ give a system with multiple signing authorities. Hence, we obtain the following.} Notice, by putting  
$\impl{UTX}$ into $\mathcal{O}$ we obtain a system with multiple signing authorities. Similarly, by putting $\spec{UTX}$ into $\mathcal{O}$ results in an ideal world in each card engages still in one session, but may use different signing authorities. 
%and hence that coarse identity is tolerated.
 \change{Now, since quasi-open bisimilarity is a congruence}, the following holds.

\begin{corollary}\label{col:bang}
	$\bang \impl{UTX} \sim \bang \spec{UTX}$, i.e. $\protocol$ is unlinkable even in the presence of multiple signing authorities.
\end{corollary}

The above means that unlinkability holds for systems with multiple signing authorities as long as we tolerate that coarse identity. That is, we permit a coarse identity, a signing authority, to exist in the system, as represented by building multiple authorities into the ideal world $\bang \spec{UTX}$, without compromising unlinkability. In particular, Corollary~\ref{col:bang} concerns the degree of unlinkability that can be established in a deployment scenario where multiple payment systems might not agree to provide a common application for unlinkable payments as discussed in Section~\ref{sec:init}, and therefore these different payment systems form a coarse identity of the card.

\change{\emph{The card has been used recently}. To clarify that the existence of cards valid for several months does not invalidate unlinkability, we consider a model of unlinkability, where cards can respond to two months at any moment. Furthermore, this model admits transitions from one month to the next, maintaining a pointer as described in Section~\ref{sec:keys}. To reflect such behaviour of cards, we build into the definition of a process modelling a card the ability to respond to two months at any time and, whenever the new month is asked, to invalidate the oldest of the two months. Notice that this requires a card to carry the state, i.e. to remember that it should respond only to the most recent two months and never respond to older months if asked. In Appendix~\ref{appx:months} we show how to employ recursion to model such behaviour and prove the following. 

\begin{customthm}{2}\label{thm:months}
$\impl{UTXMM} \sim \spec{UTXMM}$.
\end{customthm}

In the above $\impl{UTXMM}$ and $\spec{UTXMM}$ define the real and the ideal worlds in an enhanced model.
The ideal world models an infinite supply of cards that are used only once, and in that single session, may either respond to the two most recent months, or the three most recent months (the latter modelling the tolerance of cards that are one month behind and can still be updated to the current month). 
Therefore, a coarse identity of whether or not the card has already been used in a session with an up-to-date terminal in the current month can also exist in the system without compromising unlinkability. There is an additional assumption made in this model, specifically, that we do not verify the unlinkability of cards which have not been used at all in the previous month. The coarse identity of having used the card in the past month, but not in the current month, barely gives any identifying information away at all. However, a card that has not been used for over two month, is relatively easy to identify among a pool of cards that are used in a normal, more frequent, manner, since it may be tracked with high probability by observing whether it responds rather than blocks when presented with a two-month-old certificate.
%, assuming majority of cards are used at least monthly.
%
%Depending on how frequently cardholders use their cards we can 
}

\delete{This justifies the observation in Section~\ref{sec:privacyreq} where we have pointed out that unlinkability can only be achieved up to the fingerprint comprising the coarse identities of the card being revealed. In particular, Corollary~\ref{col:bang} concerns the case where multiple payment systems might not agree to provide a common application for unlinkable payments as discussed in Section~\ref{sec:init}, and therefore these different payment systems form a coarse identity of the card. Corollary~\ref{col:bang} ensures that even in this scenario $\protocol$ remains unlinkable.}

\delete{
%\subsubsection{Further unlinkability proofs}
\emph{Further unlinkability proofs}
	In the current work we restrict the analysis to the case where all cards are synchronised to execute within the same month $\texttt{MM}$, however in Appendix~\ref{appx:months}, we present the model admitting cards valid in different months as well as advancing pointers from one month to the next. The verification of this model is left as future work.}

\subsection{Related methods for proving unlinkability.}\label{sec:tools}
%To answer this question negatively we discuss existing tools and perform a small case study involving the full $\protocol$ protocol and the related, strictly simpler, key agreement phase. 
In the proofs of Theorems~\ref{thm:link}~and~\ref{thm:months} establishing the unlinkability of $\protocol$, we have constructed and checked by hand a bisimulation between the real and the ideal worlds of the respective models of the protocol. In this section we address the question of whether current tools can be used to confidently reach the same conclusion. Below we discuss existing tools and perform a small case study involving the full $\protocol$ protocol and the related, strictly simpler, key agreement phase defined in earlier work~\cite{horne2022csf}.

%The choice of equivalence between the idealised and the real-world specifications, used to define a privacy property, may crucially affect the verification outcome~\cite{arapinis2010analysing, HORNE2023113842}. For instance, a protocol may be proven correct under a coarser equivalence, while at the same time admitting a practical attack under a finer equivalence. Alternatively, a correct protocol may admit an attack having no practical interpretation, if the equivalence is so fine, that it is violated for technical reasons.

%The developers of state-of-the-art verification tools agree that \emph{observational equivalence} is the coarsest useful notion suitable for defining privacy properties. 

The widely-used tools such as Tamarin~\cite{tamarineq} and ProVerif~\cite{BLANCHET20083} offer limited support for bisimilarity checking -- they can verify so-called diff-equivalence, i.e., equivalence between two processes that differ only in the messages exchanged. Definition~\ref{def:unlinkformal} does not fall into the diff-equivalence category since the $\impl{UTX}$ and $\spec{UTX}$ processes have different structures. Hence we rule out the use of Tamarin. ProVerif, however, makes an attempt to represent the equivalence problem for arbitrary processes as a diff-equivalence problem, thus it can be considered as a candidate for verifying the unlinkability of $\protocol$. Moreover, recently, two ProVerif-related tools have been introduced with the aim of improving the verification of equivalence-based properties. In this work, we call them schematically T1 and T2. 

The T1 tool~\cite{csf2023unlink} transforms the existing ProVerif model to another model that ProVerif is more likely to verify. The T2 tool~\cite{csf2023beyond}, which improves on Ukano~\cite{ukano}, 
%capable of handling a restricted class of the protocols,
is essentially a new version of ProVerif that may verify observational equivalence (aka.\ early bisimilarity) between two arbitrary processes, lifting restrictions on diff-equivalence. Neither ProVerif nor T1 or T2 has been able to verify the unlinkability of the single-month model of $\protocol$.
This means that, for the time being, our manual proof of Theorem~\ref{thm:link} is justified.

However, to assess the reliability of T1 and T2 and out of curiosity, we performed the following test. Firstly, we attempted to verify the BDH key agreement protocol~\cite{rfc}, which has already been proven to be linkable~\cite{horne2022csf}. Secondly, we asked tools to verify the UBDH protocol, which roughly corresponds to the \emph{Initialisation} followed by the \emph{Validity check} phases of $\protocol$ and has already been proven to be unlinkable~\cite{horne2022csf}. For comparison, we also include the results delivered by basic ProVerif. We have verified two different models of both protocols -- with and without terminals. The presence of the terminals should not affect the verification results because in BDH and UBDH, cards and terminals share no common secret, as explained in Section~\ref{sec:resultunlink}. 

%The verification results are presented in the table below, where \shrug stands for the \emph{cannot be proved} conclusion and \cmark, \xmark, and \qmark \ indicates whether the outcome is, correct, incorrect or unreliable respectively.

The verification results are presented in the table below, where \shrug stands for the \emph{cannot be proved} verdict and \cmark and \xmark \ indicate whether the verdict is correct or not, given that the correct results for the protocols tested are known from related work.

%\noindent
%\begin{tabular}{ |p{1.631cm}||p{1.6cm}|p{1.25cm}|p{1.6cm}|}
%	%	\hline
%	%	\multicolumn{4}{|c|}{Operations} \\
%	\hline 
%	&PV 2.04 diff&T1 diff & T2 obs\\
%	\hline
%	BDH (no T) &\multicolumn{1}{|c|}{\shrug}&\multicolumn{1}{|c|}{\textcolor{red}{TRUE} \xmark}&\multicolumn{1}{|c|}{\shrug}\\
%	\hline
%	BDH (full) &\multicolumn{1}{|c|}{running forever}&\multicolumn{1}{|c|}{\textcolor{red}{TRUE} \xmark}& \multicolumn{1}{|c|}{\shrug} \\
%	\hline
%	UBDH (no T) &\multicolumn{1}{|c|}{\textcolor{blue}{TRUE} \cmark}&\multicolumn{1}{|c|}{\textcolor{blue}{TRUE} \cmark}&\multicolumn{1}{|c|}{\textcolor{blue}{TRUE} \cmark}\\
%	\hline
%	UBDH (full) &\multicolumn{1}{|c|}{running forever} &\multicolumn{1}{|c|}{\textcolor{blue}{TRUE} \cmark} &\multicolumn{1}{|c|}{\textcolor{blue}{TRUE} \cmark} \\
%	\hline
%	UTX &\multicolumn{1}{|c|}{running forever}&\multicolumn{1}{|c|}{fatal error}&\multicolumn{1}{|c|}{running forever}\\
%	\hline
%\end{tabular}

\bgroup
\setlength\tabcolsep{0.154cm}
\noindent
\begin{tabular}{ |p{1.631cm}||p{1.6cm}|p{1.4cm}|p{1.6cm}|}
	%	\hline
	%	\multicolumn{4}{|c|}{Operations} \\
	\hline 
	&PV 2.04 diff&T1 diff & T2 obs\\
	\hline
	BDH (no T) &\multicolumn{1}{|c|}{\shrug}&\multicolumn{1}{|c|}{\textcolor{red}{TRUE} \xmark}&\multicolumn{1}{|c|}{\shrug}\\
	\hline
	BDH (full) &\multicolumn{1}{|c|}{running forever}&\multicolumn{1}{|c|}{\textcolor{red}{TRUE} \xmark}& \multicolumn{1}{|c|}{\shrug} \\
	\hline
	UBDH (no T) &\multicolumn{1}{|c|}{\textcolor{blue}{TRUE} \cmark}&\multicolumn{1}{|c|}{\textcolor{blue}{TRUE} \cmark}&\multicolumn{1}{|c|}{\textcolor{blue}{TRUE} \cmark}\\
	\hline
	UBDH (full) &\multicolumn{1}{|c|}{running forever} &\multicolumn{1}{|c|}{\textcolor{blue}{TRUE} \cmark} &\multicolumn{1}{|c|}{\textcolor{blue}{TRUE} \cmark} \\
	\hline
	UTX &\multicolumn{1}{|c|}{running forever}&\multicolumn{1}{|c|}{fatal error}&\multicolumn{1}{|c|}{running forever}\\
	\hline
\end{tabular}
\egroup

We can immediately see that T1 is unreliable for the protocols we consider -- it claims that BDH (with and without terminals) is unlinkable, which is not true. Thus, the verification results for both versions of UBDH in the T1 column would require further examination before they can be trusted. In contrast, T2 does not make incorrect claims for the protocols tested, e.g. in approximately 17 hours the tool concludes that observational equivalence cannot be proved for the full BDH protocol, and within 4 hours it was able to prove that the full UBDH model is unlinkable. Interestingly, ProVerif was able to verify only the restricted version of UBDH which highlights the importance of compositionality since it might be the case that a tool is only able to prove the property when a smaller subsystem is in a form the tool can handle. The final row shows that tools freeze either without output or with an error when fed UTX. Based on this observation we hypothesise that it is the structure of UTX that tools cannot deal with rather than the scalability issue which justifies the claim we made in the introduction.

%Related work~\cite{HORNE2023113842} proves that, being a congruence relation, quasi-open bisimilarity can be used to show that unlinkability without terminals implies unlinkability with terminal for UBDH, by compositional reasoning similar to that employed in Section~\ref{sec:corollary}.

%\SM{Maybe mention explicitly that there is indeed a simple attack on BDH with only two sessions? At least that seems to be the implicit message in the following paragraph.}
At the same time, no tool in the table is able to discover the known simple attack on BDH which requires only two sessions with the same card. This holds even if we restrict the tools so they consider exactly two sessions instead of using replication, in which case the outcome is identical to the first line of the table, differing only in running time.
The repository~\cite{repoutx} contains files corresponding to each cell of the table.
We should mention here, that the \textit{noname} tool~\cite{alphabeta}, implementing a promising parallel approach to modelling privacy called \emph{alpha-beta privacy}, can discover the simple attack on BDH that eludes equivalence checkers.
% which allows discovering attacks automatically for a bounded number of sessions.
%In this work, we aim to reason about $\protocol$ without any restrictions on the number of sessions.  

%	Theorem~\ref{thm:link} holds for an unbounded number of sessions.

%	The standard version of ProVerif is, unfortunately, freezes for any protocol with more then one role.

%The last line of the table above reflects that neither ProVerif nor T1 or T2 is able to even start verifying the full scale $\protocol$. 
%Indeed, the fact that this problem is not even attempted, suggests that it is the form of UTX is out of scope of current tools. If the problem were simply too large to explore, then progress would have been reported by the tools. We expect these observations and our manual proofs can stimulate future work on tools.

%	\SY{check may testing inclusion like in the og paper}

%It would be also relatively easy to adapt the proof of Theorem~\ref{thm:link} to show that unlinkability holds for low-value contactless payments, even if all terminals are fully compromised (by excluding them from the model and making the channel with the bank public). 

\subsection{Authentication in UTX}\label{sec:securityanalysis}

%\SY{this section is substantially updated as well as the respective code in the repository. Events are removed from the specification and moved to appendix.}

Our security definition supporting the requirements identified in Section~\ref{sec:securityrequirements} relies on an authentication property called \emph{injective agreement}~\cite{lowe1997auth}. 
%that has been employed to analyse the original EMV~\cite{jorge2020emv, jorge2021usenix}.
A party \textsf{X} injectively agrees with the parties \textsf{Y} and \textsf{Z} whenever if \textsf{X} thinks it has authenticated \textsf{Y} and \textsf{Z}, then \textsf{Y} and \textsf{Z} executed the protocol exchanging the same messages as \textsf{X} (\emph{agreement}), and each run of \textsf{X} corresponds to a unique run of \textsf{Y} and \textsf{Z} (\emph{injectivity}).

% Injective agreement between parties \textsf{X} and \textsf{Y} ensures firstly, that when \textsf{X} thinks it has authenticated \textsf{Y}, then \textsf{Y} executed the protocol exchanging the same messages as \textsf{X} (\emph{agreement}), and secondly, that each run of \textsf{X} corresponds to a unique run of \textsf{Y} (\emph{injectivity}).

%\SY{for verification is as strong as for jorge emv standard inj agree from literature}

To verify injective agreement in $\protocol$ we have already included events in role specifications in Fig.~\ref{fig:specification} marking certain stages reached by processes during the execution of the protocol and then evaluate correspondence assertions~\cite{blanchet-corresp} between events listed in Fig.~\ref{fig:security} using the ProVerif tool~\cite{blanchet2001efficient}. Appendix~\ref{sec:appxproverif} contains further details regarding using ProVerif.

%Appendix~\ref{sec:appxproverif} contains the UTX specification updated with the events involved and additional details regarding using ProVerif. 

\begin{figure}[h!]
	\centering
	%	\resizebox{.8\linewidth}{!}{\includegraphics[width=\linewidth]{msc/accert.pdf}}
	\[
	\begin{array}{l}
		\multicolumn{1}{l}{\bef{The terminal agrees with the card \change{(before contacting bank)}}} \\
		\texttt{TComC}(z_1, z_2, \textit{ec}, \textit{emc}, \textit{etx}, \textit{eac}) \Rightarrow \\ \ \ \texttt{CRun}(z_1, z_2, \textit{ec}, \textit{emc}, \textit{etx}, \textit{eac}) \\[3pt]
		\multicolumn{1}{l}{\bef{The terminal agrees wih the bank and the card}} \\
		\texttt{TComBC}(\textit{req}, \textit{resp}, z_1, z_2, \textit{ec}, \textit{emc}, \textit{etx}, \textit{eac}) \Rightarrow \\ \ \ \texttt{BRunT}(\textit{req}, \textit{resp}) \ \wedge \ \texttt{CRun}(z_1, z_2, \textit{ec}, \textit{emc}, \textit{etx}, \textit{eac}) \\[3pt]
		\multicolumn{1}{l}{\bef{The bank agrees with the terminal and the card}} \\
		\texttt{BComTC}(\textit{req}) \Rightarrow \\  \ \ \texttt{TRunBC}(\textit{req}, z_1, z_2, \textit{ec}, \textit{emc}, \textit{etx}, \textit{eac}) \ \wedge \\  \ \ \texttt{CRun}(z_1, z_2, \textit{ec}, \textit{emc}, \textit{etx}, \textit{eac}))\\[3pt]
		\multicolumn{1}{l}{\bef{Bank agrees with the card on the encrypted cryptogram}} \\
		\texttt{BComC}(\textsf{EAC}) \Rightarrow  \texttt{CRunB}(\textsf{EAC})
	\end{array}
	\]  
	\caption{Injective agreement correspondences in $\protocol$.}\label{fig:security}
\end{figure}

The events in Fig.~\ref{fig:security} are parametrised by the messages the card, the terminal and the bank exchange, i.e. $z_1$ and $z_2$ stand for the ephemeral terminal's key and the blinded card's public key, \textit{ec}, \textit{emc}, \textit{etx}, \textit{eac} represent the messages the card exchanges with the terminal and \textit{req}, \textit{resp} the message the terminal exchanges with the bank. Finally, \textsf{EAC} represents the encrypted cryptogram $\enc{\nlist{\textsf{AC}, \textsf{AC}_{hmac}}}{\key{cb}}$.
%Role specifications updated with exact locations where we place the events are presented in Appendix~\ref{sec:appxproverif}.

The first three assertions in Fig.~\ref{fig:security} are straightforward -- whenever the terminal or the bank thinks it has executed the session with the rest of the agents, they have exchanged the same messages, thereby agreeing on crucial data such as the derived keys, transaction details, the cryptogram, etc. The last assertion, representing the agreement between the bank and the card, ensures that an honest card was involved in low-value contactless payment even if terminals are fully compromised. In this scenario the terminals can be omitted in the specification as explained in the related work~\cite{thesis}.

\emph{Security under compromised terminals.} Using ProVerif, we support the point made at the end of Section~\ref{sec:acgen} that even if a terminal neglects to perform the checks required to authenticate the card, the bank is still ensured that a valid card is executing a transaction. To model that, we remove the Verheul signature verification in the terminal's process. In that case, the first property that the terminal authenticates the card fails as expected, while others are preserved. 
%The respective code is provided in the directory \emph{compromised} of~\cite{repoutx}. 

\emph{Security under compromised $\chi_{\texttt{MM}}$.} Another scenario in which the terminal accepts a potentially fake card is when the key $\chi_{\texttt{MM}}$ is leaked, allowing attackers to manufacture cards passing the terminal's check by producing valid Verheul signatures. The verification outcome in this case is similar -- the terminal-card agreement fails, making offline transactions insecure,
%as the terminal cannot verify the cryptogram, 
while online transactions are still safe, i.e. the injective agreement involving the bank holds. Therefore, the payment system should notify terminal owners to stop accepting offline payments if $\chi_{\texttt{MM}}$ has been compromised.

The repository~\cite{repoutx} contains the code specifying the injective agreement in the $\protocol$ protocol and the expected secrecy of the private data. All properties are successfully verified within 100 minutes. The code verifying additional scenarios described above is provided in the directory \emph{compromised}. 
%\emph{Security under compromised terminals.} Finally, using ProVerif we support the point made at the end of Section~\ref{sec:acgen} that if a terminal neglects to perform the checks required to authenticate the card, then it is possible that the bank can decline the transaction even after the card has approved it. 
%%the bank will
%%
%% fail to authorise the payment. 
%%
%%the terminal cannot be reimbursed for the honest cryptogram replayed in a session with an unauthenticated device. 
%To model that we remove the Verheul signature verification in the terminal's process. In that case, the first property that the card is authenticated by the terminal fails as expected as well as the property that the terminal is being paid by the bank while the property that the card is authenticated by the bank is preserved.

\change{
\subsubsection{Remark on replay protection.}
We explain here a small difference between the model used to verify unlinkability and authentication concerning replay protection. Recall that replay protection is enforced by the uniqueness by the bank of the triple $\nlist{\text{PAN}, \textsf{TX}, a}$, as specified in Section~\ref{sec:acgen} and in Fig.~\ref{fig:protocol}. This check is an essential ingredient for authentication, without which authentication could not be verified. Hence replay protection is accounted for in the threat model used in Section~\ref{sec:securityanalysis}, specifically in line 165 of the ProVerif model.

In contrast, the threat model we use for unlinkability simplifies this aspect by allowing terminals to replay cryptograms, that is the bank skips the uniqueness check. This does not introduce problems for the following two reasons. Firstly, the bank in the $\protocol$ protocol sends no message intended for the card, hence there is no way for the terminal to probe cards with such a message in an attempt to track them. Secondly, an observable \texttt{auth} output by the terminal that reveals whether the cryptogram was accepted by the bank introduces no issues regardless of the presence of the check. With replay protection, any attempt to replay the message from the terminal to the bank, i.e. the cryptogram with auxiliary data would result in the absence of \texttt{auth}, while, without replay protection, the replay would result in the message \texttt{auth} being always present. In both cases it is impossible for an attacker to link the presence or the absence of the \texttt{auth} with any session other than the one in which the cryptogram was created, and hence it cannot be used to link two sessions with the same card.
}

\subsection{Estimation of the runtime performance}\label{sec:runtime}
Concluding the analysis, we give a rough estimate of the runtime performance of the $\protocol$ protocol focusing on the card operations. Indeed since the terminal is a more powerful device than the card we expect its contribution to the runtime to be minuscule. We make our assessment based on the estimations reported in~\cite{malina2018assessment, dzurenda2017performance} for the Multos Card ML3 supporting ECC scalar multiplication. The table below summarises the amount of time for individual operations performed by the card in the $\protocol$ protocol. As we expect the equality check and forming $n$-tuples operations to be negligible, we omit them in our calculation. Overall the numbers add up to 700ms per on-card computation per session. 
%Indeed this number does not include the computation time taken by the terminal, however, being a more powerful device than the card, we expect the terminal's contribution to be minuscule. 
We expect that further optimisation and using more recent smart card platforms would lower this number within the current 500 \textit{ms} recommendation~\cite{emvcontactless}. 

%hence a concept implementation and the study of users' behaviour are needed to clarify that $\protocol$ is fast enough to be usable contactlessly.

%Based on the number of on-card operations and the known performance assessment~\cite{malina2018assessment, dzurenda2017performance} of cryptographic functions on smart cards. 

\noindent
\begin{tabular}{ |p{1.6cm}||p{0.7cm}|p{0.7cm}|p{1.0cm}|p{0.7cm}|p{1.55cm}|  }
%	\hline
%	\multicolumn{4}{|c|}{Operations} \\
	\hline
	Operation&$\smult{\cdot}{\cdot}$&$\hash{\cdot}$&$\dec{\cdot}{\cdot}$&$\enc{\cdot}{\cdot}$&$\checksig{\cdot}{\cdot}$\\
	\hline
	$\#$ of ops&6&3&2&3&1\\
	\hline
	\textit{ms} per op&61&11&13&13&228\\
	\hline
\end{tabular}

%The table above presents the evaluation of the Multos Card ML3 platform chosen because it directly supports ECC scalar multiplication (in contrast to the popular Java Card platform). 
The numbers from the third line correspond to the 256-bit security level for $\phi$ and $\texttt{check}$ operations, which are evaluated using the Barreto-Naehrig pairing-friendly curve since Verheul signatures are pairing-based, and ECDSA, respectively. To the best of our knowledge, there is no credible source for 256-bit security assessment for the rest, hence we use the available benchmarks -- $\texttt{dec}$ and $\{ \ \}$ are evaluated using 128-bit key AES in CBC mode on 128-bit message, and, finally, $\texttt{h}$ has been tested using SHA-256 on 128-bit message.

%Given the data, we calculate that on-card computation can be estimated within 700 \textit{ms} per transaction. Indeed this number does not include the computation time taken by the terminal
%%~\footnote{\footnotesize As well as the response time from the bank, which can take up to several seconds in the current EMV in the case of online transactions.}, 
%however, being a more powerful device than the card, we expect the terminal's contribution to be minuscule. 

%This estimate exceeds the current 500 \textit{ms} requirement~\cite{emvcontactless}, hence a concept implementation and the study of users' behaviour are needed to clarify that $\protocol$ is fast enough to be usable contactlessly.

	\section{Conclusion}

%\SY{PIN low value to appx, recursive in appx}

%\SY{lessons from this design that can generalize to other design problems}
%
%\SY{Active privacy is an issue We need to develop methods for proving the privacy}
%
%\SY{Stretching assumption on limitation of attacker model. It is impossible to have unlinkability in the 2nd phase}

In this paper, we have identified in Section~\ref{sec:building} the requirements for a smartcard-based payments protocol, and have demonstrated that at least one protocol satisfying these requirements exists -- the $\protocol$ protocol  presented in Fig.~\ref{fig:protocol}. 
We strengthen the initial security of EMV as explained in Section~\ref{sec:requirements}. In particular, we request that the application cryptogram is secret and can only be processed by a legitimate acquiring bank. This requirement is addressed in $\protocol$ by using the certified bank's public key that the card obtains at the beginning of each transaction and uses to encrypt the cryptogram as we have explained in Sections~\ref{sec:connectingterminal},~\ref{sec:acgen}. Fig.~\ref{fig:security} summarises how we have proved in ProVerif that $\protocol$ satisfies all security requirements we have identified.
%\SM{The last sentence has a grammatical problem.}

%We have chosen unlinkability as the main privacy requirement since it is among the strongest standardised in ISO 15408 and supports the right attacker.
% realistic active attacker who can force the card to interact. 
We explain how ISO 15408 supports targeting unlinkability as our privacy requirement in Section~\ref{sec:privacyreq}, and highlight that the fingerprint of the card, comprising coarse identities of a card that permit groups of cards to be tracked, should be minimised.
Since strong identities compromise unlinkability, we have hidden any strong identity of the card by utilising Verheul signatures to make the validity signature distinct in every session, as explained in Section~\ref{sec:mutualauth}, and by encrypting the cryptogram that contains the PAN to hide it from the terminal, as explained in Section~\ref{sec:acgen}.
We have minimised the card's fingerprint by introducing  certificates that reveal that the card is valid for the current and previous months without revealing the expiry date, as explained in Sections~\ref{sec:keys},~\ref{sec:mutualauth}. 
If payment systems agree on a common certification authority we may reduce the card's fingerprint further by introducing the $\psu$ application as explained in Section~\ref{sec:init}.
%and the key agreement at the start of the transaction are also among the 
Theorem~\ref{thm:link} proves that these measures indeed achieve unlinkability in $\protocol$.

We provide precisely three modes which agents should implement to process $\protocol$ payments. The modes of payment should be standardised and be common to all cards supporting $\protocol$. This avoids cards being distinguished by implementation differences. This contrasts to the current EMV standard, which has many different modes of operation, defined in 2000 pages split into several books; thus the variety of \emph{implementations} serve\change{s} as a coarse identity of the card. Moreover, having a concise, coherent, and linear presentation can improve the reliability of the system. Our message sequence chart in Fig.~\ref{fig:protocol} and the applied $\pi$-calculus specification of $\protocol$ in Fig.~\ref{fig:specification} go some way towards this aim.

\after{Roll-out of the $\protocol$ protocol is feasible. The software of banks and terminals can be updated in advance across a region so both accept unlinkable payments, while continuing the support of old payment methods. Then cards supporting unlinkable payments can be issued. Of course, new cards must implement only one application to avoid attacks that downgrade cards to EMV.}

%The $\protocol$ example demonstrates that the safest way is to present the protocol in a specification language such as applied $\pi$-calculus.

%We hypothesise that the existence of insecure configurations in EMV is rooted in the way it is written: over 2000 pages split into several books with a heavy cross-referencing. 
%
%The existence of research papers that \emph{explain} how EMV works~\cite{van2016emv, el2018overview} backs up our claim. 
%
%The $\protocol$ example demonstrates that the safest way is to present the protocol in a specification language such as applied $\pi$-calculus.

%\before{As mentioned in Section~\ref{sec:attackermodel} high-value online transactions can be compromised via relay attacks when the PIN is exposed. Hence introducing relay protection~\cite{radu2022practical} can mitigate such attacks while maintaining its unlinkability. Other future work includes a proof of concept implementation clarifying that $\protocol$ is fast enough to be usable contactlessly.}  

Regarding the protocol design future work includes introducing relay protection~\cite{radu2022practical}. Firstly, it should mitigate the situation where a high-value online transaction is compromised via a relay attack if the PIN is exposed, as we mention in Section~\ref{sec:attackermodel}. Secondly, it is essential for the protocol that supports PIN tries counter, which limits the number of incorrect attempts to enter the PIN. An active attacker can exploit PIN tries counter by relaying messages from the honest terminal waiting to process an online high-value transaction to the card and entering the PIN incorrectly enough times to exceed the limit, thereby blocking the card from any online transactions. Then to identify such cards, an attacker should yet again relay communication between the card and an online terminal -- transactions would be declined with an explicit reason of the PIN tries exceeded. Relay protection would mitigate this scenario, making it impossible to enter the PIN remotely since the user should be physically close and aware of someone entering the PIN. 
%Notice that an active attacker could not exceed PIN tries for offline high-value purchases since such transactions are always contact (and the user is close), as we explain at the beginning of Section~\ref{sec:protocol}.

%Regarding the verification future work includes developing automated methods for proving the privacy of security protocols which would lower the analysis effort and improve its reliability -- the proof of the Theorem~\ref{thm:link} in Appendix~\ref{sec:appx} illustrates the high cost of the manual analysis of a single protocol \change{which cannot be verified using existing tools as we have demonstrated in Section~\ref{sec:tools}.}

Regarding the verification future work includes developing automated methods for proving the privacy of security protocols which would lower the analysis effort \delete{and improve its reliability} -- the proof of the Theorem~\ref{thm:link} in Appendix~\ref{sec:appx} illustrates the high cost of the manual analysis of a single protocol. \change{Regardless of the proof method, hand or computer-assisted, we consider checking proofs essential to improve the reliability of the result since even tools occasionally cannot be trusted as we have demonstrated in Section~\ref{sec:tools}.}

	\bibliographystyle{ACM-Reference-Format}
	\bibliography{ref}
	
	\appendix
	\clearpage 
\section{Key establishment unlinkability}\label{app:unlinkkey}
For completeness, we provide the definition that only considers the authenticated key establishment part. Notice that the Def.~\ref{def:unlinkformal} in Section~\ref{sec:resultunlink} of unlinkability of the $\protocol$ protocol contains processes specifying the terminal's behaviour and the bank, while the definition below concerns only cards.

\begin{definition}\emph{(unlinkability for authenticated key establishment)~\cite{horne2022csf}}
	We say that a card process scheme $C$ specifying authenticated key establishment is unlinkable whenever the following processes are quasi-open bisimilar.
	
\begin{gather*}
	\begin{aligned}
		&\nu \, \user, s, si, \chi_{\texttt{MM}}. \cout{out}{\pks{s}}.\cout{out}{\pkv{ \chi_{\texttt{MM}}}}.\mathbf{\bang} \nu \text{PIN}, mk, c, \text{PAN}.\\ 
		&\begin{array}{l}	
			\textcolor{red}{\bang}\nw \ch.\cout{card}{\ch}.C(\ch, c, \pk{s}, \sigv{\chi_{\texttt{MM}}}{\pkg{c}}, \text{PAN}, mk, \text{PIN})
		\end{array}
	\end{aligned}
	\\
	\sim 
	\\
	\begin{aligned}
		&\nu \, \user, s, si, \chi_{\texttt{MM}}. \cout{out}{\pks{s}}.\cout{out}{\pkv{ \chi_{\texttt{MM}}}}.\mathbf{\bang} \nu \text{PIN}, mk, c, \text{PAN}.\\ 
		&\begin{array}{l}	
			\nw \ch.\cout{card}{\ch}.C(\ch, c, \pk{s}, \sigv{\chi_{\texttt{MM}}}{\pkg{c}}, \text{PAN}, mk, \text{PIN})
		\end{array}	
	\end{aligned}
\end{gather*}	
\end{definition}

	\section{Theorem~\ref{thm:link}}\label{sec:appx}
\def\spcard{
	\begin{equation}
		\notag
		\begin{aligned}
			%			& \text{\textcolor{blue}{$\ctail_0(\cparams)$}} \triangleq  \nu ch.\cout{term}{ch}.\ctail_1(\cparams)\\
			C(&\ch, c, pk_s, \texttt{vsig}_\texttt{MM}, \text{PAN}, \mk, \text{PIN}) \triangleq \\
			& \text{\textcolor{blue}{$\ctail_1(\cparams)$}} \\
			& \cin{\ch}{z_1}.\\
			& \text{\textcolor{blue}{$\ctail_2(\cparams, z_1)$}} \\
			& \nw a.\cout{\ch}{\smult{a}{\pkg{c}}}.\\
			& \text{\textcolor{blue}{$\ctail_3(\cparams, z_1, a)$}} \\
			& \lett \, \key{c} \coloneqq \hash{\smult{\mult{a}{c}}{z_1}} \inn \\
			& \cin{\ch}{m}.\\ 
			& \text{\textcolor{blue}{$\ctail_4(\cparams, \key{c}, a, m)$}} \\
			& \lett \, \nlist{\textsf{MC}, \textsf{MC}_s} \coloneqq \dec{m}{\key{c}} \inn \\
			& \ifff \ \checksig{pk_s}{\textsf{MC}_s} = \textsf{MC} \ \thenn \\
			& \ifff \ \proj{1}{\textsf{MC}} =\texttt{MM} \ \thenn \\
			& \cout{\ch}{\enc{\nlist{\smult{a}{\pkg{c}}, \smult{a}{\texttt{vsig}_\texttt{MM}}}}{\key{c}}}.\\
			& \text{\textcolor{blue}{$\ctail_5(\cparams, \key{c}, a, m)$}} \\
			& \cin{\ch}{x}.\\
			& \text{\textcolor{blue}{$\ctail_6(\cparams, \key{c}, a, m, x)$}} \\
			% low-value online, low-value offline, high-value online
			& \lett \, \nlist{\textsf{TX},\textsf{uPin}} \coloneqq \dec{x}{\key{c}} \inn\\
			& \lett \, \textsf{AC} \coloneqq \nlist{a, \text{PAN}, \textsf{TX}} \inn\\
			& \lett \, \textsf{AC}^{\texttt{ok}} \coloneqq \nlist{a, \text{PAN}, \textsf{TX}, \texttt{ok}} \inn\\
			& \lett \, \textsf{AC}^{\fail} \coloneqq \nlist{a, \text{PAN}, \textsf{TX}, \fail} \inn\\
			& \lett \, \key{cb} \coloneqq \hash{\smult{\mult{a}{c}}{\proj{3}{\proj{1}{\dec{m}{\key{c}}}}}} \inn \\
			& \ifff \ \textsf{uPin} = \fail \ \thenn \\
			& \cout{\ch}{\enc{\enc{\nlist{\textsf{AC},\hash{\nlist{\textsf{AC}, \mk}}}}{\key{cb}}}{\key{c}}} \\
			& \ifff \ \textsf{uPin} =\text{PIN} \ \thenn \\
			& \cout{\ch}{\enc{\nlist{\enc{\nlist{\textsf{AC}^{\texttt{ok}},\hash{\nlist{\textsf{AC}^{\texttt{ok}}, \mk}}}}{\key{cb}}, \texttt{ok}}}{\key{c}}} \\
			& \elsee \ \cout{\ch}{\enc{\nlist{\enc{\nlist{\textsf{AC}^{\fail},\hash{\nlist{\textsf{AC}^{\fail}, \mk}}}}{\key{cb}}, \fail}}{\key{c}}}\\
			& \text{\textcolor{blue}{$\ctail_7 \triangleq 0$}} \\
		\end{aligned}
	\end{equation}
}

\def\spton{
	\begin{equation}
		\notag
	\begin{aligned}
		%			& \text{\textcolor{blue}{$\ttonhi_0(user, \tparams)$}} \triangleq  \nu ch.\cout{term}{ch}.\ttonhi_1(user, \tparams)\\
		T_{\texttt{onhi}}(&\user, \ch, pk_\texttt{MM}, \textsf{crt}, \kbt) \triangleq \\
		& \text{\textcolor{blue}{$\ttonhi_1(user, \tparams)$}} \\
		& \nw \, \mathsf{TXdata}. \lett \, \textsf{TX} \coloneqq \nlist{\mathsf{TXdata}, \hi} \inn \\
		& \nw t.\cout{\ch}{\pkg{t}}. \\
		& \text{\textcolor{blue}{$\ttonhi_2(user, \tparams, t, \textsf{TX})$}} \\
		& \cin{\ch}{z_2}. \\
		& \text{\textcolor{blue}{$\ttonhi_3(user, \tparams, t, \textsf{TX}, z_2)$}} \\
		& \lett \, \key{t} \coloneqq \hash{\smult{t}{z_2}} \inn \\
		& \cout{\ch}{\enc{\textsf{crt}}{\key{t}}}. \\
		& \text{\textcolor{blue}{$\ttonhi_4(user, \tparams, t, \textsf{TX}, z_2)$}} \\
		& \cin{\ch}{n}.\\
		& \text{\textcolor{blue}{$\ttonhi_5(user, \tparams, t, \textsf{TX}, z_2, n)$}} \\
		& \lett \, \nlist{\textsf{B}, \textsf{B}_s} \coloneqq \dec{n}{\key{t}} \inn \\
		& \ifff \ \checksigv{pk_\texttt{MM}}{\textsf{B}_s} = \textsf{B} \ \thenn \\
		& \ifff \ \textsf{B} = z_2 \ \thenn \\
		& \cin{\user}{\textsf{uPIN}}.\\
		& \text{\textcolor{blue}{$\ttonhi_6(user, \tparams, t, \textsf{TX}, z_2, n, \textsf{uPIN})$}} \\
		& \cout{\ch}{\enc{\nlist{\textsf{TX}, \fail}}{\key{t}}}. \\
		& \text{\textcolor{blue}{$\ttonhi_7(user, \tparams, t, \textsf{TX}, z_2, n, \textsf{uPIN})$}} \\
		& \cin{\ch}{y}.\\
		& \text{\textcolor{blue}{$\ttonhi_8(user, \tparams, t, \textsf{TX}, z_2, n, \textsf{uPIN}, y)$}} \\
		& \cout{\ch}{\enc{\nlist{\textsf{TX}, z_2, \dec{y}{\key{t}}, \textsf{uPIN}}}{\kbt}}. \\
		& \text{\textcolor{blue}{$\ttonhi_9(user, \tparams, t, \textsf{TX}, z_2, n, \textsf{uPIN}, y)$}} \\
		& \cin{\ch}{r}. \\
		& \text{\textcolor{blue}{$\ttonhi_{10}(user, \tparams, t, \textsf{TX}, z_2, n, \textsf{uPIN}, y, r)$}} \\
		& \ifff \ \proj{1}{\dec{r}{\kbt}} = \textsf{TX} \ \thenn \\
		& \ifff \ \proj{2}{\dec{r}{\kbt}} = \texttt{accept} \ \thenn \\
		&  \cout{\ch}{\texttt{auth}} \text{\textcolor{blue}{$\ttonhi_{11} \triangleq 0$}} \\
	\end{aligned}
	\end{equation}
}

\def\sptof{
	\begin{equation}
		\notag
		\begin{aligned}
			%			& \text{\textcolor{blue}{$\ttofhi_0(user, \tparams)$}} \triangleq  \nu ch.\cout{term}{ch}.\ttofhi_1(user, \tparams)\\
			T_{\texttt{offhi}}(&\user, \ch, pk_\texttt{MM}, \textsf{crt},  \kbt) \triangleq \\
			& \text{\textcolor{blue}{$\ttofhi_1(user, \tparams)$}} \\
			& \nw \, \mathsf{TXdata}. \lett \, \textsf{TX} \coloneqq \nlist{\mathsf{TXdata}, \hi} \inn \\
			& \nw t.\cout{\ch}{\pkg{t}}. \\
			& \text{\textcolor{blue}{$\ttofhi_2(user, \tparams, t, \textsf{TX})$}} \\
			& \cin{\ch}{z_2}. \\
			& \text{\textcolor{blue}{$\ttofhi_3(user, \tparams, t, \textsf{TX}, z_2)$}} \\
			& \lett \, \key{t} \coloneqq \hash{\smult{t}{z_2}} \inn \\
			& \cout{\ch}{\enc{\nlist{\textsf{crt}}}{\key{t}}}. \\
			& \text{\textcolor{blue}{$\ttofhi_4(user, \tparams, t, \textsf{TX}, z_2)$}} \\
			& \cin{\ch}{n}.\\
			& \text{\textcolor{blue}{$\ttofhi_5(user, \tparams, t, \textsf{TX}, z_2, n)$}} \\
			& \lett \, \nlist{\textsf{B}, \textsf{B}_s} \coloneqq \dec{n}{\key{t}} \inn \\
			& \ifff \ \checksigv{pk_\texttt{MM}}{\textsf{B}_s} = \textsf{B} \ \thenn \\
			& \ifff \ \textsf{B} = z_2 \ \thenn \\
			& \cin{\user}{\textsf{uPIN}}.\\
			& \text{\textcolor{blue}{$\ttofhi_6(user, \tparams, t, \textsf{TX}, z_2, n, \textsf{uPin})$}} \\
			& \cout{\ch}{\enc{\nlist{\textsf{TX}, \textsf{uPin}}}{\key{t}}}. \\
			& \text{\textcolor{blue}{$\ttofhi_7(user, \tparams, t, \textsf{TX}, z_2, n, \textsf{uPin})$}} \\
			& \cin{\ch}{y}.\\
			& \text{\textcolor{blue}{$\ttofhi_8(user, \tparams, t, \textsf{TX}, z_2, n, \textsf{uPin}, y)$}} \\
			& \ifff \ \proj{2}{\dec{y}{\key{t}}} = \texttt{ok} \ \thenn \\
			& \cout{\ch}{\texttt{auth}}. \\
			& \text{\textcolor{blue}{$\ttofhi_9(user, \tparams, t, \textsf{TX}, z_2, n, \textsf{uPin}, y)$}} \\
			& \cout{\ch}{\enc{\nlist{\textsf{TX}, z_2, \textsf{EAC}, \fail}}{\kbt}} \\
			& \text{\textcolor{blue}{$\ttofhi_{10}(user, \tparams, t, \textsf{TX}, z_2, n, \textsf{uPin}, y)$}} \\
			& \ifff \ \proj{2}{\dec{y}{\key{t}}} \neq \texttt{ok} \ \thenn \\
			& \cout{ch}{\enc{\nlist{\textsf{TX}, z_2, \textsf{EAC}, \fail}}{\kbt}} \text{\textcolor{blue}{$\ttofhi_{11} \triangleq 0$}} 
		\end{aligned}
	\end{equation}
}

\def\sptlo{
	\begin{equation}
		\notag
			\begin{aligned}
			%			& \text{\textcolor{blue}{$\ttlo_0(\tparams)$}} \triangleq  \nu ch.\cout{term}{ch}.\ttlo_1(\tparams)\\
			T_{\texttt{lo}} (&\ch, pk_\texttt{MM}, \textsf{crt}, \kbt) \triangleq \\
			& \text{\textcolor{blue}{$\ttlo_1(\tparams)$}} \\
			& \nw \, \mathsf{TXdata}. \lett \, \textsf{TX} \coloneqq \nlist{\mathsf{TXdata}, \lo} \inn \\
			& \nw t.\cout{\ch}{\pkg{t}}. \\
			& \text{\textcolor{blue}{$\ttlo_2(\tparams, t, \textsf{TX})$}} \\
			& \cin{\ch}{z_2}. \\
			& \text{\textcolor{blue}{$\ttlo_3(\tparams, t,\textsf{TX}, z_2)$}} \\
			& \lett \, \key{t} \coloneqq \hash{\smult{t}{z_2}} \inn \\
			& \cout{\ch}{\enc{\textsf{crt}}{\key{t}}}. \\
			& \text{\textcolor{blue}{$\ttlo_4(\tparams, t, \textsf{TX}, z_2)$}} \\
			& \cin{\ch}{n}.\\
			& \text{\textcolor{blue}{$\ttlo_5(\tparams, t, \textsf{TX}, z_2, n)$}} \\
			& \lett \, \nlist{\textsf{B}, \textsf{B}_s} \coloneqq \dec{n}{\key{t}} \inn \\
			& \ifff \ \checksigv{pk_\texttt{MM}}{\textsf{B}_s} = \textsf{B} \ \thenn \\
			& \ifff \ \textsf{B} = z_2 \ \thenn \\
			& \cout{\ch}{\enc{\nlist{\textsf{TX}, \fail}}{\key{t}}}. \\
			& \text{\textcolor{blue}{$\ttlo_6(\tparams, t, \textsf{TX}, z_2, n)$}} \\
			& \cin{\ch}{y}.\\
			& \text{\textcolor{blue}{$\ttlo_7(\tparams, t, \textsf{TX}, z_2, n, y)$}} \\
			& \cout{\ch}{\enc{\nlist{\textsf{TX}, z_2, \dec{y}{\key{t}}, \fail}}{\kbt}}. \\
			& \text{\textcolor{blue}{$\ttlo_8(\tparams, t, \textsf{TX}, z_2, n, y)$}} \\
			& \cin{\ch}{r}. \\
			& \text{\textcolor{blue}{$\ttlo_9(\tparams, t, \textsf{TX}, z_2, n, y, r)$}} \\
			& \ifff \ \proj{1}{\dec{r}{\kbt}} = \textsf{TX} \ \thenn \\
			& \ifff \ \proj{2}{\dec{r}{\kbt}} = \texttt{accept} \ \thenn \\
			& \cout{\ch}{\texttt{auth}} \text{\textcolor{blue}{$\ttlo_{10} \triangleq 0$}} \\
		\end{aligned}
	\end{equation}
}

\def\spbank{
	\begin{equation}
		\notag
		\begin{aligned}
			B(&\ch, si, \kbt, b_t) \triangleq \\
			& \text{\textcolor{blue}{$\btail_1(\bparams)$}} \\
			& \cin{\ch}{x}.\\
			& \text{\textcolor{blue}{$\btail_2(\bparams, x)$}} \\
			& \lett \, dx \coloneqq \dec{x}{\kbt} \inn \\
			& \lett \, \text{PAN} \coloneqq \proj{2}{\proj{1}{\dec{\proj{3}{dx}}{\smult{b_t}{\proj{2}{dx}}}}} \inn \\
			& \cin{\nlist{si, \text{PAN}}}{y}. \\
			& \text{\textcolor{blue}{$\btail_3(\bparams, x, y)$}} \\
			& \lett \, \nlist{\textsf{TX}', z_2, \textsf{EAC}, \textsf{uPIN}} \coloneqq \dec{x}{\kbt} \\
			& \lett \, \key{bc} \coloneqq \hash{\smult{b_t}{z_2}} \inn \\
			& \lett \, \nlist{\textsf{AC}, \textsf{AC}_{hmac}} = \dec{\textsf{EAC}}{\key{bc}} \inn \\
			& \lett \, \nlist{\text{PIN}, mk, pk_{c}} \coloneqq y \\
			& \ifff \ \hash{\nlist{\textsf{AC}, mk}} = \textsf{AC}_{hmac} \ \thenn \\
			& \ifff \ \proj{3}{\textsf{AC}} = \textsf{TX}' \ \thenn \\
			& \ifff \ \smult{\proj{1}{\textsf{AC}}}{pk_{c}}= z_2 \\
			%			& \textcolor{blue}{?\nlist{\text{PAN}, \textsf{TX}, a} is \ unique ?}\\
			& \lett \, r \coloneqq \nlist{\textsf{TX}', \texttt{accept}} \inn \\
			% low-value offline, low-value online
			& \ifff \ \proj{2}{TX'} = \lo \ \thenn \\
			& \cout{\ch}{\enc{r}{\kbt}} \\
			& \ifff \ \proj{2}{\textsf{TX}'} = \hi \ \thenn \\
			% high-value offline		
			& \ifff \ \proj{4}{\textsf{AC}} = \texttt{ok} \ \thenn \\ 
			& \cout{\ch}{\enc{r}{\kbt}} \\
			% high-value online	
			& \elsee \ \ifff \ \textsf{uPIN} = \text{PIN} \ \thenn \\
			& \cout{\ch}{\enc{r}{\kbt}} \\
			& \text{\textcolor{blue}{$\btail_4 \triangleq 0 $}} \\
		\end{aligned}
	\end{equation}
}
This appendix contains the proof of unlinkability of the $\protocol$ protocol for the single-month model, i.e. of the Theorem~\ref{thm:link} from Section~\ref{sec:resultunlink}.
\begin{customthm}{1}The \begin{upshape}$\protocol$\end{upshape} protocol is unlinkable. 
\begin{proof}

By Def.~\ref{def:unlinkformal} we must prove that $\spec{UTX} \sim \impl{UTX}$. To do so we provide a relation $\mathfrak{R}$, s.t. $\spec{UTX} \, \mathfrak{R} \, \impl{UTX}$ and check that it is a quasi-open bisimulation (Def.~6 in~\cite{horne2021quasi}). The program is as follows. Construct $\mathfrak{R}$, check that $\mathfrak{R}$ is a bisimulation, i.e. each state can match each other's actions, that $\mathfrak{R}$ is \emph{open} as in Def.~5 in~\cite{horne2021quasi}, and that any related states are \emph{statically equivalent} as in Def.~4 in~\cite{horne2021quasi}.

First, let us define the following three parameter lists $\squeezespaces{1}\cparams \coloneqq (ch, c, s, \chi_\texttt{MM}, \text{PAN}, \mk, \text{PIN})$, $\squeezespaces{1}\tparams \coloneqq (\ch, pk_\texttt{MM}, \textsf{crt},  \kbt)$, and finally $\squeezespaces{1}\bparams \coloneqq (ch, si, \ch, b_t)$. 
$\cparams$, that always comes without vector, should not be confused by the reader with the private key $\chi_{\texttt{MM}}$, that always comes with a subscript $\texttt{MM}$.
Then we define ``tail'' subprocesses representing different stages of the execution for each role specification.
%in Fig.~\ref{fig:subproccardbank}~\ref{fig:subprocterminal}. 
A process highlighted in \textcolor{blue}{blue} defines the process starting from the next line, i.e. each tail subprocess defines actions left to complete the protocol. For instance, $\squeezespaces{1}\ctail_2(\cparams, z_1) \triangleq \nw a.\cout{\ch}{\smult{a}{\pkg{c}}}.\ctail_3(\cparams, z_1, a)$. For terminal tails to be properly defined we include the output of the $\texttt{auth}$ message when the respective terminal accepts the transaction.

\begin{figure*}[t!]
	\captionsetup[subfigure]{labelformat=empty}
	\begin{subfigure}[t]{0.33\linewidth}
		\centering
		\squeezespaces{1}
		\begin{equation}
			\notag
			\scalebox{1}{\parbox{.5\linewidth}{\[
					\begin{aligned}
						%			& \text{\textcolor{blue}{$\ctail_0(\cparams)$}} \triangleq  \nu ch.\cout{term}{ch}.\ctail_1(\cparams)\\
						C(&\ch, c, pk_s, \texttt{vsig}_\texttt{MM}, \text{PAN}, \mk, \text{PIN}) \triangleq \\
						& \text{\textcolor{blue}{$\ctail_1(\cparams)$}} \\
						& \cin{\ch}{z_1}.\\
						& \text{\textcolor{blue}{$\ctail_2(\cparams, z_1)$}} \\
						& \nw a.\cout{\ch}{\smult{a}{\pkg{c}}}.\\
						& \text{\textcolor{blue}{$\ctail_3(\cparams, z_1, a)$}} \\
						& \lett \, \key{c} \coloneqq \hash{\smult{\mult{a}{c}}{z_1}} \inn \\
						& \cin{\ch}{m}.\\ 
						& \text{\textcolor{blue}{$\ctail_4(\cparams, \key{c}, a, m)$}} \\
						& \lett \, \nlist{\textsf{MC}, \textsf{MC}_s} \coloneqq \dec{m}{\key{c}} \inn \\
						& \ifff \ \checksig{pk_s}{\textsf{MC}_s} = \textsf{MC} \ \thenn \\
						& \ifff \ \proj{1}{\textsf{MC}} =\texttt{MM} \ \thenn \\
						& \cout{\ch}{\enc{\nlist{\smult{a}{\pkg{c}}, \smult{a}{\texttt{vsig}_\texttt{MM}}}}{\key{c}}}.\\
						& \text{\textcolor{blue}{$\ctail_5(\cparams, \key{c}, a, m)$}} \\
						& \cin{\ch}{x}.\\
						& \text{\textcolor{blue}{$\ctail_6(\cparams, \key{c}, a, m, x)$}} \\
						% low-value online, low-value offline, high-value online
						& \lett \, \nlist{\textsf{TX},\textsf{uPin}} \coloneqq \dec{x}{\key{c}} \inn\\
						& \lett \, \textsf{AC}^{\fail} \coloneqq \nlist{a, \text{PAN}, \textsf{TX}} \inn\\
						& \lett \, \textsf{AC}^{\texttt{ok}} \coloneqq \nlist{a, \text{PAN}, \textsf{TX}, \texttt{ok}} \inn\\
						& \lett \, \textsf{AC}^{\texttt{no}} \coloneqq \nlist{a, \text{PAN}, \textsf{TX}, \texttt{no}} \inn\\
						& \lett \, \key{cb} \coloneqq \hash{\smult{\mult{a}{c}}{\proj{3}{\proj{1}{\dec{m}{\key{c}}}}}} \inn \\
						& \ifff \ \textsf{uPin} = \fail \ \thenn \\
						& \cout{\ch}{\enc{\nlist{\enc{\nlist{\textsf{AC}^{\fail},\hash{\nlist{\textsf{AC}^{\fail}, \mk}}}}{\key{cb}}}, \fail, \textsf{TX}}{\key{c}}} \\
						& \ifff \ \textsf{uPin} =\text{PIN} \ \thenn \\
						& \cout{\ch}{\enc{\nlist{\enc{\nlist{\textsf{AC}^{\texttt{ok}},\hash{\nlist{\textsf{AC}^{\texttt{ok}}, \mk}}}}{\key{cb}}, \texttt{ok}, \textsf{TX}}}{\key{c}}} \\
						& \elsee \ \cout{\ch}{\enc{\nlist{\enc{\nlist{\textsf{AC}^{\texttt{no}},\hash{\nlist{\textsf{AC}^{\texttt{no}}, \mk}}}}{\key{cb}}, \texttt{no}, \textsf{TX}}}{\key{c}}}\\
						& \text{\textcolor{blue}{$\ctail_7 \triangleq 0$}} \\
					\end{aligned}
					\]}}
		\end{equation}
	\end{subfigure}
	\begin{subfigure}[t]{0.65\linewidth}
		\centering
		\squeezespaces{1}
		\begin{equation}
			\notag
			\scalebox{1}{\parbox{.5\linewidth}{\[
					\begin{aligned}
						B(&\ch, si, \kbt, b_t) \triangleq \\
						& \text{\textcolor{blue}{$\btail_1(\bparams)$}} \\
						& \cin{\ch}{x}.\\
						& \text{\textcolor{blue}{$\btail_2(\bparams, x)$}} \\
						& \lett \, dx \coloneqq \dec{x}{\kbt} \inn \\
						& \lett \, \text{PAN} \coloneqq \proj{2}{\proj{1}{\dec{\proj{3}{dx}}{\smult{b_t}{\proj{2}{dx}}}}} \inn \\
						& \cin{\nlist{si, \text{PAN}}}{y}. \\
						& \text{\textcolor{blue}{$\btail_3(\bparams, x, y)$}} \\
						& \lett \, \nlist{\textsf{TX}', z_2, \textsf{EAC}, \textsf{uPIN}} \coloneqq \dec{x}{\kbt} \\
						& \lett \, \key{bc} \coloneqq \hash{\smult{b_t}{z_2}} \inn \\
						& \lett \, \nlist{\textsf{AC}, \textsf{AC}_{hmac}} = \dec{\textsf{EAC}}{\key{bc}} \inn \\
						& \lett \, \nlist{\text{PIN}, mk, pk_{c}} \coloneqq y \\
						& \ifff \ \hash{\nlist{\textsf{AC}, mk}} = \textsf{AC}_{hmac} \ \thenn \\
						& \ifff \ \proj{3}{\textsf{AC}} = \textsf{TX}' \ \thenn \\
						& \ifff \ \smult{\proj{1}{\textsf{AC}}}{pk_{c}}= z_2 \\
						%			& \textcolor{blue}{?\nlist{\text{PAN}, \textsf{TX}, a} is \ unique ?}\\
						& \lett \, r \coloneqq \nlist{\textsf{TX}', \texttt{accept}} \inn \\
						% low-value offline, low-value online
						& \ifff \ \proj{2}{TX'} = \lo \ \thenn \\
						& \cout{\ch}{\enc{r}{\kbt}} \\
						& \ifff \ \proj{2}{\textsf{TX}'} = \hi \ \thenn \\
						% high-value offline		
						& \ifff \ \proj{4}{\textsf{AC}} = \texttt{ok} \ \thenn \\ 
						& \cout{\ch}{\enc{r}{\kbt}} \\
						% high-value online	
						& \elsee \ \ifff \ \textsf{uPIN} = \text{PIN} \ \thenn \\
						& \cout{\ch}{\enc{r}{\kbt}} \\
						& \text{\textcolor{blue}{$\btail_4 \triangleq 0 $}} \\
					\end{aligned}
					\]}}
		\end{equation}
	\end{subfigure}
	\caption{Subprocesses defining the execution stages in the $\protocol$ protocol for the card and the bank.} \label{fig:subproccardbank}
\end{figure*}

\begin{figure*}
		\begin{subfigure}[t]{0.33\linewidth}
		\centering
		\squeezespaces{1}
			\begin{equation}
			\notag
			\scalebox{1}{\parbox{.5\linewidth}{\[
					\begin{aligned}
						%			& \text{\textcolor{blue}{$\ttonhi_0(user, \tparams)$}} \triangleq  \nu ch.\cout{term}{ch}.\ttonhi_1(user, \tparams)\\
						T_{\texttt{onhi}}(&\user, \ch, pk_\texttt{MM}, \textsf{crt}, \kbt) \triangleq \\
						& \text{\textcolor{blue}{$\ttonhi_1(user, \tparams)$}} \\
						& \nw \, \mathsf{TXdata}. \lett \, \textsf{TX} \coloneqq \nlist{\mathsf{TXdata}, \hi} \inn \\
						& \nw t.\cout{\ch}{\pkg{t}}. \\
						& \text{\textcolor{blue}{$\ttonhi_2(user, \tparams, t, \textsf{TX})$}} \\
						& \cin{\ch}{z_2}. \\
						& \text{\textcolor{blue}{$\ttonhi_3(user, \tparams, t, \textsf{TX}, z_2)$}} \\
						& \lett \, \key{t} \coloneqq \hash{\smult{t}{z_2}} \inn \\
						& \cout{\ch}{\enc{\textsf{crt}}{\key{t}}}. \\
						& \text{\textcolor{blue}{$\ttonhi_4(user, \tparams, t, \textsf{TX}, z_2)$}} \\
						& \cin{\ch}{n}.\\
						& \text{\textcolor{blue}{$\ttonhi_5(user, \tparams, t, \textsf{TX}, z_2, n)$}} \\
						& \lett \, \nlist{\textsf{B}, \textsf{B}_s} \coloneqq \dec{n}{\key{t}} \inn \\
						& \ifff \ \checksigv{pk_\texttt{MM}}{\textsf{B}_s} = \textsf{B} \ \thenn \\
						& \ifff \ \textsf{B} = z_2 \ \thenn \\
						& \cin{\user}{\textsf{uPIN}}.\\
						& \text{\textcolor{blue}{$\ttonhi_6(user, \tparams, t, \textsf{TX}, z_2, n, \textsf{uPIN})$}} \\
						& \cout{\ch}{\enc{\nlist{\textsf{TX}, \fail}}{\key{t}}}. \\
						& \text{\textcolor{blue}{$\ttonhi_7(user, \tparams, t, \textsf{TX}, z_2, n, \textsf{uPIN})$}} \\
						& \cin{\ch}{y}.\\
						& \text{\textcolor{blue}{$\ttonhi_8(user, \tparams, t, \textsf{TX}, z_2, n, \textsf{uPIN}, y)$}} \\
						& \lett \, \nlist{\textsf{EHAC}, \textsf{pinV}, \textsf{tx}} \coloneqq \dec{y}{\key{t}} \inn \\
						& \ifff \ \textsf{tx} = \textsf{TX} \ \thenn \\
						& \cout{\ch}{\enc{\nlist{\textsf{TX}, z_2, \textsf{EHAC}, \textsf{uPIN}}}{\kbt}}. \\
						& \text{\textcolor{blue}{$\ttonhi_9(user, \tparams, t, \textsf{TX}, z_2, n, \textsf{uPIN}, y)$}} \\
						& \cin{\ch}{r}. \\
						& \text{\textcolor{blue}{$\ttonhi_{10}(user, \tparams, t, \textsf{TX}, z_2, n, \textsf{uPIN}, y, r)$}} \\
						& \ifff \ \proj{1}{\dec{r}{\kbt}} = \textsf{TX} \ \thenn \\
						& \ifff \ \proj{2}{\dec{r}{\kbt}} = \texttt{accept} \ \thenn \\
						&  \cout{\ch}{\texttt{auth}} \text{\textcolor{blue}{$\ttonhi_{11} \triangleq 0$}} \\
					\end{aligned}
					\]}}
		\end{equation}
	\end{subfigure}
	\begin{subfigure}[t]{0.33\linewidth}
		\centering
		\squeezespaces{1}
		\begin{equation}
			\notag
			\scalebox{1}{\parbox{.5\linewidth}{\[
					\begin{aligned}
						%			& \text{\textcolor{blue}{$\ttofhi_0(user, \tparams)$}} \triangleq  \nu ch.\cout{term}{ch}.\ttofhi_1(user, \tparams)\\
						T_{\texttt{offhi}}(&\user, \ch, pk_\texttt{MM}, \textsf{crt},  \kbt) \triangleq \\
						& \text{\textcolor{blue}{$\ttofhi_1(user, \tparams)$}} \\
						& \nw \, \mathsf{TXdata}. \lett \, \textsf{TX} \coloneqq \nlist{\mathsf{TXdata}, \hi} \inn \\
						& \nw t.\cout{\ch}{\pkg{t}}. \\
						& \text{\textcolor{blue}{$\ttofhi_2(user, \tparams, t, \textsf{TX})$}} \\
						& \cin{\ch}{z_2}. \\
						& \text{\textcolor{blue}{$\ttofhi_3(user, \tparams, t, \textsf{TX}, z_2)$}} \\
						& \lett \, \key{t} \coloneqq \hash{\smult{t}{z_2}} \inn \\
						& \cout{\ch}{\enc{\nlist{\textsf{crt}}}{\key{t}}}. \\
						& \text{\textcolor{blue}{$\ttofhi_4(user, \tparams, t, \textsf{TX}, z_2)$}} \\
						& \cin{\ch}{n}.\\
						& \text{\textcolor{blue}{$\ttofhi_5(user, \tparams, t, \textsf{TX}, z_2, n)$}} \\
						& \lett \, \nlist{\textsf{B}, \textsf{B}_s} \coloneqq \dec{n}{\key{t}} \inn \\
						& \ifff \ \checksigv{pk_\texttt{MM}}{\textsf{B}_s} = \textsf{B} \ \thenn \\
						& \ifff \ \textsf{B} = z_2 \ \thenn \\
						& \cin{\user}{\textsf{uPIN}}.\\
						& \text{\textcolor{blue}{$\ttofhi_6(user, \tparams, t, \textsf{TX}, z_2, n, \textsf{uPin})$}} \\
						& \cout{\ch}{\enc{\nlist{\textsf{TX}, \textsf{uPin}}}{\key{t}}}. \\
						& \text{\textcolor{blue}{$\ttofhi_7(user, \tparams, t, \textsf{TX}, z_2, n, \textsf{uPin})$}} \\
						& \cin{\ch}{y}.\\
						& \text{\textcolor{blue}{$\ttofhi_8(user, \tparams, t, \textsf{TX}, z_2, n, \textsf{uPin}, y)$}} \\
						& \lett \, \nlist{\textsf{EHAC}, \textsf{pinV}, \textsf{tx}} \coloneqq \dec{y}{\key{t}} \inn \\
						& \ifff \ \textsf{tx} = \textsf{TX}\ \thenn \\
						& \ifff \ \textsf{pinV} = \texttt{ok} \ \thenn \\
						& \cout{\ch}{\texttt{auth}}. \\
						& \text{\textcolor{blue}{$\ttofhi_9(user, \tparams, t, \textsf{TX}, z_2, n, \textsf{uPin}, y)$}} \\
						& \cout{\ch}{\enc{\nlist{\textsf{TX}, z_2, \textsf{HEAC}, \fail}}{\kbt}} \\
						& \text{\textcolor{blue}{$\ttofhi_{10}(user, \tparams, t, \textsf{TX}, z_2, n, \textsf{uPin}, y)$}} \\
						& \ifff \ \textsf{pinV} \neq \texttt{ok} \ \thenn \\
						& \cout{ch}{\enc{\nlist{\textsf{TX}, z_2, \textsf{HEAC}, \fail}}{\kbt}} \text{\textcolor{blue}{$\ttofhi_{11} \triangleq 0$}} 
					\end{aligned}
					\]}}
		\end{equation}
	\end{subfigure}
	\begin{subfigure}[t]{0.33\linewidth}
		\centering
		\squeezespaces{1}
			\begin{equation}
			\notag
			\scalebox{1}{\parbox{.5\linewidth}{\[
					\begin{aligned}
						%			& \text{\textcolor{blue}{$\ttlo_0(\tparams)$}} \triangleq  \nu ch.\cout{term}{ch}.\ttlo_1(\tparams)\\
						T_{\texttt{lo}} (&\ch, pk_\texttt{MM}, \textsf{crt}, \kbt) \triangleq \\
						& \text{\textcolor{blue}{$\ttlo_1(\tparams)$}} \\
						& \nw \, \mathsf{TXdata}. \lett \, \textsf{TX} \coloneqq \nlist{\mathsf{TXdata}, \lo} \inn \\
						& \nw t.\cout{\ch}{\pkg{t}}. \\
						& \text{\textcolor{blue}{$\ttlo_2(\tparams, t, \textsf{TX})$}} \\
						& \cin{\ch}{z_2}. \\
						& \text{\textcolor{blue}{$\ttlo_3(\tparams, t,\textsf{TX}, z_2)$}} \\
						& \lett \, \key{t} \coloneqq \hash{\smult{t}{z_2}} \inn \\
						& \cout{\ch}{\enc{\textsf{crt}}{\key{t}}}. \\
						& \text{\textcolor{blue}{$\ttlo_4(\tparams, t, \textsf{TX}, z_2)$}} \\
						& \cin{\ch}{n}.\\
						& \text{\textcolor{blue}{$\ttlo_5(\tparams, t, \textsf{TX}, z_2, n)$}} \\
						& \lett \, \nlist{\textsf{B}, \textsf{B}_s} \coloneqq \dec{n}{\key{t}} \inn \\
						& \ifff \ \checksigv{pk_\texttt{MM}}{\textsf{B}_s} = \textsf{B} \ \thenn \\
						& \ifff \ \textsf{B} = z_2 \ \thenn \\
						& \cout{\ch}{\enc{\nlist{\textsf{TX}, \fail}}{\key{t}}}. \\
						& \text{\textcolor{blue}{$\ttlo_6(\tparams, t, \textsf{TX}, z_2, n)$}} \\
						& \cin{\ch}{y}.\\
						& \text{\textcolor{blue}{$\ttlo_7(\tparams, t, \textsf{TX}, z_2, n, y)$}} \\
						& \lett \, \nlist{\textsf{EHAC}, \textsf{pinV}, \textsf{tx}} \coloneqq \dec{y}{\key{t}} \inn \\
						& \ifff \ \textsf{tx} = \textsf{TX}\ \thenn \\
						& \cout{\ch}{\enc{\nlist{\textsf{TX}, z_2, \textsf{EHAC}, \fail}}{\kbt}}. \\
						& \text{\textcolor{blue}{$\ttlo_8(\tparams, t, \textsf{TX}, z_2, n, y)$}} \\
						& \cin{\ch}{r}. \\
						& \text{\textcolor{blue}{$\ttlo_9(\tparams, t, \textsf{TX}, z_2, n, y, r)$}} \\
						& \ifff \ \proj{1}{\dec{r}{\kbt}} = \textsf{TX} \ \thenn \\
						& \ifff \ \proj{2}{\dec{r}{\kbt}} = \texttt{accept} \ \thenn \\
						& \cout{\ch}{\texttt{auth}} \text{\textcolor{blue}{$\ttlo_{10} \triangleq 0$}} \\
					\end{aligned}
					\]}}
		\end{equation}
	\end{subfigure}
\caption{Subprocesses defining the execution stages in the $\protocol$ protocol for the terminal.} \label{fig:subprocterminal}
\end{figure*}

The idea behind forming $\mathfrak{R}$ is to pair all \emph{reachable} states based on the number of sessions, yet ignoring the number of existing cards. If not specified otherwise, below we talk about \emph{started} sessions, i.e. the ones with the announced channel, hence we define the following sets of sessions: $\squeezespaces{1}\mathcal{D} \coloneqq \{1 \hdots D \}$ for cards, $\squeezespaces{1}\mathcal{FG} \coloneqq \{1 \hdots F + G \}$ for terminals, and $\squeezespaces{1}\mathcal{FM} \coloneqq \{1 \hdots F + M \}$ for the bank, where $F$ is the number of bank-terminal sessions with a shared secret key $\kbt$. 
These reachable states are defined by partitions of the session sets, where the element of the partition defines all sessions at a certain stage. We consider the following partitions. $\squeezespaces{1}A \coloneqq \{\alpha_1 \hdots \alpha_7\}$ of $\mathcal{D}$,  $\squeezespaces{1}\Gamma \coloneqq \{\gamma^{\on}_1 \hdots \gamma^{\on}_{11}, \gamma^{\of}_1 \hdots \gamma^{\of}_11, \gamma^{\lo}_1 \hdots \gamma^{\lo}_{10}\}$ of $\mathcal{FG}$, and $\squeezespaces{1}B \coloneqq \{\beta_1 \hdots \beta_4\}$ of $\mathcal{FM}$. Here, e.g. $\alpha_2$ defines all sessions where the actions left are defined by the process of the form $\ctail_2(\cdot)$, or, similarly, $\gamma^{\of}_5$ defines all offline high-value terminal sessions where the actions left are defined by $\ttofhi_5(\cdot)$.

We also define the list of global parameters $\vec{\epsilon} \coloneqq (user, s, si, \chi_{\texttt{MM}})$, introduce the shorthand for different transaction types $\textsf{TX}_i \coloneqq \nlist{\mathsf{TXdata}_i, \lo} \ \text{or} \ \nlist{\mathsf{TXdata}_i, \hi}$, and define the numbers $E \coloneqq \bigcup_{i=3}^{7}\alpha_i$ and $L \coloneqq \bigcup_{i=2}^{11}\gamma^{\on}_i \cup \bigcup_{i=2}^{11}\gamma^{\of}_i \cup \bigcup_{i=2}^{10}\gamma^{\lo}_i$ standing for the number of session where fresh blinding factor $a_i$ or fresh ephemeral private key $t_i$ has already been generated. 

The following processes are also required to make the definition of $\mathfrak{R}$ less bulky. 
\[
	\begin{aligned}
	&\spec{PC} \triangleq \nu \text{PIN}, mk, c, \text{PAN}.( \\
	&
	\begin{array}{l}
		\nw \ch.\cout{card}{ch}.C(ch, c, \pk{s}, \sigv{\chi_{\texttt{MM}}}{\pkg{c}}, \text{PAN}, mk, \text{PIN}) \cpar \\
		\bang \cout{user}{\text{PIN}} \cpar \\
		\bang	\cout{\nlist{si, \text{PAN}}}{\nlist{\text{PIN}, mk, \pkg{c}}})
	\end{array}
\end{aligned}
\]
\[
\begin{aligned}
	& \impl{PC} \triangleq \ \nu \text{PIN}, mk, c, \text{PAN}.\bang( \\
	&
 \begin{array}{l}
		\nw \ch.\cout{card}{\ch}.C(\ch, c, \pk{s}, \sigv{\chi_{\texttt{MM}}}{\pkg{c}}, \text{PAN}, mk, \text{PIN}) \cpar \\
		\cout{\user}{\text{PIN}} \cpar \\
		\cout{\nlist{si, \text{PAN}}}{\nlist{\text{PIN}, mk, \pkg{c}}})
	\end{array}
\end{aligned}
\]
\[
\begin{aligned}
	\textit{PBT} \triangleq \ \nu b_t.\bang \nu &\kbt.( \\
	&
	\begin{array}{l}
		\nu \ch.\cout{bank}{\ch}.B(ch, si, \ch, b_t) \cpar \\ 
		\lett \, \textsf{crt} \coloneqq \nlist{\nlist{\texttt{MM}, \pkg{b_t}}, \sig{s}{\nlist{\texttt{MM}, \pkg{b_t}}}} \inn \\
		\nu \ch.\cout{term}{\ch}.T_{\texttt{onhi}}(\user, \ch, \pkv{\chi_{\texttt{MM}}}, \textsf{crt}, \kbt) \ + \\ \nu \ch.\cout{term}{\ch}.T_{\texttt{offhi}}(\user, \ch, \pkv{\chi_{\texttt{MM}}}, \textsf{crt}, \kbt) \ +  \\ \nu \ch.\cout{term}{\ch}.T_{\texttt{lo}}(\ch, \pkv{\chi_{\texttt{MM}}}, \textsf{crt}, \kbt))
	\end{array}
\end{aligned}
\]

%\begin{equation}
%	\notag
%	\begin{split}
%		& \begin{aligned}
%			\spec{PC} \triangleq & \nu \text{PIN}, mk, c, \text{PAN}.( \\
%			&
%			\quad \begin{array}{l}
%			\nw \ch.\cout{card}{ch}.C(ch, c, \pk{s}, \sigv{\chi_{\texttt{MM}}}{\pkg{c}}, \text{PAN}, mk, \text{PIN}) \cpar \\
%			 \bang \cout{user}{\text{PIN}} \cpar \\
%			 \bang	\cout{\nlist{si, \text{PAN}}}{\nlist{\text{PIN}, mk, \pkg{c}}})
%			\end{array}
%		\end{aligned}
%%		\\[0.01ex]
%\\[10pt]
%		& \begin{aligned}
%			\impl{PC} \triangleq \ & \nu \text{PIN}, mk, c, \text{PAN}.\bang( \\
%			&
%			\quad \begin{array}{l}
%			\nw \ch.\cout{card}{\ch}.C(\ch, c, \pk{s}, \sigv{\chi_{\texttt{MM}}}{\pkg{c}}, \text{PAN}, mk, \text{PIN}) \cpar \\
%			 \cout{\user}{\text{PIN}} \cpar \\
%		     \cout{\nlist{si, \text{PAN}}}{\nlist{\text{PIN}, mk, \pkg{c}}})
%		     \end{array}
%		\end{aligned}
%\\[10pt]
%	& \begin{aligned}
%		\textit{PBT} \triangleq \ & \nu b_t.\bang \nu \kbt.( \\
%		&
%		\quad \begin{array}{l}
%			\nu \ch.\cout{bank}{\ch}.B(ch, si, \ch, b_t) \cpar \\ 
%			\lett \, \textsf{crt} \coloneqq \nlist{\nlist{\texttt{MM}, \pkg{b_t}}, \sig{s}{\nlist{\texttt{MM}, \pkg{b_t}}}} \inn \\
%			\nu \ch.\cout{term}{\ch}.T_{\texttt{onhi}}(\user, \ch, \pkv{\chi_{\texttt{MM}}}, \textsf{crt}, \kbt) \ + \\ \nu \ch.\cout{term}{\ch}.T_{\texttt{offhi}}(\user, \ch, \pkv{\chi_{\texttt{MM}}}, \textsf{crt}, \kbt) \ +  \\ \nu \ch.\cout{term}{\ch}.T_{\texttt{lo}}(\ch, \pkv{\chi_{\texttt{MM}}}, \textsf{crt}, \kbt))
%		\end{array}
%	\end{aligned}
%	\end{split}
%\end{equation}

We define the processes corresponding to different stages of the bank's and card's executions, and the processes corresponding to the user entering the PIN, and the bank looking up for the card's details. In what follows $k^c_i(a, x) \coloneqq  \hash{\smult{\mult{a}{c_i}}{x}}$. Also, let $K$ be the total number of card sessions for which the channel is not yet announced in the spec world, $H$ be the number of cards in the impl world, the list $\squeezespaces{1}\vec{K} \coloneqq \{K_1, \hdots, K_H\}$ be s.t. $K_h$ defines the number of sessions with the card $h$ for which the channel is not yet announced in the impl world, $\squeezespaces{1}\Lambda \coloneqq \{\lambda_1 \hdots \lambda_H\}$ be the partition of $\mathcal{D}$, where $\lambda_h$ is the set of sessions with the card $h$ in the impl world, and finally $D_h \coloneqq |\lambda_h|$. To keep track of inputs in the card, bank, and terminal sessions we use matricies $\underset{D \times 3}{X}$, $\underset{G \times 1}{Y}$ and $\underset{F \times 4}{Z}$ respectively; the element $(i, j)$ defines $j$th input in the $i$th session.
\[
B^\rho_{i} \triangleq 
\left\{
\begin{array}{lr}
	\nu ch.\cout{bank}{ch}.B(ch, si, \kbt_i, b_t) & \mbox{if $i \in \mathcal{FG}\setminus\mathcal{FM}$}
	\\
	\btail_1(\bparams_i) & \mbox{if $i \in \beta_1$}
	\\
	\btail_2(\bparams_i, Y_1^i\rho) & \mbox{if $i \in \beta_2$}
	\\
	\btail_3(\bparams_i, Y_1^i\rho, \text{DB}) & \mbox{if $i \in \beta_3$}
	\\
	\btail_4 & \mbox{if $i \in \beta_4$}		
\end{array}
\right.
\]
\[
C_i \triangleq 
\left\{
\begin{array}{lr}
	\parbox[r]{4.0cm}{$\nw ch.\cout{card}{ch}. \\ C(ch, c_i, \pk{s}, \sigv{\chi_{\texttt{MM}}}{\pkg{c}}, \text{PAN}_i, mk_i, \text{PIN}_i)$} & \mbox{\begin{minipage}{2cm}\vspace{-7pt}if $i  \notin \mathcal{D}$, $i \leq K$\end{minipage}} 
	\\
	\ctail_1(\cparams_i) & \mbox{if $i  \in \alpha_1$} 
	\\
	\ctail_2(\cparams_i, X^1_i\sigma) & \mbox{if $i  \in \alpha_2 $} 
	\\
	\ctail_3(\cparams_i, X^1_i\sigma, a_i) & \mbox{if $i  \in \alpha_3 $}
	\\
	\ctail_4(\cparams_i, k^c_i(a_i, X^1_i\sigma), a_i, X^2_i\sigma) & \mbox{if $i  \in \alpha_4 $}
	\\
	\ctail_5(\cparams_i, k^c_i(a_i, X^1_i\sigma), a_i, X^2_i\sigma) & \mbox{if $i  \in \alpha_5 $}
	\\
	\ctail_6(\cparams_i, k^c_i(a_i, X^1_i\sigma), a_i, X^2_i\sigma, X^3_i\sigma) & \mbox{if $i  \in \alpha_6$}
	\\
	\ctail_7 & \mbox{if $i  \in \alpha_7$}	
\end{array}
\right.
\]
\[
C^j_i \triangleq 
\left\{
\begin{array}{lr}
	\parbox[r]{4.0cm}{$\nw ch.\cout{card}{ch}. \\ C(ch, c_j, \pk{s}, \sigv{\chi_{\texttt{MM}}}{\pkg{c}}, \text{PAN}_j, mk_j, \text{PIN}_j)$} & \mbox{\begin{minipage}{2cm}\vspace{-7pt}if $i  \notin \mathcal{D}$, $i \leq K_j$\end{minipage}}
	\\
	\ctail_1(\cparams_j) & \mbox{if $i  \in \alpha_1 \cap \lambda_j$} 
	\\
	\ctail_2(\cparams_j, X^1_i\theta) & \mbox{if $i  \in \alpha_2 \cap \lambda_j $} 
	\\
	\ctail_3(\cparams_j, X^1_i\theta, a_i) & \mbox{if $i  \in \alpha_3 \cap \lambda_j$}
	\\
	\ctail_4(\cparams_j, k^c_j(a_i, X^1_i\theta), a_i, X^2_i\theta) & \mbox{if $i  \in \alpha_4 \cap \lambda_j$}
	\\
	\ctail_5(\cparams_j, k^c_j(a_i, X^1_i\theta), a_i, X^2_i\theta) & \mbox{if $i  \in \alpha_5 \cap \lambda_j$}
	\\
	\ctail_6(\cparams_j, k^c_j(a_i, X^1_i\theta), a_i, X^2_i\theta, X^3_i\theta) & \mbox{if $i  \in \alpha_6 \cap \lambda_j$}
	\\
	\ctail_7 & \mbox{if $i  \in \alpha_7 \cap \lambda_j$}	
\end{array}
\right. 
\]
\[
	U^j_i \triangleq 
\left\{
\begin{array}{ll}
	0 & \parbox[r]{4.6cm}{if $j \leq H$ and $\exists l \in \bigcup_{t=6}^{11}\gamma^{\on}_t$ or $ l \in \bigcup_{t=6}^{9}\gamma^{\of}_t$ s.t. $\text{PIN}_j$ is consumed in session $l$}
	\\
	\cout{user}{\text{PIN}_j} & \mbox{if $j \leq H$ and else}	
\end{array}
\right.
\]
\[
DB^j_i \triangleq 
\left\{
\begin{array}{ll}
	0 & \hspace{-3.0cm}\parbox[r]{4.6cm}{if $j \leq H$ and $\exists l \in \beta_3 \cup \beta_4$, s.t. parameters of the card $j$ are consumed in the session $l$}
	\\
	\cout{\nlist{si, \text{PAN}_j}}{\nlist{\text{PIN}_j, mk_j, \pkg{c_j}}} & \hspace{-0.2cm}\mbox{if $j \leq H$ and else}	
\end{array}
\right.
\]

We define the processes corresponding to different stages of the terminal's execution. Firstly, if $i \in \mathcal{FM}\setminus\mathcal{FG}$.
\[
T^\rho_i =
\begin{array}{lr}
	\begin{array}{l} 
		\nu ch.\cout{term}{ch}.T_{\texttt{onhi}}(user, ch, \pkv{\chi_{\texttt{MM}}}, \textsf{crt}, \kbt_i) \, + \\
		\nu ch.\cout{term}{ch}.T_{\texttt{offhi}}(user, ch, \pkv{\chi_{\texttt{MM}}}, \textsf{crt}, \kbt_i) \, + \\
		\nu ch.\cout{term}{ch}.T_{\texttt{lo}}(ch, \pkv{\chi_{\texttt{MM}}}, \textsf{crt}, \kbt)_i 
	\end{array}
\end{array}
\]
\[
\arraycolsep=1.4pt
\squeezespaces{0.9}
T^\rho_{i} \triangleq 
	\left\{
	\begin{array}{lr}
		\ttonhi_1(user, \tparams_i) & \mbox{if $i \in \gamma^{\on}_1$}
		\\
		\ttonhi_2(user, \tparams_i, t_i, \text{TX}_i) & \mbox{if $i \in \gamma^{\on}_2$}
		\\
		\ttonhi_3(user, \tparams_i, t_i, \text{TX}_i, Z^1_i\rho) & \mbox{if $i \in \gamma^{\on}_3$}
		\\
		\ttonhi_4(user, \tparams_i, t_i, \text{TX}_i, Z^1_i\rho) & \mbox{if $i \in \gamma^{\on}_4$}
		\\
		\ttonhi_5(user, \tparams_i, t_i, \text{TX}_i, Z^1_i\rho, Z^2_i\rho) & \mbox{if $i \in \gamma^{\on}_5$}
		\\
		\ttonhi_6(user, \tparams_i, t_i, \text{TX}_i, Z^1_i\rho, Z^2_i\rho, \text{uPIN}) & \mbox{if $i \in \gamma^{\on}_6 $}
		\\
		\ttonhi_7(user, \tparams_i, t_i, \text{TX}_i, Z^1_i\rho, Z^2_i\rho, \text{uPIN}) & \mbox{if $i \in \gamma^{\on}_7$}
		\\
		\ttonhi_8(user, \tparams_i, t_i, \text{TX}_i, Z^1_i\rho, Z^2_i\rho, \text{uPIN}, Z^3_i\rho) & \mbox{if $i \in \gamma^{\on}_8$}
		\\
		\ttonhi_9(user, \tparams_i, t_i, \text{TX}_i, Z^1_i\rho, Z^2_i\rho, \text{uPIN}, Z^3_i\rho) & \mbox{if $i \in \gamma^{\on}_9$}
		\\
		\ttonhi_{10}(user, \tparams_i, t_i, \text{TX}_i, Z^1_i\rho, Z^2_i\rho, \text{uPIN}, Z^3_i\rho, \text{R}) & \mbox{if $i \in \gamma^{\on}_{10}$}
		\\
		\ttonhi_{11} & \mbox{if $i \in \gamma^{\on}_{11}$}
	\end{array}
	\right.
\]
\[
\arraycolsep=1.4pt
\squeezespaces{0.9}
T^\rho_{i} \triangleq 
\left\{
\begin{array}{lr}
	\ttofhi_1(user, \tparams_i) & \mbox{if $i \in \gamma^{\of}_1$}
	\\
	\ttofhi_2(user, \tparams_i, t_i, \text{TX}_i) & \mbox{if $i \in \gamma^{\of}_2$}
	\\
	\ttonhi_3(user, \tparams_i, t_i, \text{TX}_i, Z^1_i\rho) & \mbox{if $i \in \gamma^{\of}_3$}
	\\
	\ttonhi_4(user, \tparams_i, t_i, \text{TX}_i, Z^1_i\rho) & \mbox{if $i \in \gamma^{\of}_4$}
	\\
	\ttonhi_5(user, \tparams_i, t_i, \text{TX}_i, Z^1_i\rho, Z^2_i\rho) & \mbox{if $i \in \gamma^{\of}_5$}
	\\
	\ttonhi_6(user, \tparams_i, t_i, \text{TX}_i, Z^1_i\rho, Z^2_i\rho, \text{uPIN}) & \mbox{if $i \in \gamma^{\of}_6$}
	\\
	\ttonhi_7(user, \tparams_i, t_i, \text{TX}_i, Z^1_i\rho, Z^2_i\rho, \text{uPIN}) & \mbox{if $i \in \gamma^{\of}_7$}
	\\
	\ttonhi_8(user, \tparams_i, t_i, \text{TX}_i, Z^1_i\rho, Z^2_i\rho, \text{uPIN}, Z^3_i\rho) & \mbox{if $i \in \gamma^{\of}_8$}
	\\
	\ttonhi_9(user, \tparams_i, t_i, \text{TX}_i, Z^1_i\rho, Z^2_i\rho, \text{uPIN}, Z^3_i\rho) & \mbox{if $i \in \gamma^{\of}_9$}
	\\
	\ttonhi_{10}(user, \tparams_i, t_i, \text{TX}_i, Z^1_i\rho, Z^2_i\rho, \text{uPIN}, Z^3_i\rho) & \mbox{if $i \in \gamma^{\of}_{10}$}
	\\
	\ttonhi_{11} & \mbox{if $i \in \gamma^{\of}_{11}$}
\end{array}
\right.
\]
\[
\arraycolsep=1.4pt
\squeezespaces{0.9}
T^\rho_{i} \triangleq 
\left\{
\begin{array}{lr}
	\ttlo_1(\tparams_i) & \mbox{if $i \in \gamma^{\lo}_1$}
	\\
	\ttlo_2(\tparams_i, t_i, \text{TX}_i) & \mbox{if $i \in \gamma^{\lo}_2$}
	\\
	\ttlo_3(\tparams_i, t_i, \text{TX}_i, Z^1_i\rho) & \mbox{if $i \in \gamma^{\lo}_3$}
	\\
	\ttlo_4(\tparams_i, t_i, \text{TX}_i, Z^1_i\rho) & \mbox{if $i \in \gamma^{\lo}_4$}
	\\
	\ttlo_5(\tparams_i, t_i, \text{TX}_i, Z^1_i\rho, Z^2_i\rho) & \mbox{if $i \in \gamma^{\lo}_5$}
	\\
	\ttlo_6(\tparams_i, t_i, \text{TX}_i, Z^1_i\rho, Z^2_i\rho) & \mbox{if $i \in \gamma^{\lo}_6$}
	\\
	\ttlo_7(\tparams_i, t_i, \text{TX}_i, Z^1_i\rho, Z^2_i\rho, Z^3_i\rho) & \mbox{if $i \in \gamma^{\lo}_7$}
	\\
	\ttlo_8(\tparams_i, t_i, \text{TX}_i, Z^1_i\rho, Z^2_i\rho, Z^3_i\rho) & \mbox{if $i \in \gamma^{\lo}_8$}
	\\
	\ttlo_9(\tparams_i, t_i, \text{TX}_i, Z^1_i\rho, Z^2_i\rho, Z^3_i\rho, \text{R}) & \mbox{if $i \in \gamma^{\lo}_9$}
	\\
	\ttlo_{10} & \mbox{if $i \in \gamma^{\lo}_{10}$}
\end{array}
\right.
\]

Using the notation introduced above we define the relation $\mathfrak{R}$ as the least symmetric open relation satisfying the conditions in Fig.~\ref{fig:relationutx}, where the generic states in the spec world $\specc{(K, F, A, \Gamma, B)}(X,Y,Z)$ and the generic state in the impl world $\impll{(\vec{K}, F, A, \Gamma, B, \Lambda)}(X,Y,Z)$ are defined in Fig.~\ref{fig:genspec},~\ref{fig:genimpl} respectively. Notice that the generic impl state is additionally parametrised by the partition $\Lambda$ which elements track all sessions with a particular card.

\begin{figure}[h!]
		\[\arraycolsep=4.3pt
		\squeezespaces{1}
	\begin{array}{rcl}
			\spec{UTX} &  \mathfrak{R} & \impl{UTX} 
			\\[16pt]
			\begin{array}{r} 
				\spec{UTX}^1 \ \triangleq \\
				\nw \vec{\epsilon}.(\sub{pk_s}{\pks{s}} \cpar \\
				\cout{out}{\pkv{\chi_{\texttt{MM}}}}. \\ (\bang \spec{PC} \cpar \nw b_t.\bang \textit{PBT}))
			\end{array} 
			& \mathfrak{R} &
			\begin{array}{l}
				\impl{UTX}^1 \ \triangleq \\
				\nw \vec{\epsilon}.(\sub{pk_s}{\pks{s}} \cpar \\
				\cout{out}{\pkv{\chi_{\texttt{MM}}}}. \\ (\bang \impl{PC} \cpar \nw b_t.\bang \textit{PBT}))
			\end{array} 
			\\[32pt]
			\begin{array}{r}
			\spec{UTX}^2 \ \triangleq \\ \squeezespaces{0.9}{\nw \vec{\epsilon}.(\sigma_0 \cpar \bang \spec{PC} \cpar \nw b_t.\bang \textit{PBT})}
			\end{array} & \mathfrak{R} & \begin{array}{l} \impl{UTX}^2 \ \triangleq \\ \squeezespaces{0.9}{\nw \vec{\epsilon}.(\sigma_0 \cpar \bang \impl{PC} \cpar \nw b_t.\bang \textit{PBT})} \end{array}
			\\[16pt]
		\squeezespaces{0.9}{\specc{(K, F, A, \Gamma, B)}(X,Y,Z)} & \mathfrak{R} & \squeezespaces{0.9}{\impll{(\vec{K}, F, A, \Gamma, B, \Lambda)}(X,Y,Z)}
\end{array}
		\]
	\caption{Defining conditions for the relation $\mathfrak{R}$.}
	\label{fig:relationutx}
\end{figure}

\begin{figure}[h!]
	\[
	\begin{array}{l}
		\specc{(K, F, A, \Gamma, B)}(X,Y,Z) \triangleq \\
		\begin{array}{l}
			\mathopen{\nu \, \vec{\epsilon}, \text{PIN}_{1 \hdots D+K}, mk_{1 \hdots D+K}, c_{1 \hdots D+K}, \text{PAN}_{1 \hdots D+K}}, 
			\\[4pt]
			\dot{ch}_{1 \hdots D}, a_{1 \hdots E}, b_t, 
%			\ch_{1 \hdots F+G+M}, 
			\ddot{ch}_{1 \hdots F+G},
			\\[4pt]
			\dddot{ch}_{1 \hdots F+M}, t_{1 \hdots L},  \textsf{TX}_{1 \hdots L}.(~\sigma \cpar  
			\\[4pt]
			\;\;
			\begin{array}[t]{l}
				C_1 \cpar \hdots \cpar 0 \cpar \bang \cout{user}{\text{PIN}_1} \cpar
				\\[4pt]
				\quad \hdots \cpar 0 \cpar \bang 	\cout{\nlist{si, \text{PAN}_1}}{\nlist{\text{PIN}_1, mk_1, \pkg{c_1}}}) \cpar
				\\[4pt]
				\hdots
				\\[4pt]
				C_i \cpar \hdots \cpar 0 \cpar \bang \cout{user}{\text{PIN}_i} \cpar
				\\[4pt]
				\quad \hdots \cpar 0 \cpar \bang 	\cout{\nlist{si, \text{PAN}_i}}{\nlist{\text{PIN}_i, mk_i, \pkg{c_i}}}) \cpar
				\\[4pt]
				\hdots
				\\[4pt]
				C_{D+K} \cpar \hdots \cpar 0 \cpar \bang \cout{user}{\text{PIN}_{D+K}} \cpar
				\\[4pt]
				\quad \hdots \cpar 0 \cpar \bang 	\cout{\nlist{si, \text{PAN}_{D+K}}}{\nlist{\text{PIN}_{D+K}, mk_{D+K}, \pkg{c_{D+K}}}}) \cpar
				\\[4pt]
				\bang \spec{PC} \cpar
				\\[4pt]
				B^\sigma_1 \cpar T^\sigma_1 \cpar 
				\\[4pt]
				\hdots \cpar 
				\\[4pt]
				B^\sigma_j \cpar T^\sigma_j \cpar 
				\\[4pt]
				\hdots \cpar 
				\\[4pt]
				B^\sigma_{F+G+M} \cpar T^\sigma_{F+G+M} \cpar \bang \textit{PBT})
			\end{array}
		\end{array}
	\end{array}
	\]
	\caption{The generic state in the spec world.}\label{fig:genspec}
\end{figure}

\begin{figure}[h!]
	\[
	\begin{array}{l}
		\impll{(\vec{K}, F, A, \Gamma, B, \Lambda)}(X,Y,Z) \triangleq \\
		\begin{array}{l}
			\mathopen{\nu \vec{\epsilon}, \text{PIN}_{1 \hdots H}, mk_{1 \hdots H}, c_{1 \hdots H}, \text{PAN}_{1 \hdots H}},
			\dot{ch}_{1 \hdots D}, 
			\\[3pt]
			a_{1 \hdots E}, b_t,  
%			\ch_{1 \hdots F+G+M}, 
			\ddot{ch}_{1 \hdots F+G}, \dddot{ch}_{1 \hdots F+M}
			\\[3pt]
			t_{1 \hdots L},  \textsf{TX}_{1 \hdots L}.(~\theta \cpar 
			\\
			\begin{array}[t]{l}
				C^1_1 \cpar U^1_1 \cpar \textit{DB}^1_1 \cpar  
				\\[-0pt]
				\hdots
				\\[-0pt]
				C^1_{i_1} \cpar U^1_{i_1} \cpar \textit{DB}^1_{i_1} \cpar
				\\[-0pt]
				\hdots
				\\[-0pt]
				C^1_{D_1+K_1} \cpar U^1_{D_1+K_1} \cpar \textit{DB}^1_{D_1+K_1} \cpar
				\\[3pt]
				\bang(\nw ch.\cout{card}{ch}. 
				\\ 
				\begin{array}{l}
				C(ch, c_j, \pk{s}, \sigv{\chi_{\texttt{MM}}}{\pkg{c}}, \text{PAN}_j, mk_j, \text{PIN}_j) \cpar 
				\\[3pt]
				\cout{user}{\text{PIN}_1} \cpar \textit{DB}(si, \text{PAN}_1, mk_1, \text{PIN}_1)) \cpar \end{array}
				\\[-0pt]
				\hdots
				\\[-0pt]
				C^h_{D_{h-1} + K_{h-1}+1} \cpar U^h_{D_{h-1} + K_{h-1}+1} \cpar \textit{DB}^h_{D_{h-1} + K_{h-1} +1} \cpar  
				\\[-0pt]
				\hdots
				\\[-0pt]
				C^h_{i_h} \cpar U^h_{i_h} \cpar \textit{DB}^h_{i_h} \cpar
				\\[-0pt]
				\hdots
				\\[-0pt]
				C^h_{D_{h-1} + K_{h-1} + D_h+K_h} \cpar U^h_{D_{h-1} + K_{h-1} + D_h+K_h} \cpar 
				\\[3pt]
				\quad \textit{DB}^h_{D_{h-1} + K_{h-1} + D_h+K_h} \cpar
				\\[3pt]
				\bang(\nw ch.\cout{card}{ch}. 
				\\ 
				\begin{array}{l} C(ch, c_h, \pk{s}, \sigv{\chi_{\texttt{MM}}}{\pkg{c}}, \text{PAN}_h, mk_h, \text{PIN}_h) \cpar 
				\\[3pt]
				\cout{user}{\text{PIN}_h} \cpar \textit{DB}(si, \text{PAN}_h, mk_h, \text{PIN}_h)) \cpar \end{array}
				\\[-0pt]
				\hdots
				\\[-0pt]
				C^H_{D_{H-1}+K_{H-1}+1} \cpar U^H_{D_{H-1}+K_{H-1}+1} \cpar \textit{DB}^H_{D_{H-1}+K_{H-1}+1} \cpar 
				\\[-0pt]
				\hdots
				\\[-0pt]
				C^H_{i_H} \cpar U^H_{i_H} \cpar \textit{DB}^H_{i_H} \cpar
				\\[-0pt]
				\hdots
				\\[-0pt]
				C^H_{D_{H-1}+K_{H-1} + D_H+K_H} \cpar U^H_{D_{H-1}+K_{H-1} + D_H+K_H} \cpar
				\\[3pt]
				\quad  \textit{DB}^H_{D_{H-1}+K_{H-1} + D_H+K_H} \cpar
				\\[3pt]
				\bang(\nw ch.\cout{card}{ch}. 
				\\ 
				\begin{array}{l}
				C(ch, c_H, \pk{s}, \sigv{\chi_{\texttt{MM}}}{\pkg{c}}, \text{PAN}_H, mk_H, \text{PIN}_H) \cpar 
				\\[3pt]
				\cout{user}{\text{PIN}_H} \cpar \textit{DB}(si, \text{PAN}_H, mk_H, \text{PIN}_H)) \cpar \end{array}
				\\[3pt]
				\bang \impl{PC} \cpar
				\\[3pt]
				B^\theta_1 \cpar T^\theta_1 \cpar 
				\\[-0pt]
				\hdots \cpar 
				\\[-0pt]
				B^\theta_j \cpar T^\theta_j \cpar 
				\\[-0pt]
				\hdots \cpar 
				\\[-0pt]
				B^\theta_{F+G+M} \cpar T^\theta_{F+G+M} \cpar \bang \textit{PBT})
			\end{array}
		\end{array}	 
	\end{array}
	\]
	\caption{The generic state in the impl world.}\label{fig:genimpl}
\end{figure}

To conclude the definition of $\mathfrak{R}$ we should clarify which messages are available to an attacker at a given state, i.e. to define the substitutions $\sigma_0$, $\sigma$ and $\theta$. The definition of $\sigma_0$ is straightforward. $$\sigma_0 = \sub{pk_s, pk_\texttt{MM}}{\pks{s}, \pkv{\chi_{\texttt{MM}}}}$$

To define of $\sigma$ and $\theta$ we introduce the following shorthand for messages. Firstly we define the card's public key $\texttt{pk}_c \coloneqq \pkg{c}$, the transaction $\textsf{TX} = \proj{1}{\dec{z}{\hash{\smult{\mult{a}{c}}{x}}}}$ and the following cryptograms: $\textsf{AC}^\fail = \nlist{a, \text{PAN}, \textsf{TX}}$, $\textsf{AC}^{\texttt{ok}} = \nlist{a, \text{PAN}, \textsf{TX}, \texttt{ok}}$, and the failure cryptogram $\textsf{AC}^{\texttt{no}} = \nlist{a, \text{PAN}, \textsf{TX}, \texttt{no}}$. We will also need $\widehat{\mathbf{kbt}} = \proj{3}{\proj{1}{\dec{y}{\hash{\smult{\mult{a}{c}}{x}}}}}$ and the list of parameters $\vec{e} \coloneqq (a, c, mk, \text{PAN}, x,  y,  z)$

\begin{itemize}[leftmargin=0pt]
	\item[] $\textit{ecert}(\hat{\mathbf{t}}, x) = \enc{\nlist{\nlist{\texttt{MM}, \pkg{b_t}}, \sig{s}{\nlist{\texttt{MM}, \pkg{b_t}}}}}{\hash{\smult{\hat{\mathbf{t}}}{x}}}$ 
	\item[] $\textit{emcert}(a, c, x) = \enc{\nlist{\smult{a}{\texttt{pk}_c}, \smult{a}{\sigv{\chi_{\texttt{MM}}}{\texttt{pk}_c}}}}{\hash{\smult{\mult{a}{c}}{x}}}$
	\item[] $\textit{etx}(t, tx, x) = \enc{\nlist{tx, \fail}}{\hash{\smult{t}{x}}}$ 
	\item[] $\textit{etxpin}(t, tx, x, \text{uPIN}) = \enc{\nlist{tx, \text{uPIN}}}{\hash{\smult{t}{x}}}$ 
	\item[] $\textit{eaclo}(\vec{e}) = \enc{\nlist{\enc{\nlist{\textsf{AC}^{\fail},\hash{\nlist{\texttt{AC}^{\fail}, \mk}}}}{\hash{\smult{\mult{a}{c}}{\widehat{\mathbf{kbt}}}}}},\fail, \textsf{TX}}{\hash{\smult{\mult{a}{c}}{x}}}$  
	%	& \quad \text{ where } \textsf{AC} = \nlist{a, \text{PAN}, \proj{1}{\dec{z}{\hash{\smult{\mult{a}{c}}{x}}}}} \\
	\item[] $\squeezespaces{0.3}\textit{eachi}(\vec{e}) = \enc{\nlist{\enc{\nlist{\textsf{AC}^{\texttt{ok}},\hash{\nlist{\texttt{AC}^{\texttt{ok}}, \mk}}}}{\hash{\smult{\mult{a}{c}}{\widehat{\mathbf{kbt}}}}}},\texttt{ok}, \textsf{TX}}{\hash{\smult{\mult{a}{c}}{x}}}$ 
	%	& \quad \text{ where } \textsf{AC}^{\texttt{ok}} = \nlist{a, \text{PAN}, \proj{1}{\dec{z}{\hash{\smult{\mult{a}{c}}{x}}}}, \texttt{ok}} \\	
	\item[] $\squeezespaces{0.01}\textit{eacfail}(\vec{e}) = \enc{\nlist{\enc{\nlist{\textsf{AC}^{\texttt{no}},\hash{\nlist{\textsf{AC}^{\texttt{no}}, \mk}}}}{\hash{\smult{\mult{a}{c}}{\widehat{\mathbf{kbt}}}}}},\texttt{no}, \textsf{TX}}{\hash{\smult{\mult{a}{c}}{x}}}$ 
\end{itemize}

%	\[
%	\begin{aligned}
%	& \textit{ecert}(\hat{\mathbf{t}}, x) = \enc{\nlist{\nlist{\texttt{MM}, \pkg{b_t}}, \sig{s}{\nlist{\texttt{MM}, \pkg{b_t}}}}}{\hash{\smult{\hat{\mathbf{t}}}{x}}} \\
%	& \textit{emcert}(a, c, x) = \enc{\nlist{\smult{a}{\texttt{pk}_c}, \smult{a}{\sigv{\chi_{\texttt{MM}}}{\texttt{pk}_c}}}}{\hash{\smult{\mult{a}{c}}{x}}} \\
%	& \textit{etx}(t, tx, x) = \enc{\nlist{tx, \fail}}{\hash{\smult{t}{x}}} \\
%	& \textit{etxpin}(t, tx, x, \text{uPIN}) = \enc{\nlist{tx, \text{uPIN}}}{\hash{\smult{t}{x}}} \\
%	& \textit{eaclo}(\vec{e}) = \enc{\enc{\nlist{\textsf{AC}, \hash{\nlist{\textsf{AC}, \mk}}}}{\hash{\smult{\mult{a}{c}}{\widehat{\mathbf{kbt}}}}}}{\hash{\smult{\mult{a}{c}}{x}}} \\
%%	& \quad \text{ where } \textsf{AC} = \nlist{a, \text{PAN}, \proj{1}{\dec{z}{\hash{\smult{\mult{a}{c}}{x}}}}} \\
%	& \textit{eachi}(\vec{e}) = \enc{\nlist{\enc{\nlist{\textsf{AC}^{\texttt{ok}},\hash{\nlist{\texttt{AC}^{\texttt{ok}}, \mk}}}}{\hash{\smult{\mult{a}{c}}{\widehat{\mathbf{kbt}}}}}},\texttt{ok}}{\hash{\smult{\mult{a}{c}}{x}}} \\
%%	& \quad \text{ where } \textsf{AC}^{\texttt{ok}} = \nlist{a, \text{PAN}, \proj{1}{\dec{z}{\hash{\smult{\mult{a}{c}}{x}}}}, \texttt{ok}} \\	
%	&\textit{eacfail}(\vec{e}) = \enc{\nlist{\enc{\nlist{\textsf{AC}^{\fail},\hash{\nlist{\textsf{AC}^{\fail}, \mk}}}}{\hash{\smult{\mult{a}{c}}{\widehat{\mathbf{kbt}}}}}},\fail}{\hash{\smult{\mult{a}{c}}{x}}} \\
%	\end{aligned}
%	\]

To present $\sigma$ and $\theta$ in Fig.~\ref{fig:sigmatheta} we introduce the index function $\ind: \{\sigma, \theta\} \rightarrow \mathcal{D}$ defined for a substitution $\rho \in \{\sigma, \theta\}$ as follows:  $\ind(\sigma) = i, \, \ind(\theta)=j$.

The aliases for the messages output in session $i$, and available to the attacker are as follows. Terminal's, card's, and bank's channels are labelled as $cht_i$, $chc_i$, and $chb_i$ respectively. Terminal's messages are labelled as $ua_i$, $ub_i$, $uc_i$, $ud_i$, and $ue_i$. Card's messages as $va_i$, $vb_i$, $vc_i$. Bank's reply as $wa_i$. 

We start with a natural freshness conditions, i.e. that  message labels from $\dom{\sigma}$, $\dom{\theta}$ cannot refer to neither bound nor free names. Firstly, for any $i,k \in \mathcal{D} \cup \mathcal{F} \cup \mathcal{G}, l \in \{1 \hdots 3\}, m \in \{1 \hdots 4\}$ we have $pk_s$, $pk_\texttt{MM}$, $cht_i$, $chc_i$, $chb_i$, $ua_i$, $va_i$, $ub_i$, $vb_i$, $uc_i$, $vc_i$, $ud_i$, $wa_i$, $ue_i$, $X^l_i$, $Y^1_i$, $Z^m_i$ $\fresh{}$ $\{card, term, \texttt{ok}, \fail, \texttt{accept}, \texttt{auth}, \lo,\hi\}$~$\cup$ $\{user, s, si, \chi_{\texttt{MM}}, b_t, \text{PIN}_k, mk_k, c_k, \dot{ch}_k, a_k, \ddot{ch}_k, \dddot{ch}_k, t_k, \textsf{TX}_k\}$. Then for the inputs we have $\fv{X^1_i} \ \fresh{} \ \{va_i, vb_i, vc_i\}$, $\fv{X^2_i} \ \fresh{} \ \{vb_i, vc_i\} $, $\fv{X^3_i} \ \fresh{} \ \{vc_i\}$, $\fv{Z^1_i} \ \fresh{} \ \{ub_i, uc_i, ud_i, ue_i\} $, $	\fv{Z^2_i} \ \fresh{}\ \{uc_i, ud_i, ue_i\}$, $\fv{Z^3_i} \ \fresh{} \ \{ud_i, ue_i\} $, $\fv{Z^4_i} \ \fresh{} \ \{ue_i\}$, $\fv{Y^1_i} \ \fresh{} \ \{wa_i\}$.

\begin{figure*}
	\[
	\arraycolsep=1.7pt
	\begin{array}{ll}
		cht_i \rho = \ddot{ch}_i & \mbox{if $i  \in\mathcal{FG} $}
		\\[10pt]
		chc_i \rho = \dot{ch}_i & \mbox{if $i  \in \mathcal{D} $}
		\\[10pt]
		chb_i \rho = \dddot{ch}_i & \mbox{if $i  \in \mathcal{FM} $}
		\\[10pt]
		ua_i \rho = \smult{t_i}{\gen} & \mbox{if $i  \in \bigcup_{l=2}^{11} \gamma^{\on}_l$ or $\bigcup_{l=2}^{11} \gamma^{\of}_l$ or $\bigcup_{l=2}^{10} \gamma^{\lo}_l$}
		\\[10pt]
		va_i \rho = \smult{a_i}{\pkg{c_{\ind(\rho)}}} & \mbox{if $i  \in \bigcup_{l=3}^{7} \alpha_l$ (and, if $\rho=\theta$, $i \in \lambda_j$)}
		\\[10pt]
		ub_i \rho = \textit{ecert}(t_i, Z^i_1\rho) & \mbox{if $i  \in \bigcup_{l=4}^{11} \gamma^{\on}_l$ or $\bigcup_{l=4}^{11} \gamma^{\of}_l$ or $ \bigcup_{l=4}^{10} \gamma^{\lo}_l$}
		\\[10pt]
		vb_i \rho = \textit{emcert}(a_i, c_{\ind(\rho)}, X^i_1\rho) & \parbox{18cm}{\squeezespaces{1}if $i  \in \bigcup_{l=5}^{7} \alpha_l$ (and, if $\rho=\theta$, $i \in \lambda_j$) and \\
			$\checksig{\pks{s}}{\proj{2}{\dec{X^2_i\rho}{\hash{\smult{\mult{a_i}{c_{\ind(\rho)}}}{X^1_i\rho}}}}} =$ \\ $\hspace*{0.3cm}=\proj{1}{\dec{X^2_i\rho}{\hash{\smult{\mult{a_i}{c_{\ind(\rho)}}}{X^1_i\rho}}}}$ and \\
			$\proj{1}{\proj{1}{\dec{X^2_i\rho}{\hash{\smult{\mult{a_i}{c_{\ind(\rho)}}}{X^1_i\rho}}}}} = \texttt{MM}$} 
		\\[10pt]
			uc_i \rho = \textit{etxpin}(t_i, tx_i, Z^i_1\rho, \text{uPIN}) & \parbox{15cm}{\squeezespaces{1}if $i  \in \bigcup_{l=7}^{11} \gamma^{\of}_l$ and \\ $\checksigv{\pkv{\chi_{\texttt{MM}}}}{\proj{2}{\dec{Z^2_i\rho}{\hash{\smult{t_i}{Z^1_i\rho}}}}} =$\\$\hspace*{0.3cm}=\proj{1}{\dec{Z^2_i\rho}{\hash{\smult{t_i}{Z^1_i\rho}}}}$ and \\$\proj{1}{\dec{Z^2_i\rho}{\hash{\smult{t_i}{Z^1_i\rho}}}} = Z^i_1\rho$}
			\\[30pt]
			\ \, \qquad = \textit{etx}(t_i, tx_i, Z^i_1\rho) & \parbox{15cm}{\squeezespaces{1}if $i  \in \bigcup_{l=7}^{11} \gamma^{\on}_l$ or $ \bigcup_{l=6}^{10} \gamma^{\lo}_l$ and 
				\\ 
				$\checksigv{\pkv{\chi_{\texttt{MM}}}}{\proj{2}{\dec{Z^2_i\rho}{\hash{\smult{t_i}{Z^1_i\rho}}}}} =$ \\ $\hspace*{0.3cm}= \proj{1}{\dec{Z^2_i\rho}{\hash{\smult{t_i}{Z^1_i\rho}}}}$ and 
				\\
				$\proj{1}{\dec{Z^2_i\rho}{\hash{\smult{t_i}{Z^1_i\rho}}}} = Z^i_1\rho$}
			\\[30pt]
			vc_i \rho = \textit{eaclo}(a_i, c_{\ind(\rho)}, mk_i, \text{PAN}_i, X^i_1\rho,  X^i_2\rho,  X^i_3\rho) & \parbox{15cm}{if $i  \in \alpha_7$ (and, if $\rho=\theta$, $i \in \lambda_j$) and $\proj{2}{\dec{X^3_i\rho}{\hash{\smult{\mult{a_i}{c_{\ind(\rho)}}}{X^1_i\rho}}}} = \fail$}	
			\\[10pt]
			\ \, \qquad = \textit{eachi}(a_i, c_{\ind(\rho)}, mk_i, \text{PAN}_i, X^i_1\rho,  X^i_2\rho,  X^i_3\rho) & \parbox{15cm}{if $i  \in \alpha_7 $ (and, if $\rho=\theta$, $i \in \lambda_j$) and \\ $\proj{2}{\dec{X^3_i\rho}{\hash{\smult{\mult{a_i}{c_{\ind(\rho)}}}{X^1_i\rho}}}} = \text{PIN}_i$}	
			\\[10pt]
			\ \, \qquad = \textit{eacfail}(a_i, c_{\ind(\rho)}, mk_i, \text{PAN}_i, X^i_1\rho,  X^i_2\rho,  X^i_3\rho) & \mbox{if $i  \in \alpha_7$ (and, if $\rho=\theta$, $i \in \lambda_j$) and else}
			\\[10pt]
			ud_i \rho = \enc{\nlist{\textsf{TX}_i, Z_1^i\rho, \proj{1}{\dec{Z_3^i\rho}{\hash{\smult{t_i}{Z_1^i\rho}}}}, \text{uPIN}}}{\kbt_i} & \mbox{if $i  \in \bigcup_{l=9}^{11} \gamma^{\on}_l$ and $\proj{3}{\dec{Z_3^i\rho}{\hash{\smult{t_i}{Z_1^i\rho}}}} = \textsf{TX}_i$} 
			\\[10pt]
			\ \ \qquad =  \enc{\nlist{\textsf{TX}_i, Z_1^i\rho, \proj{1}{\dec{Z_3^i\rho}{\hash{\smult{t_i}{Z_1^i\rho}}}}, \fail}}{\kbt_i} & \mbox{if $i  \in \gamma^{\of}_{11}$ and $\proj{3}{\dec{Z_3^i\rho}{\hash{\smult{t_i}{Z_1^i\rho}}}} = \textsf{TX}_i$} 
			\\[10pt]
			\ \ \qquad =  \enc{\nlist{\textsf{TX}_i, Z_1^i\rho, \proj{1}{\dec{Z_3^i\rho}{\hash{\smult{t_i}{Z_1^i\rho}}}}, \fail}}{\kbt_i} & \mbox{if $i  \in \bigcup_{l=8}^{10} \gamma^{\lo}_l$ and $\proj{3}{\dec{Z_3^i\rho}{\hash{\smult{t_i}{Z_1^i\rho}}}} = \textsf{TX}_i$} 
			\\[13pt]
			wa_i\rho = \enc{\nlist{\proj{1}{dy_i}, \texttt{accept}}}{\kbt_i} & \parbox{18.5cm}{\squeezespaces{1}let $dy_i = \dec{Y^1_i\rho}{\kbt_i}$ and $\nlist{\text{PIN}_j, mk_j, \pkg{c_j}} = \text{DB}$ \\ if $i  \in \beta_4$ and $\exists j$, s.t. $j \in \alpha_7$ and
				%	\\$\text{PAN}_j = \proj{2}{\proj{1}{\dec{\proj{3}{dy_i}}{\hash{\smult{b_t}{\proj{2}{dy_i}}}}}}$ and 
				\\$\squeezespaces{1}\hash{\nlist{\proj{1}{\dec{\proj{3}{dy_i}}{\hash{\smult{b_t}{\proj{2}{dy_i}}}}}, mk_j}}=$\\$\hspace*{0.3cm}=\proj{2}{\dec{\proj{3}{dy_i}}{\hash{\smult{b_t}{\proj{2}{dy_i}}}}}$ and 
				\\$\proj{3}{\proj{1}{\dec{\proj{3}{dy_i}}{\hash{\smult{b_t}{\proj{2}{dy_i}}}}}}= \proj{1}{dy_i}$ and 
				\\$\smult{\proj{1}{\proj{1}{\dec{\proj{3}{dy_i}}{\hash{\smult{b_t}{\proj{2}{dy_i}}}}}}}{\pkg{c_j}}= \proj{2}{dy_i}$ and 
				\\ $\squeezespaces{1}\bigl(\proj{2}{\proj{1}{dy_i}} = \lo \ \text{or} \ \proj{2}{\proj{1}{dy_i}}= \hi \\ \text{and} \ \proj{4}{\proj{1}{\dec{\proj{3}{dy_i}}{\hash{\smult{b_t}{\proj{2}{dy_i}}}}}} = \texttt{ok} \ 
				\text{or else if} \ \proj{4}{dy_i} = \text{PIN}_j\bigl)$}
			\\[55pt]
			ue_i\rho = \texttt{auth} &  \parbox{15cm}{if $i \in \gamma^{\of}_9 \cup \gamma^{\of}_{11}$ and $\proj{2}{\dec{Z^3_i}{\hash{\smult{t_i}{Z^1_i}}}} = \texttt{ok}$ or\\
				if $i \in \gamma^{\on}_{11}$ or $i \in \gamma^{\lo}_{10}$ and $\squeezespaces{0.8}\proj{1}{\dec{Z_4^i}{\kbt_i}} = \textsf{TX}_i$ and $\squeezespaces{0.7}\proj{2}{\dec{Z_4^i}{\kbt_i}} = \texttt{accept}$}	
	\end{array}
	\]
	\caption{The definition of active substitutions $\sigma$ and $\theta$.}\label{fig:sigmatheta}
	\end{figure*}

\textit{Bisimulation}. Now, when the relation is defined, we can start consider all possible moves each side can make. Since we have defined $\mathfrak{R}$ as a symmetric relation, we consider only the cases when the spec process starts first. 

\textit{Case} 1. $\co{out}(pk_s)$, $\spec{UP} \ \mathfrak{R} \ \impl{UP}$. 

The process $\spec{UP}$ can make the transition $\co{out}(pk_s)$ to the state $\spec{UP}^1$. There is a state $\impl{UP}^1$ to which the process $\impl{UP}$ can make the transition $\co{out}(pk_s)$. By the definition of $\mathfrak{R}$ we have $\spec{UP}^1 \ \mathfrak{R} \ \impl{UP}^1$.

\textit{Case} 2. $\co{out}(pk_\texttt{MM})$, $\spec{UP}^1 \ \mathfrak{R} \ \impl{UP}^1$. 

Identical to Case 1. 

\textit{Case} 3. $\co{card}(chc_{D+1})$. 
%$\specc{(K, F, A, \Gamma, B)}(X,Y,Z) \ \mathfrak{R} \ \impll{(\vec{K}, F, A, \Gamma, B, \Lambda)}(X,Y,Z)$. 

From now on, we will only track the effect of the transition on the parameters defining the state, hence the parameters not affected by the transition are omitted. 

The spec process either creates a new card and outputs the channel, hence transits to $\alpha_1 \cup \{D+1\}$ or outputs a channel for the waiting card evolving to $K-1, \alpha_1 \cup \{D+1\}$. The impl process can match by either creating a new card and announcing the channel $\alpha_1 \cup \{D+1\}, \Lambda \cup \{D+1\}$, starting new session for the existing card $h$ making the transition to $\alpha_1 \cup \{D+1\}, \lambda_h \cup \{D+1\}$ or outputting the channel for the waiting card $h$ evolving to $K_h-1, \alpha_1 \cup \{D+1\}, \lambda_h \cup \{D+1\}$. In either case the resulting states are related by $\mathfrak{R}$. 

\textit{Case} 4. $\co{term}(cht_{F+G+1})$. 

Either the spec process starts a new $\on$, $\of$ or $\lo$ terminal session by also creating a symmetric bank-terminal symmetric key transiting to respectively $\gamma^{\on}_1 \cup \{F+G+1\}$, $\gamma^{\of}_1 \cup \{F+G+1\}$ or $\gamma^{\lo}_1 \cup \{F+G+1\}$. Or the spec process starts a new $\on$, $\of$ or $\lo$ terminal session for the existing bank-terminal key transiting to respectively $F+1, \gamma^{\on}_1 \cup \{F+G+1\}$, $F+1, \gamma^{\of}_1 \cup \{F+G+1\}$ or $F+1, \gamma^{\lo}_1 \cup \{F+G+1\}$. The impl process can always match and the resulting states are related by $\mathfrak{R}$. 

\textit{Case} 5. $\co{bank}(chb_{F+G+1})$. 

Identical to Case 4. Notice that in cases 3-5, as new card terminal or bank sessions started, the input matrices $X$, $Y$ and $Z$ also grow by one row to accommodate future inputs.

\textit{Case} 6. $\co{cht_i}(ua_i)$, $i \in \gamma^{\on}_1$, $i \in \gamma^{\of}_1$ or $i \in \gamma^{\lo}_1$. 

The spec process moves to the respective state $\gamma^{\on}_1 \setminus \{i\}, \gamma^{\on}_2 \cup \{i\}$, $\gamma^{\of}_1 \setminus \{i\}, \gamma^{\of}_2 \cup \{i\}$ or $\gamma^{\lo}_1 \setminus \{i\}, \gamma^{\lo}_2 \cup \{i\}$. The impl process can always match by transiting to the state parametrised identically, hence the resulting states are related by $\mathfrak{R}$.

\textit{Case} 7. $chc_i \, X^1_i$, $i \in \alpha_1$.

The spec process moves to the respective state $\alpha_1 \setminus \{i\}, \alpha_2 \cup \{i\}$ where the $(i, 1)$ element of the matrix $X$ is replaced by $X^1_i$. The impl process can always match by transiting to the state parametrised identically, hence the resulting states are related by $\mathfrak{R}$.

\textit{Case} 8. $\co{chc_i}(va_i)$, $i \in \alpha_2$.

The spec process moves to the respective state $\alpha_2 \setminus \{i\}, \alpha_3 \cup \{i\}$. The process on the right can always match, it is parametrised by the same partition, hence the resulting states are related by $\mathfrak{R}$.

\textit{Case} 9. $cht_i \, Z^1_i$, $i \in \gamma^{\on}_2$, $i \in \gamma^{\of}_2$ or $i \in \gamma^{\lo}_2$.

The spec process moves to the respective state $\gamma^{\on}_2 \setminus \{i\}, \gamma^{\on}_3 \cup \{i\}$, $\gamma^{\of}_2 \setminus \{i\}, \gamma^{\of}_3 \cup \{i\}$ or $\gamma^{\lo}_2 \setminus \{i\}, \gamma^{\lo}_3 \cup \{i\}$ where the $(i, 1)$ element of the matrix $Z$ is replaced by  $Z^1_i$. The impl process can always match by transiting to the state parametrised identically, hence the resulting states are related by $\mathfrak{R}$.

\textit{Case} 10. $\co{cht_i}(ub_i)$, $i \in \gamma^{\on}_3$, $i \in \gamma^{\of}_3$ or $i \in \gamma^{\lo}_3$.

The spec process moves to the respective state $\gamma^{\on}_3 \setminus \{i\}, \gamma^{\on}_4 \cup \{i\}$, $\gamma^{\of}_3 \setminus \{i\}, \gamma^{\of}_4 \cup \{i\}$ or $\gamma^{\lo}_3 \setminus \{i\}, \gamma^{\lo}_4 \cup \{i\}$. The impl process can always match by transiting to the state parametrised identically, hence the resulting states are related by $\mathfrak{R}$.

\textit{Case} 11. $chc_i \, X^2_i$, $i \in \alpha_3$.

The spec process moves to the state $\alpha_3 \setminus \{i\}, \alpha_4 \cup \{i\}$ where the $(i, 2)$ element of the matrix $X$ is replaced by $X^2_i$. The impl process can always match by transiting to the state parametrised identically, hence the resulting states are related by $\mathfrak{R}$.

\textit{Case} 12. $\co{chc_i}(vb_i)$, $i \in \alpha_4$ and the conditions for $vb_i\sigma$ in the definition of $\sigma$ in Fig.~\ref{fig:sigmatheta} are satisfied given the inputs $(X,Y,Z)$.

Let $\alpha_4 \setminus \{i\}, \alpha_5 \cup \{i\}$ define the resulting state in which the spec process can make the transition. Consider the second condition. We can rewrite this condition as $\squeezespaces{1}X^2_i\sigma = \enc{\nlist{\nlist{\texttt{MM}, M_1}, M_2}}{\hash{\smult{\mult{a_i}{c_i}}{X^1_i\sigma}}}$, where $M_1$ and $M_2$ are arbitrary messages. As $\texttt{MM}$ is a global constant and $M_1$ and $M_2$ are arbitrary attacker's inputs, the term $\nlist{\nlist{\texttt{MM}, M_1}, M_2}$ can always be produced. Since at the point of input of $X^2_i$ the only message on the network that refers to (some multiple) of $a_i$ is $\smult{a_i}{\pkg{c_i}}$ available through the alias $va_i$, the key is either of the form $\hash{va_i}$ or $\hash{\smult{\hat{t}}{va_i}}$ for some $\hat{t}$ before $\sigma$ is applied. In either case we then conclude that $X^2_i\theta = \enc{\nlist{\nlist{\texttt{MM}, M_1}, M_2}}{\hash{\smult{\mult{a_i}{c_h}}{X^1_i\theta}}}$ holds since $va_i\theta = \smult{a_i}{\pkg{c_h}}$ for some card $h$ and we have established that the second condition holds also in the impl case. The case when the spec process stats first is similar. Now notice, from the second equation we have the following $\squeezespaces{1}\dec{X^2_i\sigma}{\hash{\smult{\mult{a_i}{c_i}}{X^i_1\sigma}}} = \dec{X^2_i\theta}{\hash{\smult{\mult{a_i}{c_h}}{X^i_1\sigma}}}$ $= \nlist{\nlist{\texttt{MM}, M_1}, M_2}$, and the first condition becomes $$\squeezespaces{1}\checksig{\proj{2}{\nlist{\nlist{\texttt{MM}, M_1}, M_2}}}{\pks{s}} = \proj{1}{\nlist{\nlist{\texttt{MM}, M_1}, M_2}}$$ which is independent of $\sigma$ and $\theta$ and trivially holds in the impl case. We conclude that the impl process can always match by transiting to the state parametrised by $\alpha_4 \setminus \{i\}, \alpha_5 \cup \{i\}$ and the resulting states are related by $\mathfrak{R}$.

\textit{Case} 13.
$cht_i \, Z_2^i$, $i \in \gamma^{\on}_4$, $i \in \gamma^{\of}_4$ or $i \in \gamma^{\lo}_4$

The spec process moves to the state $\gamma^{\on}_4 \setminus \{i\}, \gamma^{\on}_5 \cup \{i\}$, $\gamma^{\of}_4 \setminus \{i\}, \gamma^{\of}_5 \cup \{i\}$ or $\gamma^{\lo}_4 \setminus \{i\}, \gamma^{\lo}_5 \cup \{i\}$ and the element $(i, 2)$ of the matrix $Z$ is replaced by $Z^2_i$. The impl process can always match by transiting to the state parametrised identically, hence the resulting states are related by $\mathfrak{R}$.

\textit{Case} 14.
$\tau$, $i \in \gamma^{\on}_5$ or $i \in \gamma^{\of}_5$ and the conditions for $uc_i\sigma$ in the definition of $\sigma$ are satisfied given the inputs $(X,Y,Z)$. 

Let $\gamma^{\on}_5 \setminus \{i\}, \gamma^{\on}_6 \cup \{i\}$ or $\gamma^{\of}_5 \setminus \{i\}, \gamma^{\of}_6 \cup \{i\}$ define the resulting state in which the spec process can make the transition. Similarly to Case 12 consider the second condition. We can rewrite it as $\squeezespaces{1}Z^2_i\sigma = \enc{\nlist{Z^i_1\sigma, M}}{\hash{\smult{t_i}{Z^1_i\sigma}}}$ for some $M$. Since at the point of input of $X^2_i$ the only message on the network that refers to a multiple of $t_i$ is $\smult{t_i}{\gen}$ available through the alias $ua_i$, the key is either of the form $\hash{ua_i}$ or $\hash{\smult{\hat{t}}{ua_i}}$ for some $\hat{t}$ before $\sigma$ is applied. In either case we conclude that $\squeezespaces{1}Z^2_i\theta = \enc{\nlist{Z^i_1\theta, M}}{\hash{\smult{t_i}{Z^1_i\theta}}}$ holds since $ua_i\theta = \smult{t_i}{\gen}$, so we have established that the second condition holds also in the impl case. The case when the spec process stats first is similar. Now notice that if follows from the second equation that $\squeezespaces{1}\dec{Z^2_i\sigma}{\hash{\smult{t_i}{Z^1_i\sigma}}}=\nlist{Z^1_i\sigma, M}$ and the first condition becomes $\squeezespaces{1} \checksigv{\pkv{\chi_{\texttt{MM}}}}{M}= Z^1_i\sigma$, i.e. the input $Z^1_i\sigma$ is signed with the signing key $\chi_{\texttt{MM}}$, hence can only be the multiple of some $\pkg{c_l}$ for some $l$. The only message on the network of this form is $\smult{a_l}{\pkg{c_l}}$ available through the alias $va_l$, hence the input $Z^1_i$ is either $va_l$ or $\smult{\hat{s}}{va_l}$ which, under $\theta$ also give signed inputs in the impl case. We conclude that the impl process can always match by transiting to the state parametrised by $\gamma^{\on}_5 \setminus \{i\}, \gamma^{\on}_6 \cup \{i\}$ or $\gamma^{\of}_5 \setminus \{i\}, \gamma^{\of}_6 \cup \{i\}$ and the resulting states are related by $\mathfrak{R}$. Notice that there is no $\text{PIN}$ check at this point and either ``right'' or ``wrong'' $\text{PIN}$s are always available, hence whenever the user enters a $\text{PIN}$ on the spec/impl side, the same type of $\text{PIN}$ can always be entered on the impl/spec side respectively.

\textit{Case} 15.
$\co{cht_i}(uc_i)$, $i \in \gamma^{\on}_6$, $i \in \gamma^{\of}_6$, or $i \in \gamma^{\lo}_5$ \emph{and} the conditions for $uc_i\sigma$ in the definition of $\sigma$ are satisfied given the inputs $(X,Y,Z)$.

The spec process moves to the respective state $\gamma^{\on}_6 \setminus \{i\}, \gamma^{\on}_7 \cup \{i\}$, $\gamma^{\of}_6 \setminus \{i\}, \gamma^{\of}_7 \cup \{i\}$ or $\gamma^{\lo}_5 \setminus \{i\}, \gamma^{\lo}_6 \cup \{i\}$. Notice that the $\lo$ case is similar to Case 14. The impl process can always match by transiting to the state parametrised identically, hence the resulting states are related by $\mathfrak{R}$.

\textit{Case} 16.
$chc_i \, X^3_i$, $i \in \alpha_5$.

The spec process moves to the state $\alpha_5 \setminus \{i\}, \alpha_6 \cup \{i\}$ where the $(i, 3)$ element of the matrix $X$ is replaced by $X^3_i$. The impl process can always match by transiting to the state parametrised identically, hence the resulting states are related by $\mathfrak{R}$. 

\textit{Case} 17.
$\co{chc_i}(vc_i)$, $i \in \alpha_6$ and the conditions for $vc_i\sigma$ in the definition of $\sigma$ are satisfied given the inputs $(X,Y,Z)$.

Let $\alpha_6 \setminus \{i\}, \alpha_7 \cup \{i\}$ define the resulting state in which the spec process can make the transition. Consider either the case when either the $\fail$ or $\text{PIN}_i$ is the element of the received input. We can rewrite these guards as follows $\squeezespaces{1}X^3_i\sigma = \enc{\nlist{M, N}}{\hash{\smult{\mult{a_i}{c_i}}{X^1_i\sigma}}}$ where $M$ is arbitrary, $N \in \{\fail, \text{PIN}_i\}$ and apply the argument from Case 12. The \texttt{else} branch is identical to Case 19.4. Hence the impl process can always match by moving to the state parametrised by $\alpha_6 \setminus \{i\}, \alpha_7 \cup \{i\}$ and the resulting states are related by $\mathfrak{R}$.

\textit{Case} 18.
$cht_i \, Z_3^i$, $i \in \gamma^{\on}_7$, $i \in \gamma^{\of}_7$ or $i \in \gamma^{\lo}_6$.

The spec process moves to the state $\gamma^{\on}_7 \setminus \{i\}, \gamma^{\on}_8 \cup \{i\}$ or $\gamma^{\of}_7 \setminus \{i\}, \gamma^{\of}_8 \cup \{i\}$ or $\gamma^{\lo}_6 \setminus \{i\}, \gamma^{\lo}_7 \cup \{i\}$ and the element $(i, 3)$ of the matrix $Z$ is replaced by $Z^3_i$. The impl process can always match by transiting to the state parametrised identically, hence the resulting states are related by $\mathfrak{R}$.

\textit{Case} 19.1.
$\co{cht_i}(ud_i)$, $i \in \gamma^{\on}_8$.

The spec process moves to the state $\gamma^{\on}_8 \setminus \{i\}, \gamma^{\on}_9 \cup \{i\}$ as online terminal just adds the entered $\text{PIN}$ and sends the cryptogram to the bank. The impl process can always match by moving to the state parametrised identically, then the resulting states are related by $\mathfrak{R}$. 

\textit{Case} 19.2.``Right'' $\text{PIN}$.
$\co{cht_i}(ue_i)$, $i \in \gamma^{\of}_8$ and the condition for $ue_i\sigma$ that the card replied affirmatively in the definition of $\sigma$ is satisfied given the inputs $(X,Y,Z)$.

Let $\gamma^{\of}_8 \setminus \{i\}, \gamma^{\of}_9 \cup \{i\}$ define the resulting state in which the spec process can make the transition. The check that the condition holds in the impl case for the substitution $\theta$ is identical to Case 12. We conclude that the impl process can always match by transiting to the state parametrised by $\gamma^{\of}_8 \setminus \{i\}, \gamma^{\of}_9 \cup \{i\}$ and the resulting states are related by $\mathfrak{R}$.

\textit{Case} 19.3. ``Right'' $\text{PIN}$.
$\co{cht_i}(ud_i)$, $i \in \gamma^{\of}_9$.

Let $\gamma^{\of}_9 \setminus \{i\}, \gamma^{\of}_{11} \cup \{i\}$ define the resulting state in which the spec process can make the transition as offline terminal simply forwards the cryptogram to the bank. The impl process can always match by transiting to the state parametrised identically, hence the resulting states are related by $\mathfrak{R}$. 

\textit{Case} 19.4. ``Wrong'' $\text{PIN}$.
$\co{cht_i}(ud_i)$, $i \in \gamma^{\of}_10$ and the condition for $ue_i\sigma$ that the card replied affirmatively in the definition of $\sigma$ does \emph{not hold} given the inputs $(X,Y,Z)$.

Let $\gamma^{\of}_10 \setminus \{i\}, \gamma^{\of}_{11} \cup \{i\}$ define the resulting state in which the spec process can make the transition. By the Def. 2 in~\cite{horne2021quasi} the inequality holds whenever there is no unifying the left and the right part substitution $\rho$ that cannot refer private values in its domain or range. Consider the opposite for impl, i.e. that there is a unifying substitution $\rho$ (in case $\squeezespaces{1}D+K>H$ w.l.o.g. we assume that $\rho$ does not refer $\squeezespaces{1}\text{PIN}_i$, $mk_i$, $c_i$, $\text{PAN}_i$, $\squeezespaces{1}i \in \{H+1, \hdots, D+K\}$), i.e. $\squeezespaces{1}\proj{2}{\dec{Z^3_i\theta\rho}{\hash{\smult{t_i\rho}{Z^1_i\theta\rho}}}} = \texttt{ok}\rho$  that can be rewritten as $\squeezespaces{1}\proj{2}{\dec{Z^3_i\rho\theta}{\hash{\smult{t_i}{Z^1_i\rho\theta}}}} = \texttt{ok}\rho$. Then applying the argument from Case 12 we conclude that $\rho$ also unifies the condition for the spec state, which contradicts the initial condition. We conclude that the impl process can always match by transiting to the state parametrised by $\gamma^{\of}_10 \setminus \{i\}, \gamma^{\of}_{11} \cup \{i\}$ and the resulting states are related by $\mathfrak{R}$. 

\textit{Case} 19.5.
$\co{cht_i}(ud_i)$, $i \in \gamma^{\lo}_7$.

Identical to Case 19.1, the resulting state where the spec process can make the transition is $\gamma^{\lo}_7 \setminus \{i\}, \gamma^{\of}_8 \cup \{i\}$. The impl process can always match by transiting to the state parametrised identically, hence the resulting states are related by $\mathfrak{R}$. 

\textit{Case} 20.
$chb_i \, Y_1^i$, $i \in \beta_1$.

The spec process moves to the state $\beta_1 \setminus \{i\}, \beta_2 \cup \{i\}$ and the element $(i, 1)$ of the matrix $Y$ is replaced by $Y^1_i$. The impl process can always match by transiting to the state parametrised identically, hence the resulting states are related by $\mathfrak{R}$.

\textit{Case} 21.
$\tau$, $i \in \beta_2$.

Let $\beta_2 \setminus \{i\}, \beta_3 \cup \{i\}$ define the resulting state in which the spec process can make the transition. Notice that the database process $\squeezespaces{1}\cout{\nlist{si, \text{PAN}_j}}{\nlist{\text{PIN}_j, mk_j, \pkg{c_j}}}$ can only be accessed in case that there is a terminal in session $i \in \gamma^{on}_9 \cup \gamma^{of}_{11} \cup \gamma^{lo}_8$ (and using $\kbt_i$) and a corresponding card's session $j$ where the cryptogram containing legitimate $\text{PAN}_j$ has been sent to the terminal. Also, since the card's data retrieved by the bank from the database is obtained privately using the secret $si$, and the legitimate $\text{PAN}_j$ it is always ``right'' in contrast to the $\text{PIN}$ entered to the terminal (the ``wrong'' $\text{PIN}$ can be entered). Hence the subsequent integrity checks for the cryptogram always using the correct data for the received cryptogram. The impl process can always match by transiting to the state parametrised by $\beta_2 \setminus \{i\}, \beta_3 \cup \{i\}$ and the resulting states are related by $\mathfrak{R}$.

\textit{Case} 22.
$\co{chb_i}(wa_i)$, $i \in \beta_3$ and the conditions for $wa_i\sigma$ hold given the inputs $(X,Y,Z)$. 

Let $\beta_3 \setminus \{i\}, \beta_4 \cup \{i\}$  define the resulting state in which the spec process can make the transition. Since the communication between the terminal and the bank is private as discussed in Case 20, the conditions for $wa_i\theta$ also hold since they depend only on whether the input $Y^1_i\theta$ can be successfully decrypted. Also notice that since high-value terminals had infinite supply of both right and wrong PINs, the respective transactions are either accepted or declined (by not passing the $\text{PIN}$ guard) simultaneously by spec and impl processes. The impl process can always match by transiting to the state parametrised by $\co{chb_i}(wa_i)$, $i \in \beta_3$ and the resulting states are related by $\mathfrak{R}$.

\textit{Case} 23.
$cht_i \, Z^4_i$, $i \in \gamma^{\on}_9$ or $i \in \gamma^{\lo}_8$.

The spec process moves to the state $\gamma^{\on}_9 \setminus \{i\}, \gamma^{\on}_{10} \cup \{i\}$ or $\gamma^{\of}_8 \setminus \{i\}, \gamma^{\of}_9 \cup \{i\}$ and the element $(i, 4)$ of the matrix $Z$ is replaced by $Z^4_i$. The impl process can always match by moving to the state parametrised identically, thus the resulting states are related by $\mathfrak{R}$.

\textit{Case} 24.
$\co{cht_i}(ue_i)$, $i \in \gamma^{\on}_{10}$ or $i \in \gamma^{\lo}_9$ and the conditions for $ue_i\sigma$ hold given the inputs $(X,Y,Z)$. 

Identical to Case 22 since the communication between the bank and the terminal is private. The resulting state where the spec process can make the transition is $\gamma^{\on}_{10} \setminus \{i\}, \gamma^{\on}_{11} \cup \{i\}$ or $\gamma^{\lo}_9 \setminus \{i\}, \gamma^{\lo}_{10} \cup \{i\}$ The impl process can always match by transiting to the state parametrised identically, hence the resulting states are related by $\mathfrak{R}$.

\textit{Openness}. 
Intuitively, the relation is open if an attacker with the power to manipulate free variables by applying a substitution $\varphi$ that cannot refer to the variables bound by $\nw$, and the message aliases from the $\dom{\sigma}=\dom{\theta}$, and with the power to extend the environment by declaring some free variables private (and out messages referring to them on the network) cannot reach the pair of states that are not related as formally stated in the respective definitions in~\cite{horne2021quasi}. 

The relation $\mathfrak{R}$ is open by definition, and it is straightforward that manipulating free variables that include $\textit{out}$, $\textit{card}$, $\textit{term}$, $\textit{bank}$, $\texttt{MM}$, $\texttt{auth}$, $\fail$, $\texttt{ok}$, $\lo$, $\hi$, and, possibly, free variables that inputs may contain, introduce neither new transitions not considered above nor affects static equivalence (see below) as these variables distributed symmetrically in the related states.

\newcommand{\nf}[1]{\mathopen{#1}\downharpoonright}
\newcommand{\norm}[2]{\ensuremath{\nf{{#1}}^{\vec{{#2}}}_E}}

\textit{Static equivalence}. We proceed by proving that any two states related by $\mathfrak{R}$ are statically equivalent. Static equivalence trivially holds when frames are both empty or both are $\sigma_0$, hence we proceed with a general case. Firstly, however, we introduce the necessary terminology. 

Recall, that for a state $\nw \vec{z}.\left(\rho \cpar P\right)$ we call $\nw \vec{z}.\rho$ a \emph{frame}. A \emph{recipe} in the context of a given frame $\nw \vec{z}.\rho$ is a message term that is not referencing any variables in $\vec{z}$. We call the recipe $M$ \emph{non-trivial} under $\nw \vec{z}.\rho$ if it does reference message aliases from $\dom{\rho}$.

\begin{figure*}[t!]
	\[
	\begin{array}{ll}
		pk_s\sigma_0 = \pks{s}
		\\[8pt]
		pk_\texttt{MM} \sigma_0 =  \pkv{\chi_{\texttt{MM}}}
		\\[8pt]
		cht_i \rho = \ddot{ch}_i & \mbox{if $i  \in\mathcal{FG} $}
		\\[8pt]
		chc_i \rho = \dot{ch}_i & \mbox{if $i  \in \mathcal{D} $}
		\\[8pt]
		chb_i \rho = \dddot{ch}_i & \mbox{if $i  \in \mathcal{FM} $}
		\\[8pt]
		ua_i \rho = \smult{t_i}{\gen} & \mbox{if $i  \in \bigcup_{l=2}^{11} \gamma^{\on}_l$ or $\bigcup_{l=2}^{11} \gamma^{\of}_l$ or $\bigcup_{l=2}^{10} \gamma^{\lo}_l$}
		\\[8pt]
		va_i \rho = \smult{\mult{a_i}{c_{\ind(\rho)}}}{\gen} & \mbox{if $i  \in \bigcup_{l=3}^{7} \alpha_l$ (and, if $\rho=\theta$, $i \in \lambda_j$)}
		\\[8pt]
		\proj{1}{\proj{1}{\dec{ub_i}{\hash{\smult{B^1_i}{ua_i}}}}} \rho = \texttt{MM} & \mbox{\textcolor{teal}{if $Z^1_i = \smult{B^1_i}{\gen}$} and $i  \in \bigcup_{l=4}^{11} \gamma^{\on}_l$ or $\bigcup_{l=4}^{11} \gamma^{\of}_l$ or $ \bigcup_{l=4}^{10} \gamma^{\lo}_l$}
		\\[8pt]
		\proj{2}{\proj{1}{\dec{ub_i}{\hash{\smult{B^1_i}{ua_i}}}}} \rho = \pkg{\bt} & \mbox{}
		\\[8pt]
		\proj{2}{\dec{ub_i}{\hash{\smult{B^1_i}{ua_i}}}} \rho = \sig{s}{\nlist{\texttt{MM}, \pkg{b_t}}} & \mbox{}
		\\[8pt]
		\proj{1}{\checksig{pk_s}{\proj{2}{\dec{ub_i}{\hash{\smult{B^1_i}{ua_i}}}}}} \rho = \texttt{MM} & \mbox{}
		\\[8pt]
		\proj{2}{\checksig{pk_s}{\proj{2}{\dec{ub_i}{\hash{\smult{B^1_i}{ua_i}}}}}} \rho = \pkg{\bt} & \mbox{}
		\\[8pt]		
		ub_i \rho = \textit{ecert}(t_i, Z^i_1\rho) \textcolor{gray}{\ =\enc{-}{-}}
		& \mbox{\textcolor{teal}{if $Z^1_i \neq \smult{B^1_i}{\gen}$} and $i  \in \bigcup_{l=4}^{11} \gamma^{\on}_l$ or $\bigcup_{l=4}^{11} \gamma^{\of}_l$ or $ \bigcup_{l=4}^{10} \gamma^{\lo}_l$}
		\\[-30pt]
		\proj{1}{\dec{vb_i}{\hash{\smult{A^i_1}{va_i}}}} \rho = \smult{\mult{a_i}{c_{\ind(\rho)}}}{\gen} & \begin{minipage}{10cm}\vspace{1.7cm}\parbox{9.5cm}{\squeezespaces{1} \textcolor{teal}{if $X^1_i = \smult{A^1_i}{\gen}$} and if $i  \in \bigcup_{l=5}^{7} \alpha_l$ (and, if $\rho=\theta$, $i \in \lambda_j$) and
				\\$\checksig{\pks{s}}{\proj{2}{\dec{X^2_i\rho}{\hash{\smult{\mult{a_i}{c_{\ind(\rho)}}}{X^1_i\rho}}}}} =$ \\ \hspace*{0.3cm} $=\proj{1}{\dec{X^2_i\rho}{\hash{\smult{\mult{a_i}{c_{\ind(\rho)}}}{X^1_i\rho}}}}$ and 
				\\$\proj{1}{\proj{1}{\dec{X^2_i\rho}{\hash{\smult{\mult{a_i}{c_{\ind(\rho)}}}{X^1_i\rho}}}}} = \texttt{MM}$}\end{minipage}
		\\[-40pt]
		\proj{2}{\dec{vb_i}{\hash{\smult{A^i_1}{va_i}}}} \rho = \smult{\mult{a_i}{c_{\ind(\rho)}}}{\sigv{\chi_{\texttt{MM}}}{\gen}} & \parbox{9.5cm}{ }
		\\[8pt]
		\checksigv{pk_\texttt{MM}}{\proj{2}{\dec{vb_i}{\hash{\smult{A^i_1}{va_i}}}}} \rho = \smult{\mult{a_i}{c_{\ind(\rho)}}}{\gen} & \parbox{9.5cm}{ }
	\end{array}
	\]
	\caption{Normalisations $\norm{\sigma}{x}$ and $\norm{\theta}{y}$ (1/2).} \label{fig:normalisation1}
\end{figure*}

In what follows we call \emph{$m$-atomic} a message $M$ if there are no such $M_1$, $M_2$, s.t. $M =_E \mult{M_1}{M_2}$, and \emph{$\phi$-atomic} if there are no such $M_1$, $M_2$, s.t. $M =_E \smult{M_1}{M_2}$. A subterm $N$ of $M$ is an \emph{immediate $m$-factor} if it is $m$-atomic and there is a message term $K$, s.t. $\mult{N}{K} = M$. 

To prove that $	\specc{(K, F, A, \Gamma, B)}(X,Y,Z)$ is statically equivalent to $	\impll{(\vec{K}, F, A, \Gamma, B, \Lambda)}(X,Y,Z)$ we will identify building blocks for messages available to an attacker and present these building blocks in a unique form up to multiplication. We call this form a \emph{weak normal form}~\cite{horne2022csf} and define it inductively as follows

\begin{itemize}
	\item $M=\gen$ or $M$ is a variable, then $\nf{M} = M$.
	\item $M= \mult{M_1}{M_2}$, then $\nf{M} = \mult{\nf{M_1}}{\nf{M_2}}$.
	\item $M = \smult{M_1}{M_2}$, then $\nf{M} = \smult{\nf{M_1}}{\nf{M_2}}$ if $\nf{M_2}$ is $\phi$-atomic. Or else $\squeezespaces{0.8}\nf{M} = \smult{\mult{\nf{M_1}}{\nf{M_2'}}}{\nf{M_2''}}$, where $\squeezespaces{0.8}M_2 =_E \smult{M_2'}{M_2''}$ and $\nf{M_2''}$ is $\phi$-atomic.
	\item $M = \enc{M_1}{M_2}$, then $\nf{M} = \enc{\nf{M_1}}{\nf{M_2}}$.
	\item $M=\nlist{M_1, \hdots, M_n}$, then $\nf{M} = \nlist{\nf{M_1}, \hdots, \nf{M_2}}$.
	\item $M=\hash{M_1}$,  $M=\pks{M_1}$, or $M=\pkv{M_1}$ then $\nf{M} = \hash{\nf{M_1}}$, $\nf{M} = \pks{\nf{M_1}}$, or $\nf{M} = \pkv{\nf{M_1}}$. 
	\item $M = \sig{M_2}{M_1}$, then $\nf{M} = \sig{\nf{M_2}}{\nf{M_1}}$.
	\item $\squeezespaces{1}M = \sigv{M_2}{M_1}$, then $\squeezespaces{1}\nf{M} = \sigv{\nf{M_2}}{\nf{M_1}}$ if $\nf{M_1}$ is $\phi$-atomic. Otherwise $\squeezespaces{1}\nf{M} = \smult{\nf{M_1'}}{\sigv{\nf{M_2}}{\nf{M_1''}}}$, where $\squeezespaces{1}M_1 = \smult{M_1'}{M_1''}$ and $\squeezespaces{1}\nf{M_1''}$ is $\phi$-atomic.
	\item $\squeezespaces{0.3}M = \checksig{\pks{M_2}}{\sig{M_2}{M_1}}$, then $\squeezespaces{0.3}\nf{M} = \nf{M_1}$.
	\item $\squeezespaces{1}M = \checksigv{\pkv{M_2}}{\sigv{M_2}{M_1}}$, then $\squeezespaces{1}\nf{M} = \nf{M_1}$.
	\item $M = \proj{k}{\nlist{M_1, \hdots, M_n}}$ then $\nf{M} = \nf{M_k}$.
	\item $M = \dec{M_2}{\enc{M_1}{M_2}}$, then $\nf{M} = \nf{M_1}$.
	\item Otherwise $\nf{M}=M$.
\end{itemize}

The notion of the \emph{normalisation} of a frame $\nw\vec{z}.\rho$ (denoted as $\norm{\rho}{z}$) with respect to the equational theory $E$ captures the saturation of the range of $\rho$ with weak normal forms of messages that have recipes under $\nw\vec{z}.\rho$ and is defined by the following procedure. 

\begin{enumerate}
	\item $u\rho=M$ for any $u \in \dom{\rho}$ is replaced by $u\rho = \nf{M}$.
	\item If $u\rho = \mult{K_1}{K_2}$ and there is a recipe $M_1$ for an immediate $m$-factor $K_1$, then $M_1\rho$ is added to the normalisation. If there is a recipe $M_2$ for an immediate $m$-factor $K_2$, then $M_2\rho$ is also added to the normalisation. 
	\item If $u\rho = \nlist{K_1, \hdots, K_n}$, then $u\rho$ is replaced by $\proj{i}{u\rho} = K_i$, $1 \leq i \leq n$.
	\item If $u\rho = \enc{K_1}{K_2}$ and there is a recipe $M_2$ for $K_2$, then $u\rho$ is replaced by $\dec{M_2}{u}\rho = K_1$.
	\item If $u\rho = \sig{N_2}{N_1}$ and there is a recipe $M_2$ for $N_2$, then $u\rho$ is replaced by $\checksig{\pks{M_2}}{u}\rho = N_1$. 
	\item If $u\rho = \sig{N_2}{N_1}$ and there is a recipe $M_2$ for $\pks{N_2}$, then $\checksig{M_2}{u}\rho = N_1$ is added to the normalisation.
	\item If $u\rho = \sigv{N_2}{N_1}$ and there is a recipe $M_2$ for $N_2$, then $u\rho$ is replaced by $\checksigv{\pks{M_2}}{u}\rho = N_1$. 
	\item If $u\rho = \sigv{N_2}{N_1}$ and there is a recipe $M_2$ for $\pkv{N_2}$, then $\checksigv{M_2}{u}\rho = N_1$ is added to the normalisation.
	%	\item Otherwise a message term remains untouched.
\end{enumerate}

We are now ready to define in Fig.~\ref{fig:normalisation1}, ~\ref{fig:normalisation2} the normalisations $\norm{\sigma}{x}$ and $\norm{\theta}{y}$ of the frames $\nw \vec{x}.\sigma$ and $\nw \vec{y}.\theta$, where $\vec{x}$ and $\vec{y}$ define the sets of bound names in the state $\specc{(K, F, A, \Gamma, B)}(X,Y,Z)$ and the state $\impll{(\vec{K}, F, A, \Gamma, B, \Lambda)}(X,Y,Z)$ respectively. As before we use the index function $\ind: \{\sigma, \theta\} \rightarrow \mathcal{D}$ defined as $\ind(\sigma) = i, \, \ind(\theta)=j$. We also define denote an attacker's input as $A^1_i$, $B^1_i$. 

\begin{figure*}[h]
	\[
	\arraycolsep=1.4pt
	\begin{array}{ll}
		vb_i \rho = \textit{emcert}(a_i, c_{\ind(\rho)}, X^i_1\rho) \textcolor{gray}{\ =\enc{-}{-}} & \parbox{9.5cm}{\squeezespaces{1} \textcolor{teal}{if $X^1_i \neq \smult{A^1_i}{\gen}$} and if $i  \in \bigcup_{l=5}^{7} \alpha_l$ (and, if $\rho=\theta$, $i \in \lambda_j$) and
			\\$\checksig{\pks{s}}{\proj{2}{\dec{X^2_i\rho}{\hash{\smult{\mult{a_i}{c_{\ind(\rho)}}}{X^1_i\rho}}}}} =$ \\ \hspace*{0.3cm} $=\proj{1}{\dec{X^2_i\rho}{\hash{\smult{\mult{a_i}{c_{\ind(\rho)}}}{X^1_i\rho}}}}$ and 
			\\$\proj{1}{\proj{1}{\dec{X^2_i\rho}{\hash{\smult{\mult{a_i}{c_{\ind(\rho)}}}{X^1_i\rho}}}}} = \texttt{MM}$}
		\\[20pt]
		uc_i \rho = \textit{etxpin}(t_i, tx_i, Z^i_1\rho, \text{uPIN}) \textcolor{gray}{\ =\enc{-}{-}} & \parbox{10.4cm}{\squeezespaces{1}if $i  \in \bigcup_{l=7}^{11} \gamma^{\of}_l$ and
			\\$\checksigv{\pkv{\chi_{\texttt{MM}}}}{\proj{2}{\dec{Z^2_i\rho}{\hash{\smult{t_i}{Z^1_i\rho}}}}} =$\\ \hspace*{0.3cm} $=\proj{1}{\dec{Z^2_i\rho}{\hash{\smult{t_i}{Z^1_i\rho}}}}$ and 
			\\$\proj{1}{\dec{Z^2_i\rho}{\hash{\smult{t_i}{Z^1_i\rho}}}} = Z^i_1\rho$}
		\\[20pt]
		\ \, \qquad = \textit{etx}(t_i, tx_i, Z^i_1\rho) \textcolor{gray}{\ =\enc{-}{-}} & \parbox{10.3cm}{\squeezespaces{1}if $i  \in \bigcup_{l=7}^{11} \gamma^{\on}_l$ or $ \bigcup_{l=6}^{10} \gamma^{\lo}_l$ and 
			\\ 
			$\checksigv{pk_\texttt{MM}}{\proj{2}{\dec{Z^2_i\rho}{\hash{\smult{t_i}{Z^1_i\rho}}}}} =$\\ $\hspace*{0.3 cm} = \proj{1}{\dec{Z^2_i\rho}{\hash{\smult{t_i}{Z^1_i\rho}}}}$ and 
			\\
			$\proj{1}{\dec{Z^2_i\rho}{\hash{\smult{t_i}{Z^1_i\rho}}}} = Z^i_1\rho$}
		\\[20pt]
		vc_i \rho = \textit{eaclo}(a_i, c_{\ind(\rho)}, mk_i, \text{PAN}_i, X^i_1\rho,  X^i_2\rho,  X^i_3\rho) \textcolor{gray}{\ =\enc{-}{-}} & \mbox{if $i  \in \alpha_7$ (and, if $\rho=\theta$, $i \in \lambda_j$) and $\proj{2}{\dec{X^3_i\rho}{\hash{\smult{\mult{a_i}{c_{\ind(\rho)}}}{X^1_i\rho}}}} = \fail$}	
		\\[8pt]
		\ \, \qquad = \textit{eachi}(a_i, c_{\ind(\rho)}, mk_i, \text{PAN}_i, X^i_1\rho,  X^i_2\rho,  X^i_3\rho) \textcolor{gray}{\ =\enc{-}{-}} & \parbox{10.4cm}{if $i  \in \alpha_7 $ (and, if $\rho=\theta$, $i \in \lambda_j$) and \\ $\proj{2}{\dec{X^3_i\rho}{\hash{\smult{\mult{a_i}{c_{\ind(\rho)}}}{X^1_i\rho}}}} = \text{PIN}_i$}	
		\\[8pt]
		\ \, \qquad = \textit{eacfail}(a_i, c_{\ind(\rho)}, mk_i, \text{PAN}_i, X^i_1\rho,  X^i_2\rho,  X^i_3\rho) \textcolor{gray}{\ =\enc{-}{-}} & \mbox{if $i  \in \alpha_7$ (and, if $\rho=\theta$, $i \in \lambda_j$) and else}
		\\[8pt]
		ud_i \rho = \enc{\nlist{\textsf{TX}_i, Z_1^i\rho, \proj{1}{\dec{Z_3^i\rho}{\hash{\smult{t_i}{Z_1^i\rho}}}}, \text{uPIN}}}{\kbt_i} & \mbox{if $i  \in \bigcup_{l=9}^{11} \gamma^{\on}_l$ and $\proj{3}{\dec{Z_3^i\rho}{\hash{\smult{t_i}{Z_1^i\rho}}}} = \textsf{TX}_i$} 
		\\[8pt]
		\ \ \qquad = \enc{\nlist{\textsf{TX}_i, Z_1^i\rho, \proj{1}{\dec{Z_3^i\rho}{\hash{\smult{t_i}{Z_1^i\rho}}}}, \fail}}{\kbt_i} & \mbox{if $i  \in \gamma^{\of}_{11}$ and $\proj{3}{\dec{Z_3^i\rho}{\hash{\smult{t_i}{Z_1^i\rho}}}} = \textsf{TX}_i$} 
		\\[8pt]
		\ \ \qquad =  \enc{\nlist{\textsf{TX}_i, Z_1^i\rho, \proj{1}{\dec{Z_3^i\rho}{\hash{\smult{t_i}{Z_1^i\rho}}}}, \fail}}{\kbt_i} & \mbox{if $i  \in \bigcup_{l=8}^{10} \gamma^{\lo}_l$ and $\proj{3}{\dec{Z_3^i\rho}{\hash{\smult{t_i}{Z_1^i\rho}}}} = \textsf{TX}_i$}
		\\[20pt]
		wa_i\rho = \enc{\nlist{\proj{1}{dy_i}, \texttt{accept}}}{\kbt_i} & \parbox{11.1cm}{\squeezespaces{1}let $dy_i = \dec{Y^1_i\rho}{\kbt_i}$ and $\nlist{\text{PIN}_j, mk_j, \pkg{c_j}} = \text{DB}$ \\ if $i  \in \beta_4$ and $\exists j$, s.t. $j \in \alpha_7$ and
			%	\\$\text{PAN}_j = \proj{2}{\proj{1}{\dec{\proj{3}{dy_i}}{\hash{\smult{b_t}{\proj{2}{dy_i}}}}}}$ and 
			\\$\squeezespaces{1}\hash{\nlist{\proj{1}{\dec{\proj{3}{dy_i}}{\hash{\smult{b_t}{\proj{2}{dy_i}}}}}, mk_j}}=$\\$\hspace*{0.3cm}=\proj{2}{\dec{\proj{3}{dy_i}}{\hash{\smult{b_t}{\proj{2}{dy_i}}}}}$ and \\
			$\proj{3}{\proj{1}{\dec{\proj{3}{dy_i}}{\hash{\smult{b_t}{\proj{2}{dy_i}}}}}}= \proj{1}{dy_i}$ and 
			\\$\smult{\proj{1}{\proj{1}{\dec{\proj{3}{dy_i}}{\hash{\smult{b_t}{\proj{2}{dy_i}}}}}}}{\pkg{c_j}}= \proj{2}{dy_i}$ and 
			\\ $\bigl(\proj{2}{\proj{1}{dy_i}} = \lo$ or 
			\\$\proj{2}{\proj{1}{dy_i}}= \hi$ and $\proj{4}{\proj{1}{\dec{\proj{3}{dy_i}}{\hash{\smult{b_t}{\proj{2}{dy_i}}}}}} = \texttt{ok}$ 
			\\or else if $\proj{4}{dy_i} = \text{PIN}_j\bigl)$}
		\\[55pt]
		ue_i\rho = \texttt{auth} &  \parbox{7cm}{if $i \in \gamma^{\of}_9 \cup \gamma^{\of}_{11}$ and $\proj{2}{\dec{Z^3_i}{\hash{\smult{t_i}{Z^1_i}}}} = \texttt{ok}$ or\\
			if $i \in \gamma^{\on}_{11}$ or $i \in \gamma^{\lo}_{10}$ and $\proj{1}{\dec{Z_4^i}{\kbt_i}} = \textsf{TX}_i$ and $\proj{2}{\dec{Z_4^i}{\kbt_i}} = \texttt{accept}$}	
	\end{array}
	\]
	\caption{Normalisations $\norm{\sigma}{x}$ and $\norm{\theta}{y}$ (2/2).} \label{fig:normalisation2}
\end{figure*}

Notice that $\vec{y} \supset \vec{x}$, hence it is enough to prove that for all messages $M$ and $N$, s.t. $\vec{x} \text{ \fresh{} } M, N$, we have $M\sigma =_E N\sigma$ iff $M\theta =_E N\theta$. We conduct the proof of static equivalence by induction on the structure of the weak normal form of $N\sigma$. We always start from the equation in the frame $\nw\vec{x}.\sigma$, since the converse case is similar. In what follows $M_i$, $N_i$ are recipes, i.e. are always fresh for $\vec{x}$.

\textit{Case} 1. $N\sigma =_E \gen$. 

\textit{Case} 1.1. $N = \gen$. If $M$ is a recipe for $\gen$, then $M = \gen$, since there is no non-trivial recipe for $\gen$ under $\norm{\sigma}{x}$ and we have $\gen\sigma =_E \gen\sigma$ iff $\gen\theta =_E \gen\theta$ as required. 

\textit{Case} 1.2. $N \neq \gen$. There is nothing to prove, since there is no non-trivial recipe for $\gen$ under $\norm{\sigma}{x}$. 

\textit{Case} 2. $N\sigma =_E z$, $z$ is a variable. 

\textit{Case} 2.1. $N = z$. If $M$ is a recipe for $z$, then $M = z$, since there is no non-trivial recipe for $z$ under $\norm{\sigma}{x}$ and we have $z\sigma =_E z\sigma$ if and only if $z\theta =_E z\theta$ as required. 

\textit{Case} 2.2. $\squeezespaces{0.5}N\sigma =_E ch_i$, where $\squeezespaces{0.5}ch_i \in \{\dot{ch}_i, \ddot{ch}_i, \dddot{ch}_i,\}$. Since $N$ is fresh for $\vec{x}$, $\squeezespaces{0.5}N \in \{chc_i, cht_i, chb_i\}$, and in either case there is a unique recipe $\squeezespaces{0.5}M \in \{chc_i, cht_i, chb_i\}$ for $ch_i$ under $\norm{\sigma}{x}$, and we have $M\sigma =_E N\sigma$ iff $M\theta =_E N\theta$ as required.  

\textit{Case} 3. $N\sigma =_E \mult{K_1}{K_2}$.

Notice that all message terms in the range of $\norm{\sigma}{x}$ are $m$-atomic, hence no message is an immediate $m$-factor of another message. Therefore $N\sigma$ is generated by $m$-factors which have a recipe under $\norm{\sigma}{x}$. 

\textit{Case} 3.1. $\squeezespaces{0.5}N = N_1^{\epsilon_1} \cdot \, \hdots \, \cdot N_k^{\epsilon_k}$, and we have $\squeezespaces{0.5}N\sigma = N_1^{\epsilon_1}\sigma \cdot \, \hdots \, \cdot N_k^{\epsilon_k}\sigma$. By the induction hypothesis suppose that for all recipes $M_i$ for an $m$-factor $N_i\sigma$ of $N\sigma$, we have $M_i\sigma =_E N_i\sigma$ iff $M_i\theta =_E N_i\theta$ for $i \in \{1, \hdots, k\}$. By applying multiplication, we have $M_1^{\epsilon_1}\theta \cdot \, \hdots \, \cdot M_k^{\epsilon_k}\theta = (M_1^{\epsilon_1} \cdot \, \hdots \,\cdot M_k^{\epsilon_k})\theta =_E (N_1^{\epsilon_1} \cdot \, \hdots \, \cdot N_k^{\epsilon_k})\theta = N_1^{\epsilon_1}\theta \cdot \, \hdots \, \cdot N_k^{\epsilon_k}\theta$ as required, and $N_i\theta$ is an $m$-factor of $N\theta$.

\textit{Case} 4. $N\sigma =_E \smult{K_1}{K_2}$.

We have several recipes of the form $\smult{\cdot}{\cdot}$ in the domain of $\norm{\sigma}{x}$.

\begin{equation}\notag
	\begin{array}{ll}
		V_1 \coloneqq ua_i, V_2 \coloneqq va_i \\[2pt]
		V_3 \coloneqq \proj{2}{\proj{1}{\dec{ub_i}{\hash{\smult{B^1_i}{ua_i}}}}} \\[2pt]
		V_4 \coloneqq  \proj{2}{\checksig{pk_s}{\proj{2}{\dec{ub_i}{\hash{\smult{B^1_i}{ua_i}}}}}} \\[2pt]
		V_5 \coloneqq  	\proj{1}{\dec{vb_i}{\hash{\smult{A^i_1}{va_i}}}} \\[2pt]
		V_6 \coloneqq  	\proj{2}{\dec{vb_i}{\hash{\smult{A^i_1}{va_i}}}} \\[2pt]
		V_7 \coloneqq  	\checksigv{pk_\texttt{MM}}{\proj{2}{\dec{vb_i}{\hash{\smult{A^i_1}{va_i}}}}} \\[2pt]
	\end{array}
\end{equation}

\textit{Case} 4.1. $N\sigma =_E \pkg{t_i}$. Since $N$ is fresh for $\vec{x}$, $N = V_1$. Let $M$ be a recipe for $\pkg{t_i}$, then $M = V_1$ and we have $V_1\sigma =_E V_1\sigma$ iff $V_1\theta =_E V_1\theta$ as required.

\textit{Case} 4.2. $N\sigma =_E \smult{\mult{a_i}{c_{i}}}{\gen}$ and $X^1_i = \smult{A^1_i}{\gen}$. Since $N$ is fresh for $\vec{x}$, $N \in \{V_2, V_5, V_7\}$. Let $M$ be a recipe for $\smult{\mult{a_i}{c_{i}}}{\gen}$, then $M \in \{V_2, V_5, V_7\}$ and we have $M\sigma =_E N\sigma$ iff $M\theta =_E N\theta$ for any $N$ and $M$ as required. If $X^1_i \neq \smult{A^1_i}{\gen}$, $N=V_2$, there is only one recipe $M_1 = V_1$ and the argument is the same. 

\textit{Case} 4.3. $N\sigma =_E \pkg{\bt}$ and $Z^1_i = \smult{B^1_i}{\gen}$. Since $N$ is fresh for $\vec{x}$, $N \in \{V_3, V_4\}$. Let $M$ be a recipe for $\pkg{\bt}$, then $M \in \{V_3, V_4\}$ and the argument is identical to Case 4.1. If $Z^1_i \neq \smult{B^1_i}{\gen}$ there is no recipe for $\pkg{\bt}$ and there is nothing to prove. 

\textit{Case} 4.4. $N\sigma =_E \smult{\mult{a_i}{c_{i}}}{\sigv{\chi_{\texttt{MM}}}{\gen}}$ when $X^1_i = \smult{A^1_i}{\gen}$. Identical to Case 4.1, where $N=M=V_6$, and there is nothing to prove if $X^1_i \neq \smult{A^1_i}{\gen}$.

\textit{Case} 4.5. $N = \smult{N_1}{N_2}$, $N_2 \in \{V_1, \hdots V_7\}$ for  $Z^1_i = \smult{B^1_i}{\gen}$ and $X^1_i = \smult{A^1_i}{\gen}$. By the induction hypothesis  suppose that for all recipes $M_1$ for $N_1\sigma$ , we have $M_1\sigma =_E N_1\sigma$ iff $M_1\theta =_E N_1\theta$, then by multiplying $N_2$ by $M_1$ we get $\smult{M_1\theta}{N_2\theta} = \smult{M_1}{N_2}\theta =_E \smult{N_1}{N_2}\theta = \smult{N_1\theta}{N_2\theta}$ for any $N_2$ as required. In case $Z^1_i \neq \smult{B^1_i}{\gen}$ and $X^1_i = \smult{A^1_i}{\gen}$ we have $N_2 \in \{\hdots \widehat{V}_3, \widehat{V}_4 \hdots\}$; in case $Z^1_i = \smult{B^1_i}{\gen}$ and $X^1_i \neq \smult{A^1_i}{\gen}$ we have $N_2 \in \{\hdots \widehat{V}_5, \widehat{V}_6, \widehat{V}_7\}$; and in case $Z^1_i \neq \smult{B^1_i}{\gen}$ and $X^1_i \neq \smult{A^1_i}{\gen}$ we have $N_2 \in \{V_1, V_2\}$, and the argument is the same. 

\textit{Case} 4.6. $N = \texttt{sig}(\hdots \, \sig{N_2}{N_1} \, \hdots \, ,N_k)$, $N_1 \in \{V_1, \hdots,  V_7\}$ for  $Z^1_i = \smult{B^1_i}{\gen}$ and $X^1_i = \smult{A^1_i}{\gen}$. By the induction hypothesis suppose that for all recipes $M_i$ for $N_i\sigma$ we have $M_i\sigma =_E N_i\sigma$ iff $M_i\theta =_E N_i\theta$ for any $i \in \{2, \hdots, k\}$. By applying the $\sigv{\cdot}{\cdot}$ function to $N_1$, we have 
\begin{equation}\notag
	\begin{array}{ll}
		\texttt{vsig}(\hdots \, \sigv{M_2}{N_1} \, \hdots \, ,M_k)\theta = \\[2pt]
		\texttt{vsig}(\hdots \, \sigv{M_2\theta}{N_1\theta} \, \hdots \, ,M_k\theta) =_E \\[2pt]
		\texttt{vsig}(\hdots \, \sigv{N_2\theta}{N_1\theta} \, \hdots \, ,N_k\theta) = \\[2pt]
		\texttt{vsig}(\hdots \, \sigv{N_2}{N_1} \, \hdots \, ,N_k)\theta
	\end{array}
\end{equation}
as required. In case $Z^1_i \neq \smult{B^1_i}{\gen}$ and $X^1_i = \smult{A^1_i}{\gen}$ we have $N_1 \in \{\hdots \widehat{V}_3, \widehat{V}_4 \hdots\}$; in case $Z^1_i = \smult{B^1_i}{\gen}$ and $X^1_i \neq \smult{A^1_i}{\gen}$ we have $N_1 \in \{\hdots \widehat{V}_5, \widehat{V}_6, \widehat{V}_7\}$; and in case $Z^1_i \neq \smult{B^1_i}{\gen}$ and $X^1_i \neq \smult{A^1_i}{\gen}$ we have $N_1 \in \{V_1, V_2\}$, and the argument is the same. 

\textit{Case} 4.7. $N = \smult{N_1}{N_2}$. Similar to Case 3.1 for $\squeezespaces{0.7}\epsilon_1=\epsilon_2=1$, $k=2$. 

\textit{Case} 5. $N\sigma =_E \pair{K_1}{K_2}$.

The range of $\norm{\sigma}{x}$ contains no pair, hence the only option is $N = \pair{N_1}{N_2}$, which is identical to Case 4.7.

\textit{Case} 6. $N\sigma =_E \hash{K_1}$. Identical to Case 3.1, where $\epsilon_1=1$, $k=1$. 

\textit{Case} 7. $N\sigma =_E \pks{K_1}$. 

\textit{Case} 7.1. $N\sigma =_E \pks{s}$. Then $N=pk_s$, since $N$ is fresh for $\vec{x}$. There is a unique recipe $M = pk_s$ for $\pks{s}$ and we have $pk_s\sigma =_E pk_s\sigma$ if and only if $pk_s\theta =_E pk_s\theta$ as required.

\textit{Case} 7.2. $N = \pks{N_1}$. Identical to Case 6.

\textit{Case} 8. $N\sigma =_E \pkv{K_1}$. Identical to Case 7, since there is a unique recipe $pk_\texttt{MM}$ in the range of $\norm{\sigma}{x}$.

\textit{Case} 9. $N\sigma =_E \sigv{K_2}{K_1}$. Identical to Case 5.

\textit{Case} 10. $N\sigma =_E \sig{K_2}{K_1}$. Identical to Case 7, since there is a unique recipe $\proj{2}{\dec{ub_i}{\hash{\smult{B^1_i}{ua_i}}}}$ in the range of $\norm{\sigma}{x}$ if $Z^1_i=\smult{B^1_i}{\gen}$, and there is nothing to prove if $Z^1_i \neq \smult{B^1_i}{\gen}$.

\textit{Case} 11. $N\sigma =_E \enc{K_1}{K_2}$. 

Cases 11.1-11.4 are identical to Case 2.2, however we list all possibilities for the sake of completeness. Let us define $\textit{ENK} = \{uc_i\sigma, vc_i\sigma, ud_i\sigma, wa_i\sigma\}$.

\textit{Case} 11.1. $N\sigma =_E \textit{enk}$,  where $\textit{enk} \in \textit{ENK}$, and $Z^1_i = \smult{B^1_i}{\gen}$ and $X^1_i = \smult{A^1_i}{\gen}$.

\textit{Case} 11.2. $N\sigma =_E \textit{enk}$, where $\textit{enk} \in \textit{ENK} \cup \{ub_i\sigma\}$, and $Z^1_i \neq \smult{B^1_i}{\gen}$ and $X^1_i = \smult{A^1_i}{\gen}$.

\textit{Case} 11.2. $N\sigma =_E \textit{enk}$, where $\textit{enk} \in \textit{ENK} \cup \{vb_i\sigma\}$, and $Z^1_i = \smult{B^1_i}{\gen}$ and $X^1_i \neq \smult{A^1_i}{\gen}$.

\textit{Case} 11.4. $N\sigma =_E \textit{enk}$, where $\textit{enk} \in \textit{ENK} \cup \{ub_i\sigma, vb_i\sigma\}$, and $Z^1_i \neq \smult{B^1_i}{\gen}$ and $X^1_i \neq \smult{A^1_i}{\gen}$.

\textit{Case} 11.5. $N = \enc{N_1}{N_2}$. identical to Case 5.

\textit{Case} 12. $N\sigma =_E \texttt{MM}$.

\textit{Case} 12.1. $N = \texttt{MM}$. Identical to Case 1.1.

\textit{Case} 12.2. $\squeezespaces{0.9}N\neq \texttt{MM}$, and $\squeezespaces{0.9}N\sigma =_E \texttt{MM}$. Similar to 2.2 for $\squeezespaces{0.9}Z^1_i = \smult{B^1_i}{\gen}$.

\textit{Case} 13. $N\sigma \in \{\fail, \texttt{ok}, \texttt{accept}, \texttt{auth}, \texttt{lo}, \texttt{hi} \}$. Identical to Case 1. 

\end{proof}
\end{customthm}

	\section{Contactless Low-value payments are unlinkable even if the PIN is compromised}\label{apx:lowval}

Here we expand on the point made at the end of Section~\ref{sec:attackermodel} -- assuming that an attacker cannot execute contact sessions with a card, unlinkability of low-value contactless payments is preserved even if the PIN, the card's strong identity, is compromised. The key observation is that low-value terminals do not require the PIN, i.e. no input on the private channel $\user$ is expected. 

To model the situation, we drop high-value terminals $T_{\texttt{onhi}}$ and $T_{\texttt{offhi}}$ from the picture and simplify how banks communicate with honest terminals, i.e. instead of session-specific bank-terminal symmetric key $\kbt$, as in Fig.~\ref{fig:system}, we use one global shared symmetric key. Fig.~\ref{fig:systemlo} contains the real-world and the idealised-unlinkable versions of the system with low-value payments only. Let us call such reduced version of the protocol UTXL ($\protocol$ Low).

\begin{figure}[h]
	\begin{subfigure}[t]{1.0\columnwidth}
		\caption{The real protocol specification $\impl{UTXL}$.}
		\begin{equation}
			\notag
			\begin{aligned}
				&\nu \, s, si, \chi_{\texttt{MM}}, b_t, \kbt. \cout{\out_s}{\pks{s}}.\Bigl( \\ 
				&
				\;\; 
				\begin{array}{l}
					\mathbf{\bang} \nu\text{PIN}, mk, c, \text{PAN}. \bigl( \\[4pt]
					\;\;
					\begin{array}{l}
						\cout{\opin}{\text{PIN}}.\\[4pt]
						\lett \, \textsf{crtC} \coloneqq  \sigv{\chi_{\texttt{MM}}}{\pkg{c}} \inn \\[4pt]
						\textcolor{red}{\bang}\nw \ch.\cout{card}{\ch}.C(\ch, c, \pk{s}, \textsf{crtC}, \text{PAN}, mk, \text{PIN}) \cpar \\[4pt]
						! \cout{\nlist{si, \text{PAN}}}{\nlist{\text{PIN}, mk, \pkg{c}}} \;\bigl) \; \cpar \; \\[4pt]
						\bang \nu \ch.\cout{bank}{\ch}.B(\ch, si, \kbt, b_t) \cpar \\[4pt]
						\lett \, \textsf{crt} \coloneqq \nlist{\nlist{\texttt{MM}, \pkg{b_t}}, \sig{s}{\nlist{\texttt{MM}, \pkg{b_t}}}} \inn \\[4pt]
						\cout{\out_v}{\pkv{ \chi_{\texttt{MM}}}}.\\[4pt]
						\cout{\out_c}{\textsf{crt}}. \\[4pt]
						\bang \nu \ch.\cout{term}{\ch}.T_{\texttt{lo}}(\ch, \pkv{\chi_{\texttt{MM}}}, \textsf{crt}, \kbt)~\Bigl)
					\end{array}
				\end{array}
			\end{aligned}
		\end{equation} 
%		\vspace{0.1cm}
	\end{subfigure} 
	
	\begin{subfigure}[t]{1.0\columnwidth}
		\caption{The ideal unlinkable protocol specification $\spec{UTXL}$.}
		\begin{equation}
			\notag
			\begin{aligned}
				&\nu \, s, si, \chi_{\texttt{MM}}, b_t, \kbt. \cout{\out_s}{\pks{s}}.\Bigl( \\ 
				&
				\;\; 
				\begin{array}{l}
					\mathbf{\bang} \nu\text{PIN}, mk, c, \text{PAN}. \bigl( \\[4pt]
					\;\;
					\begin{array}{l}
						\cout{\opin}{\text{PIN}}.\\[4pt]
						\lett \, \textsf{crtC} \coloneqq  \sigv{\chi_{\texttt{MM}}}{\pkg{c}} \inn \\[4pt]
						\nw \ch.\cout{card}{\ch}.C(\ch, c, \pk{s}, \textsf{crtC}, \text{PAN}, mk, \text{PIN}) \cpar \\[4pt]
						! \cout{\nlist{si, \text{PAN}}}{\nlist{\text{PIN}, mk, \pkg{c}}} \;\bigl) \; \cpar \; \\[4pt]
						\bang \nu \ch.\cout{bank}{\ch}.B(\ch, si, \kbt, b_t) \cpar \\[4pt]
						\lett \, \textsf{crt} \coloneqq \nlist{\nlist{\texttt{MM}, \pkg{b_t}}, \sig{s}{\nlist{\texttt{MM}, \pkg{b_t}}}} \inn \\[4pt]
						\cout{\out_v}{\pkv{ \chi_{\texttt{MM}}}}.\\[4pt]
						\cout{\out_c}{\textsf{crt}}. \\[4pt]
						\bang \nu \ch.\cout{term}{\ch}.T_{\texttt{lo}}(\ch, \pkv{\chi_{\texttt{MM}}}, \textsf{crt}, \kbt)~\Bigl)
					\end{array}
				\end{array}
			\end{aligned}
		\end{equation}
	\end{subfigure} 
	\caption{Specifications for the real UTXL protocol and its ideal unlinkable version.}\label{fig:systemlo}
\end{figure}

Besides the differences with the full system in Fig.~\ref{fig:system} emphasised above, in the system with low-value transactions only, we explicitly output PINs on the public channel $\opin$, and the public information that an attacker may use to construct a low-value accepting terminal -- the public key to verify the card, and the bank's certificate -- in the channels $\out_v$, $\out_c$ next to the terminal's process (($\cout{\out_v}{\pkv{ \chi_{\texttt{MM}}}}$, $\cout{\out_c}{\textsf{crt}}$)). 
To verify the unlinkability of UTXL, we should show that $\impl{UTXL} \sim \spec{UTXL}$, however in that case, we can reduce the amount of work needed for verification using compositionality and verify a strictly smaller system.

Let us consider in Fig.~\ref{fig:systemlosmall} the respective real-world and idealised subsystems of UTXL called SUTXL (Small UTXL) comprising only cards and banks.

\begin{figure}[h]
	\begin{subfigure}[t]{1.0\columnwidth}
		\caption{The real protocol specification $\impl{SUTXL}$.}
			\begin{equation}
			\notag
			\begin{aligned}
				&\nu \, s, si, \chi_{\texttt{MM}}, b_t. \\[4pt]
				&\lett \, \textsf{crt} \coloneqq \nlist{\nlist{\texttt{MM}, \pkg{b_t}}, \sig{s}{\nlist{\texttt{MM}, \pkg{b_t}}}} \inn \\[4pt]
				&\cout{\out_v}{\pkv{ \chi_{\texttt{MM}}}}.\\[4pt]
				&\cout{\out_c}{\textsf{crt}}.\\[4pt]
				&\cout{\out_s}{\pks{s}}.\Bigl( \\ 
				&
				\;\;
				\begin{array}{l}
					\mathbf{\bang} \nu\text{PIN}, mk, c, \text{PAN}. \bigl( \\[4pt]
					\cout{\opin}{\text{PIN}}.\\[4pt]
					\lett \, \textsf{crtC} \coloneqq  \sigv{\chi_{\texttt{MM}}}{\pkg{c}} \inn \\[4pt]
					\textcolor{red}{\bang}\nw \ch.\cout{card}{\ch}.C(\ch, c, \pk{s}, \textsf{crtC}, \text{PAN}, mk, \text{PIN}) \cpar \\[4pt]
					! \cout{\nlist{si, \text{PAN}}}{\nlist{\text{PIN}, mk, \pkg{c}}} \;\bigl) \; \cpar \; \\[4pt]
					\bang \nu \ch.\cout{bank}{\ch}.B(\ch, si, \kbt, b_t)~\Bigl)
				\end{array}
			\end{aligned}
		\end{equation}
%		\vspace{0.1cm}
	\end{subfigure} 
	
	\begin{subfigure}[t]{1.0\columnwidth}
		\caption{The ideal unlinkable protocol specification $\spec{SUTXL}$.}
		\begin{equation}
			\notag
			\begin{aligned}
				&\nu \, s, si, \chi_{\texttt{MM}}, b_t. \\[4pt]
				&\lett \, \textsf{crt} \coloneqq \nlist{\nlist{\texttt{MM}, \pkg{b_t}}, \sig{s}{\nlist{\texttt{MM}, \pkg{b_t}}}} \inn \\[4pt]
				&\cout{\out_v}{\pkv{ \chi_{\texttt{MM}}}}.\\[4pt]
				&\cout{\out_c}{\textsf{crt}}.\\[4pt]
				&\cout{\out_s}{\pks{s}}.\Bigl( \\ 
				&
				\;\;
				\begin{array}{l}
					\mathbf{\bang} \nu\text{PIN}, mk, c, \text{PAN}. \bigl( \\[4pt]
					\cout{\opin}{\text{PIN}}.\\[4pt]
					\lett \, \textsf{crtC} \coloneqq  \sigv{\chi_{\texttt{MM}}}{\pkg{c}} \inn \\[4pt]
					\nw \ch.\cout{card}{\ch}.C(\ch, c, \pk{s}, \textsf{crtC}, \text{PAN}, mk, \text{PIN}) \cpar \\[4pt]
					! \cout{\nlist{si, \text{PAN}}}{\nlist{\text{PIN}, mk, \pkg{c}}} \;\bigl) \; \cpar \; \\[4pt]
					\bang \nu \ch.\cout{bank}{\ch}.B(\ch, si, \kbt, b_t)~\Bigl)
				\end{array}
			\end{aligned}
		\end{equation}
	\end{subfigure} 
	\caption{Subsystem specifications for SUTXL.}\label{fig:systemlosmall}
\end{figure}

In SUTXL we not only assume that all transactions are low-value and executed with any unauthorised device constructed using public information, but we also allow the bank to process any transactions received, as the variable $\kbt$ in SUTXL specification is not bound (in contrast to UTXL). To justify that it is enough to verify that $\impl{SUTXL} \sim \spec{SUTXL}$, we give the following context relating SUTXL and UTXL.
\[
				\begin{aligned}
					\mathcal{L}\{\cdot\} \ \triangleq \ &\nw \, out_v, out_c, \kbt. \left( \{\cdot\}  \cpar \right. \\[2pt]
					& \cin{\out_v}{\textit{pk}_{\texttt{MM}}}.\cin{\out_c}{\textsf{crt}}. \\[2pt]
					& \cout{\out_v'}{\textit{pk}_{\texttt{MM}}}.\cout{\out_c'}{\textsf{crt}}.\\[2pt]
					& \left.\bang \nu \ch.\cout{term}{\ch}.T_{\texttt{lo}}(\ch, \textit{pk}_{\texttt{MM}}, \textsf{crt}, \kbt) \right)
				\end{aligned}
\]
		
In contrast to the context $\mathcal{O}$ used in Corollary~\ref{col:bang} after we plug in the context $\mathcal{L}\{\cdot\}$ either $\impl{SUTXL}$ or $\spec{SUTXL}$, it takes two $\tau$ transitions and the application of the substitution $\sub{\out_v', \out_c'}{\out_v,\out_c}$ (since quasi-open bisimilarity is closed under substitutions) to obtain the initial bigger system ($\impl{UTXL}$ and $\spec{UTXL}$ respectively). 

To conclude that the system where all payments are low-value is still left to verify that the subsystem represented by the SUTXL protocol in Fig.~\ref{fig:systemlosmall} is unlinkable. We formulate this claim separately as a hypothesis since we are leaving the proof for future work. However, we expect such proof to be quite close to the proof of the Theorem~\ref{thm:link} in Appendix~\ref{sec:appx} since it considers a more general case. 

\begin{hypothesis}
$\impl{SUTXL} \sim \spec{SUTXL}$.
\end{hypothesis}

Finally, the directory \emph{UTXL} in the repository~\cite{repoutx} contains ProVerif code verifying injective agreement for both UTXL and SUTXL real-world specifications.
	\section{Theorem~\ref{thm:months}}\label{appx:months}

%Here we introduce the model that admits transitions from one month to the next. We will call an enhanced model UTXMM. To reflect such behaviour of cards, we require each card to respond to two months at any time and, whenever the new month is asked, to invalidate the oldest of two months. Notice that this requires a card to carry the state, i.e. to ``remember'' that it should respond only to the recent month and never return responding to the older months if asked. Below we show how we can employ recursion to model such behaviour. 
This appendix contains the details about the $\protocol$ model UTXMM admitting cards valid in different months and the proof Theorem~\ref{thm:months} from Section~\ref{sec:resultunlink}.

Without loss of generality, we restrict the model to three months $\texttt{M1}$, $\texttt{M2}$, $\texttt{M3}$, and populate the world with two types of cards -- responding to $\texttt{M1}$, $\texttt{M2}$ or to $\texttt{M2}$, $\texttt{M3}$. Notice that a card can advance its pointer to the next month only in the real-world system, where it can participate in multiple transactions. In contrast, in the idealised scenario, where cards are disposable, no change in the state of a given card is required. This requires us to have two different role specifications for cards.

In Fig.~\ref{fig:role_card_month_real}, we give the specification for the card's role in the real-world system. 
\begin{figure}[t!]
	\begin{equation}
		\squeezespaces{0.92}
		\notag
		\begin{aligned}
			\textcolor{blue}{Crw^{rec}}&(c, pk_s, \texttt{vsig}_\texttt{M1}, \texttt{vsig}_\texttt{M2}, \texttt{vsig}_\texttt{M3}, \text{PAN}, \mk, \text{PIN}) \triangleq \\
			& \textcolor{red}\nw \ch.\cout{card}{\ch}. \\
			& \cin{\ch}{z_1}. \\
			& \nw a.\; \lett z_2 \coloneqq \smult{a}{\pkg{c}} \; \inn \; \\ 
			%			& C^{rec}(\ch, c, pk_s, \texttt{vsig}_\texttt{M1}, \texttt{vsig}_\texttt{M2}, \texttt{vsig}_\texttt{M3}, \text{PAN}, \mk, \text{PIN}) + \\
			& \cout{\ch}{z_2}.\\
			& \lett \, \key{c} \coloneqq \hash{\smult{\mult{a}{c}}{z_1}} \inn \\
			%			& C^{rec}(\ch, c, pk_s, \texttt{vsig}_\texttt{M1}, \texttt{vsig}_\texttt{M2}, \texttt{vsig}_\texttt{M3}, \text{PAN}, \mk, \text{PIN})+ \\
			& \cin{\ch}{m}.\ \ \textcolor{purple}{*} \\ 
			& \lett \, \nlist{\nlist{\textsf{MM},y_B}, \textsf{MC}_s} \coloneqq \dec{m}{\key{c}} \; \inn \\
			%			& C^{rec}(\ch, c, pk_s, \texttt{vsig}_\texttt{M1}, \texttt{vsig}_\texttt{M2}, \texttt{vsig}_\texttt{M3}, \text{PAN}, \mk, \text{PIN})+ \\			
			& \ifff \ \checksig{pk_s}{\textsf{MC}_s} = \nlist{\textsf{MM},y_B} \  \\
			& \ifff \ \textsf{MM} = \texttt{M1} \ \thenn \\
			& \quad \textcolor{magenta}{Cont^{rec}}(\ch, c, pk_s, \texttt{vsig}_\texttt{M1}, \texttt{vsig}_\texttt{M2}, \texttt{vsig}_\texttt{M3}, \text{PAN}, \mk, \text{PIN}) \\
			& \elsee \; \ifff \ \textsf{MM} = \texttt{M2} \ \thenn \\
			& \quad \textcolor{magenta}{Cont^{rec}}(\ch, c, pk_s, \texttt{vsig}_\texttt{M1}, \texttt{vsig}_\texttt{M2}, \texttt{vsig}_\texttt{M3}, \text{PAN}, \mk, \text{PIN}) \\
			& \elsee \; \ifff \ \textsf{MM} = \texttt{M3} \ \thenn \\
			& \quad \nw \chi_{\texttt{M4}}. \lett \, \textsf{crtC4} \coloneqq  \sigv{\chi_{\texttt{M4}}}{\pkg{c}} \inn \\
			& \quad \textcolor{magenta}{Cont^{rec}}(\ch, c, pk_s, \texttt{vsig}_\texttt{M2}, \texttt{vsig}_\texttt{M3}, \textsf{crtC4}, \text{PAN}, \mk, \text{PIN})
			\\
			& \elsee \\
			& \quad \textcolor{blue}{Crw^{rec}}(c, pk_s, \texttt{vsig}_\texttt{M1}, \texttt{vsig}_\texttt{M2}, \texttt{vsig}_\texttt{M3}, \text{PAN}, \mk, \text{PIN})	
			\\
			%	& \ifff \ \proj{1}{\textsf{MC}} =\texttt{MM} \ \thenn \\
			\\
			\textcolor{magenta}{Cont^{rec}}&(\ch, c, pk_s, \texttt{vsig}_\texttt{M1}, \texttt{vsig}_\texttt{M2}, \texttt{vsig}_\texttt{M3}, \text{PAN}, \mk, \text{PIN}) \triangleq \\
			& \cout{\ch}{\enc{\nlist{\smult{a}{\pkg{c}}, \smult{a}{\texttt{vsig}_\texttt{MM}}}}{\key{c}}}.\\
			%			& C^{rec}(\ch, c, pk_s, \texttt{vsig}_\texttt{M1}, \texttt{vsig}_\texttt{M2}, \texttt{vsig}_\texttt{M3}, \text{PAN}, \mk, \text{PIN}) + \\
			& \cin{\ch}{x}.\\
			% low-value online, low-value offline, high-value online
			& \lett \, \nlist{\textsf{TX},\textsf{uPin}} \coloneqq \dec{x}{\key{c}} \inn\\
			& \lett \, \textsf{AC}^\fail \coloneqq \nlist{a, \text{PAN}, \textsf{TX}} \; \inn\\
			& \lett \, \textsf{AC}^{\texttt{ok}} \coloneqq \nlist{a, \text{PAN}, \textsf{TX}, \texttt{ok}} \; \inn\\
			& \lett \, \textsf{AC}^{\texttt{no}} \coloneqq \nlist{a, \text{PAN}, \textsf{TX}, \texttt{no}} \; \inn\\
			& \lett \, \key{cb} \coloneqq \hash{\smult{\mult{a}{c}}{y_B}} \; \inn \\
			%			& C^{rec}(\ch, c, pk_s, \texttt{vsig}_\texttt{M1}, \texttt{vsig}_\texttt{M2}, \texttt{vsig}_\texttt{M3}, \text{PAN}, \mk, \text{PIN}) + \\ 
			& \ifff \ \textsf{uPin} = \fail \ \thenn \\
			& \quad \cout{\ch}{\enc{\enc{\textsf{AC}^{\fail},\h{\textsf{AC}^{\fail}, \mk}}{\key{cb}}, \fail, \textsf{TX}}{\key{c}}}. \\
			& \quad \textcolor{blue}{Crw^{rec}}(c, pk_s, \texttt{vsig}_\texttt{M1}, \texttt{vsig}_\texttt{M2}, \texttt{vsig}_\texttt{M3}, \text{PAN}, \mk, \text{PIN})
			\\			& \elsee \, \ifff \ \textsf{uPin} =\text{PIN} \ \thenn \\			
			& \quad \cout{\ch}{\enc{\enc{\textsf{AC}^{\texttt{ok}},\h{\textsf{AC}^{\texttt{ok}}, \mk}}{\key{cb}}, \texttt{ok}, \textsf{TX}}{\key{c}}}. \\
			& \quad \textcolor{blue}{Crw^{rec}}(c, pk_s, \texttt{vsig}_\texttt{M1}, \texttt{vsig}_\texttt{M2}, \texttt{vsig}_\texttt{M3}, \text{PAN}, \mk, \text{PIN})
			\\			& \elsee \, \\
			&\quad \cout{\ch}{\enc{\enc{\textsf{AC}^{\texttt{no}},\h{\textsf{AC}^{\texttt{no}}, \mk}}{\key{cb}}, \texttt{no}, \textsf{TX}}{\key{c}}}. \\
			& \quad \textcolor{blue}{Crw^{rec}}(c, pk_s, \texttt{vsig}_\texttt{M1}, \texttt{vsig}_\texttt{M2}, \texttt{vsig}_\texttt{M3}, \text{PAN}, \mk, \text{PIN})			
		\end{aligned}	
	\end{equation}
%	\vspace*{0mm}
	\caption{The real-world specification of the card role in UTXMM.} \label{fig:role_card_month_real}
\end{figure}
The specification 
%of the card's behaviour in the real-world scenario 
is split into two parts. In the initial part, represented by the process $Crw^{rec}$, the card decides if it needs to advance the pointer to the next month, and the rest, represented by $Cont^{rec}$. The card is initially set up to respond for months $\texttt{M1}$, $\texttt{M2}$. Whenever one of the two is asked, the process $Cont^{rec}$ at the end of the transaction refers to $Crw^{rec}$ with the same parameters, but if the month asked is $\texttt{M3}$, the process $Cont^{rec}$ at the end of the run calls the process $Crw^{rec}$ with a ``shifted'' list of month signatures: $\texttt{vsig}_\texttt{M2}, \texttt{vsig}_\texttt{M3}, \textsf{crtC4}$. The possibility to complete the session responding to the month $\texttt{M1}$ is now lost for the card $c$ -- it simply aborts the protocol by restarting the session (the else branch in the last line of $Crw^{rec}$).

The card's role in the idealised world is specified in Fig.~\ref{fig:role_card_month_ideal1}. There is no recursion in the card's role in contrast to the real-world spec. The card simply continues the run replying to any month asked, and then is getting disposed of. 

\begin{figure}[t]
	\begin{equation}
		\squeezespaces{0.92}
		\notag
		\begin{aligned}
			Cid&(\ch, c, pk_s, \texttt{vsig}_\texttt{M1}, \texttt{vsig}_\texttt{M2}, \texttt{vsig}_\texttt{M3}, \text{PAN}, \mk, \text{PIN}) \triangleq \\
			& \cin{\ch}{z_1}. \\
			& \nw a.\; \lett z_2 \coloneqq \smult{a}{\pkg{c}} \; \inn \; \\ 
			& \cout{\ch}{z_2}.\\
			& \lett \, \key{c} \coloneqq \hash{\smult{\mult{a}{c}}{z_1}} \; \inn \\
			& \cin{\ch}{m}. \ \ \textcolor{purple}{*} \\ 
			& \lett \, \nlist{\nlist{\textsf{MM},y_B}, \textsf{MC}_s} \coloneqq \dec{m}{\key{c}} \; \inn\; \\
			& \ifff \ \checksig{pk_s}{\textsf{MC}_s} = \nlist{\textsf{MM},y_B} \  \\
			& \ifff \ \textsf{MM} = \texttt{M1} \ \thenn \\
			& \quad \textcolor{magenta}{Cont}(\ch, c, pk_s, \texttt{vsig}_\texttt{M1}, \texttt{vsig}_\texttt{M2}, \texttt{vsig}_\texttt{M3}, \text{PAN}, \mk, \text{PIN}) \\
			& \elsee \; \ifff \ \textsf{MM} = \texttt{M2} \ \thenn \\
			& \quad \textcolor{magenta}{Cont}(\ch, c, pk_s, \texttt{vsig}_\texttt{M1}, \texttt{vsig}_\texttt{M2}, \texttt{vsig}_\texttt{M3}, \text{PAN}, \mk, \text{PIN}) \\
			& \elsee \; \ifff \ \textsf{MM} = \texttt{M3} \ \thenn \\
			& \quad \nw \chi_{\texttt{M4}}. \lett \, \textsf{crtC4} \coloneqq  \sigv{\chi_{\texttt{M4}}}{\pkg{c}} \inn \\
			& \quad \textcolor{magenta}{Cont}(\ch, c, pk_s, \texttt{vsig}_\texttt{M2}, \texttt{vsig}_\texttt{M3}, \textsf{crtC4}, \text{PAN}, \mk, \text{PIN})
			\\
			%	& \ifff \ \proj{1}{\textsf{MC}} =\texttt{MM} \ \thenn \\
			\\
			\textcolor{magenta}{Cont}&(\ch, c, pk_s, \texttt{vsig}_\texttt{M1}, \texttt{vsig}_\texttt{M2}, \texttt{vsig}_\texttt{M3}, \text{PAN}, \mk, \text{PIN}) \triangleq \\
			& \cout{\ch}{\enc{\nlist{\smult{a}{\pkg{c}}, \smult{a}{\texttt{vsig}_\texttt{MM}}}}{\key{c}}}.\\
			& \cin{\ch}{x}.\\
			% low-value online, low-value offline, high-value online
			& \lett \, \nlist{\textsf{TX},\textsf{uPin}} \coloneqq \dec{x}{\key{c}} \; \inn\\
			& \lett \, \textsf{AC}^\fail \coloneqq \nlist{a, \text{PAN}, \textsf{TX}} \; \inn\\
			& \lett \, \textsf{AC}^{\texttt{ok}} \coloneqq \nlist{a, \text{PAN}, \textsf{TX}, \texttt{ok}} \; \inn\\
			& \lett \, \textsf{AC}^{\texttt{no}} \coloneqq \nlist{a, \text{PAN}, \textsf{TX}, \texttt{no}} \; \inn\\
			& \lett \, \key{cb} \coloneqq \hash{\smult{\mult{a}{c}}{y_B}} \; \inn \\
			& \ifff \ \textsf{uPin} = \fail \ \thenn \\
			& \quad \cout{\ch}{\enc{\enc{\textsf{AC}^{\fail},\h{\textsf{AC}^{\fail}, \mk}}{\key{cb}}, \fail, \textsf{TX}}{\key{c}}} \\
			& \elsee \, \ifff \ \textsf{uPin} =\text{PIN} \ \thenn \\			
			& \quad \cout{\ch}{\enc{\enc{\textsf{AC}^{\texttt{ok}},\h{\textsf{AC}^{\texttt{ok}}, \mk}}{\key{cb}}, \texttt{ok}, \textsf{TX}}{\key{c}}} \\
			& \elsee \, \\
			&\quad \cout{\ch}{\enc{\enc{\textsf{AC}^{\texttt{no}},\h{\textsf{AC}^{\texttt{no}}, \mk}}{\key{cb}}, \texttt{no}, \textsf{TX}}{\key{c}}}
		\end{aligned}	
	\end{equation}
%	\vspace*{-2mm}
	\caption{The ideal-world specification of the card role in UTXMM.} \label{fig:role_card_month_ideal1}
\end{figure}

Finally, we define the spec and imp worlds of UTXMM in Fig.~\ref{fig:systemUTXMM}. 

\begin{figure}[h!]
	\begin{subfigure}[t]{1.0\columnwidth}
		\caption{The real protocol specification $\impl{UTXMM}$.}
		\begin{equation*}
			\notag
			\begin{aligned}
				&\nu \, \user, s, si, \chi_{\texttt{M1}}, \chi_{\texttt{M2}}, \chi_{\texttt{M3}}.\\ &\cout{out}{\pks{s}}.\cout{out}{\pkv{ \chi_{\texttt{M1}}}}.\cout{out}{\pkv{ \chi_{\texttt{M2}}}}.\cout{out}{\pkv{ \chi_{\texttt{M3}}}}.\Bigl( \\ 
				&
				\;\; 
				\begin{array}{l}
					\mathbf{\bang} \nu \text{PIN}, mk, c, \text{PAN}. \bigl(\\
					\quad \lett \, \textsf{crtC1} \coloneqq  \sigv{\chi_{\texttt{M1}}}{\pkg{c}} \inn \\
					\quad \lett \, \textsf{crtC2} \coloneqq  \sigv{\chi_{\texttt{M2}}}{\pkg{c}} \inn \\
					\quad \lett \, \textsf{crtC3} \coloneqq  \sigv{\chi_{\texttt{M3}}}{\pkg{c}} \inn \\
					\quad 
					\begin{array}[t]{l}
						\;\; Crw^{rec}(c, \pk{s}, \textsf{crtC1}, \textsf{crtC2}, \textsf{crtC3}, \text{PAN}, mk, \text{PIN}) + \\
						\;\; \nw \chi_{\texttt{M4}}. \lett \, \textsf{crtC4} \coloneqq  \sigv{\chi_{\texttt{M4}}}{\pkg{c}} \inn \\
						\;\; Crw^{rec}(c, \pk{s}, \textsf{crtC2}, \textsf{crtC3}, \textsf{crtC4}, \text{PAN}, mk, \text{PIN}) 
						\\
						\cpar \; !\cout{\user}{\text{PIN}} \cpar \; ! \cout{\nlist{si, \text{PAN}}}{\nlist{\text{PIN}, mk, \pkg{c}}} \;\bigl) \; \cpar \; \\
						
						%					\textit{DB}(si, \text{PAN}, mk, \text{PIN}, \pkg{c})~\textcolor{green}{\Bigl)} \cpar
					\end{array}
					\\
					\nu b_t.\bang \nu \kbt. \textcolor{cyan}{\bigl(}
					\\
					\;\; 
					\begin{array}[t]{l}
						\nu \ch.\cout{bank}{\ch}.B(\ch, si, \kbt, b_t) \cpar \\
						\lett \, \textsf{crt1} \coloneqq \nlist{\nlist{\texttt{M1}, \pkg{b_t}}, \sig{s}{\nlist{\texttt{M1}, \pkg{b_t}}}} \inn \\
						\lett \, \textsf{crt2} \coloneqq \nlist{\nlist{\texttt{M2}, \pkg{b_t}}, \sig{s}{\nlist{\texttt{M2}, \pkg{b_t}}}} \inn \\
						\lett \, \textsf{crt3} \coloneqq \nlist{\nlist{\texttt{M3}, \pkg{b_t}}, \sig{s}{\nlist{\texttt{M3}, \pkg{b_t}}}} \inn \\
						\nu \ch.\cout{term}{\ch}.T(user, \ch, \pkv{\chi_{\texttt{M1}}}, \textsf{crt1}, \kbt)+\\
						\nu \ch.\cout{term}{\ch}.T(user, \ch, \pkv{\chi_{\texttt{M2}}}, \textsf{crt2}, \kbt)+\\
						\nu \ch.\cout{term}{\ch}.T(user, \ch, \pkv{\chi_{\texttt{M3}}}, \textsf{crt3}, \kbt)
						\textcolor{cyan}{\bigl)} ~\Bigl)
					\end{array}
				\end{array}
			\end{aligned}
		\end{equation*}
%		\vspace{0.5cm}
	\end{subfigure} 
	
	\begin{subfigure}[t]{1.0\columnwidth}
		\caption{The ideal unlinkable protocol specification $\spec{UTXMM}$.}
		\begin{equation*}
			\notag
			\begin{aligned}
				&\nu \, \user, s, si, \chi_{\texttt{M1}}, \chi_{\texttt{M2}}, \chi_{\texttt{M3}}.\\ &\cout{out}{\pks{s}}.\cout{out}{\pkv{ \chi_{\texttt{M1}}}}.\cout{out}{\pkv{ \chi_{\texttt{M2}}}}.\cout{out}{\pkv{ \chi_{\texttt{M3}}}}.\Bigl( \\ 
				&
				\;\; 
				\begin{array}{l}
					\mathbf{\bang} \nu \text{PIN}, mk, c, \text{PAN}. \bigl(\\
					\quad \lett \, \textsf{crtC1} \coloneqq  \sigv{\chi_{\texttt{M1}}}{\pkg{c}} \inn \\
					\quad \lett \, \textsf{crtC2} \coloneqq  \sigv{\chi_{\texttt{M2}}}{\pkg{c}} \inn \\
					\quad \lett \, \textsf{crtC3} \coloneqq  \sigv{\chi_{\texttt{M3}}}{\pkg{c}} \inn \\
					\quad 
					\begin{array}[t]{l}
						\;\; \textcolor{red}\nw \ch.\cout{card}{\ch}. \\ \;\; Cid(\ch, c, \pk{s}, \textsf{crtC1}, \textsf{crtC2}, \textsf{crtC3}, \text{PAN}, mk, \text{PIN}) + \\
						\;\; \nw \chi_{\texttt{M4}}. \lett \, \textsf{crtC4} \coloneqq  \sigv{\chi_{\texttt{M4}}}{\pkg{c}} \inn \\
						\;\; \textcolor{red}\nw \ch.\cout{card}{\ch}. \\ \;\; Cid(\ch, c, \pk{s}, \textsf{crtC2}, \textsf{crtC3}, \textsf{crtC4}, \text{PAN}, mk, \text{PIN}) 
						\\
						\cpar \; !\cout{\user}{\text{PIN}} \cpar \; ! \cout{\nlist{si, \text{PAN}}}{\nlist{\text{PIN}, mk, \pkg{c}}} \;\bigl) \; \cpar \; \\
						
						%					\textit{DB}(si, \text{PAN}, mk, \text{PIN}, \pkg{c})~\textcolor{green}{\Bigl)} \cpar
					\end{array}
					\\
					\nu b_t.\bang \nu \kbt. \textcolor{cyan}{\bigl(}
					\\
					\;\; 
					\begin{array}[t]{l}
						\nu \ch.\cout{bank}{\ch}.B(\ch, si, \kbt, b_t) \cpar \\
						\lett \, \textsf{crt1} \coloneqq \nlist{\nlist{\texttt{M1}, \pkg{b_t}}, \sig{s}{\nlist{\texttt{M1}, \pkg{b_t}}}} \inn \\
						\lett \, \textsf{crt2} \coloneqq \nlist{\nlist{\texttt{M2}, \pkg{b_t}}, \sig{s}{\nlist{\texttt{M2}, \pkg{b_t}}}} \inn \\
						\lett \, \textsf{crt3} \coloneqq \nlist{\nlist{\texttt{M3}, \pkg{b_t}}, \sig{s}{\nlist{\texttt{M3}, \pkg{b_t}}}} \inn \\
						\nu \ch.\cout{term}{\ch}.T(user, \ch, \pkv{\chi_{\texttt{M1}}}, \textsf{crt1}, \kbt)+\\
						\nu \ch.\cout{term}{\ch}.T(user, \ch, \pkv{\chi_{\texttt{M2}}}, \textsf{crt2}, \kbt)+\\
						\nu \ch.\cout{term}{\ch}.T(user, \ch, \pkv{\chi_{\texttt{M3}}}, \textsf{crt3}, \kbt)
						\textcolor{cyan}{\bigl)} ~\Bigl)
					\end{array}
				\end{array}
			\end{aligned}
		\end{equation*}
	\end{subfigure} 
	\caption{Specifications for the real UTXMM protocol and its ideal unlinkable version.}\label{fig:systemUTXMM}
\end{figure}

We would like to highlight two crucial differences with the respective specification of UTX presented previously in Fig.~\ref{fig:system}. Firstly, since we consider two types of cards that respond either to $\texttt{M1}$, $\texttt{M2}$ or to $\texttt{M2}$, $\texttt{M3}$ we are taking care of populating the system with both types. Right at the start, there are cards with the pointer already advanced, represented by the second branch in the choice in the card's part of the specification, i.e. $\nw \chi_{\texttt{M4}}. \lett \, \textsf{crtC4} \coloneqq \hdots$. Secondly, the replication in the $\impl{UTXMM}$ is now implicit since the process $Crw^{rec}$ is recursive. This makes the relation witnessing unlinkability in the proof of the following theorem surprisingly compact.
%We expect an enhanced model of the UTX protocol described in this section to be also unlinkable. 
%\begin{theorem}
%%	\textup{UTXLMM} is unlinkable, i.e. 
%	$\impl{UTXMM} \sim \spec{UTXMM}$.
%\end{theorem}

\begin{customthm}{2}$\impl{UTXMM} \sim \spec{UTXMM}$.
\begin{proof}
The proof of the claim above in full detail requires at least the same amount of work as the proof of Theorem~\ref{thm:link}, hence we explain here how to adapt the comprehensive proof from Appendix~\ref{sec:appx} with the main focus on the ingenious part -- defining the relation $\mathfrak{S}$, such that $\spec{UTXMM} \, \mathfrak{S} \, \impl{UTXMM}$, which the reader can verify against quasi-open bisimilarity definition.
%as we did in the proof of Theorem~\ref{thm:link} in Appendix~\ref{sec:appx}. 
To define $\mathfrak{S}$ we reuse the notation from the proof of Theorem~\ref{thm:link} and introduce some additional notation below.  

As before, firstly, we define the parameter list for the card process $\squeezespaces{1}\cparams{\texttt{MM}} \coloneqq (c, pk_s, \texttt{vsig}_\texttt{MM}, \texttt{vsig}_{\texttt{MM}+1}, \texttt{vsig}_{\texttt{MM}+2},  \text{PAN}, \mk, \text{PIN})$, and the parameter list for the terminal process $\squeezespaces{1}\tparamsmm \coloneqq (\ch, pk_\texttt{MM}, \textsf{crt}_\texttt{MM},  \kbt)$. Then we define ``tail'' subprocesses to track different stages of the execution in Fig.~\ref{fig:subproccard}. Notice that only the card role specification requires an upgrade with parametrising the ``tails'' --  the terminal and the bank role specifications remain untouched in the UTXMM model; we can reuse the notation already introduced in the proof of Theorem~\ref{thm:link} (the only change is the use of $\tparamsmm$ instead of $\tparams$).

\begin{figure*}[b!]
	\captionsetup[subfigure]{labelformat=empty}
	\begin{subfigure}[t]{0.33\linewidth}
		\centering
		\squeezespaces{1}
		\begin{equation}
			\squeezespaces{1}
			\notag
			\begingroup
			\addtolength{\jot}{-1.2pt}
			\begin{aligned}
				Crw^{rec}(&\cparams{\texttt{M1}}) \triangleq \\
				& \text{\textcolor{blue}{$\ctaili_0(\cparams{\texttt{M1}})$}} \\
				& \nw \ch.\cout{card}{\ch}. \\
				& \text{\textcolor{blue}{$\ctaili_1(\ch, \cparams{\texttt{M1}})$}} \\
				& \cin{\ch}{z_1}. \\
				& \text{\textcolor{blue}{$\ctaili_2(\ch, \cparams{\texttt{M1}}, z_1)$}} \\
				& \nw a.\; \lett z_2 \coloneqq \smult{a}{\pkg{c}} \; \inn \; \\ 
				%			& C^{rec}(\ch, c, pk_s, \texttt{vsig}_\texttt{M1}, \texttt{vsig}_\texttt{M2}, \texttt{vsig}_\texttt{M3}, \text{PAN}, \mk, \text{PIN}) + \\
				& \cout{\ch}{z_2}.\\
				& \text{\textcolor{blue}{$\ctaili_3(\ch, \cparams{\texttt{M1}}, z_1, a)$}} \\
				& \lett \, \key{c} \coloneqq \hash{\smult{\mult{a}{c}}{z_1}} \inn \\
				%			& C^{rec}(\ch, c, pk_s, \texttt{vsig}_\texttt{M1}, \texttt{vsig}_\texttt{M2}, \texttt{vsig}_\texttt{M3}, \text{PAN}, \mk, \text{PIN})+ \\
				& \cin{\ch}{m}.\\ 
				& \text{\textcolor{blue}{$\ctaili_4(\ch, \cparams{\texttt{M1}}, \key{c}, a, m)$}} \\
				& \lett \, \nlist{\nlist{\textsf{MM},y_B}, \textsf{MC}_s} \coloneqq \dec{m}{\key{c}} \; \inn \\
				%			& C^{rec}(\ch, c, pk_s, \texttt{vsig}_\texttt{M1}, \texttt{vsig}_\texttt{M2}, \texttt{vsig}_\texttt{M3}, \text{PAN}, \mk, \text{PIN})+ \\			
				& \ifff \ \checksig{pk_s}{\textsf{MC}_s} = \nlist{\textsf{MM},y_B} \  \\
				& \ifff \ \textsf{MM} = \texttt{M1} \ \thenn \\
				& \quad Cont^{rec}(\ch, \cparams{\texttt{M1}}) \\
				& \elsee \; \ifff \ \textsf{MM} = \texttt{M2} \ \thenn \\
				& \quad Cont^{rec}(\ch, \cparams{\texttt{M1}}) \\
				& \elsee \; \ifff \ \textsf{MM} = \texttt{M3} \ \thenn \\
				& \quad \nw \chi_{\texttt{M4}}. \lett \, \texttt{vsig}_\texttt{M4} \coloneqq  \sigv{\chi_{\texttt{M4}}}{\pkg{c}} \inn \\
				& \quad Cont^{rec}(\ch, \cparams{\texttt{M2}})
				\\
				& \elsee \\
				& \quad Crw^{rec}(\cparams{\texttt{M1}})	
				\\
				%	& \ifff \ \proj{1}{\textsf{MC}} =\texttt{MM} \ \thenn \\
				\\
				Cont^{rec}(& \ch, \cparams{\texttt{M1}}) \triangleq \\
				& \cout{\ch}{\enc{\nlist{\smult{a}{\pkg{c}}, \smult{a}{\texttt{vsig}_\texttt{MM}}}}{\key{c}}}.\\
				& \text{\textcolor{blue}{$\ctaili_5(\ch, \cparams{\texttt{M1}}, \key{c}, a, m)$}} \\
				%			& C^{rec}(\ch, c, pk_s, \texttt{vsig}_\texttt{M1}, \texttt{vsig}_\texttt{M2}, \texttt{vsig}_\texttt{M3}, \text{PAN}, \mk, \text{PIN}) + \\
				& \cin{\ch}{x}.\\
				& \text{\textcolor{blue}{$\ctaili_6(\ch, \cparams{\texttt{M1}}, \key{c}, a, m, x)$}} \\
				% low-value online, low-value offline, high-value online
				& \lett \, \nlist{\textsf{TX},\textsf{uPin}} \coloneqq \dec{x}{\key{c}} \inn\\
				& \lett \, \textsf{AC}^\fail \coloneqq \nlist{a, \text{PAN}, \textsf{TX}} \; \inn\\
				& \lett \, \textsf{AC}^{\texttt{ok}} \coloneqq \nlist{a, \text{PAN}, \textsf{TX}, \texttt{ok}} \; \inn\\
				& \lett \, \textsf{AC}^{\texttt{no}} \coloneqq \nlist{a, \text{PAN}, \textsf{TX}, \texttt{no}} \; \inn\\
				& \lett \, \key{cb} \coloneqq \hash{\smult{\mult{a}{c}}{y_B}} \; \inn \\
				%			& C^{rec}(\ch, c, pk_s, \texttt{vsig}_\texttt{M1}, \texttt{vsig}_\texttt{M2}, \texttt{vsig}_\texttt{M3}, \text{PAN}, \mk, \text{PIN}) + \\ 
				& \ifff \ \textsf{uPin} = \fail \ \thenn \\
				& \quad \cout{\ch}{\enc{\enc{\textsf{AC}^{\fail},\h{\textsf{AC}^{\fail}, \mk}}{\key{cb}}, \fail, \textsf{TX}}{\key{c}}}. \\
				& \quad Crw^{rec}(\cparams{\texttt{M1}})
				\\			& \elsee \, \ifff \ \textsf{uPin} =\text{PIN} \ \thenn \\			
				& \quad \cout{\ch}{\enc{\enc{\textsf{AC}^{\texttt{ok}},\h{\textsf{AC}^{\texttt{ok}}, \mk}}{\key{cb}}, \texttt{ok}, \textsf{TX}}{\key{c}}}. \\
				& \quad Crw^{rec}(\cparams{\texttt{M1}})
				\\			& \elsee \, \\
				&\quad \cout{\ch}{\enc{\enc{\textsf{AC}^{\texttt{no}},\h{\textsf{AC}^{\texttt{no}}, \mk}}{\key{cb}}, \texttt{no}, \textsf{TX}}{\key{c}}}. \\
				& \quad Crw^{rec}(\cparams{\texttt{M1}})			
			\end{aligned}	
		\endgroup
		\end{equation}
	\end{subfigure}
	\begin{subfigure}[t]{0.65\linewidth}
		\centering
		\squeezespaces{1}
		\begin{equation}
			\squeezespaces{1}
			\notag
			\begingroup
			\addtolength{\jot}{-1.2pt}
			\begin{aligned}
				Cid(&\ch, \cparams{\texttt{M1}}) \triangleq \\
				& \text{\textcolor{blue}{$\ctails_1(\ch, \cparams{\texttt{M1}})$}} \\
				& \cin{\ch}{z_1}. \\
				& \text{\textcolor{blue}{$\ctails_2(\ch, \cparams{\texttt{M1}}, z_1)$}} \\
				& \nw a.\; \lett z_2 \coloneqq \smult{a}{\pkg{c}} \; \inn \; \\ 
				& \cout{\ch}{z_2}.\\
				& \text{\textcolor{blue}{$\ctails_3(\ch, \cparams{\texttt{M1}}, z_1, a)$}} \\
				& \lett \, \key{c} \coloneqq \hash{\smult{\mult{a}{c}}{z_1}} \; \inn \\
				& \cin{\ch}{m}.\\ 
				& \text{\textcolor{blue}{$\ctails_4(\ch, \cparams{\texttt{M1}}, \key{c}, a, m)$}} \\
				& \lett \, \nlist{\nlist{\textsf{MM},y_B}, \textsf{MC}_s} \coloneqq \dec{m}{\key{c}} \; \inn\; \\
				& \ifff \ \checksig{pk_s}{\textsf{MC}_s} = \nlist{\textsf{MM},y_B} \  \\
				& \ifff \ \textsf{MM} = \texttt{M1} \ \thenn \\
				& \quad Cont(\ch, \cparams{\texttt{M1}}) \\
				& \elsee \; \ifff \ \textsf{MM} = \texttt{M2} \ \thenn \\
				& \quad Cont(\ch, \cparams{\texttt{M1}}) \\
				& \elsee \; \ifff \ \textsf{MM} = \texttt{M3} \ \thenn \\
				& \quad \nw \chi_{\texttt{M4}}. \lett \, \texttt{vsig}_\texttt{M4} \coloneqq  \sigv{\chi_{\texttt{M4}}}{\pkg{c}} \inn \\
				& \quad Cont(\ch, \cparams{\texttt{M2}})
				\\
				%	& \ifff \ \proj{1}{\textsf{MC}} =\texttt{MM} \ \thenn \\
				\\
				Cont(&\ch, \cparams{\texttt{M1}}) \triangleq \\
				& \cout{\ch}{\enc{\nlist{\smult{a}{\pkg{c}}, \smult{a}{\texttt{vsig}_\texttt{MM}}}}{\key{c}}}.\\
				& \text{\textcolor{blue}{$\ctails_5(\ch, \cparams{\texttt{M1}}, \key{c}, a, m)$}} \\
				& \cin{\ch}{x}.\\
				& \text{\textcolor{blue}{$\ctails_6(\ch, \cparams{\texttt{M1}}, \key{c}, a, m, x)$}} \\
				% low-value online, low-value offline, high-value online
				& \lett \, \nlist{\textsf{TX},\textsf{uPin}} \coloneqq \dec{x}{\key{c}} \; \inn\\
				& \lett \, \textsf{AC}^\fail \coloneqq \nlist{a, \text{PAN}, \textsf{TX}} \; \inn\\
				& \lett \, \textsf{AC}^{\texttt{ok}} \coloneqq \nlist{a, \text{PAN}, \textsf{TX}, \texttt{ok}} \; \inn\\
				& \lett \, \textsf{AC}^{\texttt{no}} \coloneqq \nlist{a, \text{PAN}, \textsf{TX}, \texttt{no}} \; \inn\\
				& \lett \, \key{cb} \coloneqq \hash{\smult{\mult{a}{c}}{y_B}} \; \inn \\
				& \ifff \ \textsf{uPin} = \fail \ \thenn \\
				& \quad \cout{\ch}{\enc{\enc{\textsf{AC}^{\fail},\h{\textsf{AC}^{\fail}, \mk}}{\key{cb}}, \fail, \textsf{TX}}{\key{c}}} \\
				& \elsee \, \ifff \ \textsf{uPin} =\text{PIN} \ \thenn \\			
				& \quad \cout{\ch}{\enc{\enc{\textsf{AC}^{\texttt{ok}},\h{\textsf{AC}^{\texttt{ok}}, \mk}}{\key{cb}}, \texttt{ok}, \textsf{TX}}{\key{c}}} \\
				& \elsee \, \\
				&\quad \cout{\ch}{\enc{\enc{\textsf{AC}^{\texttt{no}},\h{\textsf{AC}^{\texttt{no}}, \mk}}{\key{cb}}, \texttt{no}, \textsf{TX}}{\key{c}}} \\
				& \text{\textcolor{blue}{$\ctails_0$}} \triangleq 0\\
			\end{aligned}
		\endgroup	
		\end{equation}
	\end{subfigure}
	\caption{Execution stages for the card processes in UTXMM model.} \label{fig:subproccard}
\end{figure*}

Let us then define the list of global parameters as follows $\squeezespaces{1}\vec{\epsilon}_{\texttt{MM}} \coloneqq (\user, s, si, \chi_{\texttt{M1}}, \chi_{\texttt{M2}}, \chi_{\texttt{M3}})$ and the initial set of messages on the network $\squeezespaces{1}\sigma_0^{\texttt{MM}} \coloneqq \sub{pk_s}{\pks{s}} \sub{pk_\texttt{M1}}{\pkv{\chi_{\texttt{M1}}}} 
\sub{pk_\texttt{M2}}{\pkv{\chi_{\texttt{M2}}}} \sub{pk_\texttt{M3}}{\pkv{\chi_{\texttt{M3}}}}$. Finally, to define the relation $\mathfrak{S}$ in Fig.~\ref{fig:relationutxmm} we also introduce the processes $\spec{PCMM}$, $\impl{PCMM}$ and $\textit{PBTMM}$.

\[
\begin{aligned}
	&\spec{PCMM} \triangleq \nu \text{PIN}, mk, c, \text{PAN}. \bigl(\\
	&
	\begin{array}{l}
		\quad \lett \, \textsf{crtC1} \coloneqq  \sigv{\chi_{\texttt{M1}}}{\pkg{c}} \inn \\
		\quad \lett \, \textsf{crtC2} \coloneqq  \sigv{\chi_{\texttt{M2}}}{\pkg{c}} \inn \\
		\quad \lett \, \textsf{crtC3} \coloneqq  \sigv{\chi_{\texttt{M3}}}{\pkg{c}} \inn \\
		\quad 
		\begin{array}[t]{l}
			\;\; \nw \ch.\cout{card}{\ch}. \\ \;\; Cid(\ch, c, \pk{s}, \textsf{crtC1}, \textsf{crtC2}, \textsf{crtC3}, \text{PAN}, mk, \text{PIN}) + \\
			\;\; \nw \chi_{\texttt{M4}}. \lett \, \textsf{crtC4} \coloneqq  \sigv{\chi_{\texttt{M4}}}{\pkg{c}} \inn \\
			\;\; \nw \ch.\cout{card}{\ch}. \\ \;\; Cid(\ch, c, \pk{s}, \textsf{crtC2}, \textsf{crtC3}, \textsf{crtC4}, \text{PAN}, mk, \text{PIN}) 
			\\
			\cpar \; !\cout{\user}{\text{PIN}} \cpar \; ! \cout{\nlist{si, \text{PAN}}}{\nlist{\text{PIN}, mk, \pkg{c}}} \;\bigl)
		\end{array}
	\end{array}
\end{aligned}
\]

\[
\begin{aligned}
	& \impl{PCMM} \triangleq \ \text{PIN}, mk, c, \text{PAN}. \bigl( \\
	&
	\begin{array}{l}
		\quad \lett \, \textsf{crtC1} \coloneqq  \sigv{\chi_{\texttt{M1}}}{\pkg{c}} \inn \\
		\quad \lett \, \textsf{crtC2} \coloneqq  \sigv{\chi_{\texttt{M2}}}{\pkg{c}} \inn \\
		\quad \lett \, \textsf{crtC3} \coloneqq  \sigv{\chi_{\texttt{M3}}}{\pkg{c}} \inn \\
		\quad
		\begin{array}[t]{l}
			\;\; Crw^{rec}(c, \pk{s}, \textsf{crtC1}, \textsf{crtC2}, \textsf{crtC3}, \text{PAN}, mk, \text{PIN}) + \\
			\;\; \nw \chi_{\texttt{M4}}. \lett \, \textsf{crtC4} \coloneqq  \sigv{\chi_{\texttt{M4}}}{\pkg{c}} \inn \\
			\;\; Crw^{rec}(c, \pk{s}, \textsf{crtC2}, \textsf{crtC3}, \textsf{crtC4}, \text{PAN}, mk, \text{PIN}) 
			\\
			\cpar \; !\cout{\user}{\text{PIN}} \cpar \; ! \cout{\nlist{si, \text{PAN}}}{\nlist{\text{PIN}, mk, \pkg{c}}} \;\bigl)
		\end{array}
	\end{array}
\end{aligned}
\]

\[
\begin{aligned}
	\textit{PBTMM} & \triangleq\ \nu b_t.\bang \nu \kbt.( \\
	&
	\begin{array}[t]{l}
		\nu \ch.\cout{bank}{\ch}.B(\ch, si, \kbt, b_t) \cpar \\
		\lett \, \textsf{crt1} \coloneqq \nlist{\nlist{\texttt{M1}, \pkg{b_t}}, \sig{s}{\nlist{\texttt{M1}, \pkg{b_t}}}} \inn \\
		\lett \, \textsf{crt2} \coloneqq \nlist{\nlist{\texttt{M2}, \pkg{b_t}}, \sig{s}{\nlist{\texttt{M2}, \pkg{b_t}}}} \inn \\
		\lett \, \textsf{crt3} \coloneqq \nlist{\nlist{\texttt{M3}, \pkg{b_t}}, \sig{s}{\nlist{\texttt{M3}, \pkg{b_t}}}} \inn \\
		\nu \ch.\cout{term}{\ch}.T(user, \ch, \pkv{\chi_{\texttt{M1}}}, \textsf{crt1}, \kbt) + \\
		\nu \ch.\cout{term}{\ch}.T(user, \ch, \pkv{\chi_{\texttt{M2}}}, \textsf{crt2}, \kbt) + \\
		\nu \ch.\cout{term}{\ch}.T(user, \ch, \pkv{\chi_{\texttt{M3}}}, \textsf{crt3}, \kbt)
		\textcolor{cyan}{\bigl)} ~\Bigl)
	\end{array}
\end{aligned}
\]

In Fig.~\ref{fig:relationutxmm} we again use partitions to parametrise the related states. Since the UTXMM model admits several month, there are two updates over the relation $\mathfrak{R}$ from the proof of Theorem~\ref{thm:link}. Firstly, we introduce a new partition $\Lambda^\texttt{MM} = \{\lambda^{\texttt{M1}}, \lambda^\texttt{M2}\}$ of the set of all card's sessions $\mathcal{D}$, where, e.g. $\lambda^{\texttt{M2}}$ is the set of all sessions where the oldest month the card can reply is $\texttt{M2}$. Secondly, we introduce a new partition $\Gamma^\texttt{MM} \coloneqq \{\gamma^{\texttt{M1}}, \gamma^{\texttt{M3}}, \gamma^{\texttt{M3}}\}$ of the set of sessions with a terminal to indicate which month certificate is requested in the respective session. Generic states for the specification and the implementation worlds are defined in Fig.~\ref{fig:utxmmgenericideal},~\ref{fig:utxmmgenericreal}.
%Fig.~\ref{fig:genspecmm} and Fig.~\ref{fig:genimplmm} respectively. 

\begin{figure}[h!]
	\[\arraycolsep=3.43pt
	\squeezespaces{1}
	\begin{array}{rcl}
		\spec{UTXMM} &  \mathfrak{R} & \impl{UTXMM} 
		\\[10pt]
		\begin{array}{r} 
			%			\spec{UTXMM}^1 \ \triangleq \\
			\nw \vec{\epsilon}_{\texttt{MM}}.(\sub{pk_s}{\pks{s}} \cpar \\
			\cout{out}{\pkv{\chi_{\texttt{M1}}}}. \\ \cout{out}{\pkv{\chi_{\texttt{M2}}}}.\\ \cout{out}{\pkv{\chi_{\texttt{M3}}}}. \\ (\bang \spec{PCMM} \cpar \nw b_t.\bang \textit{PBTMM}))
		\end{array} 
		& \mathfrak{R} &
		\begin{array}{l}
			%			\impl{UTXMM}^1 \ \triangleq \\
			\nw \vec{\epsilon}_{\texttt{MM}}.(\sub{pk_s}{\pks{s}} \cpar \\
			\cout{out}{\pkv{\chi_{\texttt{M1}}}}. \\ \cout{out}{\pkv{\chi_{\texttt{M2}}}}.\\ \cout{out}{\pkv{\chi_{\texttt{M3}}}}.\\ (\bang \impl{PCMM} \cpar \nw b_t.\bang \textit{PBTMM}))
		\end{array} 
		\\[35pt]
		\begin{array}{r} 
			%	\spec{UTXMM}^1 \ \triangleq \\
			\nw \vec{\epsilon}_{\texttt{MM}}.(\sub{pk_s}{\pks{s}} \\ \sub{pk_\texttt{M1}}{\pkv{\chi_{\texttt{M1}}}}\cpar \\
			\cout{out}{\pkv{\chi_{\texttt{M2}}}}.\\ \cout{out}{\pkv{\chi_{\texttt{M3}}}}. \\ (\bang \spec{PCMM} \cpar \nw b_t.\bang \textit{PBTMM}))
		\end{array} 
		& \mathfrak{R} &
		\begin{array}{l}
			%	\impl{UTXMM}^1 \ \triangleq \\
			\nw \vec{\epsilon}_{\texttt{MM}}.(\sub{pk_s}{\pks{s}} \\ \sub{pk_\texttt{M1}}{\pkv{\chi_{\texttt{M1}}}}\cpar \\
			\cout{out}{\pkv{\chi_{\texttt{M2}}}}.\\ \cout{out}{\pkv{\chi_{\texttt{M3}}}}.\\ (\bang \impl{PCMM} \cpar \nw b_t.\bang \textit{PBTMM}))
		\end{array} 
		\\[35pt]
		\begin{array}{r} 
			%	\spec{UTXMM}^1 \ \triangleq \\
			\nw \vec{\epsilon}_{\texttt{MM}}.(\sub{pk_s}{\pks{s}} \\ \sub{pk_\texttt{M1}}{\pkv{\chi_{\texttt{M1}}}} \\
			\sub{pk_\texttt{M2}}{\pkv{\chi_{\texttt{M2}}}} \cpar \\ \cout{out}{\pkv{\chi_{\texttt{M3}}}}. \\ (\bang \spec{PCMM} \cpar \nw b_t.\bang \textit{PBTMM}))
		\end{array} 
		& \mathfrak{R} &
		\begin{array}{l}
			%	\impl{UTXMM}^1 \ \triangleq \\
			\nw \vec{\epsilon}_{\texttt{MM}}.(\sub{pk_s}{\pks{s}} \\ \sub{pk_\texttt{M1}}{\pkv{\chi_{\texttt{M1}}}} \\
			\sub{pk_\texttt{M2}}{\pkv{\chi_{\texttt{M2}}}} \cpar \\ \cout{out}{\pkv{\chi_{\texttt{M3}}}}.\\ (\bang \impl{PCMM} \cpar \nw b_t.\bang \textit{PBTMM}))
		\end{array} 
		\\[35pt]
		\begin{array}{r} 
			%	\spec{UTXMM}^1 \ \triangleq \\
			\nw \vec{\epsilon}_{\texttt{MM}}.(\sigma_0^{\texttt{MM}} \cpar \\ 
			\bang \spec{PCMM} \cpar \nw b_t.\bang \textit{PBTMM})
		\end{array} 
		& \mathfrak{R} &
		\begin{array}{l}
			%	\impl{UTXMM}^1 \ \triangleq \\
			\nw \vec{\epsilon}_{\texttt{MM}}.(\sigma_0^{\texttt{MM}} \cpar \\ \bang \impl{PCMM} \cpar \nw b_t.\bang \textit{PBTMM})
		\end{array} 
		\\[12pt]
		%		\begin{array}{r}
		%			\spec{UTX}^2 \ \triangleq \\ \nw \vec{\epsilon}_{MM}.(\sigma_0 \cpar \bang \spec{PC} \cpar \nw b_t.\bang \textit{PBT}) 
		%		\end{array} & \mathfrak{R} & \begin{array}{l} \impl{UTX}^2 \ \triangleq \\ \nw \vec{\epsilon}_{MM}.(\sigma_0 \cpar \bang \impl{PC} \cpar \nw b_t.\bang \textit{PBT}) \end{array}
		\specc{(K, \vec{\Sigma})}(X,Y,Z) & \mathfrak{R} & \impll{(\vec{K}, \vec{\Sigma}, \Lambda, \Lambda^\texttt{MM})}(X,Y,Z)
	\end{array}
	\]
	Where $\vec{\Sigma} = \{F, A, \Gamma, \Gamma^{\texttt{MM}}, B\}$
	\caption{Defining conditions for the relation $\mathfrak{S}$.}
	\label{fig:relationutxmm}
\end{figure}

%\begin{figure*}[b!]
\begin{figure}
		\centering
		\squeezespaces{1}
		\[
		\begin{array}{l}
			\specc{(K, F, A, \Gamma, \Gamma^{\texttt{MM}}, B)}(X,Y,Z) \triangleq \\
			\begin{array}{l}
				\mathopen{\nu \, \vec{\epsilon}_{\texttt{MM}}, \text{PIN}_{1 \hdots D+K}, mk_{1 \hdots D+K}, c_{1 \hdots D+K}, \text{PAN}_{1 \hdots D+K}}, 
				\\[4pt]
				\dot{ch}_{1 \hdots D}, a_{1 \hdots E}, b_t, 
				%			\ch_{1 \hdots F+G+M}, 
				\ddot{ch}_{1 \hdots F+G},
				\\[4pt]
				\dddot{ch}_{1 \hdots F+M}, t_{1 \hdots L},  \textsf{TX}_{1 \hdots L}.(~\sigma^\texttt{MM} \cpar  
				\\[4pt]
				\;\;
				\begin{array}[t]{l}
					C_1 \cpar \hdots \cpar 0 \cpar \bang \cout{user}{\text{PIN}_1} \cpar
					\\[4pt]
					\quad \hdots \cpar 0 \cpar \bang 	\cout{\nlist{si, \text{PAN}_1}}{\nlist{\text{PIN}_1, mk_1, \pkg{c_1}}}) \cpar
					\\[3pt]
					\hdots
					\\[3pt]
					C_i \cpar \hdots \cpar 0 \cpar \bang \cout{user}{\text{PIN}_i} \cpar
					\\[4pt]
					\quad \hdots \cpar 0 \cpar \bang 	\cout{\nlist{si, \text{PAN}_i}}{\nlist{\text{PIN}_i, mk_i, \pkg{c_i}}}) \cpar
					\\[3pt]
					\hdots
					\\[3pt]
					C_{D+K} \cpar \hdots \cpar 0 \cpar \bang \cout{user}{\text{PIN}_{D+K}} \cpar
					\\[4pt]
					\quad \hdots \cpar 0 \cpar \bang 	\cout{\nlist{si, \text{PAN}_{D+K}}}{\nlist{\text{PIN}_{D+K}, mk_{D+K}, \pkg{c_{D+K}}}}) \cpar
					\\[4pt]
					\bang \spec{PCMM} \cpar
					\\[4pt]
					B^{\sigma^\texttt{MM}}_1 \cpar T^{\sigma^\texttt{MM}}_1 \cpar 
					\\[3pt]
					\hdots \cpar 
					\\[3pt]
					B^{\sigma^\texttt{MM}}_j \cpar T^{\sigma^\texttt{MM}}_j \cpar 
					\\[3pt]
					\hdots \cpar 
					\\[3pt]
					B^{\sigma^\texttt{MM}}_{F+G+M} \cpar T^{\sigma^\texttt{MM}}_{F+G+M} \cpar \bang \textit{PBTMM})
				\end{array}
			\end{array}
		\end{array}
		\]
	\caption{The generic of the idealised UTXMM world.} \label{fig:utxmmgenericideal}
\end{figure}

%\begin{figure*}[b!]
\begin{figure}
		\centering
		\squeezespaces{1}
		\[
		\begin{array}{l}
			\impll{(\vec{K}, F, A, \Gamma, \Gamma^{\texttt{MM}}, B, \Lambda, \Lambda^\texttt{MM})}(X,Y,Z) \triangleq \\
			\begin{array}{l}
				\mathopen{\nu \vec{\epsilon}_{\texttt{MM}}, \text{PIN}_{1 \hdots H}, mk_{1 \hdots H}, c_{1 \hdots H}, \text{PAN}_{1 \hdots H}},
				\dot{ch}_{1 \hdots D}, 
				\\[4pt]
				a_{1 \hdots E}, b_t,  
				%			\ch_{1 \hdots F+G+M}, 
				\ddot{ch}_{1 \hdots F+G}, \dddot{ch}_{1 \hdots F+M}
				\\[4pt]
				t_{1 \hdots L},  \textsf{TX}_{1 \hdots L}.(~\theta^\texttt{MM} \cpar 
				\\[4pt]
				\begin{array}[t]{l}
					\crec^1_{i_1} \cpar \hdots \cpar 0 \cpar \bang \cout{user}{\text{PIN}_1} \cpar 
					\\[4pt]
					\quad \hdots \cpar 0 \cpar \bang 	\cout{\nlist{si, \text{PAN}_1}}{\nlist{\text{PIN}_1, mk_1, \pkg{c_1}}}) \cpar
					\\[3pt]
					\hdots
					\\[3pt]
					\crec^h_{i_h} \cpar \hdots \cpar 0 \cpar \bang \cout{user}{\text{PIN}_h} \cpar  
					\\[4pt]
					\quad \hdots \cpar 0 \cpar \bang 	\cout{\nlist{si, \text{PAN}_h}}{\nlist{\text{PIN}_h, mk_h, \pkg{c_h}}}) \cpar
					\\[3pt]
					\hdots
					\\[3pt]
					\crec^H_{i_H} \cpar \hdots \cpar 0 \cpar \bang \cout{user}{\text{PIN}_H} \cpar  
					\\[4pt]
					\quad \hdots \cpar 0 \cpar \bang 	\cout{\nlist{si, \text{PAN}_H}}{\nlist{\text{PIN}_H, mk_H, \pkg{c_H}}}) \cpar
					\\[4pt]
					\bang \impl{PCMM} \cpar
					\\[4pt]
					B^{\theta^\texttt{MM}}_1 \cpar T^{\theta^\texttt{MM}}_1 \cpar 
					\\[3pt]
					\hdots \cpar 
					\\[3pt]
					B^{\theta^\texttt{MM}}_j \cpar T^{\theta^\texttt{MM}}_j \cpar 
					\\[3pt]
					\hdots \cpar 
					\\[3pt]
					B^{\theta^\texttt{MM}}_{F+G+M} \cpar T^{\theta^\texttt{MM}}_{F+G+M} \cpar \bang \textit{PBTMM})
				\end{array}
			\end{array}	 
		\end{array}
		\]
		Where $i_h \in \{D_{h-1} + K_{h-1}+1 , \hdots, D_{h-1} + K_{h-1} + D_h+K_h\}$
	\caption{The generic of the real UTXMM world.} \label{fig:utxmmgenericreal}
\end{figure}

To finalise the definition it is left to define $R_i^j$ and $T^\rho_i$ which is straightforward given the definition of $C_i^j$ and $T_i$ from the proof of the Theorem~\ref{thm:link}, e.g. $R_i^j = \ctaili_5(\dot{\ch}_i, \vec{\chi}^{\texttt{\textcolor{cyan}{M1}}}_j, k^c_j(a_i, X^1_i\theta), a_i, X^2_i\theta)$ if $i \in \alpha_5 \cap \lambda_j \cap \textcolor{cyan}{\lambda^{\texttt{M1}}}$ and $R_i^j = \ctaili_5(\dot{\ch}_i, \vec{\chi}^{\texttt{\textcolor{cyan}{M2}}}_j, k^c_j(a_i, X^1_i\theta), a_i, X^2_i\theta)$ if $i \in \alpha_5 \cap \lambda_j \cap \textcolor{cyan}{\lambda^{\texttt{M2}}}$, i.e. the only difference is the explicit indication if the card accepts the bank's certificates for months $\texttt{M1}$ and $\texttt{M2}$ or $\texttt{M2}$ and $\texttt{M3}$. Similarly, the definition of $T^\rho_i$ is straightforward to update, e.g. $T^\rho_i = \ttonhi_5(user, \vec{\psi}^{\textcolor{cyan}{\texttt{MM}}}_i, t_i, \text{TX}_i, Z^1_i\rho, Z^2_i\rho)$ if $i \in \gamma^{\of}_5 \cap \textcolor{cyan}{\gamma^{\texttt{MM}}}$, where $\texttt{MM} \in \{\texttt{M1},\texttt{M2},\texttt{M3} \}$. The definitions of $\sigma^\texttt{MM}$ and $\theta^\texttt{MM}$ are obtained analogously from the definitions of $\sigma$ and $\theta$ in Fig.~\ref{fig:sigmatheta} by involving partitions $\Lambda^{\texttt{MM}}$ and $\Gamma^{\texttt{MM}}$ whenever a message that depends on the month $\texttt{MM}$ appears on the network -- either explicitly, as, e.g. $\textit{ecert}$ or is guarded by month-dependant check at any point in the past as, e.g. $\textit{etx}$. 
\end{proof}
\end{customthm}

	\section{Security analysis details}\label{sec:appxproverif}

Here we present the details on specifying injective agreement explained in Section~\ref{sec:securityanalysis} in $\protocol$ using ProVerif. The respective code is provided in the dedicated repository~\cite{repoutx}.

The specification of UTX we use in ProVerif to verify the security properties of UTX is given in Fig.~\ref{fig:proverifsystem}. It differs from the one presented in Fig.~\ref{fig:system}a. in two aspects -- firstly, we drop session channels and use common open public channels $\textit{card}$, $\textit{bank}$ and $\textit{term}$ in role specifications, and secondly, we allow all three types of terminals to run in parallel since the non-deterministic choice operator is not supported in ProVerif. 

\begin{figure}
	\begin{equation}
		\notag
		\begin{aligned}
			&\nu \, \user, s, si, \chi_{\texttt{MM}}. \cout{out}{\pks{s}}.\cout{out}{\pkv{ \chi_{\texttt{MM}}}}.\Bigl( \\ 
			&
			\;\; 
			\begin{array}{l}
				\mathbf{\bang} \nu \text{PIN}, mk, c, \text{PAN}. \bigl(\\
				\quad \lett \, \textsf{crtC} \coloneqq  \sigv{\chi_{\texttt{MM}}}{\pkg{c}} \inn \\
				\quad 
				\begin{array}[t]{l}
					\;\; \bang C(\textit{card}, c, \pk{s}, \textsf{crtC}, \text{PAN}, mk, \text{PIN}) \\
					\cpar \; !\cout{\user}{\text{PIN}} \cpar \; ! \cout{\nlist{si, \text{PAN}}}{\nlist{\text{PIN}, mk, \pkg{c}}} \;\bigl) \; \cpar \; \\
					
					%					\textit{DB}(si, \text{PAN}, mk, \text{PIN}, \pkg{c})~\textcolor{green}{\Bigl)} \cpar
				\end{array}
				\\
				\nu b_t.\bang \nu \kbt. \textcolor{cyan}{\bigl(}
				\\
				\;\; 
				\begin{array}[t]{l}
					B(\textit{bank}, si, \kbt, b_t) \cpar \\
					\lett \, \textsf{crt} \coloneqq \nlist{\nlist{\texttt{MM}, \pkg{b_t}}, \sig{s}{\nlist{\texttt{MM}, \pkg{b_t}}}} \inn \\
					T_{\texttt{onhi}}(\user, \textit{term}, \pkv{\chi_{\texttt{MM}}}, \textsf{crt}, \kbt) \cpar \\
					T_{\texttt{offhi}}( \user, \textit{term}, \pkv{\chi_{\texttt{MM}}}, \textsf{crt},  \kbt) \cpar \\
					T_{\texttt{lo}} (\textit{term}, \pkv{\chi_{\texttt{MM}}}, \textsf{crt}, \kbt)
					\textcolor{cyan}{\bigl)} ~\Bigl)
				\end{array}
			\end{array}
		\end{aligned}
	\end{equation}
	\caption{The ProVerif real-world model of UTX.}\label{fig:proverifsystem}
\end{figure}

The syntax and semantics of correspondence assertions used in Fig.~\ref{fig:security} is as follows. A (simplified) correspondence assertion is a formula of the form $\Phi_0 \Rightarrow \Phi_1$, where $\Phi_0$ is a conjunction of events, and $\Phi_1$ is a conjunction of disjunctions of events. A protocol specification satisfies such a formula if, for any execution trace of the protocol where the events in $\Phi_0$ are true, it is the case that one of the events in each conjunct of $\Phi_1$ is true. The variables of $\Phi_0$ have an implicit universal quantifier, while the variables in $\Phi_1$ (that are not in $\Phi_0$) are quantified existentially.  
For example, $A(x,y)\Rightarrow B(x,z)\vee C(y,w)$ expresses that, for any $x$ and $y$, whenever the event $A(x,y)$ occurs, then either the event $B(x,z)$ should occur, for some $z$, or the event $C(y,w)$ should occur, for some $w$. 

%The processes in Fig.~\ref{fig:eventscard},~\ref{fig:eventsterminal},~\ref{fig:eventsbank} specify the exact locations for events from Fig.~\ref{fig:security} in the specification of the card's, bank's, and the terminal's processes respectively. 

	The message theory from Fig.~\ref{fig:syntax}, that faithfully models the cryptographic primitives used in the \protocol\; protocol, is currently outside the
scope of ProVerif. There are two sources of complexity in the theory. One is scalar multiplication of an unbounded number of terms with an element of the elliptic curve group, as modelled by the first three equations in Fig.~\ref{fig:syntax}. We handle this problem by extending a standard abstraction used in ProVerif to model the Diffie-Hellman based key agreement (where exponentiation plays the role of scalar multiplication). Instead of the first three equations from Fig.~\ref{fig:syntax}, we have the following two equations. 
\[
\begin{array}{rcl}
	\smult{M}{\smult{N}{\gen}} &=_{E}& \smult{M}{\smult{N}{\gen}} \\[1pt]
	\smult{L}{\smult{M}{\smult{N}{\gen}}} &=_{E}& \smult{M}{\smult{L}{\smult{N}{\gen}}}
\end{array}
\]
These equations together cover all permutations of three scalar multipliers on
top of the group generator $\gen$ and are sufficient for modelling
the blinded Diffie-Hellman key agreement as used in our protocol: the
three scalars are the secret key of the card, the blinding factor
chosen by the card, and the scalar chosen by the terminal in the first
message to the card. A second source of complexity in Fig.~\ref{fig:syntax} is the last equation, which allows us to homomorphically push scalar multiplication inside a signature. We handle this second problem by replacing the homomorphic equation $$\smult{M}{\sigv{K}{N}} =_E \sigv{K}{\smult{M}{N}}$$ with the equation allowing to verify a multiplied signature directly without pushing the multiplication inside the signature. $$\checksigv{\pkv{K}}{\smult{M}{\sigv{K}{N}}} =_{E} \smult{M}{N}$$

\iffalse

\begin{figure}[h]
	\begingroup
	\addtolength{\jot}{1pt}
	\begin{equation}
		\squeezespaces{1}
		\notag
		\begin{aligned}
			C(& \ch, c, pk_s, \texttt{vsig}_\texttt{MM}, \text{PAN}, \mk, \text{PIN}) \triangleq \\ 
			& \cin{\ch}{z_1}. \\
			& \nw a.\; \lett z_2 \coloneqq \smult{a}{\pkg{c}} \; \inn \\ 
			&\cout{\ch}{z_2}.\\
			& \lett \, \key{c} \coloneqq \hash{\smult{\mult{a}{c}}{z_1}} \inn \\
			%		& \event{CRunning}{\text{PAN},\key{c},z_1,z_2}. \\
			& \cin{\ch}{m}.\\ 
			& \lett \, \nlist{\nlist{\textsf{MM},y_B}, \textsf{MC}_s} \coloneqq \dec{m}{\key{c}}  \inn \\
			&  \ifff \ \checksig{pk_s}{\textsf{MC}_s} = \nlist{\textsf{MM},y_B} \ \thenn \\
			%	& \ifff \ \proj{1}{\textsf{MC}} =\texttt{MM} \ \thenn \\
			& \lett \, \textit{emc} = \enc{\nlist{\smult{a}{\pkg{c}}, \smult{a}{\texttt{vsig}_\texttt{MM}}}}{\key{c}}\; \inn \\
			& \cout{\ch}{\textit{emc}}.\\
			& \cin{\ch}{x}.\\
			% low-value online, low-value offline, high-value online
			& \lett \, \nlist{\textsf{TX},\textsf{uPin}} \coloneqq \dec{x}{\key{c}} \; \inn\\
			& \lett \, \textsf{AC} \coloneqq \nlist{a, \text{PAN}, \textsf{TX}} \; \inn\\
			& \lett \, \textsf{AC}^{\texttt{ok}} \coloneqq \nlist{a, \text{PAN}, \textsf{TX}, \texttt{ok}} \; \inn\\
			& \lett \, \textsf{AC}^{\fail} \coloneqq \nlist{a, \text{PAN}, \textsf{TX}, \fail} \; \inn\\
			& \lett \, \key{cb} \coloneqq \hash{\smult{\mult{a}{c}}{y_B}} \inn \\
			& \ifff \ \textsf{uPin} = \fail \ \thenn \\ 
			%		& \quad \event{CAccept}{\sf{PAN,TX}}.\\
			& \quad \lett \, \textit{eac} \coloneqq \enc{\enc{\nlist{\textsf{AC},\hash{\nlist{\textsf{AC}, \mk}}}}{\key{cb}}}{\key{c}} \; \inn \\
			& \quad \event{\textcolor{teal}{CRunningWithB}}{\textit{eac}} \\
			& \quad \event{\textcolor{blue}{CR}\textcolor{red}{unn}\textcolor{violet}{ing}}{z_1, z_2, m, \textit{emc}, x, \textit{eac}} \\
			& \quad \cout{\ch}{\textit{eac}} \\
			& \elsee \, \ifff \ \textsf{uPin} =\text{PIN} \ \thenn \\
			& \quad \lett \, \textit{eac} \coloneqq \enc{\nlist{\enc{\nlist{\textsf{AC}^{\texttt{ok}},\h{\nlist{\textsf{AC}^{\texttt{ok}}, \mk}}}}{\key{cb}}, \texttt{ok}}}{\key{c}} \; \inn \\
			& \quad \event{\textcolor{teal}{CRunningWithB}}{\enc{\nlist{\textsf{AC}^{\texttt{ok}},\h{\nlist{\textsf{AC}^{\texttt{ok}}, \mk}}}}{\key{cb}}} \\
			& \quad \event{\textcolor{blue}{CR}\textcolor{red}{unn}\textcolor{violet}{ing}}{z_1, z_2, m, \textit{emc}, x, \textit{eac}} \\
			%		& \quad \event{CAccept}{\sf{PAN,TX}}. \\
			& \quad \cout{\ch}{\textit{eac}} \\
			& \elsee \, \\
			& \quad \lett \, \textit{eac} \coloneqq \enc{\nlist{\enc{\nlist{\textsf{AC}^{\fail}, \h{\nlist{\textsf{AC}^{\fail}, \mk}}}}{\key{cb}}, \fail}}{\key{c}} \; \inn \\
			& \quad \event{\textcolor{teal}{CRunningWithB}}{\enc{\nlist{\textsf{AC}^{\fail},\h{\nlist{\textsf{AC}^{\fail}, \mk}}}}{\key{cb}}} \\
			& \quad \event{\textcolor{blue}{CR}\textcolor{red}{unn}\textcolor{violet}{ing}}{z_1, z_2, m,\textit{emc}, x, \textit{eac}} \\
			&\quad \cout{\ch}{\textit{eac}}
		\end{aligned}
	\end{equation}
	\endgroup
	\caption{Events in the card's role.}\label{fig:eventscard}
\end{figure}

	\begin{figure}[h]
			\begingroup
		\addtolength{\jot}{1pt}
	\begin{equation}
		\squeezespaces{1}
		\notag
		\begin{aligned}
			B(&\ch, si, \kbt, b_t) \triangleq \\
			& \cin{\ch}{x}.\\
			& \lett \, \mbox{\squeezespaces{1}$\nlist{\textsf{TX}', z_2, \textsf{EAC}, \textsf{uPIN}} \coloneqq \dec{x}{\kbt}$} \\
			%		& \event{BRunning}{\kbt,\textsf{TX'}}. \\
			& \lett \, \key{bc} \coloneqq \hash{\smult{b_t}{z_2}} \inn \\
			& \lett \, \mbox{\squeezespaces{0.5}$\nlist{\textsf{AC}, \textsf{AC}_{hmac}} = \dec{\textsf{EAC}}{\key{bc}}$} \inn \\
			& \lett \, \nlist{x_a,\textsf{PAN,TX,pinV}} = \textsf{AC} \; \inn \\
			%			& \cin{si}{=\text{PAN},\text{PIN}, mk, pk_{c}}. \\
			&\cin{\nlist{si, \text{PAN}}}{\text{PIN}, mk, pk_{c}}. \\
			& \ifff \ \hash{\nlist{\textsf{AC}, mk}} = \textsf{AC}_{hmac} \ \thenn \\
			& \ifff \ \textsf{TX} = \textsf{TX'} \ \thenn \\
			%		& \event{BCommit}{\kbt,\textsf{TX}, \text{PAN}}. \\
			& \ifff \ \smult{x_a}{pk_{c}}= z_2 \\
			%	& \lett \, r \coloneqq \nlist{\textsf{TX'}, \texttt{accept}} \inn \\
			% low-value offline, low-value online
			& \lett \ \nlist{\textsf{TXdata,TXtype}}  \coloneqq \textsf{TX'} \ \inn \\
			& \ifff \; \textsf{TXtype}=\lo \; \thenn \\
			& \quad \event{\textcolor{teal}{BCommitWithC}}{\textsf{EAC}} \\
			& \quad \event{\textcolor{red}{BRunningWithT}}{x, \enc{\nlist{\textsf{TX'}, \texttt{accept}}}{\kbt}} \\
			%		& \quad\event{BAccept}{\textsf{PAN,TX}} \\
			& \quad \event{\textcolor{violet}{BCommitWithTC}}{x} \\
			& \quad\cout{\ch}{\enc{\nlist{\textsf{TX'}, \texttt{accept}}}{\kbt}} \\
			& \elsee \; \ifff \; \textsf{TXtype} = \hi \ \thenn \\
			% high-value offline		
			& \quad \mbox{\squeezespaces{1}$\ifff \ (\textsf{pinV} = \texttt{ok}) \vee (\textsf{uPIN} = \text{PIN}) \ \thenn$} \\ 
			%		& \quad \quad \event{BAccept}{\text{PAN},\textsf{TX}}. \\
			& \quad \quad \event{\textcolor{teal}{BCommitWithC}}{\textsf{EAC}} \\
			& \quad \quad \event{\textcolor{red}{BRunningWithT}}{x, \enc{\nlist{\textsf{TX}', \texttt{accept}}}{\kbt}} \\
			& \quad \quad \event{\textcolor{violet}{BCommitWithTC}}{x} \\
			& \quad \quad \cout{\ch}{\enc{\nlist{\textsf{TX}', \texttt{accept}}}{\kbt}} \\
			% high-value online	
			%& \quad \elsee \ \ifff \ \textsf{uPIN} = \text{PIN} \ \thenn \\
			%& \quad \quad \event{BAccept}{\textsf{PAN,TX}} \\
			%& \quad \quad \cout{\ch}{\enc{\textsf{TX}', \texttt{accept}}{\kbt}} \\
			& \quad \elsee \\
			& \quad \quad \event{\textcolor{brown}{BReject}}{\kbt, \textsf{TX'}} \\
			& \quad \quad \event{\textcolor{teal}{BCommitWithC}}{\textsf{EAC}} \\
			& \quad \quad \event{\textcolor{red}{BRunningWithT}}{x, \enc{\nlist{\textsf{TX'}, \texttt{reject}}}{\kbt}} \\
			& \quad \quad \event{\textcolor{violet}{BCommitWithTC}}{x} \\
			& \quad \quad  \cout{\ch}{\enc{\nlist{\textsf{TX'}, \texttt{reject}}}{\kbt}}. \\ 
			%		&  \quad \quad \event{BReject}{\textsf{PAN,TX}}. \\
		\end{aligned}
	\end{equation}
\endgroup
	\caption{Events in the bank's's role.}\label{fig:eventsbank}
\end{figure}

\begin{figure}[h]
	\begingroup
	\addtolength{\jot}{1.0pt}
	\begin{equation}
		\squeezespaces{1}
		\notag
		\begin{aligned}
			T_{\texttt{onhi}}(&\user, \ch, pk_\texttt{MM}, \textsf{crt}, \kbt)  \triangleq \\
			& \nw \, \textsf{TXdata}. \\ 
			& \lett \, \textsf{TX} \coloneqq \nlist{\textsf{TXdata}, \hi} \; \inn \\
			& \nw t. \lett z_1 \coloneqq \pkg{t} \inn \; \\ & \cout{\ch}{z_1}. \\
			& \cin{\ch}{z_2}. \\
			& \lett \, \key{t} \coloneqq \hash{\smult{t}{z_2}} \inn \\
			%		& \event{TRunning}{\kbt,\key{t},z_1,z_2, \textsf{TX}}. \\
			& \cout{\ch}{\enc{\textsf{crt}}{\key{t}}}. \\
			& \cin{\ch}{n}.\\
			& \lett \, \nlist{\textsf{B}, \textsf{B}_s} \coloneqq \dec{n}{\key{t}} \; \inn \\
			& \ifff \ \checksigv{pk_\texttt{MM}}{\textsf{B}_s} = \textsf{B} \ \thenn \\
			& \ifff \ \textsf{B} = z_2 \ \thenn \\
			%& \event{TCommit}{\key{t},z_1,z_2, \textsf{TX}}. \\
			& \cin{user}{\textsf{uPIN}}.\\
			& \cout{\ch}{\enc{\nlist{\textsf{TX}, \fail}}{\key{t}}}. \\
			& \cin{\ch}{y}. \\
			& \event{\textcolor{blue}{TCommitWithC}}{z_1, z_2, \enc{\textsf{crt}}{\key{t}}, n, \enc{\nlist{\textsf{TX}, \fail}}{\key{t}}, y} \\
			& \lett \, \textit{req} = \enc{\nlist{\textsf{TX}, z_2, \dec{y}{\key{t}}, \textsf{uPIN}}}{\kbt} \; \inn \\
			& \event{\textcolor{violet}{TRunningWithBC}}{\textit{req}, z_1, z_2, \enc{\textsf{crt}}{\key{t}}, n, \enc{\nlist{\textsf{TX}, \fail}}{\key{t}}, y}\\
			& \cout{\ch}{\textit{req}}. \\
			& \cin{\ch}{r}. \\
			& \ifff \; \dec{r}{\kbt} = \nlist{\textsf{TX},\texttt{rtype}} \ \thenn \\
			& \quad \event{\textcolor{red}{TCommitWithBC}}{\textit{req}, r, z_1, z_2, \enc{\textsf{crt}}{\key{t}}, n, \enc{\nlist{\textsf{TX}, \fail}}{\key{t}}, y} \\ 
			%		& \quad \event{TCommit}{\kbt,\key{t},z_1,z_2, \textsf{TX}}. \\
			&  \quad \ifff \; \texttt{rtype} = \texttt{accept} \;\thenn\\ 
			%		& \quad \quad  \; \event{TAccept}{\kbt,\TX}. \\
			& 	\quad \quad \event{\textcolor{brown}{TAccept}}{\kbt,\textsf{TX}} \\
			&   \quad \quad \cout{\ch}{\texttt{auth}} \\
		\end{aligned}
	\end{equation}
	\endgroup
	\caption{Events in the terminal's role.}\label{fig:eventsterminal}
\end{figure}

\fi

\end{document}